\documentclass[preprint,11pt]{elsarticle}
\usepackage[bookmarks=true,colorlinks=true,linkcolor=blue]{hyperref}

\pdfoutput=1

\biboptions{sort&compress}

\usepackage{color}

\usepackage{soul}

\usepackage[usenames,dvipsnames,svgnames,table]{xcolor}

\usepackage{amssymb,amsmath}
\usepackage{amsthm}
\usepackage{multirow}

\usepackage{calc}
\usepackage[margin=1.in]{geometry}

\renewcommand{\url}[1]{}
\newcommand{\citeCount}[1]{}
\newtheorem{comment}{Comment}

\DeclareMathOperator*{\erfc}{erfc}

\newcommand{\dx}{\Delta x}

\newcommand{\dt}{\Delta t}

\newcommand{\bogus}[1]{{}}

\newenvironment{myIndent}%
 {\list{}{\leftmargin=0.1in\rightmargin=0.1in}\item[]}%
  {\endlist}

{\noindent\textbf{Procedure~}{#1}\begin{myIndent}\em}
{\end{myIndent}}

\newcommand{\bv}{\mathbf{ b}}

\newcommand{\gv}{\mathbf{ g}}

\newcommand{\uv}{\mathbf{ u}}
\newcommand{\vv}{\mathbf{ v}}

\newcommand{\xv}{\mathbf{ x}}
\newcommand{\yv}{\mathbf{ y}}
\newcommand{\zv}{\mathbf{ z}}
\newcommand{\Av}{\mathbf{ A}}
\newcommand{\Bv}{\mathbf{ B}}

\newcommand{\Ev}{\mathbf{ E}}

\newcommand{\Gv}{\mathbf{ G}}

\newcommand{\Iv}{\mathbf{ I}}
\newcommand{\Jv}{\mathbf{ J}}

\newcommand{\Uv}{\mathbf{ U}}
\newcommand{\Vv}{\mathbf{ V}}

\newcommand{\zerov}{\mathbf{0}}

\newcommand{\Ac}{{\mathcal A}}

\newcommand{\Dc}{{\mathcal D}}

\newcommand{\Jc}{{\mathcal J}}

\newcommand{\Oc}{{\mathcal O}}
\newcommand{\Pc}{{\mathcal P}}

\newcommand{\Sc}{{\mathcal S}}

\newcommand{\Vc}{{\mathcal V}}
\newcommand{\Psiv}{\boldsymbol{\Psi}}
\newcommand{\Phiv}{\boldsymbol{\Phi}}

\newcommand{\omegav}{\boldsymbol{\omega}}

\newcommand{\grad}{\nabla}

\newcommand{\tableFont}{\scriptsize}

\newcommand{\num}[2]{#1e#2} 

\clearpage

\newcommand{\red}{\color{red}}

\newcommand{\strutt}{\rule{0pt}{9pt}}

\usepackage{tikz}
\usetikzlibrary{calc,patterns,angles,quotes}
\usepackage{pgfplots}

\newlength{\tfwidth}
\newlength{\tfheight}
\newlength{\tfxa}
\newlength{\tfxb}
\newlength{\tfya}
\newlength{\tfyb}
\newcommand{\trimFigWithBox}[6]{%
\setlength\fboxsep{0pt}%
\setlength\fboxrule{1.0pt}
\fbox{\includegraphics[width=#2, clip, trim=#3 #4 #5 #6]{#1}}%
}
\newcommand{\trimFigNoBox}[6]{%
\setlength\fboxsep{1pt}
\setlength\fboxrule{0.0pt}
\fbox{\includegraphics[width=#2, clip, trim=#3 #4 #5 #6]{#1}}%
}
\newcommand{\trimFigHeightWithBox}[6]{%
\setlength\fboxsep{0pt}%
\setlength\fboxrule{1.0pt}
\fbox{\includegraphics[height=#2, clip, trim=#3 #4 #5 #6]{#1}}%
}
\newcommand{\trimFigHeightNoBox}[6]{%
\setlength\fboxsep{1pt}
\setlength\fboxrule{0.0pt}
\fbox{\includegraphics[height=#2, clip, trim=#3 #4 #5 #6]{#1}}%
}

\newcommand{\trimFig}[6]{%
\setlength{\tfwidth}{(#2+#2*\real{#3})+#2*\real{#4}}
\setlength{\tfheight}{(#2+#2*\real{#5})+#2*\real{#6}}%
\setlength{\tfxa}{\tfwidth*\real{#3}}%
\setlength{\tfxb}{\tfwidth*\real{#4}}%
\setlength{\tfya}{\tfheight*\real{#5}}%
\setlength{\tfyb}{\tfheight*\real{#6}}%
\trimFigNoBox{#1}{#2}{\tfxa}{\tfya}{\tfxb}{\tfyb}%
}

\newsavebox\figBox

\newcommand{\trimw}[6]{%
\sbox\figBox{\includegraphics{#1}}
\setlength{\tfwidth}{\the\wd\figBox}
\setlength{\tfheight}{\the\ht\figBox}
\setlength{\tfxa}{\tfwidth*\real{#3}}%
\setlength{\tfxb}{\tfwidth*\real{#4}}%
\setlength{\tfya}{\tfheight*\real{#5}}%
\setlength{\tfyb}{\tfheight*\real{#6}}%
\trimFigNoBox{#1}{#2}{\tfxa}{\tfya}{\tfxb}{\tfyb}%
}

\newcommand{\trimwb}[6]{%

\sbox\figBox{\includegraphics{#1}}
\setlength{\tfwidth}{\the\wd\figBox}
\setlength{\tfheight}{\the\ht\figBox}
\setlength{\tfxa}{\tfwidth*\real{#3}}%
\setlength{\tfxb}{\tfwidth*\real{#4}}%
\setlength{\tfya}{\tfheight*\real{#5}}%
\setlength{\tfyb}{\tfheight*\real{#6}}%
\trimFigWithBox{#1}{#2}{\tfxa}{\tfya}{\tfxb}{\tfyb}%
}

\newcommand{\trimh}[6]{%
\sbox\figBox{\includegraphics{#1}}
\setlength{\tfwidth}{\the\wd\figBox}
\setlength{\tfheight}{\the\ht\figBox}
\setlength{\tfxa}{\tfwidth*\real{#3}}%
\setlength{\tfxb}{\tfwidth*\real{#4}}%
\setlength{\tfya}{\tfheight*\real{#5}}%
\setlength{\tfyb}{\tfheight*\real{#6}}%
\trimFigHeightNoBox{#1}{#2}{\tfxa}{\tfya}{\tfxb}{\tfyb}%
}

\newcommand{\trimhb}[6]{%

\sbox\figBox{\includegraphics{#1}}
\setlength{\tfwidth}{\the\wd\figBox}
\setlength{\tfheight}{\the\ht\figBox}
\setlength{\tfxa}{\tfwidth*\real{#3}}%
\setlength{\tfxb}{\tfwidth*\real{#4}}%
\setlength{\tfya}{\tfheight*\real{#5}}%
\setlength{\tfyb}{\tfheight*\real{#6}}%
\trimFigHeightWithBox{#1}{#2}{\tfxa}{\tfya}{\tfxb}{\tfyb}%
}
\newcommand{\logLogSlopeTriangleInv}[5]
{

    \pgfplotsextra
    {
        \pgfkeysgetvalue{/pgfplots/xmin}{\xmin}
        \pgfkeysgetvalue{/pgfplots/xmax}{\xmax}
        \pgfkeysgetvalue{/pgfplots/ymin}{\ymin}
        \pgfkeysgetvalue{/pgfplots/ymax}{\ymax}

        \pgfmathsetmacro{\xArel}{#1}
        \pgfmathsetmacro{\yArel}{#3}
        \pgfmathsetmacro{\xBrel}{#1-#2}
        \pgfmathsetmacro{\yBrel}{\yArel}
        \pgfmathsetmacro{\xCrel}{\xArel}

        \pgfmathsetmacro{\lnxB}{\xmin*(1-(#1-#2))+\xmax*(#1-#2)} 
        \pgfmathsetmacro{\lnxA}{\xmin*(1-#1)+\xmax*#1} 
        \pgfmathsetmacro{\lnyA}{\ymin*(1-#3)+\ymax*#3} 
        \pgfmathsetmacro{\lnyC}{\lnyA-#4*(\lnxA-\lnxB)}
        \pgfmathsetmacro{\yCrel}{\lnyC-\ymin)/(\ymax-\ymin)} 

        \coordinate (A) at (rel axis cs:\xArel,\yArel);
        \coordinate (B) at (rel axis cs:\xBrel,\yBrel);
        \coordinate (C) at (rel axis cs:\xCrel,\yCrel);

        \draw[#5]   (A)-- node[pos=0.5,anchor=south] {1}
                    (B)-- 
                    (C)-- node[pos=0.5,anchor=west] {#4}
                    cycle;
    }
}

\usepackage{algorithm}
\usepackage{algpseudocode} 

\begin{document}

\small

\begin{frontmatter}
\title{
    An adaptive scalable fully implicit algorithm based on stabilized finite element for reduced visco-resistive MHD
}

\author[lanl]{Qi Tang\corref{cor1}\fnref{DOEThanks}}
\ead{qtang@lanl.gov}

\author[lanl]{Luis Chac\'on}
\ead{chacon@lanl.gov}

\author[llnl]{Tzanio V.~Kolev}
\ead{tzanio@llnl.gov}

\author[snl,unm]{John N.~Shadid}
\ead{jnshadi@sandia.gov}

\author[lanl]{Xian-Zhu Tang}
\ead{xtang@lanl.gov}

\address[lanl]{Los Alamos National Laboratory, Los Alamos, New Mexico 87545.}

\address[llnl]{Lawrence Livermore National Laboratory, Livermore, California 94550.}

\address[snl]{Sandia National Laboratories, Albuquerque, New Mexico 87123.}

\address[unm]{Department of Mathematics and Statistics, University of New Mexico, Albuquerque, NM 87131.}

\cortext[cor1]{Corresponding author}

\fntext[DOEThanks]
{This work was jointly supported by the U.S. Department
of Energy through the Fusion Theory Program of the Office
of Fusion Energy Sciences and the SciDAC partnership on
Tokamak Disruption Simulation between the
Office of Fusion Energy Sciences and the Office of Advanced
Scientific Computing.
Los Alamos National
Laboratory is operated by Triad National Security, LLC, for the National Nuclear Security Administration
of U.S. Department of Energy (Contract No. 89233218CNA000001).
Sandia National Laboratories is a multimission laboratory managed and operated by National
Technology and Engineering Solutions of Sandia, LLC., a wholly owned subsidiary of Honeywell International,
Inc., for the U.S. Department of Energy's National Nuclear Security Administration
under grant DE-NA-0003525.
This work also performed under the auspices of the U.S. Department of Energy by Lawrence Livermore National Laboratory under Contract DE-AC52-07NA27344.
This paper describes objective technical results and analysis. Any
subjective views or opinions that might be expressed in the paper do not necessarily represent the
views of the U.S. Department of Energy or the United States Government.
}

\begin{abstract}

The  magnetohydrodynamics (MHD) equations are continuum models used in the study of a wide range of plasma physics systems, including  the evolution of complex plasma dynamics in tokamak disruptions. However, efficient numerical solution methods for MHD  are extremely challenging due to disparate time and length scales, strong hyperbolic phenomena, and nonlinearity. Therefore the development of scalable,  implicit MHD algorithms and high-resolution adaptive mesh refinement strategies is of considerable importance. 
In this work, 
we  develop a high-order stabilized finite-element algorithm for   the reduced visco-resistive MHD equations based on the MFEM finite element library (mfem.org).
The scheme is fully implicit, solved with the Jacobian-free Newton-Krylov (JFNK) method with a physics-based preconditioning strategy.
Our preconditioning strategy is a generalization of the physics-based preconditioning methods in [Chac\'on, et al, JCP 2002] to adaptive, stabilized finite elements.
Algebraic multigrid methods are used to invert {sub-block} operators to achieve scalability.
A parallel adaptive mesh refinement scheme with dynamic load-balancing is implemented to efficiently resolve the multi-scale spatial features of the system.
Our implementation uses the MFEM framework, which provides arbitrary-order polynomials and flexible adaptive conforming and non-conforming meshes capabilities.
Results demonstrate the accuracy, efficiency, and scalability of the implicit scheme in the presence of large scale disparity.
The potential of the AMR approach 
is demonstrated on an island coalescence problem in the high Lundquist-number regime ($\ge10^7$) with the successful resolution of 
plasmoid instabilities and thin current sheets.

\end{abstract}


\end{frontmatter}

\clearpage
\tableofcontents

\clearpage

\section{Introduction}  \label{sec:introduction}

Magnetohydrodynamics (MHD) models the dynamics of a quasi-neutral plasma fluid in the presence of electromagnetic fields.
It has a wide range of applications including space weather prediction, astrophysics, as well as in laboratory plasma applications such as  Z-pinch, tokamak and stellarator devices~\cite{biskamp1997nonlinear}.
The full MHD model comprises evolution equations for mass, moment and energy as well as an evolution equation for the magnetic field,
while the electric field is determined by a (generalized) Ohm's law.
The ultimate goal of this study is to develop an adaptive, scalable and high-order-accurate  simulator for the fusion devices involving complex geometries.
The resistive MHD model is commonly used due to its relative computational efficiency in contrast to kinetic descriptions,  as well as the ability to efficiently capture many important macroscopic instabilities.
In  resistive MHD, the Lundquist number, $S_L$, defined as the ratio between the resistive diffusion time scale for the magnetic field and the Alfv\'en wave transit time scale,
 is commonly used to distinguish between different physics regimes.
It is well known that the Lundquist number is very high in fusion devices, in the range of $10^6$ to $10^9$~\cite{jardin2012review},
owing to the large plasma temperature and small densities.
Therefore,  in this work, we seek to develop a high-order-accurate and scalable resistive MHD solver
 that is suitable for a wide range of  Lundquist numbers.

The resistive MHD system is a strongly coupled, highly nonlinear and very stiff system that involves multiple physical phenomena  spanning a wide range of spatial and temporal scales~\cite{chacon2002, shadid2010towards}.
In the context of fusion-device modeling, it is of interest to evolve ideal and resistive instabilities (tearing mode,  resistive kink mode, coalescence instability, etc.)~excited from a given MHD equilibrium.
However,  hyperbolic dynamics in the MHD system is typically much faster than  resistive instabilities (as suggested by the high Lundquist number), which results in numerical stiffness.
It is therefore important to consider fully implicit algorithms to efficiently step  over the stiff hyperbolic component in the system~\cite{jardin2012review}
along with scalable  preconditioning strategies.
In addition,  thin current sheets develop in the high-Lundquist-number regime, along with a secondary plasmoid instability.
The plasmoid length-scale  is comparable to the current sheet width and very small compared to the size of a typical magnetic island (a ratio of approximately $10^{-3}$ to $10^{-5}$).
Plasmoid emergence, transport, and merging is on a much faster time scale than the resistive diffusion time scale.
Due to the large length scale disparity, the problem becomes truly multi-scale in space as well.
Thus,  adaptive implicit multiscale algorithms are needed to capture these dynamics in a self-consistent fashion.

Many of the earlier implicit studies for  extended MHD  have been summarized in Ref.~\cite{jardin2012review}.
Here we only highlight studies that focus on  preconditioning strategies for the implicit MHD system.
Ref.~\cite{chacon2002} is one of the earliest studies to propose a physics-based preconditioner for 2D resistive MHD  in  a finite-difference context.
This work will serve as the starting point for our physics-based  preconditioner in the stabilized finite-element formulation.
The approach is later generalized to a 2D Hall MHD~\cite{chacon20032d}, a 3D compressible resistive MHD~\cite{chacon2004non} and a 2D two-field low-$\beta$ extended MHD~\cite{chacon2016scalable}.
The significance of this series of studies is its nearly optimal performance under grid refinement, featuring $\Oc(N\log N)$ scaling, where
$N$ is the number of degrees of freedom (dofs).
A drawback of  these studies is their reliance on low-order finite differences, which complicates their generalization and extension to higher orders of accuracy.
A stabilized finite element (FE) formulation is proposed in Ref.~\cite{shadid2010towards}, which considers the lower-order continuous FE and its preconditioning strategy.
The preconditioning is less optimal compared to Ref.~\cite{chacon2002} since it uses a fully coupled algebraic multilevel preconditioner, which does not parabolize the system as done in Ref.~\cite{chacon2002}.
This approach is later extended to 3D resistive MHD~\cite{shadid2016scalable}.
A related approach, using approximate block factorization (ABF) and approximate  Schur complements has been developed for the MHD system in~\cite{cyr2013new, phillips2014block, phillips2016block}.
More recently, a fully implicit hybridization discontinuous Galerkin approach (HDG) has been developed and analyzed in~\cite{lee2019analysis} and results for scaling of a block preconditioner based on an approximate Schur complement have been presented ~\cite{muralikrishnan2020multilevel}.
Note that the approaches developed in~\cite{shadid2010towards, muralikrishnan2020multilevel} are also suitable for high-Lundquist-number cases. 
Another class of methods~\cite{hu2017stable, ma2016robust, li2021constrained} rely on  compatible finite elements and discrete exterior calculus for the resistive MHD system.
This method  can be divergence-free in $\uv$, $\Bv$ or both without introducing the potential formulation as in this study.
For instance, the approach in Ref.~\cite{li2021constrained} uses a constrained transport method similar to Ref.~\cite{christlieb2014finite, christlieb2016high} to obtain divergence-free $\Bv$.
However, in that context, an efficient preconditioner can be even more challenging, and as pointed out~\cite{li2021constrained} it is unclear the advantage of such a method in practice
compared to the potential-based approach proposed here.

Beyond an efficient preconditioner,  high-Lundquist-number simulations require two more important ingredients. 
First, for advection-dominated  problems, it is well known that some stabilization is necessary in continuous Galerkin FE formulations
to prevent spurious oscillations~\cite{donea2003finite}. 
Here, we choose the streamline upwind Petrov--Galerkin (SUPG) approach to stabilize both  the magnetic potential and vorticity equations.
The SUPG approach is chosen because it is variationally consistent and has a simple form 
which separates the stabilization term from the advection term and thus eases the generalization of the physics-based preconditioner. 
The second necessary ingredient is the adaptive mesh refinement (AMR) algorithm to address vast length-scale disparity and capture resulting localized structures. 
It is known that when the Lundquist number is sufficiently high, the tearing mode instability breaks the thin current sheet and produces plasmoids~\cite{biskamp1996magnetic}.
This has been  studied in detail in the context of magnetic reconnection, see~\cite{bhattacharjee2009fast, huang2010scaling} for instance.  
However, these studies~\cite{bhattacharjee2009fast, huang2010scaling} typically focus on the reconnection regions, and use very refined and locally stretched grids. 
Our goal is an algorithm to dynamically capture plasmoids in practical simulations as in~\cite{lynch2016reconnection, baty2020petschek}, which requires an AMR capability that self-consistently evolves as localized plasmoids form and merge.
To the best of our knowledge, the only studies that successfully develop an implicit AMR approach for resistive MHD are~\cite{philip2008implicit, adler2011efficiency, baty2019finmhd}.
Among those, only~\cite{baty2019finmhd} was able to capture some plasmoids in the reconnection study of~\cite{baty2020petschek}. 
The highest Lundquist number considered in Ref.~\cite{baty2020petschek} is $4\times 10^5$ but the definition of Lundquist number there uses the local current sheet length, and therefore it is comparable to $\sim10^6$ in our study.

The present study complements the previous ones~\cite{chacon2002, shadid2010towards, shadid2016scalable} by providing a robust, scalable, adaptive and high-order stabilized FE formulation for resistive MHD. 
Key contributions of the current work include 
(a)  generalization of the physics-based preconditioning to the FE formulation; 
(b) generalization of the previous lower-order (finite-difference or finite-element) discretization to high-order continuous  FE;
(c)  development of an AMR algorithm along with the FE discretization;
and (d)  optimization of the implicit AMR solver for high Lundquist numbers.
The generalization of the physics-based preconditioners from  finite differences to high-order FE is by no means trivial.
While it is sufficient  in  finite differences  to introduce numerical diffusion via flux limiting in the high Lunduiqst/Reynolds number cases,
the FE formulation requires  high-order stabilization,
which can often
 introduce complicated coupling terms in the system.
We find ignoring such stabilization terms in the preconditioner degrades the performance of the solver significantly.
Furthermore, both AMR  and careful optimization of
the parameters in the full solver are found to be necessary to provide a robust and efficient algorithm for the high Lundquist-number runs.

Our algorithm is implemented with the scalable FE framework, MFEM (mfem.org),
using many of its features.
For instance, we use the equal-order continuous FE bases, supporting arbitrary order of accuracy and flexible meshes.
We adopt the AMR algorithm from MFEM, but have developed our own custom error estimator and refinement strategies.
As a result, the adaptive solver supports both conforming and non-conforming AMR, for both structured and unstructured meshes.
Although the practical runs presented in the current work only utilize the octree-based AMR and quadrilateral elements, the flexible implementation provides an easy route to leverage the current solver
to practical simulations of fusion devices in the future.

The remainder of this paper is organized as follows.
Section~\ref{sec:governing} describes the governing equations of the visco-resistive MHD equation and discusses the specific formulation we choose in this work.
Section~\ref{sec:scheme}  presents the stabilized FE formulation of the model, its linearized system and the choices of time integrators.
Section~\ref{sec:preconditioning} generalizes the physics-based preconditioning approach in Ref.~\cite{chacon2002} to the stabilized FE formulation. 
Section~\ref{sec:jfnk} outlines the Jacobian-free Newton-Krylov (JFNK) approach  and Section~\ref{sec:amr} describes the AMR approach,
 including a description of the error estimator and some details on the refinement/coarsening strategy,
followed by the implementation details of the full algorithm in Section~\ref{sec:implementation}.
Section~\ref{sec:results} presents the numerical results, including a refinement study, parallel scaling tests and simulation results of practical interest. The numerical experiments start from some simple smooth problems to verify the order of accuracy and efficiency of the scheme,
 followed by some harder problems (tearing mode) to examine both of the scheme and AMR approach.
 The final example, the high Lundquist-number island coalescence problem, is the most challenging, as plasmoid instabilities develop.
Finally, Section~\ref{sec:conclusions} concludes and outlines possible future directions.

\section{Governing equations: visco-resistive MHD } \label{sec:governing}

We consider the incompressible reduced visco-resistive MHD model in a two-dimensional space $\Omega$.
The normalized model is described by a streamfunction-vorticity description ($\Phi$--$\omega$) of the fluid quantities and a magnetic flux $\Psi$, and is given by Ref.~\cite{strauss1976nonlinear}
\begin{subequations}
\begin{align}
    \grad^2 \Phi & =\omega, \\
    \left( \partial_t +\vv \cdot \grad - \eta \grad^2\right) \Psi   &= -E_0, \\
    \left( \partial_t +\vv \cdot \grad - \nu \grad^2\right) \omega  &= \Bv \cdot \grad J-S_\omega,
\end{align}    \label{eqn:mhd2d}
\end{subequations}
with the velocity and magnetic field defined as
\begin{align*}
    \vv  & = \zv \times \grad \Phi = [- \Phi_y, \Phi_x]^T, \\
    \Bv & = \zv \times \grad \Psi =  [- \Psi_y, \Psi_x]^T ,
\end{align*}
and the current density defined as
\[
J = \Delta \Psi.
\]
Other important quantities include the (constant) magnetic resistivity $\eta$,  (constant) kinematic viscosity $\nu$, two source terms $E_0$ and $S_\omega$.
The source terms are used to balance visco-resistive dissipations of  the initial equilibrium in the study.

The reduced MHD model is originally derived for the low-$\beta$ large-aspect-ratio tokamak from the full MHD model in Ref.~\cite{strauss1976nonlinear}.
The model is derived by ordering all the quantities based on a large aspect ratio and dropping all toroidal effects.
The pressure term is also dropped due to the assumption of low $\beta$.
Here $\beta$ is the ratio of the plasma pressure  to the magnetic pressure.
Other waves such as fast and slow magnetosonic waves are asymptotically removed and only
the  Alfv\'en wave remains.
The model~\eqref{eqn:mhd2d} assumes infinite toroidal aspect ratio, which removes all toroidal terms.
Despite being significantly simpler than the full MHD model,
\eqref{eqn:mhd2d} still keeps many of the fundamental challenges in the original MHD model, particularly for the tokamak problems we are interested in,
including the strongly hyperbolic nature (supporting a stiff Alfv\'en wave),
strong nonlinear coupling between the magnetic flux and  fluid quantities, numerical challenges in the strong-convection regime, large disparity in time and length scales, etc.
Note that, as pointed out  in Ref.~\cite{strauss1976nonlinear}, the current system can be viewed as a model for a rectangular tokamak,
and thus $\Psi$, $J$ and $\omega$ are the $z$-component of the magnetic potential, current density and vorticity vectors of a full 3D model, respectively.
In this context, $\vv$ and $\Bv$ are $x$- and $y$-components of the full 3D vectors.

We note that there is an alternative formulation for the reduced MHD model  based on evolving the current density instead of $\Psi$.
(Note that the current density in the system~\eqref{eqn:mhd2d} can be viewed as an auxiliary variable, and thus one could choose  evolve either $J$ or $\Psi$.)
The alternative model is given by
\begin{align*}
    \grad^2 \Phi & =\omega, \\
    \left( \partial_t +\vv \cdot \grad - \eta \grad^2\right) J   &= {\red \Bv \cdot \grad \omega} -\Delta E_0 +g(\Psi, \Phi), \\
    \left( \partial_t +\vv \cdot \grad - \nu \grad^2\right) \omega  &= \Bv \cdot \grad J-S_\omega,
\end{align*}
where
\begin{align*}
g(\Psi, \Phi) := & 2\left[ \frac{\partial^2 \Phi}{\partial x \partial y} \left( \frac{\partial^2 \Psi}{\partial x^2} -   \frac{\partial^2 \Psi}{\partial y^2}\right)-
\frac{\partial^2 \Psi}{\partial x \partial y} \left( \frac{\partial^2 \Phi}{\partial x^2} -   \frac{\partial^2 \Phi}{\partial y^2}\right) \right], \\
\Psi =& \Delta^{-1} J.
\end{align*}
The model avoids the challenging $\Bv\cdot\grad J$ term, which involves third-order derivatives for $\Psi$, and the system was first introduced in~\cite{strauss1998adaptive, lankalapalli2007adaptive}
in order to improve the AMR performance.
It is pointed out in the AMR study in Ref.~\cite{philip2008implicit} that such an implementation helps minimize the large error around the AMR coarse-fine grid interface.
The above model has also been implemented recently in Ref.~\cite{baty2019finmhd} for a continuous finite-element implementation and
the author pointed out that the challenge of $\Bv\cdot\grad J$ in~\eqref{eqn:mhd2d} will always exist in AMR simulations except for spectral methods.
We note, however, that the implementation in Ref.~\cite{philip2008implicit} is based on a second-order finite-different algorithm
and Ref.~\cite{baty2019finmhd} uses a polynomial order of 1 or 2.
In our implementation, we observed a very similar behavior around  coarse-fine interfaces in AMR examples when the polynomial order is lower than 3.
However, when choosing the polynomial order greater than or equal to 3,
coarse-fine interfaces no longer accumulate large errors.
Employing the original model~\eqref{eqn:mhd2d} has at least two advantages:
(i) the physics-based preconditioning strategy in~\cite{chacon2002} can be easily extended into the high-order finite element discretization;
(ii) for high-Lundquist-number simulations,  the original model leads to a much simpler SUPG terms in the preconditioner  compared to the terms resulting from $g(\Psi, \Phi)$ in the system evolving $J$.
Therefore, we use the model in~\eqref{eqn:mhd2d} for our implementation.

We handle  the challenging   $\Bv\cdot\grad J$ term  by introducing an auxiliary variable $J$ and inverting a mass matrix with $\Psi$.
For challenging cases with large  Lundquist numbers   (typically $S_L>10^5$ for which AMR is necessary to capture small scales)
we also  lump the mass matrix when evaluating $J$ to minimize the oscillations in
the current density during local refinement or coarsening.
Such a strategy appears to be sufficient to capture small structures of reasonably high Lundquist number without introducing large oscillations.
We note that there is an alternative approach for the ($\Av$--$\vv$) formulation of MHD given in Ref.~\cite{shadid2010towards}, in which the authors reformulate the Lorentz force
$\Jv\times\Bv=\grad \cdot [ \Bv \Bv - \frac{\|\Bv\|^2}{2} \Iv ]$ to avoid high-order derivatives in their finite element formulations.
A straightforward extension to the current system leads to  source term for $\omega$ becoming the third component of
\begin{align*}
    \grad \times ( \Jv \times \Bv )
    &= \grad \times \Big\{ \grad \cdot [ \Bv \Bv - \frac{\|\Bv\|^2}{2} \Iv ] \Big\} \\
    &= \grad \cdot \Big\{ \grad \times [ \Bv \Bv ] \Big\} \\
    &=\grad \cdot \Big\{ (\grad \times \Bv) \Bv - \Bv \times \grad \Bv\Big\}.
\end{align*}
It is clear that it does not  help our  implementation, since there are still third-order derivatives with respect to $\Psi$ in the  source term.
Overall, we find that the  $\Bv\cdot\grad J$ term is always challenging to handle in both finite-element and finite-difference discretizations, and it will be interesting to explore other techniques to
handle this operator, particularly with AMR.

\section{Stabilized finite element formulation} \label{sec:scheme}
We introduce the FE formulation for the continuous PDE system~\eqref{eqn:mhd2d}  in this section.
The system~\eqref{eqn:mhd2d} avoids the Ladyzhenskaya--Babuska--Brezzi (LBB) stability condition~\cite{donea2003finite, gunzburger2012finite} by using the potential formulations for both
velocity and magnetic field.
Therefore, it is natural to employ equal-order polynomial spaces for all the unknowns in the system.
Given the small magnitude of resistivity and viscosity, it is all but assured that our simulations will be convection-dominated in most of the domain.
In fact, we find that, for some high-Lundquist-number cases, the mesh Peclet number (the ratio between convective strength and diffusive conductance) can be higher than 100 even with a very aggressive AMR refinement.
For such cases, relying on excessive AMR refinement is unrealistic due to the rapid increase of the total degrees of freedom and efficiency reduction of any preconditioning strategy.
Therefore, it is necessary to use a stabilized FE approach for both $\Psi$ and  $\omega$ to achieve a stable solution locally.

Artificial diffusion is one of the most commonly used approaches to stabilize the convection term.
For finite-element schemes, artificial diffusion can be introduced by the  streamline upwinding stabilization.
Such an approach, however, does not fit high-order schemes,
 since it inevitably introduces a low-order error whenever the stabilization term is turned on.
To avoid this, we  rely on a residual-based stabilization approach. These methods are variationally consistent
in the sense that the exact solution satisfies the weak formulation, and it is more suitable for high-order approximations.
We employ the standard Streamline-Upwind Petrov-Galerkin (SUPG) approach with  modifications to the stabilization parameter $\tau$   for  AMR and high-order schemes.

\newcommand{\lgl}{\langle}
\newcommand{\rgl}{\rangle}
\newcommand{\Domain}{{\Omega_h}}
\newcommand{\partialD}{\partial{\Omega_h}}

\subsection{2D visco-resistive MHD stabilized FE formulation}
A stabilized SUPG FE formulation  is presented for the 2D reduced MHD equation. The formulation is fully implicit, and
the backward Euler method is used in the following presentation for simplicity and convenience.
Unless otherwise noted, the numerical cases presented employ  either Dirichlet or periodic boundary conditions.
Therefore, as an example, we consider a Dirichlet boundary condition imposed on the Perfect Electrical Conductor (PEC) wall as
\begin{align}
\omega = 0, \quad \Psi = 0, \quad \Phi=0,  \qquad {\rm for } \; \xv \in \partial \Omega, \label{eqn:bc}
\end{align}
in the discussion. {\color{red} Note that a more general PEC boundary condition may allow $\Psi$ being spatially dependent.}

Define the current $J:=\grad^2 \Psi$ as an ``auxiliary'' variable and a FE space $\Vc$ as $\{\Uv \, | \, \Uv = [\Phi, \Psi, \omega, J]^T, \,  \Phi, \Psi, \omega\in H^1_0(\Omega_h), \, J \in H^1(\Omega_h) \}.$
Using the test functions, $\phi \in H_0^1(\Omega_h)$ and $\psi \in H^1(\Omega_h)$,
a stabilized FE backward-Euler discretization gives
\begin{subequations}
\label{eqn:aWeakForm}
    \begin{alignat}{2}
       & \int_\Domain \grad \Phi^{n+1} \cdot \grad \phi \,dx + \int_\Domain \omega^{n+1} \phi  \,dx =0, &&  \\
        & \begin{split} \int_\Domain \frac{\Psi^{n+1} - \Psi^n} \dt \phi  \,dx   +  \int_\Domain \vv^{n+1}  \cdot \grad  \Psi^{n+1}  \phi \, dx & + \int_\Domain \eta \grad \Psi^{n+1}\cdot \grad \phi  \, dx   \\
           & +  \int_\Domain \tau_\eta R_\Psi(\Psi^{n+1}) \vv^{n+1}\cdot\grad\phi \, dx =  -\int_\Domain E_0 \phi  \,dx,
          \end{split}\\
      &  \begin{split}
         \int_\Domain \frac{\omega^{n+1} - \omega^n} \dt \phi  \,dx  +  \int_\Domain \vv^{n+1}  \cdot \grad  \omega ^{n+1}  \phi \, dx  & + \int_\Domain \nu \grad \omega^{n+1} \cdot \grad \phi \, dx      \\
          -  \int_\Domain  \Bv^{n+1} \cdot \grad J^{n+1}   \phi \, dx  & + \int_\Domain  \tau_\nu R_\omega(\omega^{n+1})  \vv^{n+1}\cdot\grad \phi \, dx=-\int_\Domain S_\omega \phi  \,dx,
          \end{split}
    \end{alignat}
\end{subequations}
where the current density $J$ is determined by 
    \begin{equation}
    \label{eqn:Jeqn}
\int_\Domain J^{n+1} \psi  \,dx= - \int_\Domain \grad \Psi^{n+1}\cdot \grad \psi \, dx +  \oint_{\partialD} \frac{\partial \Psi^{n+1}} {\partial n} \psi \, ds,
    \end{equation}
and $R(\cdot)$ in the formulation stands for the residual terms for each equation, which are defined as
\begin{align*}
R_\Psi(\Psi^{n+1})&:=  \frac{\Psi^{n+1} - \Psi^n} \dt +  \vv^{n+1} \cdot \grad  \Psi ^{n+1}  - \eta \Delta \Psi^{n+1}  + E_0,  \\
R_\omega(\omega^{n+1})&:= \frac{\omega^{n+1} - \omega^n} \dt +  \vv^{n+1} \cdot \grad  \omega^{n+1} - \nu \Delta \omega^{n+1}-\Bv^{n+1} \cdot \grad J^{n+1} + S_\omega.
\end{align*}
{\red Note that the boundary condition for $J$  is not needed since the only term involving $J$ in the system~\eqref{eqn:mhd2d} is $\Bv\cdot\grad J$
while the Dirichlet boundary condition given in~\eqref{eqn:bc} implies $\Bv_\perp=0$.}
In the FE formulation, the vectors $\Bv$ and $\vv$ are the strong derivatives computed from the potentials $\Psi$ and $\Phi$, respectively.
Another advantage of the above system is that the vector fields $\Bv$  and $\vv$ are divergence-free in a strong sense.

In this initial study, the stabilization parameters $\tau$ in the formulation are taken as the standard term for a transient advection-diffusion equation~\cite{donea2003finite}, and are given by
\[
\tau_\eta := \left[ \bigg(\frac{\alpha}{\dt}\bigg)^2 + \left(\frac{2\| \vv\|}{h}\right)^2 + 9 \left(\frac{4\eta}{h^2}\right)^2\right]^{-\frac 1 2}.
\]
Note that the standard formulation for a Crank-Nicholson approximation in time for SUPG has $\alpha=2$. 
In our case, we have found this to  work well for all cases but the most challenging AMR solution for high Lundquist number (i.e. $S \ge 10^6$) island coalescence computations where we have used $\alpha=10$. This limited modification of $\alpha$ has the effect of  reducing the diffusion where the flow speed is small while still providing sufficient numerical diffusion in the large-flow-speed regions. In the future, we will consider alternate forms for the stabilization parameters.
Finally, in our implementation, the stabilization parameters $\tau$ are computed element-wise, by taking a maximum of all the quadrature points  in each element.

Since the current density can be determined from $\Psi$ through~\eqref{eqn:Jeqn}, the system~\eqref{eqn:aWeakForm} is a closed system for the three unknowns $\Psi$, $\Phi$ and $\omega$.
The preconditioning strategy will therefore focus on  these three components.
Note that other than the nonlinear SUPG residual terms, the nonlinearity of the system comes from the terms related to the convection term $\vv\cdot\grad$ and the Lorentz force term $\Bv\cdot \grad J $,
which will be the main focus when deriving the preconditioning strategy.
Although the SUPG terms do not add extra stiffness to the system,  some care is needed to achieve a scalable preconditioning strategy.

\begin{comment}
Regarding to the high-order derivative term $\Bv\cdot \grad  J$,
there are three obvious choices to implement it: $(\Bv\cdot \grad  J, \psi)_\Domain$, $( -J, \Bv\cdot \grad\psi)_\Domain$ or $ (-\Bv  J, \grad\psi)_\Domain$.
Here, the strong divergence-free condition of $\Bv$ is used.
One would expect that the latter two choices are better since the operator is implemented weakly.
However, we found that the choices of the implementation help very little to minimize the oscillations in $J$    in some high-Lundquist-number runs when the AMR refinement is very aggressive.
Part of the reason is that the integral-by-parts does not fundamentally change the nature of the operator (unlike the reformulation in~\cite{shadid2010towards}).
For instance, the second operator is simply the transpose of the first operator in our implementation.
\label{com1}
\end{comment}

\subsection{Linearized system of stabilized FE formulation}
Before presenting the preconditioning strategy, we first linearize the nonlinear system  to simplify the discussion and identify the key terms.
We begin by rewriting the system~\eqref{eqn:aWeakForm} in a block operator form.
Let us define several operators to simplify the presentation:
\begin{alignat*}{2}
& M: && \text{ mass matrix}\\
& K: && \text{ stiffness matrix} \\
& \tilde K: && \text{ fixed matrix that is a sum of stiffness and boundary operators as the RHS of~\eqref{eqn:Jeqn}}\\
& N_{\uv}: && \text{ discrete operator for the continuous operator } \uv \cdot \grad  \text{ where } \uv = [-\partial_y A, \partial_x A]^T
\end{alignat*}
{\red The nonlinear terms such as $N_{\uv} \, \hat \gv$} are implemented in a matrix-free fashion in the FE formulation,
i.e., $N_{\uv}$ is assembled directly element-by-element without forming a global matrix.
The concept is referred to as {\em partial assembly} in MFEM.
{\red Here the notations such as $\hat \gv$ stand for the vector representation of the dofs corresponding to a scalar component $g$.}  

The current density is $\hat\Jv = M ^{-1} \tilde K \hat\Psiv$ based on~\eqref{eqn:Jeqn}. Substituting it into the system~\eqref{eqn:aWeakForm} leads to a
 nonlinear equation for $\Gv(\Uv^{n+1}) = 0$ (where $\Uv:=[\hat\Phiv, \hat\Psiv, \hat\omegav]^T$):
\[
\Gv(\Uv^{n+1}):=    \begin{bmatrix}
         K  & \zerov &  M  \\
        \zerov & \Ac_\eta   & \zerov \\
        \zerov & N_\Bv  M ^{-1} \tilde K   &  \Ac_{\nu}
    \end{bmatrix}
    \begin{bmatrix}
        \hat\Phiv^{n+1} \\ \hat\Psiv^{n+1} \\ \hat\omegav^{n+1}
    \end{bmatrix} +
    \begin{bmatrix}
        \zerov\smallskip \\ \tau_{\Psi}  \tilde R_\Psi(\hat\Psiv^{n+1})\smallskip \\ \tau_{\nu}  \tilde R_\omega(\hat\omegav^{n+1})
    \end{bmatrix}
    +     \begin{bmatrix}
        \zerov \smallskip \\  -\frac 1 \dt  M   \hat\Psiv^{n} +\hat \Ev_0 \smallskip \\ -  \frac 1 \dt  M   \hat\omegav^{n}+\hat  S_\omega \smallskip
    \end{bmatrix},
\]
where $\tilde R(\cdot)$  is the discrete form of the corresponding nonlinear residual operator $R(\cdot)$, and the two convection-diffusion operators are defined as:
\begin{align*}
 \Ac_\eta&:=\frac 1 \dt  M  +  N_\vv + \eta  K, \\
 \Ac_\nu&:=\frac 1 \dt  M  +  N_\vv + \nu  K.
\end{align*}
The Dirichlet boundary condition is enforced strongly in the FE formulation by eliminating the proper boundary elements if needed.

A Jacobian-free Newton-Krylov (JFNK) method is used to solve the nonlinear problem $\Gv(\hat\Uv^{n+1}) = \zerov$.
Although not explicitly  formulated in our nonlinear solver, a reasonable approximation to the Jacobian system of $\Gv(\hat\Uv^{n+1}) = \zerov$ is
\begin{align}
    \begin{bmatrix}
         K  & \zerov &  M   \\
        - N_{\Bv_0} & \Ac_\eta  +{\tau_\eta  S} & \zerov \\
    - P _{\omega_0} &   P _{J_0} + N_{\Bv_0}  M ^{-1} \tilde K   &  \Ac_\nu  + {\tau_\nu  T}
    \end{bmatrix}
    \begin{bmatrix}
        \delta \hat\Phiv^{(k+1)} \\ \delta \hat\Psiv^{(k+1)} \\ \delta \hat\omegav^{(k+1)}
    \end{bmatrix} = -
    \begin{bmatrix}
        \Gv_{\Phiv}  (\hat\Uv^{(k)} ) \\ \Gv_{\Psiv} (\hat\Uv^{(k)})\\ \Gv_{\omegav}(\hat\Uv^{(k)})
    \end{bmatrix},
   \label{eqn:jacobian}
\end{align}
where the following continuous relations have been used to derive the Jacobian matrix approximation,
\begin{align*}
    \delta \vv \cdot \grad \Psi_0 &= - \Bv_0 \cdot \grad \delta \Phi, \\
    \delta \vv \cdot \grad \omega_0 &=-(\zv \times \grad \omega_0) \cdot \grad \delta \Phi, \\
    \delta \Bv \cdot \grad J_0 & = -(\zv \times \grad J_0)\cdot \grad \delta \Psi.
\end{align*}
 $P_{J_0} \hat\uv$ stands for the discretization of the continuous operator $(\zv \times \grad J_0) \cdot \grad u$,
and $P_{\omega_0} \hat\uv $ stands for the discretization of the continuous operator $(\zv \times \grad \omega_0) \cdot \grad u$.
The readers are referred to~\cite{chacon2002} for the detailed derivations in a finite-difference context.
Its extension to the FE formulation is rather straightforward  and thus not presented here.
The major difference between~\cite{chacon2002} and the current work is the additional stabilization terms:
\begin{alignat*}{2}
& S: &  \text{ discrete operator for }  \int_\Domain \left( \frac{y} \dt + \vv_{0}\cdot \grad  y - \eta \Delta y\right)   \vv_{0}\cdot\grad\phi \, dx, \\
& T: &   \text{ discrete operator for } \int_\Domain \left( \frac{y} \dt + \vv_{0}\cdot \grad  y - \nu \Delta y\right)  \vv_{0}\cdot\grad\phi  \, dx.
\end{alignat*}
Note that even though we only keep part of the diagonal operators  of the stabilization terms in the linearized form for the preconditioner development,
{the retained diagonal terms keep the most important term from SUPG: the streamline diffusion term, $(\vv_0\cdot\grad y, \vv_0\cdot\grad\phi)_\Domain$,
Those terms are found to be critical for the inversion of certain operators in our preconditioner.
Other linearized terms in the stabilization terms are ignored in the preconditioner. Keeping all the terms is cumbersome and ultimately unimportant.
This is a common approach in preconditioning SUPG for complicated systems, see~\cite{cyr2012stabilization} for instance.}

Finally, the resulting system~\eqref{eqn:jacobian} is non-symmetric and indefinite.
As pointed on in~\cite{chacon2002}, the system is not diagonally dominant for certain waves, particularly when the time step is large.
The stiff Alfv\'en waves are retained in the linearized system.
An efficient preconditioning strategy is necessary to achieve a scalable solver.

\subsection{Time stepping}
Using the backward Euler scheme as the building block, the time-stepping algorithm is
 generalized to high-order using diagonal implicit Runge-Kutta (DIRK) methods in the implementation.
 For instance, a two-stage second-order DIRK method based on the Butcher table (there are a few choices of $x$ with different stability properties)
 \begin{align*}
\begin{array}{c|cc}
x      &  x           &  0  \\
1 - x  &  1 - 2x      &  x  \\
\hline
       & \frac{1}{2}  & \frac{1}{2}\\
\end{array},
\end{align*}
and a three-stage 3rd order DIRK method based on ($x =0.4358665215$)
 \begin{align*}
\begin{array}{c|ccc}
x              &  x               &  0              &  0   \\
(1+x)/{2}  &  ({1-x})/{2}   &  x              &  0   \\
1              &  -3x^2/2+4x-1/4  &  3x^2/2-5x+5/4  &  x   \\
\hline
               &  -3x^2/2+4x-1/4  &  3x^2/2-5x+5/4  &  x   \\
\end{array},
\end{align*}
are fully supported, among others. The implementation details of DIRK  can be found in the MFEM documentation, and many  recent studies in numerical ODEs.
As pointed out in~\cite{owren1995alternative}, the DIRK methods provide proper damping for high-order modes, useful for AMR refinement or coarsening.
Another advantage of DIRK methods is that they are single-stage, which makes the adaptive time-stepping easy to implement.
The adaptive time-stepping is necessary for challenging simulations, for instance involving plasmoids, where the local dynamical time scale may become comparable to the Alfv\'en wave speed propagation.
When that happens, the time steps typically need to be reduced significantly to resolve the dynamics of interest.

\section{Physics-based preconditioning for stabilized finite element}
\label{sec:preconditioning}
A successful physics-based preconditioning strategy for finite differences was proposed in~\cite{chacon2002}. Here, we modify and extend it to the stabilized FE formulation.
The key steps in the extension include: replacing the finite-difference discretization  with finite-element discretization, modifying certain pieces to suit the finite-element representation better,
and incorporating the treatments for the stabilization terms.
The key idea is to first explicitly expose the stiffness in the system and then apply an approximate  Schur complement based on certain physics understanding.

\subsection{Saddle-point problem and Schur complement}
The linearized system~\eqref{eqn:jacobian} is first converted to an approximate saddle-point system,  suitable for Schur factorization.
For ease of presentation, we ignore the diagonal stabilization terms and all the superscripts in~\eqref{eqn:jacobian}.
The first equation in~\eqref{eqn:jacobian} gives
\begin{align*}
    \delta \hat\omegav & =  M ^{-1}(-  K  \delta \hat \Phiv -  \Gv_{\Phiv} ),
\end{align*}
and substituting it into the third equation gives
\[
    - P _{\omega_0} \delta \hat\Phiv +\Big[  P _{J_0}+  N_{\Bv_0}  M ^{-1}\tilde K  \Big]  \delta\hat \Psiv +
    \Big[ \frac 1 \dt  M  +  N_{\vv_0} +  \nu K  \Big]  M ^{-1} (-  K  \delta \hat \Phiv-  \Gv_{\Phiv} )
    = -\Gv_{\omegav} ,
\]
which is simplified as
\begin{align*}
    \Big[  N_{\Bv_0}  M ^{-1} \tilde K +  P _{J_0} \Big] \delta\hat\Psiv -
\Big[ \frac 1 \dt  K   +  N_{\vv_0}  M ^{-1}  K  + P _{\omega_0} +  \nu K   M ^{-1}  K \Big] \delta \hat \Phiv = -\Gv_{\omegav} - \Ac_{\nu}  M ^{-1} (-\Gv_{\Phiv}).
\end{align*}
Note that   the following approximations are convenient:
\begin{align*}
      N_{\Bv_0}  M ^{-1}  K  +  P _{J_0} & \approx  K   M ^{-1}  N_{\Bv_0}, \\
     N_{\vv_0}  M ^{-1}  K  + P _{\omega_0} & \approx  K    M ^{-1} N_{\vv_0},
\end{align*}
based on their continuous analogue,
\begin{align*}
    (\Bv_0 \cdot \grad) \grad^2-(\zv \times \grad J_0    ) \cdot \grad &\approx \grad^2 (\Bv_0 \cdot \grad), \\
    (\vv_0 \cdot \grad) \grad^2-(\zv \times \grad\omega_0) \cdot \grad &\approx \grad^2 (\vv_0 \cdot \grad).
\end{align*}
The above continuous approximations are used in~\cite{chacon2002} for finite-difference approximations.
Here, we  use their finite-element version.
Then, the third equation  becomes:
\[
 K    M ^{-1}  N_{\Bv_0} \delta\hat\Psiv -
  K  \Big[ \frac 1 \dt +  M ^{-1} N_{\vv_0} + \nu M ^{-1}  K \Big] \delta \hat \Phiv
 \approx -\Gv_{\omegav} - \Ac_{\nu}  M ^{-1} (-\Gv_{\Phiv}),
\]
which is simplified to
\[
 \Big[ \frac  M  \dt +  N_{\vv_0} +\nu  K \Big] \delta \hat \Phiv - N_{B_0} \delta\hat\Psiv
 \approx -  M  K ^{-1}[-\Gv_{\omegav} - \Ac_{\nu}  M ^{-1} (-\Gv_{\Phiv})].
\]
Recall that the second equation in~\eqref{eqn:jacobian} is
\[
 \Big[ \frac  M  \dt +  N_{\vv_0} + \eta  K \Big] \delta \hat \Psiv
 =  N_{B_0} \delta\hat\Phiv    -\Gv_{\Psiv}.
\]
When written as a system, it is a saddle-point system for $\Psi$ and $\Phi$,
\begin{align}
    \begin{bmatrix}
        \Ac_{\nu}  & - N_{\Bv_0}   \smallskip \\
        - N_{\Bv_0}  & \Ac_{\eta}
    \end{bmatrix}
\begin{bmatrix}
 \delta \hat\Phiv  \smallskip \\
       \delta \hat\Psiv
 \end{bmatrix}
 =
\begin{bmatrix}
   -  M   K ^{-1} [-\Gv_{\omegav}- \Ac_{\nu}  M ^{-1} (-\Gv_{\Phiv})] \smallskip \\
    -\Gv_{\Psiv}
 \end{bmatrix},
 \label{eqn:saddle}
\end{align}
with the Schur complement given by
\[
    \Sc = \Ac_{\eta}- N_{\Bv_0}
    \Ac_{\nu}^{-1} N_{\Bv_0}.
\]
Ideally, one would like to invert the Schur complement using scalable solvers.
However, the Schur complement is a dense matrix due to the presence of $\Ac_{\nu}^{-1}$,
 and it is not possible to assemble in practice,  requiring further simplification.
 The study in~\cite{chacon2002} uses a diagonal approximation to $\Ac_{\nu}^{-1}$, and a similar idea will be used here.
 Note  there is a physical meaning in the operator $ N_{\Bv_0}\Ac_{\nu}^{-1} N_{\Bv_0}$.
As, if we simply keep the mass matrix part of the operator $\Ac_{\nu}$,  it becomes  $\dt N_{\Bv_0} M^{-1} N_{\Bv_0}$,
 which is an approximation to the continuous Alfv\'en wave propagator $(\Bv\cdot\grad)^2$.
 Since the  Alfv\'en wave   is the stiffest wave, the most important step in the preconditioner is indeed inverting the Alfv\'en wave propagator to a sufficiently good approximation.
 This propagator is symmetric positive definite
and therefore amenable to  multigrid methods for suitable discrete approximations.

 The above preconditioner works well in~\cite{chacon2002} for finite-difference schemes, and it also works extremely well for continuous finite elements
 for problems with relatively large resistivities and viscosities
 (typically $\ge10^{-4}$).
  However, for sufficiently small resistivity and viscosity values, the stabilization terms are necessary in both of the residual and the preconditioner.
Involving the stabilization term only in the residual will degrade the overall performance of the nonlinear solver for sufficiently large flows.
When we add the diagonal stabilization term into the derivation of the preconditioner,
it is easy to show that the modified saddle-point problem becomes
 \begin{align}
    \begin{bmatrix}
        \Ac_{\nu} + { M  K ^{-1} (\tau_\nu T) M ^{-1} K } & - N_{\Bv_0}   \smallskip  \\
        - N_{\Bv_0}  & \Ac_{\eta} +{\tau_\eta  S}
    \end{bmatrix}
\begin{bmatrix}
 \delta \hat\Phiv  \smallskip \\
       \delta \hat\Psiv
 \end{bmatrix}
 =
\begin{bmatrix}
   -  M   K ^{-1} [-\Gv_{\omegav}- (\Ac_{\nu}  + {\tau_\nu T})  M ^{-1} (-\Gv_{\Phiv})] \smallskip \\
    -\Gv_{\Psiv}
 \end{bmatrix}.
 \label{eqn:saddle2}
\end{align}
The  rest of the section will focus on finding a good approximate Schur-complement preconditioner for this system.

\subsection{Approximate Schur-complement preconditioner}
The approach proposed in~\cite{chacon2002} uses a diagonal approximation for $\Ac_{\nu}$ to make the Schur complement tractable.
We use a similar idea to solve the saddle-point problem~\eqref{eqn:saddle2}, and propose two approaches for the stabilized FE formulation.

The first approach inverts the following system with another level of Jacobi iteration (its iteration is denoted by $m$ here)
\begin{align}
    \begin{bmatrix}
        \Dc_{\nu}  & - N_{\Bv_0}  \\
        - N_{\Bv_0}  & \Ac_{\eta}+{\tau_\eta  S}
    \end{bmatrix}
\begin{bmatrix}
    \delta \hat\Phiv^{(m+1)} \\
\delta \hat\Psiv^{(m+1)}
 \end{bmatrix}
 =
\begin{bmatrix}
   -  M   K ^{-1} [-\Gv_{\omegav}- (\Ac_{\nu}+{\tau_\nu T})  M ^{-1} (-\Gv_{\Phiv})] - \Oc_\nu \delta \Phiv^{(m)} \\
    -\Gv_{\Psiv}
 \end{bmatrix},
 \label{eqn:jacobi1}
\end{align}
where the off-diagonal operator is defined as $\Oc_\nu =   \Ac_{\nu} + { M  K ^{-1} (\tau_\nu T) M ^{-1} K } - \Dc_\nu$ and
the diagonal operator $\Dc_\nu$ is defined as the diagonal portion for the operator of $\Ac_\nu$.
This is a direct extension of the approach in~\cite{chacon2002}.
All the inversions of  the mass and stiffness matrices are implemented through iterative methods.

In the second approach, we note the operator $M  K ^{-1} (\tau_\nu T) M ^{-1} K $ composes the diagonal stabilization operator $\tau_\nu T$ with the stiffness and mass matrices and their inverses on both sides.
If we simply assume all those operators commute with one another, we can simplify the operator to:
\begin{equation}
\label{eqn:commOper}
M  K ^{-1} (\tau_\nu T) M ^{-1} K\approx  \tau_\nu T.
\end{equation}
Note that if $\tau_\nu T$ is replaced by a mass matrix or a stiffness matrix (or their weighted sum), the above approximation  is exact.
The operator $\tau_\nu T$ consists of the mass and stiffness matrices, but in the Petrov-Galerkin sense.
Therefore, the approximation~\eqref{eqn:commOper} appears to be a reasonable assumption.
This leads to the following Jacobi iteration for the saddle-point system,
 \begin{align}
    \begin{bmatrix}
        \tilde\Dc_{\nu}  & - N_{\Bv_0}  \\
        - N_{\Bv_0}  & \Ac_{\eta}+{\tau_\eta  S}
    \end{bmatrix}
\begin{bmatrix}
    \delta \hat\Phiv^{(m+1)} \\
\delta \hat\Psiv^{(m+1)}
 \end{bmatrix}
 =
\begin{bmatrix}
   -  M   K ^{-1} [-\Gv_{\omegav}- (\Ac_{\nu}+{\tau_\nu T})  M ^{-1} (-\Gv_{\Phiv})] - \tilde \Oc_\nu \delta \Phiv^{(m)} \\
    -\Gv_{\Psiv}
 \end{bmatrix},
 \label{eqn:jacobi2}
\end{align}
where $\tilde\Oc_\nu =   \Ac_{\nu} +  \tau_\nu T  - \tilde\Dc_\nu$ and the diagonal operator $\tilde\Dc_\nu$ is the diagonal portion of the operator $ \Ac_{\nu} +  \tau_\nu T$.
We have implemented both approaches  and have found that the second approach actually performs significantly better than the first one.
They all indicate that the stabilization terms can play a significant role in the system and the preconditioner needs to treat those terms carefully.

With a Jacobi iteration of~\eqref{eqn:jacobi1} or~\eqref{eqn:jacobi2},
the Schur complement becomes
\[
    \Sc = \Ac_{\eta} + \tau_\eta S  -  N_{\Bv_0}
    \Dc_{\nu}^{-1}  N_{\Bv_0}.
\]
In practice,  the second operator is implemented weakly ({\red i.e., by integration by parts}), i.e.,
\[
-  N_{\Bv_0} \Dc_{\nu}^{-1}  N_{\Bv_0}  \approx   N_{\Bv_0}^T \Dc_{\nu}^{-1}  N_{\Bv_0},
\]
and thus
\[
    \Sc = \Ac_{\eta} + \tau_\eta S   +  N_{\Bv_0}^T
    \Dc_{\nu}^{-1}  N_{\Bv_0}.
\]
Note the last operator is symmetric and positive semi-definite.
In the implementation, the operator $\Sc$ is assembled as a matrix through a matrix-matrix multiplication and a matrix sum,
so that we can directly call available algebraic multigrid (AMG) methods to  invert this matrix.
A fixed number of Jacobi iterations is used to make sure it converges to the original saddle-point problem.

In Ref.~\cite{chacon2002}, after  $\delta \hat\Phiv^{(k+1)}$ and $\delta \hat\Psiv^{(k+1)}$ are obtained in the Jacobi iteration for
the $k$-th Newton iteration,
the last component of the solution $\delta \hat \omegav^{(k+1)}$ is directly determined by
\[
 M \delta \hat \omegav^{(k+1)} = -\Gv_{\Phiv} -  K  \delta \hat \Phiv^{(k+1)}.
\]
Such an approach works well for the second-order finite-difference approximation.
However, for high-order finite-element methods used in this work, it is found that  it leads to oscillations
in the solution of $\delta \hat \omegav^{(k+1)}$ resulting in preconditioner performance degradation.
Instead, we propose to use the original system to solve  $\delta \hat \omegav^{(k+1)}$ again, i.e.,
\[
(\Ac_{\nu} +\tau_\nu T)\delta \hat \omegav^{(k+1)} = -\Gv_{\omegav}
    + P _{\omega_0} \delta \hat\Phiv^{(k+1)}  -  K   M ^{-1}  N_{\Bv_0} \delta\hat \Psiv^{(k+1)},
\]
after the Jacobi iteration.
Then we further update $\delta \hat\Phiv^{(k+1)}$ by
\[
  K  \delta \hat \Phiv^{(k+1)} = -\Gv_{\Phiv} - M \delta \hat \omegav^{(k+1)}.
\]
The above steps invert the convection-diffusion operator and a stiffness matrix,
and therefore it will increase the smoothness in both $\delta \hat\Phiv^{(k+1)}$ and $\delta \hat\omegav^{(k+1)}$.
To achieve scalability,  AMG solver is used through Hypre ({\tt BoomerAMG})
to invert most of the linear operators in
the preconditioning stage, except the mass matrix, which is typically inverted by conjugate gradient with a generic preconditioner.
Finally, the steps of the full physics-based preconditioner are summarized in Algorithm~\ref{alg:1}.
\begin{algorithm}
\caption{Physics-based preconditioning}
\begin{algorithmic}
\State Compute $\bv = M   K ^{-1} [\Gv_{\omegav}-(\Ac_{\nu}+\tau_\nu T)  M ^{-1} \Gv_{\Phiv}]$
\State Let $\delta \Phiv^{(0)} = 0$
\For{$m=0:N_{ja}$}
\State Update $\tilde \bv= \bv -(\Ac_{\nu}+\tau_\nu T-\Dc_{\nu})\delta \Phiv^{(m)}$
\State Solve $  \Sc \, \delta \hat\Psiv^{(m+1)} =    -\Gv_{\Psiv} +   N_{\Bv_0} \Dc_{\nu}^{-1} \tilde\bv$
\State Solve $\Dc_{\nu} \, \delta \hat\Phiv^{(m+1)} = \tilde \bv +  N_{\Bv_0}\delta \hat\Psiv^{(m+1)}$
\EndFor
\State Solve $(\Ac_{\nu}+\tau_\nu T) \, \delta \hat \omegav = -\Gv_{\omegav}
    + P _{\omega_0} \delta \hat\Phiv  -  K   M ^{-1}  N_{\Bv_0} \delta\hat \Psiv^{(N_{ja})} $
\State Solve  $ K  \delta \hat \Phiv =-\Gv_{\Phiv} - M \delta \hat \omegav$
\end{algorithmic}
\label{alg:1}
\end{algorithm}

We typically set the total number of Jacobi iterations $N_{ja}$ to 4. How tightly the sub-block matrices are inverted can play an important role in the performance of the preconditioner.
Since the mass matrix and stiffness matrix are outside of the Jacobi iteration and they are essentially used to evaluate the right hand side or update the final solution,
we invert them  tightly with a  relative tolerance of $10^{-6}$.
On the other hand, the Schur complement matrix is inverted inside the Jacobi iteration, and since we do not need a very accurate solution from Jacobi iteration, we
invert it relatively loosely with a relative tolerance of $10^{-4}$.
In some cases, we find that solving the Schur complement more tightly can harm the overall performance of the preconditioner.

The overall cost in the preconditioner consists of inverting the Schur complement $N_{ja}$ times and inverting the stiffness matrix twice.
All these inversions are performed with the scalable AMG solver.
Other costs in the preconditioner are negligible compared to this one. The cost of the preconditioner   dominates the cost of the rest of the algorithm,
including evaluation of the FE residual $\Gv (\Uv)$ and the AMR overhead.
Therefore, the overall algorithm will be  \emph{scalable}   in parallel as long as the total number of linear and nonlinear iterations do not grow when we increase the number of processors
and, correspondingly, the mesh.
This will be demonstrated later in the numerical section.

It is tempting to ignore the stabilization terms in the preconditioner since they are ``small''  compared to other operators.
However, we find ignoring those terms can severely degrade preconditioner performance when the flow speed becomes sufficiently large.
The   reason is related to the Schur complement and the convection-diffusion operator $\Ac_\nu$ being inverted in the preconditioner.
As demonstrated  in~\cite{manteuffel2018nonsymmetric}, when the viscosity is around $10^{-5}$ or smaller,
the AMG preconditioner can become very inefficient to invert the convection-diffusion operator if it is under-resolved  even when a SUPG formulation is used.
Without the SUPG term, it behaves much worse and the AMG preconditioner will fail to invert the convection-diffusion operator.
The same in our context also applies to the Schur complement. Although the Schur complement involves the elliptic operator  $(\Bv\cdot\grad)^2$,
it is found that in certain regions the flow speed will dominate  the magnetic field,  resulting in  AMG failure.
Therefore,  it is critical to keep the stabilization terms in both the residual and the preconditioner.

The current study only considers the essential boundary condition as the physical boundary condition.
To impose it in the preconditioner,
the diagonal terms are replaced with $1$ and all the off-diagonal terms are replaced with $0$ in $\Ac_{\nu}$ and $\Ac_{S_L}$,
while all the terms associated with the boundaries are replaced with $0$ in $ N_{\Bv_0}$.
We have verified that there is no boundary layer introduced by the preconditioner in all the numerical examples.
Finally, all the operators such as mass and stiffness matrices are discretized by the operator of the same polynomial spaces, and we have not explored the possibility of using the same operators from lower-order representations.
As we document later, the above preconditioning strategy works well.

\section{Preconditioned JFNK-based algorithm}
\label{sec:jfnk}

We briefly summarize the high-level steps to solve the implicit system using  JFNK  combined with  inexact Newton~\cite{dembo1982inexact} and our physics-based preconditioner.
For more details on  JFNK and inexact Newton methods, see the review paper~\cite{knoll2004jacobian}.
Assume at time $t^n$ there is a known solution of $\hat \Uv^n = [\hat\Phiv^n, \hat\Psiv^n, \hat\omegav^n]^T$, and
we are seeking $\hat \Uv^{n+1}$.
 JFNK  consists of the following four important steps:
\begin{enumerate}
    \item \emph{Define a nonlinear equation}: $ \Gv(\hat\Uv)= 0$ for $\hat\Uv^{n+1}$ using the previous solution $\hat\Uv^n$.  
    \item \emph{Perform a Newton iteration}: the $k$th Newton iteration is defined by solving
    \[
        \Jc_k \delta \hat\Uv_k  = -\Gv(\hat\Uv_k),
    \]
    where the matrix vector product is defined as
       $ \Jc_k \vv:=({\Gv(\hat\Uv_k+\epsilon\vv)-\Gv(\hat\Uv_k)})/\epsilon$
    and $\epsilon$ stands for a small perturbation in the matrix-free product.
\item \emph{Solve the Jacobian system with a Krylov solver {\red with right preconditioning:}} 
   {\red FGMRES is used in an inexact Newton method, which 
     adjusts its convergence toleranace adaptively according to $\| \Jc_k \delta \hat\Uv_k + \Gv(\hat\Uv_k)\| < \zeta \| \Gv(\hat\Uv_k)\|$,
    where $\zeta$ is the inexact Newton parameter.
    A right preconditioner is used in each linear iteration:}
    \[
        {\red(\Jc_k \Pc _k) (\Pc_k^{-1}\delta \hat \Uv_k) = -\Gv(\hat\Uv_k).}
    \]
    The product of $\Jc_k \Pc_k \vv_j$ in the $j$th Krylov iteration is provided  by
    first finding $\yv = \Pc_k \vv_j$
    and then computing $\Jc_k \yv$ by the definitions in Steps 1 and 2. {$\Pc_k$} is the physics-based preconditioner described in Algorithm~\ref{alg:1}.

\end{enumerate}
The JFNK solver provided in PETSc is used here.
A fixed convergence criterion for the Newton iteration throughout all the numerical examples is used with a relative tolerance of $10^{-4}$ and no absolute tolerance.
A fixed maximum number of 20 Newton iterations is used. If the number of Newton iterations exceed 20, we determine that Newton iteration has failed.
In this work, a reasonable   range for the matrix-free parameter $\epsilon$ is found to be  $[10^{-4}, 10^{-1}]$, which is somewhat larger than  values used in previous work~\cite{chacon2002}.
This indicates the numerical residual are slightly more noisy, which is generally true since the current work uses high-order finite-element bases and
targets  much higher Lundquist numbers.
Note that while the value of $\epsilon$ does not have an impact on the converged solutions,  it may significantly affect the convergence of FGMRES.

The number of Newton iterations is also used as an indicator of the presence of fast dynamics during the time stepping.
Since we have a very efficient physics-based preconditioner for the stiff Alfv\'en wave, a general expectation is that the Newton iteration will converge very fast,  typically only requiring  2 to 8 iterations in our implementation.
If the Newton iteration fails to converge after 20 iterations, we surmise that there are fast dynamics present in the system,  which cannot be stepped over by the implicit scheme.
This  happens, for instance, when  plasmoids  appear in the solution.
When such a scenario occurs, we reduce the time step by a factor of two.
When the Newton iteration starts to converge, we slowly increase the time step to improve the efficiency until the targeted time step is reached or the Newton iteration fails again.
For simulations involving plasmoids, this strategy is necessary to run through the entire simulation.

\section{Dynamic adaptive mesh refinement}
\label{sec:amr}
{AMR  is an important component for reduced MHD, and a number of previous AMR implementations are available in~\cite{strauss1998adaptive, lankalapalli2007adaptive, philip2008implicit, baty2019finmhd}.
A key element in all AMR algorithms is the refinement/coarsening indicator, which are of two types, error-based and feature-based.
For instance, in the finite-difference-based AMR solver in~\cite{philip2008implicit},
the refinement criterion is based on the feature in the current $J$ and vorticity $\omega$, which is {\it ad hoc},  requiring parameter tuning  and resulting in unnecessary refinement or coarsening.
A common error-based approach for finite-difference is through Richardson extrapolation, which can be cumbersome to implement.
On  the other hand, an advantage of FE-based AMR algorithms  is the availability of many {\it a posteriori} error estimators  (see the textbook~\cite{ainsworth2011posteriori} for details).
Unfortunately, for a complicated system like the reduced MHD equation, there is no {\it a posteriori} error estimator currently available.
Ref.~\cite{baty2019finmhd} instead uses a feature-based indicator for the FE formulation  of the same MHD problem.
In this work, we propose to use a combination of   feature-based   and {\it a posteriori} error estimators.
Our approach is inspired by the AMR error estimator in the deal.II package~\cite{arndt2020deal}.
In deal.II, a Kelly error estimator~\cite{kelly1983posteriori} is typically used for a complicated system if the system has an elliptic nature (or parabolic if it is time-dependent).
Although the Kelly error estimator was originally proposed for the Poisson equation, the approach is used to estimate more complicated systems in deal.II.
The estimator uses the gradient of the solution as a gradient recovery estimator.
Even though the estimator does not feature theoretical bounds, in practice it typically gives a good hint for mesh refinement.

\subsection{Error estimator}
The error estimator is based on the  superconvergent patch recovery (SPR) approach proposed in~\cite{zienkiewicz1992superconvergent1, zienkiewicz1992superconvergent2},
commonly referred to as the Zienkiewicz-Zhu estimator.
In a gradient recovery approach, the {\it a posteriori} error estimator for a small domain $K$ is simply defined as the energy norm of the difference between estimated and numerical fluxes
\[
\eta^2_K(u_h):= \int_K \| G_h[u_h]- \grad u_h\|^2 \, dx,
\]
where $G_h[u_h]$ denotes an accurate approximation to the gradient of the true solution and $\grad u_h$ is the numerical gradient (the flux for the Poisson equation).
The idea of SPR is to use a least-squares approximation to recover  a more accurate flux $G_h[u_h]$ locally,
and the superconvergence of such a recovery can be proved at certain points in the element.
Note, however, that the above energy norm for the flux only rigiuously leads to a good estimator for the Poisson equation $-\Delta u = f$.
Thus, in principle, the Zienkiewicz-Zhu estimator can only be used for $\Phi$ in the system.
However, since the approach proposed in the current work is fully implicit, one can argue that the above estimator is applicable to other components as well,
since the dominant error in the numerical solution is the component containing the highest-order derivative, which is a Laplacian operator with constant coefficients.
This is a common approach for practical problems, for instance, when extending the Zienkiewicz-Zhu estimator from $-\Delta u = f$ to a more general case $-\Delta u+c u = f$; see the discussions in~\cite{ainsworth2011posteriori} for details.

The estimated error for each element is then defined as a weighted sum of the Zienkiewicz-Zhu estimated errors of certain components.
Following~\cite{philip2008implicit}, we choose one component from the fluid variables, $\omega$, and one component from the field-related variable, $J$.
It is common in simulations that both variables contain more interesting structures than other variables.
The total error  is then defined as a simple sum of two components of the solutions, such as,
\begin{equation}
\label{eqn:error}
e_K:= \alpha \, \eta_K(J) +\eta_K(\omega),
\end{equation}
where $\alpha\in(0, 0.1)$ is taken.
Here, $\alpha$ is chosen as a small value because the current magnitude can be significantly larger than the vorticities in our simulations.
Note that our MFEM implementation can readily leverage to future error estimators,  when they become available in the package.

\subsection{Refinement strategy}
\label{sec:refine}
A successful AMR algorithm necessitates a good refinement and coarsening strategy.
It is particularly important to find balance among different pieces in the adaptive solver,
 including the refinement/coarsening frequency, the overhead due to the AMR algorithm,  effective grid resolutions, smallest length scales and fastest time scales in the system, etc.
For instance,  interpolation or restriction grid operations may lead to small local oscillations in the solution near refinement boundaries, which typically need a few time steps to damp away.
If adaptive patches are changed too frequently,  the small oscillations will accumulate and  result into  larger  estimated errors  later,
which further leads to more unnecessary refinement.
On the other hand, if adaptive grids are changed too infrequently, it leads to  under-resolved structures.
We have found under-refinement to be particularly problematic in simulations with current sheets of high aspect ratio in two ways:
(i) it leads to oscillations in the current sheets, resulting in a lot more plasmoids generated (many more than well resolved runs, indicating triggered by oscillations);
(ii)   it typically results in the failure of  the solver.

We find a refinement and coarsening frequency of once every four to ten time steps to be a good balance for the problems considered.
The refinement and coarsening stages are called in the same time step.
In order to mark elements for refinement/coarsening, we first compute the estimated errors $e_{K_i}$ for each element $K_i$ using~\eqref{eqn:error}
and then valuate the total error $e_{total}$ though a $L_2$-norm of the errors of all the elements.
The $L_2$ norm instead of $L_\infty$ is chosen to avoid the total errors being dominated by a few elements.
The element $K_i$ is refined locally if
\[
e_{K_i} > \max(\beta_r \,e_{total}, e_{lgoal}) \text{ and } e_{total}>e_{goal}.
\]
The element $K_i$ is coarsened locally if
\[
e_{K_i} < \max(\beta_c \, e_{total}, \beta_c \, e_{lgoal}).
\]
Here $e_{goal}$ and $e_{lgoal}$ are the global and local error goals for the refinement and coarsening criteria,
and they are used as safeguards to avoid refining or coarsening excessively.
We typically set $\beta_r=0.1$ and $\beta_c=0.001$.
Using the percentage of the total error is common  in many finite-element AMR solvers based on error estimators, see~\cite{peng2020, bonilla2020monotonicity} for instance.
Finally, after the mesh is updated, a load-balancing algorithm is called to preserve the parallel performance.

\section{Parallel implementation in MFEM}
\label{sec:implementation}
The described algorithm for the 2D visco-resistive MHD equation was implemented in the C++ finite element library MFEM \cite{anderson2020mfem}.
Our implementation is complicated and below we comment on several specific points.

The fully implicit solvers, along with a simple explicit solver, are implemented in both serial and parallel in MFEM.
The simple explicit solver is only implemented as a verification tool. 
All the results discussed in the current work are generated with the parallel fully implicit solver.
The parallel algorithms are implemented using an MPI-based  domain decomposition method.
The vectors and small block matrices use the parallel distributed data structure in MFEM.
PETSc~\cite{balay2019petsc} provides the nonlinear Newton solver as well as the linear solvers in the preconditioner, and Hypre \cite{falgout2002hypre} provides linear solvers in the FE residual
and also provides the preconditioner for the sub-block matrices like the Schur complement and the convection-diffusion operator to improve scalability.
The physics-based preconditioner is implemented through MFEM's PETSc interface. We take the full advantage of the {\tt pcshell} interface provided by PETSc and linked by MFEM.
In this shell interface, the input is a block vector $\xv$ consisting of three component, $\delta\Phi$, $\delta\Psi$ and $\delta\omega$,
and its output is $\yv=\Pc \xv$ where $\Pc$ is the physics-based preconditioning described in  Algorithm~\ref{alg:1}.
Sub-block matrices assembled in MFEM are inverted using PETSc inside the shell interface. 
Note that the physics-based preconditioner is created once per every Newton iteration and it is reused until the FGMRES iteration converges.

{
\newcommand{\figWidth}{8.5cm}
\newcommand{\trimfig}[2]{\trimw{#1}{#2}{.0}{.0}{.0}{.0}}
\begin{figure}[htb]
\begin{center}
\begin{tikzpicture}[scale=1]
  \useasboundingbox (0.,.75) rectangle (16,5.4);  
  \draw(-1, 0) node[anchor=south west,xshift=-4pt,yshift=+0pt] {\trimfig{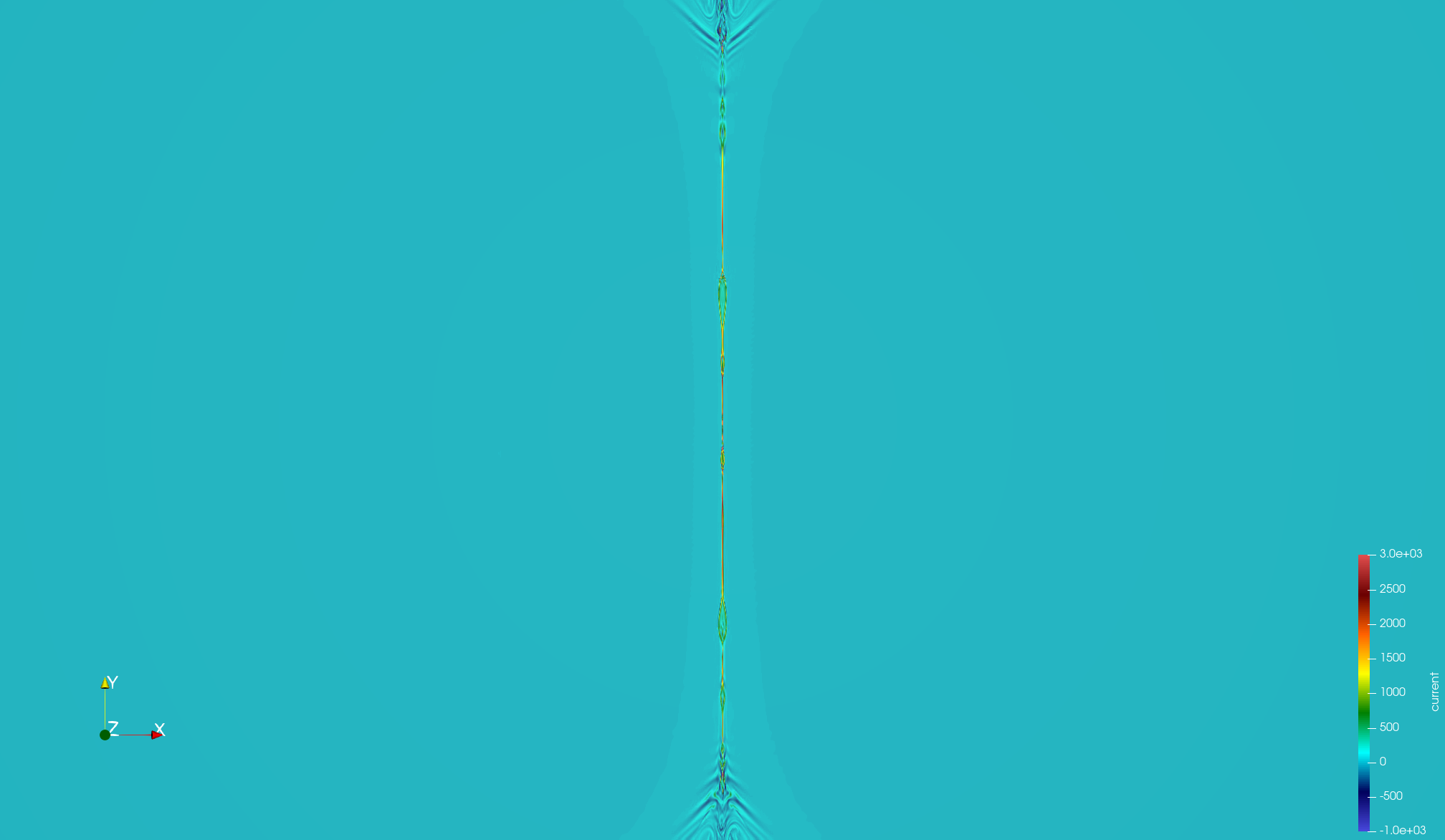}{\figWidth}};
  \draw(7.6, 0) node[anchor=south west,xshift=-4pt,yshift=+0pt] {\trimfig{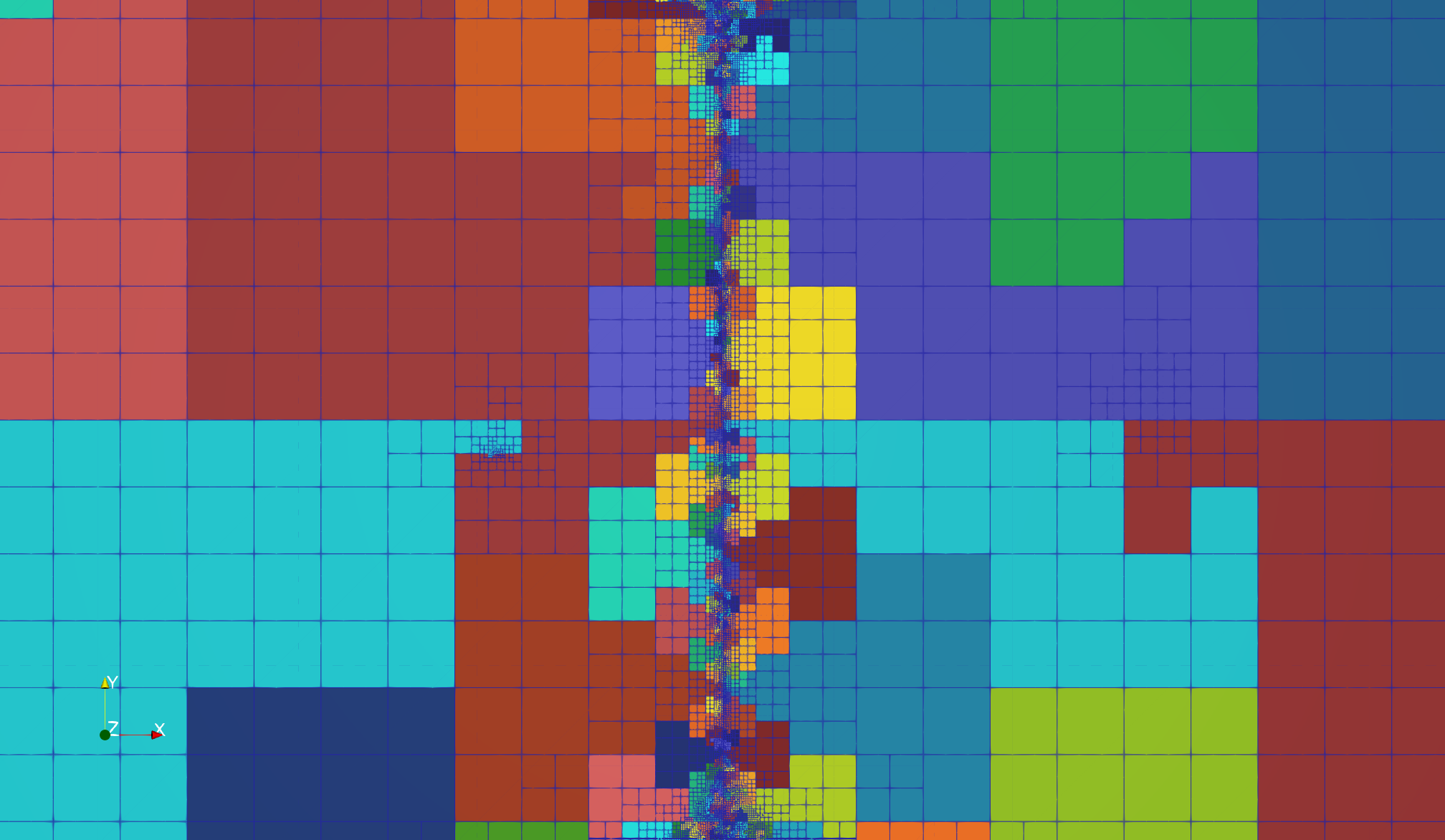}{\figWidth}};
\end{tikzpicture}
\end{center}
\caption{Island coalescence test with AMR. Left: the current sheets and plasmoids around the center region before plasmoids become unstable.
Right: the corresponding adaptive mesh and domain decomposition through dynamic load-balancing.
 Here, $\eta=\nu=10^{-7}$, $p=3$ and 7 refinement levels are used. The AMR mesh captures both current sheet and plasmoids very well. 
}
  \label{fig:amr}
\end{figure}
}

Our implementation of the algorithms is general,
supporting arbitrary order of accuracy and general meshes including triangular, quadrilateral and high-order
curvilinear meshes, taking full advantage of the MFEM capabilities.
However, in the numerical study, we focus on quadrilateral meshes and polynomial order less than or equal to four, particularly in the high-Lundquist-number runs.
For high-order polynomials such as $p=8$ or higher, the preconditioner performs considerably worse than the lower-order polynomials,
which indicates the need for further tuning. 
The AMR implementation  supports both conforming and nonconforming meshes with coarsening and refining as well as dynamic load-balancing.
In practice, we focus on the nonconforming quadrilateral meshes due to their more flexible refinement. 
A load-balancing for quadrilateral meshes is based on a space-filling curve and it leads to a continuous decomposition of the meshes throughout the simulations,
while other meshes typically cannot achieve that and the decompositions become discontinuous as time evolves.
See Figure~\ref{fig:amr} for a sample AMR mesh. 
For more details on the nonconforming AMR algorithm and its MFEM implementation, see Ref.~\cite{cerveny2019nonconforming}.

{

\begin{figure}[htb]
\begin{center}
    \begin{tikzpicture}[scale=.8]
    \useasboundingbox (-1, 0.0) rectangle (8.,6);  
\begin{scope}[xshift=0cm]
\begin{loglogaxis}[xlabel=\# of CPUs,ylabel=Time, grid=both]
\logLogSlopeTriangleInv{0.65}{0.3}{0.8}{1}{black};
\addplot coordinates {
        (1,70096)
    (2,35109)
        (4,20590)
    (8,9573.4)
    (16,5882.)
    (32,3719.9)
    (64,1730)
    (128,835.6)
    (256,397.3)
    (512,189.9)
    (1024,94.9)
    (2048,48.4)
};
\end{loglogaxis}
\end{scope}
\end{tikzpicture}
\end{center}
\caption{Strong scaling result. The test is island coalescence with a mesh size of $768\times768$ and $p=3$.  Here $\eta=\nu=10^{-3}$.   {Excellent nearly} perfect linear scaling is observed up to 2048 CPUs.
}
  \label{fig:scaling}
\end{figure}
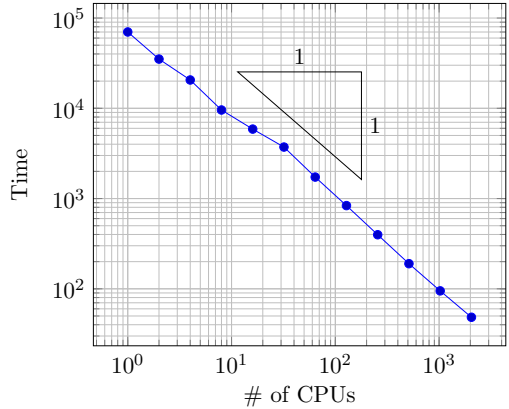
}

One significant advantage of the MFEM package is its scalability, which has been demonstrated up to hundreds of thousands of processors (as well as on GPU systems) \cite{anderson2020mfem}.
In the current work,  the majority of the cost of the algorithm comes from the preconditioner, which takes the full advantage of AMG solvers,
so it is not too challenging  to achieve  good strong scaling.
See Figure~\ref{fig:scaling} for instance, which demonstrates a perfect linear scaling for the current solver up to 2048 CPUs on a relatively coarse mesh.
(Due to the 2D nature of  problems considered, the total number of processors  typically does not exceed 4098.)
The benchmark problem is an island coalescence with a relatively small Lundquist number.
See the corresponding numerical  section for more details of the problem setup.
However, while achieving a good strong scaling is relatively easy  in MFEM, it is very challenging to achieve a good weak scaling in implicit solvers,
as it requires  algorithmic scalability. 
This will be one focus of the study in the  numerical section.

\paragraph{Reproducibility} The implementation of the solvers described above, as well as most of the numerical examples presented in this work, is freely available as an MFEM branch of {\tt tds-mhd-dev}
distributed along with MFEM's master branch on GitHub\footnote{See https://github.com/mfem/mfem/tree/tds-mhd-dev.}.
The {\tt tds-mhd-dev} branch currently contains both serial and parallel versions of the solver including a simple explicit scheme and fully implicit schemes along with physics-based preconditioning and AMR.

\section{Numerical results} \label{sec:results}

Several benchmark problems are considered in this section to demonstrate the accuracy and efficiency
of the proposed full algorithm.
The benchmark problems are organized in the order of increasing complexity.
We begin in Section~\ref{sec:wavePropagation} by considering a linear Alfv\'en wave propagation example.
This benchmark
problem provides a good first test of  the accuracy of the scheme as well as the performance of the iterative solver.
In Section~\ref{sec:Rayleigh}, we consider a second benchmark problem propagating MHD Rayleigh flow and Alfv\'en wave.
 The example is used to demonstrate the high-order spatial accuracy.
Section~\ref{sec:tearMode} considers the tearing mode instability.
Some physics scaling results are presented to validate the associated solver.
The example is also used to perform a weak scaling test to demonstrate the algorithmic scalability of the proposed solver.
This example is then used for a careful study of the performance of the AMR solver.
Attention will be devoted to the accuracy improvement of the adaptive meshes, the computational cost savings,
the performance of the iterative solver along with the AMR algorithm, etc.
Some quantitative comparisons in those aspects will be presented.
The last problem we consider in Section~\ref{sec:ic} is the island coalescence. 
The study of this example is split into two parts, steady meshes (Section~\ref{sec:ic-steady}) and adaptive meshes (Section~\ref{sec:ic-amr}).
The steady mesh results are used for validation study through a physics scaling test relating the reconnection rate with the Lundquist number.
A weak scaling demonstration up to 4096 CPUs is also presented on the uniform meshes using a moderate Lundquist number.
The robustness and efficiency of the AMR solver is then demonstrated using moderately high to high Lundquist-number setups.
In particular,  results with Lundquist number of $10^5$, $10^6$ and $10^7$ are presented.
The solutions around the current sheets show a transition from no plasmoids ($S_L =$ $10^5$), to isolated plasmoids and simple dynamics ($S_L =$ $10^6$), and finally to many colliding plasmoids
and complicated dynamics ($S_L =$ $10^7$).
Some interesting physics phenomena in the numerical solutions are observed.
The computational savings of the adaptive meshes over a uniform refined mesh is also presented.

Throughout all the tests,   a relative tolerance of $10^{-4}$ with no absolute tolerance of the Newton solver is used.
A solution update tolerance of $10^{-6}$ is added to avoid over-solving, which  only activates in the finest uniform resolution during the weak-scaling study.
The Schur complement uses the same relative tolerance of $10^{-4}$ while the advection-diffusion operator is typically solved tighter, with a relative tolerance of $10^{-6}$.
Other operators such as the mass matrix and stiffness matrix are also inverted tightly using a relative tolerance of $10^{-6}$.
The parameters associated with JFNK and inexact Newton methods use the default values of PETSc.
All the scaling tests are performed on the LANL Grizzly cluster, which consists of XEON E5-2695V4 18C 2.1GHz processors.

\subsection{Alfv\'en wave propagation}
\label{sec:wavePropagation}

We first consider a linear Alfv\'en wave propagation  test with zero dissipation ($\nu=\eta=0$).
The problem is similar to the example given in~\cite{chacon2002}, except a manufactured solution is introduced to examine the accuracy here.
Consider the following system 
\begin{align*}
    \grad^2 \Phi & =\omega, \\
    \left( \partial_t +\vv \cdot \grad \right) \Psi   &= -E_0, \\
    \left( \partial_t +\vv \cdot \grad \right) \omega  &= \Bv \cdot \grad J.
\end{align*}
in the domain of $[0, L_x]\times[0, 1]$.
A manufactured standing-wave  solution satisfies
 \begin{align*}
 \Phi &= -\alpha \sin(\pi y) \sin(k x) \sin(k t), \\
     \Psi & = -y + \alpha \sin(\pi y) \cos \left(k x\right)\cos(k t), \\
     \omega   &= \alpha (\pi^2+k^2) \sin(\pi y) \sin(k x ) \sin(k t),
\end{align*}
where the frequency is $k = 2\pi/L_x$.
To balance the equation, a time-dependent source electric field is needed $E_0 = \alpha^2 \pi k \sin(k t)\cos(k t) \sin(\pi y) \cos(\pi y)$.
A periodic boundary condition is imposed in the $x$-direction and a homogeneous Dirichlet  boundary condition is imposed in the $y$-direction.
We let $L_x = 3$ and the perturbation magnitude  $\alpha = 10^{-3}$.
The problem is essentially driven by the background magnetic field of $\Bv_0=[1, \, 0]^T$
which results into an Alfv\'en wave with speed of $1$.
Note that the system develops a standing wave in the domain.
The problem is used to examine both of the spatial and temporal accuracy of the scheme as well as the efficiency of the preconditioner.
Due to the special linear-wave structure, the problem does not need a physical or numerical dissipation.
Nevertheless, we still apply the fully implicit nonlinear solver to the system, as a way to examine overall performance of the algorithm.

\newcommand{\logLogSlopeTriangle}[5]
{
    \pgfplotsextra
    {
        \pgfkeysgetvalue{/pgfplots/xmin}{\xmin}
        \pgfkeysgetvalue{/pgfplots/xmax}{\xmax}
        \pgfkeysgetvalue{/pgfplots/ymin}{\ymin}
        \pgfkeysgetvalue{/pgfplots/ymax}{\ymax}

        \pgfmathsetmacro{\xArel}{#1}
        \pgfmathsetmacro{\yArel}{#3}
        \pgfmathsetmacro{\xBrel}{#1-#2}
        \pgfmathsetmacro{\yBrel}{\yArel}
        \pgfmathsetmacro{\xCrel}{\xArel}

        \pgfmathsetmacro{\lnxB}{\xmin*(1-(#1-#2))+\xmax*(#1-#2)} 
        \pgfmathsetmacro{\lnxA}{\xmin*(1-#1)+\xmax*#1} 
        \pgfmathsetmacro{\lnyA}{\ymin*(1-#3)+\ymax*#3} 
        \pgfmathsetmacro{\lnyC}{\lnyA+#4*(\lnxA-\lnxB)}
        \pgfmathsetmacro{\yCrel}{\lnyC-\ymin)/(\ymax-\ymin)} 

        \coordinate (A) at (rel axis cs:\xArel,\yArel);
        \coordinate (B) at (rel axis cs:\xBrel,\yBrel);
        \coordinate (C) at (rel axis cs:\xCrel,\yCrel);

        \draw[#5]   (A)-- node[pos=0.5,anchor=north] {1}
                    (B)--
                    (C)-- node[pos=0.5,anchor=west] {#4}
                    cycle;
    }
}

{
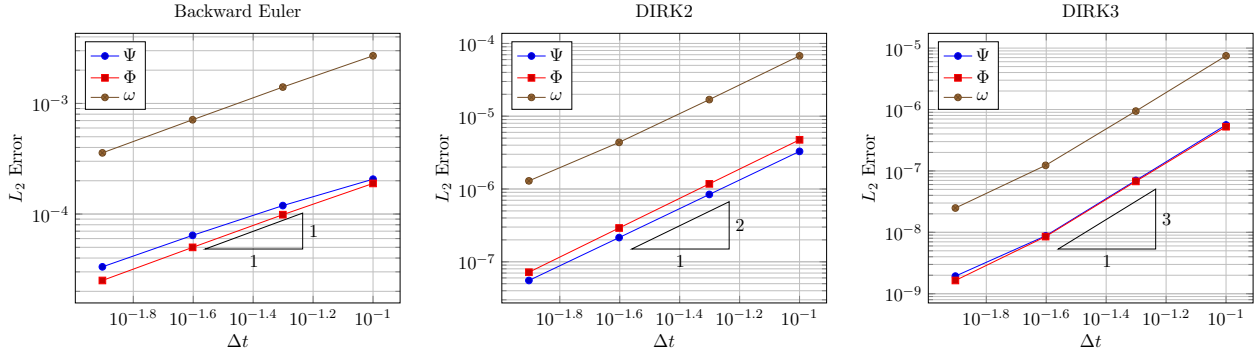
\begin{figure}[htb]
\begin{center}
\begin{tikzpicture}[scale=.63]
\useasboundingbox (0, 0.0) rectangle (24.,6.5);
\begin{loglogaxis}[
title=Backward Euler,
xlabel={$\dt$},
ylabel={$L_2$ Error},
grid=both,
legend pos=north west
]
\logLogSlopeTriangle{0.7}{0.3}{0.2}{1}{black};
\addplot+[] table[x={xColumn},y={column1}] {tables/waveTime.dat};
\addplot+[] table[x={xColumn},y={column2}] {tables/waveTime.dat};
\addplot+[] table[x={xColumn},y={column3}] {tables/waveTime.dat};
\legend{$\Psi$\\$\Phi$\\$\omega$\\}
\end{loglogaxis}

\begin{scope}[xshift=9cm]
\begin{loglogaxis}[
title=DIRK2,
xlabel={$\dt$},
ylabel={$L_2$ Error},
grid=both,
legend pos=north west
]
\logLogSlopeTriangle{0.7}{0.3}{0.2}{2}{black};
\addplot+[] table[x={xColumn},y={column1}] {tables/waveDIRK2.dat};
\addplot+[] table[x={xColumn},y={column2}] {tables/waveDIRK2.dat};
\addplot+[] table[x={xColumn},y={column3}] {tables/waveDIRK2.dat};
\legend{$\Psi$\\$\Phi$\\$\omega$\\}
\end{loglogaxis}
\end{scope}

\begin{scope}[xshift=18cm]
\begin{loglogaxis}[
title=DIRK3,
xlabel={$\dt$},
ylabel={$L_2$ Error},
grid=both,
legend pos=north west
]
\logLogSlopeTriangle{0.7}{0.3}{0.2}{3}{black};
\addplot+[] table[x={xColumn},y={column1}] {tables/waveDIRK3.dat};
\addplot+[] table[x={xColumn},y={column2}] {tables/waveDIRK3.dat};
\addplot+[] table[x={xColumn},y={column3}] {tables/waveDIRK3.dat};
\legend{$\Psi$\\$\Phi$\\$\omega$\\}
\end{loglogaxis}
\end{scope}
\end{tikzpicture}
\end{center}
\caption{Wave propagation. The refinement study for the temporal integrators are presented. A fixed computational grid of $192\times192$ is used for the first two tests,
while a computational grid of $768\times768$ is used for the third-order integrator.
The polynomial degree used in all the tests is $p=3$.
The expected temporal orders of accuracy are observed with all the integrators.
\label{fig:waveTime}}
\end{figure}
}

We have verified several time integrators implemented in the solver, including the backward Euler method, a second-order DIRK scheme
and a third-order DIRK scheme.
To avoid any potential super-convergence related to the sinusoidal waves, we compare the numerical solutions with the exact solutions at $t=1.8$.
A computational grid of $192\times192$ with polynomial order of 3 is used for the backward Euler and DIRK2 integrators while a grid of $768\times768$ is used for the DIRK3 integrator.
The numerical errors are dominated by the temporal errors for these grid resolutions.
The time step starts from $\dt=0.1$ and refined by a factor of 2.
The resulting numerical errors vs the time steps are presented in Figure~\ref{fig:waveTime}.
Here a $L_2$-norm is used to evaluate the absolute errors.
Note that the desired orders of accuracy are observed for all the time integrators.

\begin{table}[hbt]\tableFont 
\begin{center}
\caption{Wave propagation. Solver performance during the temporal refinement study (DIRK2) are presented.
The averaged number of Newton iterations and Krylov iterations in each solve are presented.
Note DIRK2 contains two backward Euler solves per time steps.
The computational grid is fixed as $192\times192$ and $p=3$.
A fixed number of CPU=32 is used for all the runs.
 \label{tableWave}}
 \medskip
\begin{tabular}{|c|c|c|c|c|c|} \hline
  \multicolumn{6}{|c|}{Wave propagation (DIRK2)} \\ \hline
\strutt  $\dt$  & Total steps &   Newton per solve     &  Krylov per solve   &    Total time (s)     &  Time/$\dt$       \\[3pt] \hline
0.1 &   18 &  2.8333 &  9.4167 &   1369.53 &   76.09           \\ \hline
0.05 &   36 & 2.7917 & 8.0972 &   1900.51 &   52.79     \\ \hline
0.025 &   72 & 2.9167 & 7.3542&  2176.62 &  30.23    \\ \hline
0.0125 &  144 & 2.4653 &  5.2743 &  2681.29 &  18.62   \\ \hline
\end{tabular}
\end{center}
\end{table}

The problem is also used to examine the performance of the preconditioner.
We use the second-order DIRK results for this study.
Performance details are summarized in Table~\ref{tableWave}.
The computational grid is fixed to $192\times192$, which corresponds to a grid spacing of $\dx= 0.015625$.
The time step of $\dt=0.1$ is therefore much larger than $\dt_{CFL}$.
Here $\dt_{CFL}$ follows the definition in Ref.~\cite{chacon2002}, which is identical to $\dx$ due to the normalized Alfv\'en wave speed of 1 and CFL of 1,  but for the high-order finite-element scheme, the stable explicit time step is typically smaller.
We also note that the number of averaged Newton and Krylov iterations only increase slightly when the time step increases.
As a result, the   large time-step run is faster than  smaller time-step runs.
The good performance of the preconditioner is consistent with the nature of the problem, as it is dominated by the Alfv\'en wave propagation.
Note that the CPU time per time-step increases when the time steps increase.
This is largely because the sub-block matrices, particularly the Schur complement,  in the preconditioner become easier to invert  for smaller time steps.
For a given targeted accuracy, the overall CPU time suggests that it is preferable to take a large time step.
This fact will be further emphasized in the next few examples.

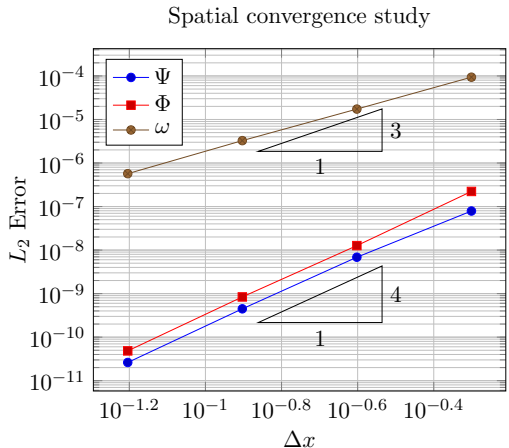
\begin{figure}[htb]
\begin{center}
\begin{tikzpicture}[scale=.8]
\useasboundingbox (0, 0.0) rectangle (20.,6.3);
\begin{scope}[xshift=6.5cm]
\begin{loglogaxis}[
title=Spatial convergence study,
xlabel={$\dx$},
ylabel={$L_2$ Error},
grid=both,
legend pos=north west
]
\logLogSlopeTriangle{0.7}{0.3}{0.7}{3}{black};
\logLogSlopeTriangle{0.7}{0.3}{0.2}{4}{black};
\addplot+[] table[x={xColumn},y={column1}] {tables/waveSpace.dat};
\addplot+[] table[x={xColumn},y={column2}] {tables/waveSpace.dat};
\addplot+[] table[x={xColumn},y={column3}] {tables/waveSpace.dat};
\legend{$\Psi$\\$\Phi$\\$\omega$\\}
\end{loglogaxis}
\end{scope}
\end{tikzpicture}
\end{center}
\caption{Wave propagation. Grid refinement study for  $p=3$.
A fixed time step of $\dt=0.001$ is used for all the simulations and the third-order DIRK integrator is used.
Fourth-order accuracy is observed for both $\Phi$ and $\Psi$ while we find order-reduction for $\omega$.
\label{fig:waveSpace}}
\end{figure}

The problem is also used to verify the spatial accuracy of the solver.
Verifying the spatial accuracy turns out to be challenging.
To make sure the numerical error is dominated by the spatial accuracy,  we use the third-order DIRK integrator with a much smaller time step $\dt=0.001$.
Due to the small time step used, 
we need to adjust the relative tolerance to $10^{-7}$ to fully expose the spatial accuracy.
We focus on $p=3$ in the convergence study.
The refinement study starts from a $6\times6$ grid and refine it by a factor of 2.
The resulting numerical errors vs.~the grid spacing are presented in Figure~\ref{fig:waveSpace}.
It is found that the expected accuracy (fourth-order) is observed for both $\Phi$ and $\Psi$ while
 $\omega$ shows order reduction, with a convergence rate around 2.5.
 The reason for the poor performance of $\omega$ is two-fold: the problem is diffusion-free,
 and the error in $\omega$ is dominated by the high-order derivative $\Bv\cdot \grad J$ term.
 More importantly, note that physics components of interest in the reduced MHD system
 are $\Phi$ and $\Psi$, since they are used to construct the velocity  and magnetic fields, respectively,
 while $\omega$ acts like an auxiliary variable in the system.
 Therefore, for the purpose of practical simulations, it is sufficient to achieve optimal accuracy for $\Phi$ and $\Psi$.

{
\newcommand{\figWidth}{5.5cm}
\newcommand{\trimfig}[2]{\trimw{#1}{#2}{.0}{.0}{.0}{.0}}
\begin{figure}[htb]
\begin{center}
\resizebox{14cm}{!}{
\begin{tikzpicture}[scale=1]
  \useasboundingbox (0.,.75) rectangle (16.,9.8);  

    \draw(-.5, 0) node[anchor=south west,xshift=-4pt,yshift=+0pt] {\trimfig{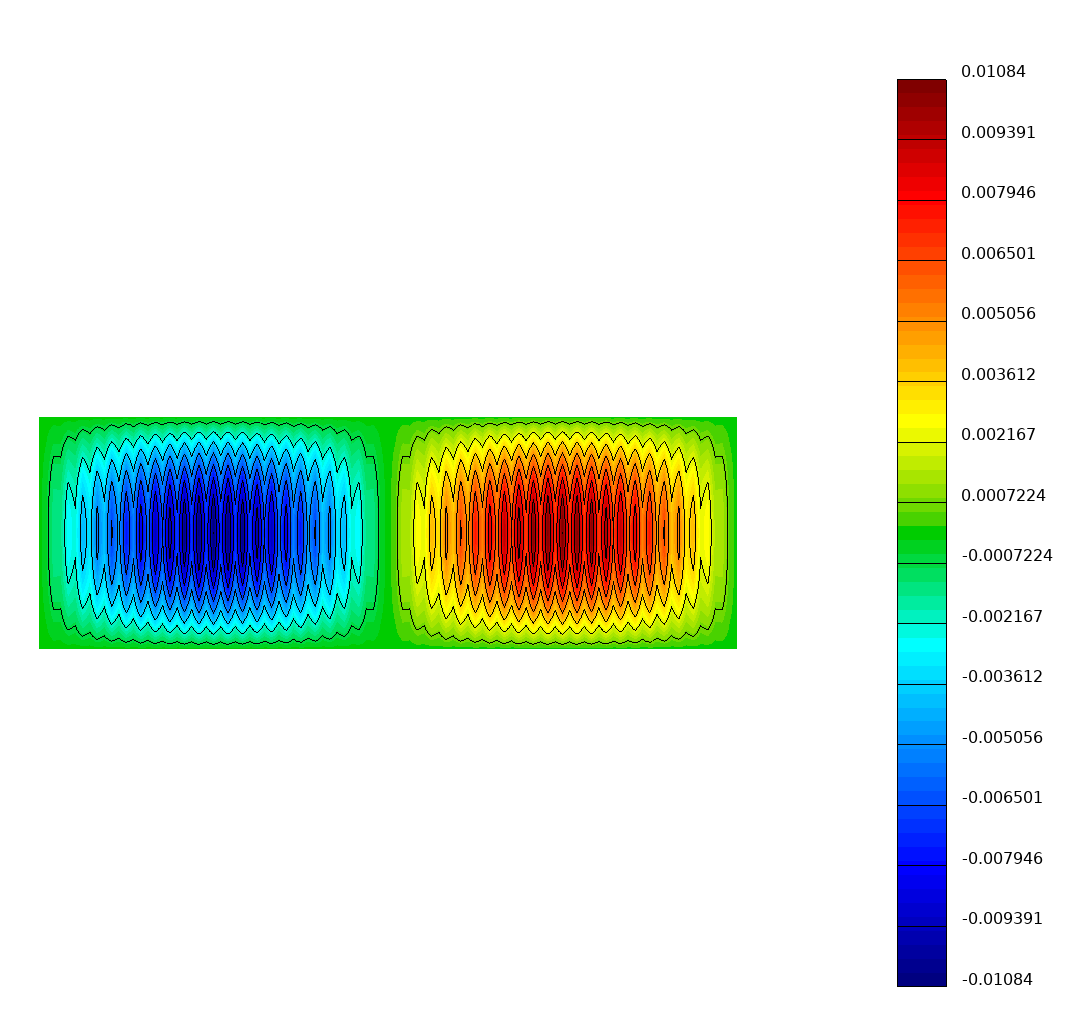}{\figWidth}};
    \draw(5, 0) node[anchor=south west,xshift=-4pt,yshift=+0pt] {\trimfig{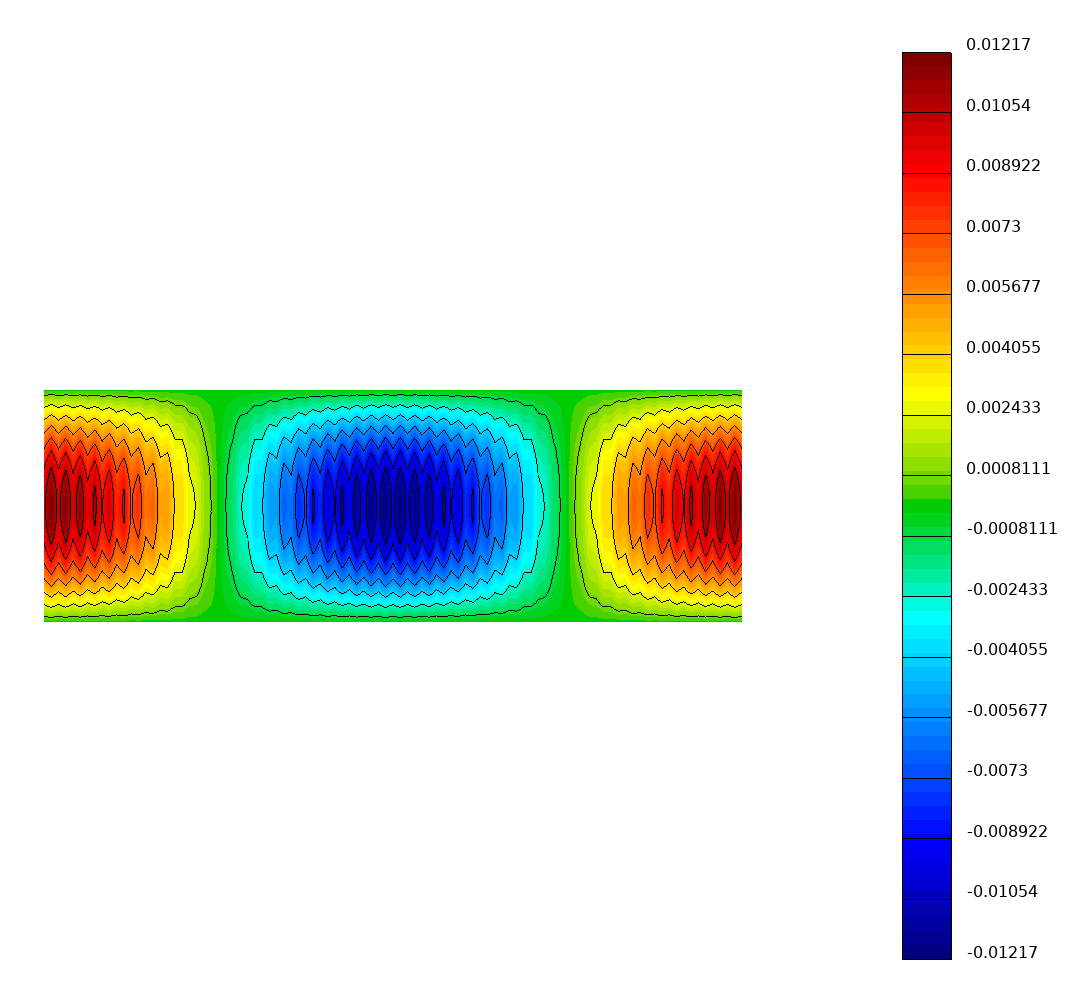}{\figWidth}};
  \draw(10.5, 0) node[anchor=south west,xshift=-4pt,yshift=+0pt] {\trimfig{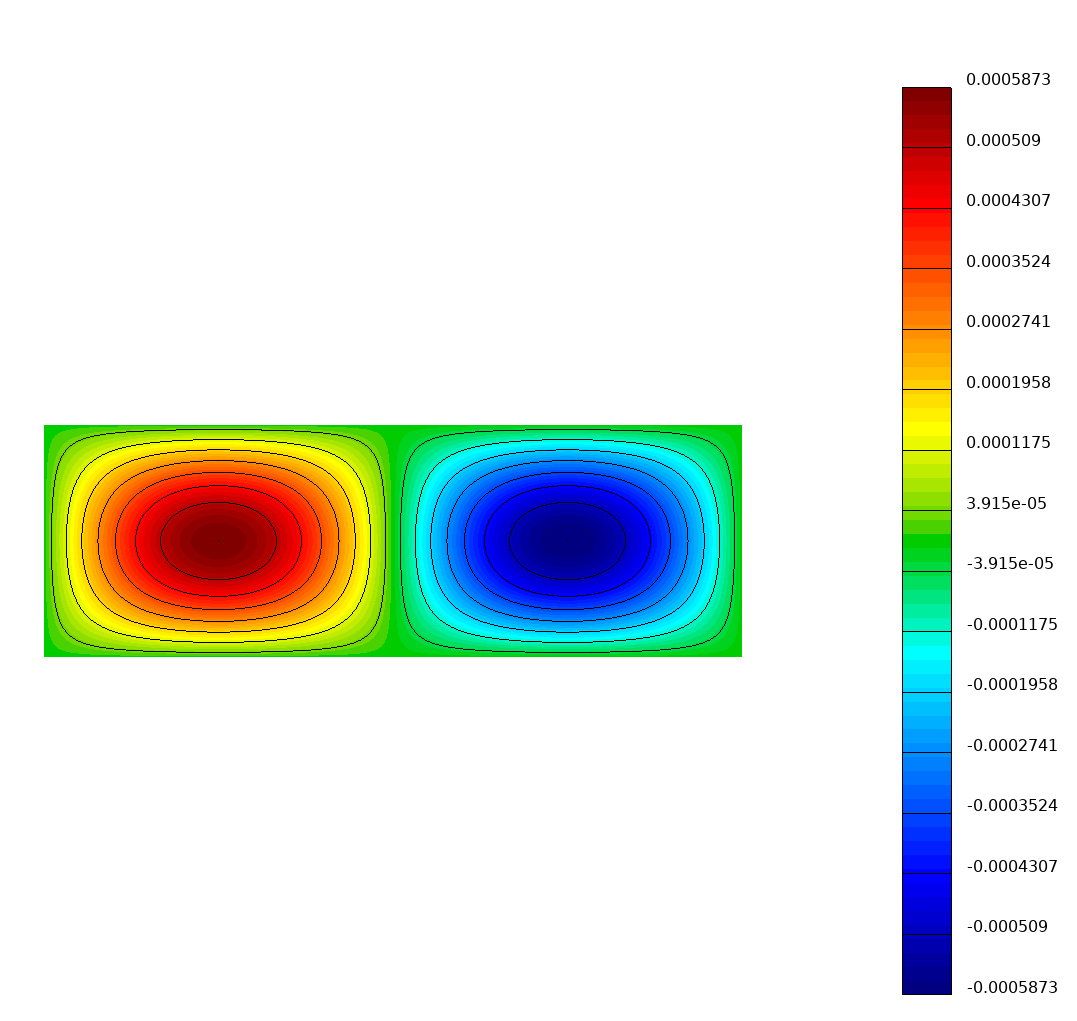}{\figWidth}};
   \draw(.0,3.6) node[draw,fill=white,anchor=west,xshift=2pt,yshift=0pt] {\small$\omega$};
    \draw(5.6,3.6) node[draw,fill=white,anchor=west,xshift=2pt,yshift=0pt] {\small$J$};
     \draw(11.1,3.6) node[draw,fill=white,anchor=west,xshift=2pt,yshift=0pt] {\small$\Phi$};
     \begin{scope}[yshift=4.5cm]
   \draw(-.5, 0) node[anchor=south west,xshift=-4pt,yshift=+0pt] {\trimfig{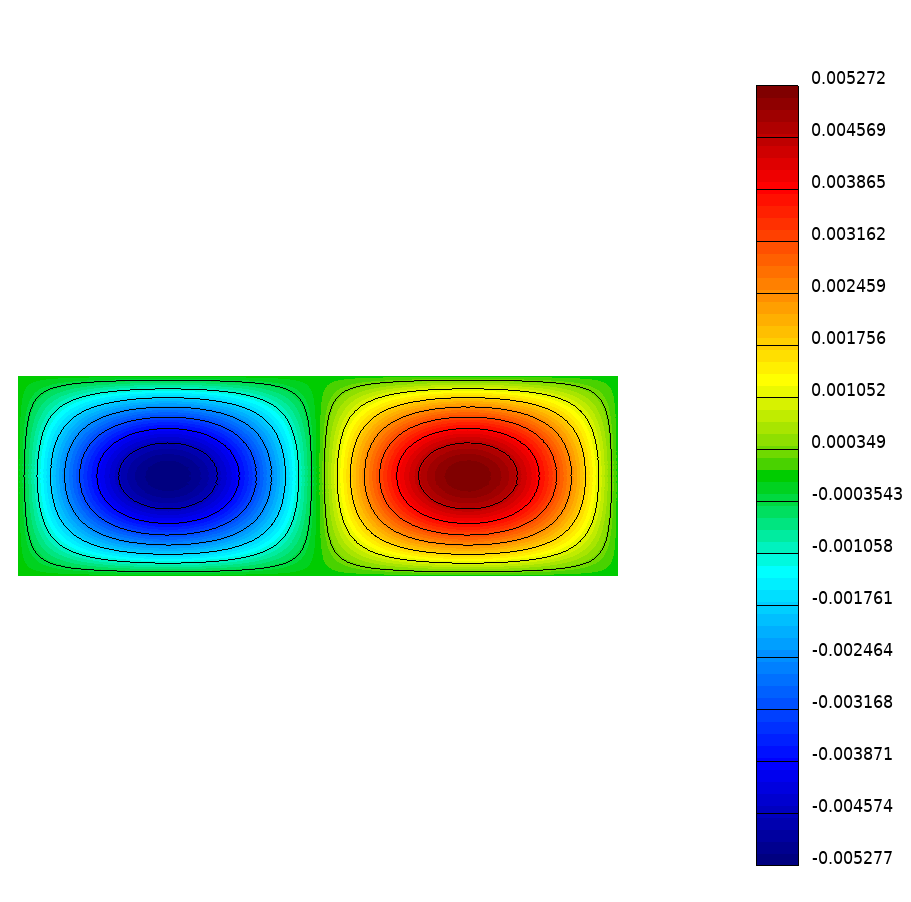}{\figWidth}};
    \draw(5, 0) node[anchor=south west,xshift=-4pt,yshift=+0pt] {\trimfig{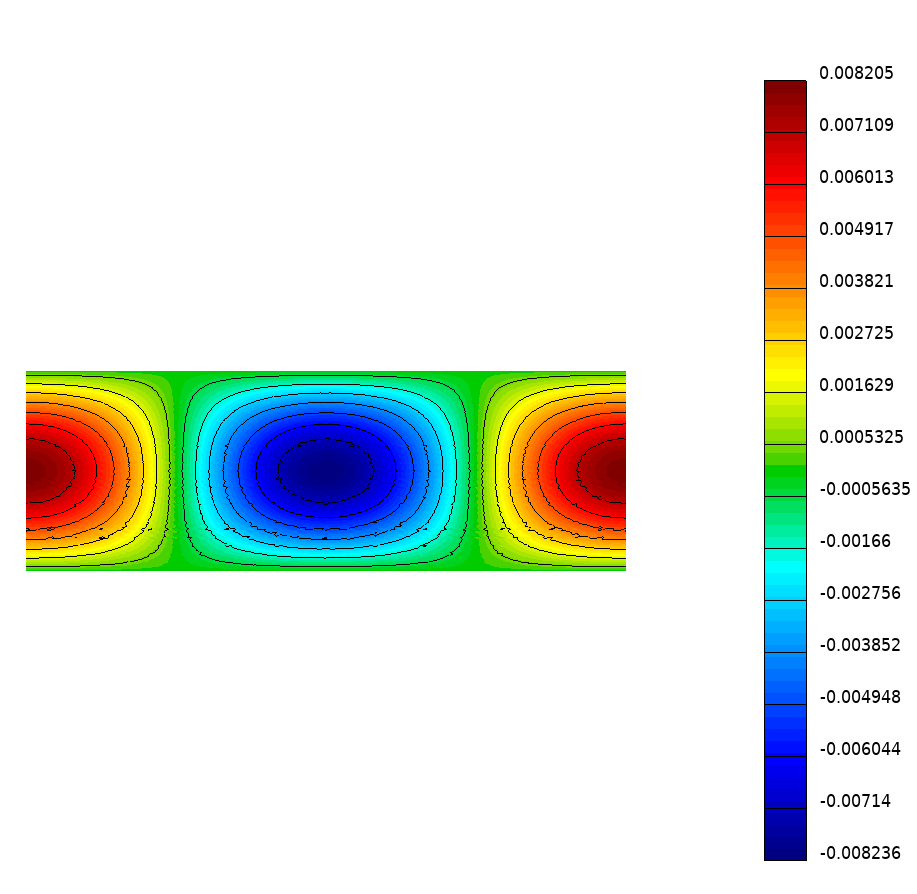}{\figWidth}};
  \draw(10.5, -.1) node[anchor=south west,xshift=-4pt,yshift=+0pt] {\trimfig{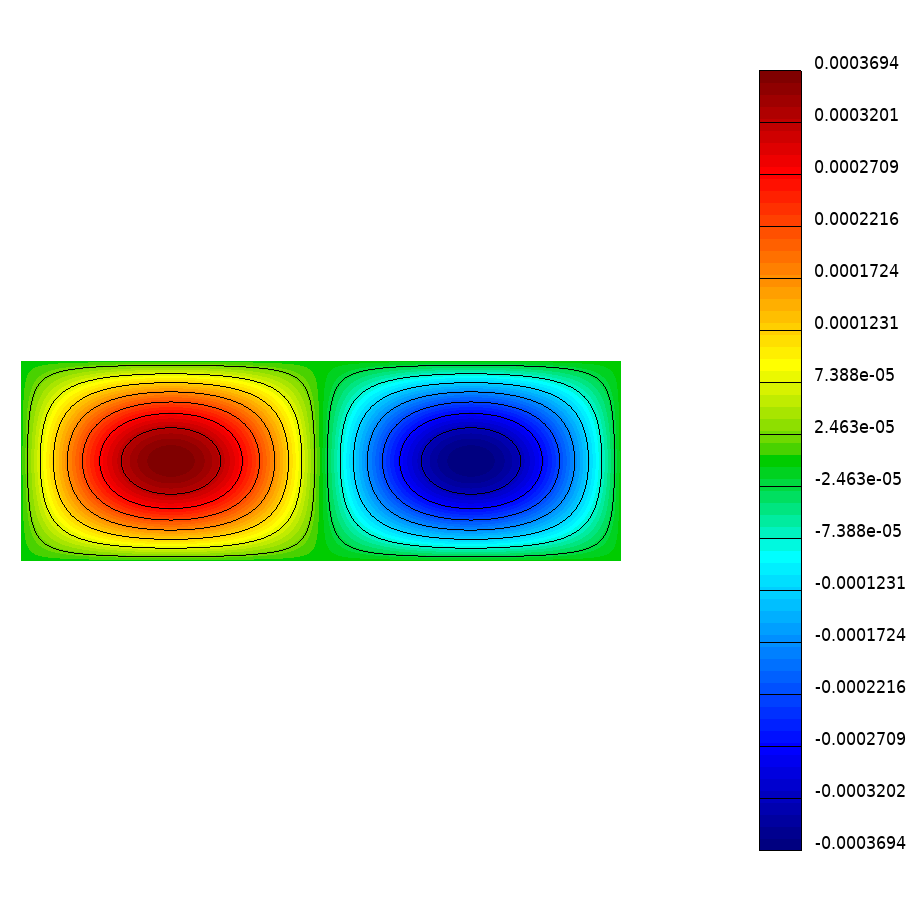}{\figWidth}};
   \draw(.0,3.6) node[draw,fill=white,anchor=west,xshift=2pt,yshift=0pt] {\small$\omega$};
    \draw(5.6,3.6) node[draw,fill=white,anchor=west,xshift=2pt,yshift=0pt] {\small$J$};
     \draw(11.1,3.6) node[draw,fill=white,anchor=west,xshift=2pt,yshift=0pt] {\small$\Phi$};
     \end{scope}
\end{tikzpicture}
}
\end{center}
\caption{Wave propagation. The numerical solutions at $t=1.8$ are presented for $p=3$ (top row) and $p=2$ (bottom row). Note that the current and vorticity become very oscillatory with $p=2$, but other
solution components remain smooth.
}
  \label{fig:p2Results}
\end{figure}
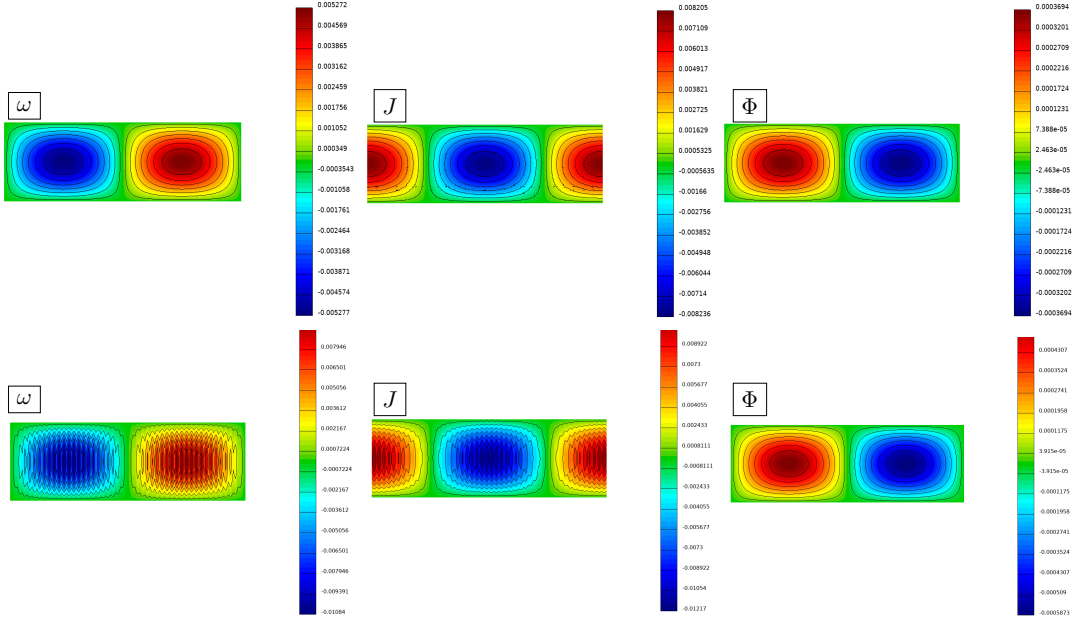
}

We  compare the results of $p=3$ and $p=2$ in Figure~\ref{fig:p2Results}.
The figure shows oscillatory behaviors in both of $\omega$ and $J$ for  $p=2$, while
other solution components, such as $\Phi$, remain smooth.
The assessment of the spatial accuracy for $p=2$ indicates $\Phi$ and $\Psi$ are second-order while the numerical error in $\omega$ decays  slower than first order.
However, as we shall see, the quality of the solution with $p=2$  improves markedly in the presence of physical dissipation.

\subsection{MHD Rayleigh flow and Alfv\'en wave propagation}
\label{sec:Rayleigh}
We consider the MHD Rayleigh flow and Alfv\'en wave propagation with a known exact solution.
An analytical solution to the MHD equation is given in~\cite{moreau2013}
and tested  in~\cite{ben1999conservative, shadid2010towards}.
The exact solution is given by $\Bv=[B_x, B_0/\sqrt{\mu_0 \rho}\,]^T$  and $\Vv=[V_x, 0]^T$ where, in our units:
\begin{align*}
B_x &= -\frac{U}{4} {e^{-{\frac {{\it A_0}\,y}{d}}}} \left( -1+{e^{{
\frac {{\it A_0}\,y}{d}}}} \right)  \left[ {\it \erfc} \left( {
\frac {-{\it A_0}\,t+y}{2 \sqrt {dt}}} \right) +{e^{{\frac {{\it A_0
}\,y}{d}}}}{\it \erfc} \left(  {\frac {{\it A_0}\,t+y}{2\sqrt {dt}}}
 \right)  \right], \\
V_x &= \frac{U}{4}  \left[{\erfc} \left( {\frac {-{\it A_0}\,t+y}{2 \sqrt {dt}}} \right)  + {\erfc} \left( {\frac {{\it A_0}\,t+y}{2 \sqrt {dt}
}} \right)
+{e^{-{\frac {{\it A_0}\,y}{d}}}}{\it \erfc} \left( {\frac {-{\it A_0}\,t+y}{2\sqrt {dt}}} \right) +{e^{{\frac {{
\it A_0}\,y}{d}}}}{\it \erfc} \left( {\frac {{\it A_0}\,t+y}{2\sqrt {
dt}}} \right)  \right],
\end{align*}
and $A_0$ here is the Alfv\'en wave speed given by $A_0 = B_0 /\sqrt{\mu_0\rho}$.
Unlike the previous test, we consider here the full SUPG formulation.

The corresponding components in the resistive MHD system considered in the current work can be worked out as
\begin{align*}
\Phi&=   \frac {U}{ 4 A_0 } \bigg[  \left( t{ A_0 }
^{2}+d{e^{-{\frac { A_0 \,y}{d}}}}- A_0 \,y+d \right)
{\rm erf} \left( {\frac { A_0 \,t-y}{2 \sqrt {dt}}}\right)+
 \left( t{ A_0 }^{2}+d{e^{{\frac { A_0 \,y}{d}}}}+ A_0
\,y+d \right) {\rm erf} \left( {\frac { A_0 \,t+y}{2\sqrt {dt}}}
\right) \\ &-d{e^{{\frac { A_0 \,y}{d}}}} +d{e^{-{\frac {{
\it A0}\,y}{d}}}}-2\, A_0 \,y \bigg]
+ {\frac {U \sqrt {dt}}{ 2 \sqrt {\pi}} \left( {e^{-{\frac { \left(  A_0 \,t-y \right) ^{2}}{4 dt}}}}   + {e^{- {\frac { \left(  A_0 \,t+y \right) ^{2}}{4dt}}}} \right) } \\
\smallskip
\Psi &={\frac {{\it B_0}\,x}{\sqrt {\mu_0 \rho}}}- {\frac {U\sqrt {dt}}{
2\sqrt {\pi}} \left( {e^{- {\frac { \left( -{\it A_0}\,t+y
 \right) ^{2}}{4 dt}}}}-{e^{- {\frac { \left( {\it A_0}\,t+y
 \right) ^{2}}{4dt}}}} \right) }- {\frac {U \left( t A_0^{2}+
d \right) }{{4\it A_0}} \left[ {\rm erf} \left( {\frac {{\it A_0}\,t-
y}{2\sqrt {dt}}}\right)-{\rm erf} \left( {\frac {{\it A_0}\,t+y}{2 \sqrt {dt}}}\right) \right] } \\
&+ {\frac {U}{{4 A_0}} \left[
 \left( d{e^{-{\frac {{\it A_0}\,y}{d}}}}+{\it A_0}\,y \right) {
\erfc} \left( {\frac {-{\it A_0}\,t+y}{2\sqrt {dt}}} \right) +
 \left( d{e^{{\frac {{\it A_0}\,y}{d}}}}-{\it A_0}\,y \right) {
\erfc} \left(  {\frac {{\it A_0}\,t+y}{2\sqrt {dt}}} \right)
 \right] } \\[4pt]
 \omega &= {\frac {UA_0}{4d} \left[ 2\,{e^{-{\frac {A_0\,y}{d}}}}-{e^{{\frac {A_0\,y}{d}}}
}{\erfc} \left(  {\frac {A_0\,t+y}{2\sqrt {dt}}} \right) - e^{-{\frac {A_0\,y}{d}}} {\erfc} \left(  {\frac {A_0\,t-y}{2\sqrt {dt}}} \right)
 \right] } \\
&+ {\frac {U}{2 \sqrt {\pi dt}} \left( {
e^{- {\frac { \left( A_0\,t-y \right) ^{2}}{4dt}}}}+{
e^{- {\frac { \left( A_0\,t+y \right) ^{2}}{4dt}}}}
 \right) }
\end{align*}
Note that due to the definition of the magnetic potential, there is a negative sign in $\Psi$ compared to the magnetic potential given in~\cite{shadid2010towards}.
It is easy to verify that the above components satisfy the MHD system~\eqref{eqn:mhd2d} with
an external  E field given by $E_0 = -U B_0/(2 \sqrt{\mu_0 \rho})$ and $d=\nu=\eta=1$.
Therefore, the magnetic Prandtl number $Pr_m = \nu/\eta = 1$ in this test.
The rest of the parameters are taken from~\cite{ben1999conservative, shadid2010towards} as $U=1.0$, $\rho = 4\times10^{-5}$,
$B_0 =1.4494\times10^{-4} $, and
$\mu_0 = 1.256636\times10^{-6} $.
The resulting Alfv\'en speed is $A_0\approx 20.44$.
The computational domain is $[0, 5]\times[0, 5]$ with a Dirichlet boundary condition applied to all the component of the system using the exact solution.
Note that unlike the case in~\cite{ben1999conservative, shadid2010towards}, the vorticity at $t=0$ has a singular boundary condition in our MHD system.
To avoid such a singularity in the boundary condition,  the exact solution is initialized at $t = 0.02$ in this test.

{
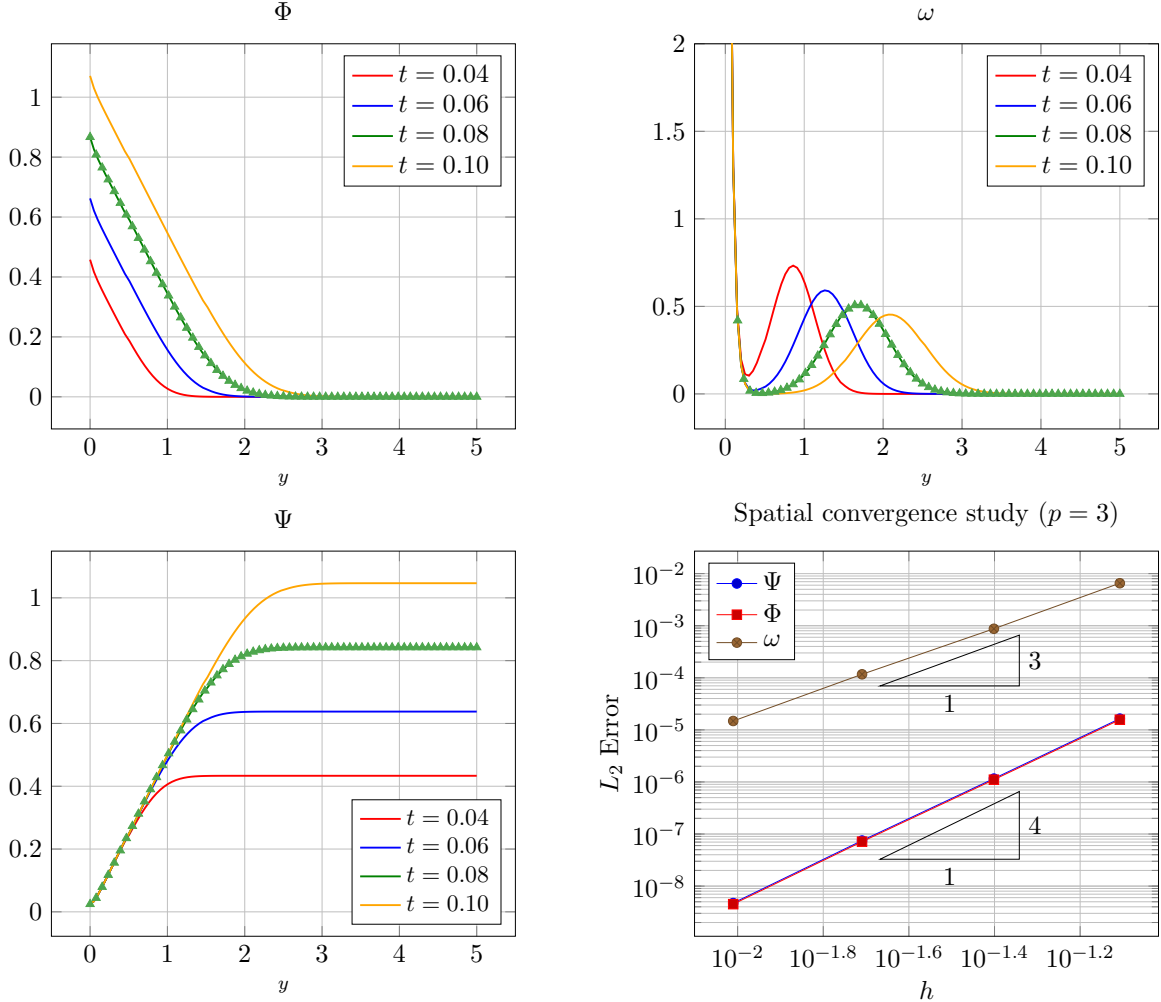
\begin{figure}[htb]
\begin{center}
\begin{tikzpicture}[scale=.9]
\useasboundingbox (0, 0.0) rectangle (16.,14);

\begin{scope}[xshift=9.5cm]
\begin{loglogaxis}[
title={Spatial convergence study ($p=3$)},
xlabel={$h$},
ylabel={$L_2$ Error},
grid=both,
legend pos=north west
]
\logLogSlopeTriangle{0.7}{0.3}{0.65}{3}{black};
\logLogSlopeTriangle{0.7}{0.3}{0.2}{4}{black};
\addplot+[] table[x={xColumn},y={psi}] {figures/rayleigh/error.txt};
\addplot+[] table[x={xColumn},y={phi}] {figures/rayleigh/error.txt};
\addplot+[] table[x={xColumn},y={omega}] {figures/rayleigh/error.txt};
\legend{$\Psi$\\$\Phi$\\$\omega$\\}
\end{loglogaxis}
\end{scope}

\begin{axis}[
title={$\Psi$},
xlabel={$y$},
label style={font=\scriptsize},
grid=both,
grid style={draw=gray!50},
legend pos=south east,
legend style={font=\small},
]
\addplot[color=red,thick] table[x={y},y={t1}] {figures/rayleigh/psi.txt};
\addplot[color=blue,thick] table[x={y},y={t2}] {figures/rayleigh/psi.txt};
\addplot[color=Green,thick] table[x={y},y={t3}] {figures/rayleigh/psi.txt};
\addplot[color=Orange,thick] table[x={y},y={t4}] {figures/rayleigh/psi.txt};
\addplot[color=Green!70, only marks, mark=triangle*] table[x={y},y={psi}] {figures/rayleigh/t0p08.txt};
\legend{$t=0.04$ , $t=0.06$ , $t=0.08$, $t=0.10$  }
\end{axis}

\begin{scope}[yshift=7.5cm]
\begin{axis}[
title={$\Phi$},
xlabel={$y$},
label style={font=\scriptsize},
grid=both,
grid style={draw=gray!50},
legend pos=north east,
]
\addplot[color=red,thick] table[x={y},y={t1}] {figures/rayleigh/phi.txt};
\addplot[color=blue,thick] table[x={y},y={t2}] {figures/rayleigh/phi.txt};
\addplot[color=Green,thick] table[x={y},y={t3}] {figures/rayleigh/phi.txt};
\addplot[color=Orange,thick] table[x={y},y={t4}] {figures/rayleigh/phi.txt};
\addplot[color=Green!70, only marks, mark=triangle*] table[x={y},y={phi}] {figures/rayleigh/t0p08.txt};
\legend{$t=0.04$ , $t=0.06$ , $t=0.08$, $t=0.10$  }
\end{axis}
\end{scope}

\begin{scope}[yshift=7.5cm, xshift=9.5cm]
\begin{axis}[
title={$\omega$},
xlabel={$y$},
label style={font=\scriptsize},
grid=both,
grid style={draw=gray!50},
legend pos=north east,
ymax=2,
]
\addplot[color=red,thick] table[x={y},y={t1}] {figures/rayleigh/omega.txt};
\addplot[color=blue,thick] table[x={y},y={t2}] {figures/rayleigh/omega.txt};
\addplot[color=Green,thick] table[x={y},y={t3}] {figures/rayleigh/omega.txt};
\addplot[color=Orange,thick] table[x={y},y={t4}] {figures/rayleigh/omega.txt};
\addplot[color=Green!70, only marks, mark=triangle*] table[x={y},y={omega}] {figures/rayleigh/t0p08.txt};
\legend{$t=0.04$ , $t=0.06$ , $t=0.08$, $t=0.10$  }
\end{axis}
\end{scope}

\end{tikzpicture}
\end{center}
\caption{MHD Rayleigh flow and Alfv\'en wave propagation.
The numerical solution at $t=0.08$ at $x=0$ are presented as green triangles in the above plots.
The exact solutions at different times are plotted as a reference.
The spatial refinement study for $p=3$ is presented at the bottom right. The expected  order of accuracy is observed in $\Phi$ and $\Psi$ while the accuracy of $\omega$ is shows
 order reduction.
\label{fig:rayleigh}}
\end{figure}
}

We present the results of $p=3$ in Figure~\ref{fig:rayleigh}.
In particular, the numerical solution at $t=0.08$ and $x=0$ solved on a grid of $64\times64$ is presented as green triangles.
The figures suggest the solution is smooth and matches well the exact solution.
A grid refinement study for $p=3$ is performed at $t = 0.08$ and its result is also presented in Figure~\ref{fig:rayleigh}.
 It is observed that an expected order of accuracy is achieved in $\Phi$ and $\Psi$,
 while  $\omega$  again shows order reduction  (although less than in the wave propagation case) with the finest resolutions showing an order of 2.8.
Here the DIRK3 time integrator is used with a very small time step.
The current test also suggests the SUPG terms do not diminish the optimal order of accuracy of the overall scheme.

\subsection{Tearing mode}
\label{sec:tearMode}
The next example we consider is the resistive tearing mode~\cite{chacon2002, glasser2004sel, philip2008implicit, lin2020krylov}.
The tearing mode is one of the most fundamental resistive instabilities. It is a spontaneous reconnection process occurring in sheared magnetic topologies~\cite{biskamp1996magnetic}.
The inverse time scale and length scale of the problem scales as $\sqrt{\eta}$ and thus the problem becomes more challenging with increasing Lundquist numbers.
 This test is used to validate the full solver through some physics scaling tests, and
 it is also used to verify the accuracy and efficiency of the adaptive solver. 

The same problem setup as described in~\cite{chacon2002} is adopted. The initial condition starts from an equilibrium given by
\begin{align*}
    \Psi_0 = \frac{1} \lambda \ln [\cosh \lambda(y-\frac 1 2)],\qquad \omega_0=0, \qquad \Phi_0&= 0,
\end{align*}
balanced with a source term of $E_0 = \eta \grad^2 \Psi_0$.
The initial magnetic field here is based on a Harris current sheet equilibrium.
The problem is excited by a perturbation on $\Psi_0$ of
\[
    \delta \Psi = 10^{-3} \sin(\pi y) \cos \left(\frac{2\pi}{L_x} x\right).
\]
The computational domain is $[0, L_x]\times [0, 1]$ and the boundary conditions are the  PEC wall (Dirichlet) along the $y$ boundaries and periodic along the $x$ direction.
The other parameters are
$L_x=3$, $\lambda =5$, and $\nu=\eta=10^{-3}$,
unless otherwise noted.

The study of this example is split into two parts in the following discussion. In the first part, the solver is verified on uniform meshes  against previous work
including some physics scaling studies, and tested in parallel up to 4096 processors.
In the second part, the solver is tested with dynamic AMR. 

\subsubsection{Uniform-mesh results}

{
\newcommand{\figWidth}{9.5cm}
\newcommand{\trimfig}[2]{\trimFig{#1}{#2}{.22}{.2}{.55}{.5}}
\begin{figure}[htb]
\begin{center}
\resizebox{16cm}{!}{
\begin{tikzpicture}[scale=1]
  \useasboundingbox (0.0,0) rectangle (18,6.2);  
  \draw(-.5,3.) node[anchor=south west,xshift=0pt,yshift=+0pt] {\trimfig{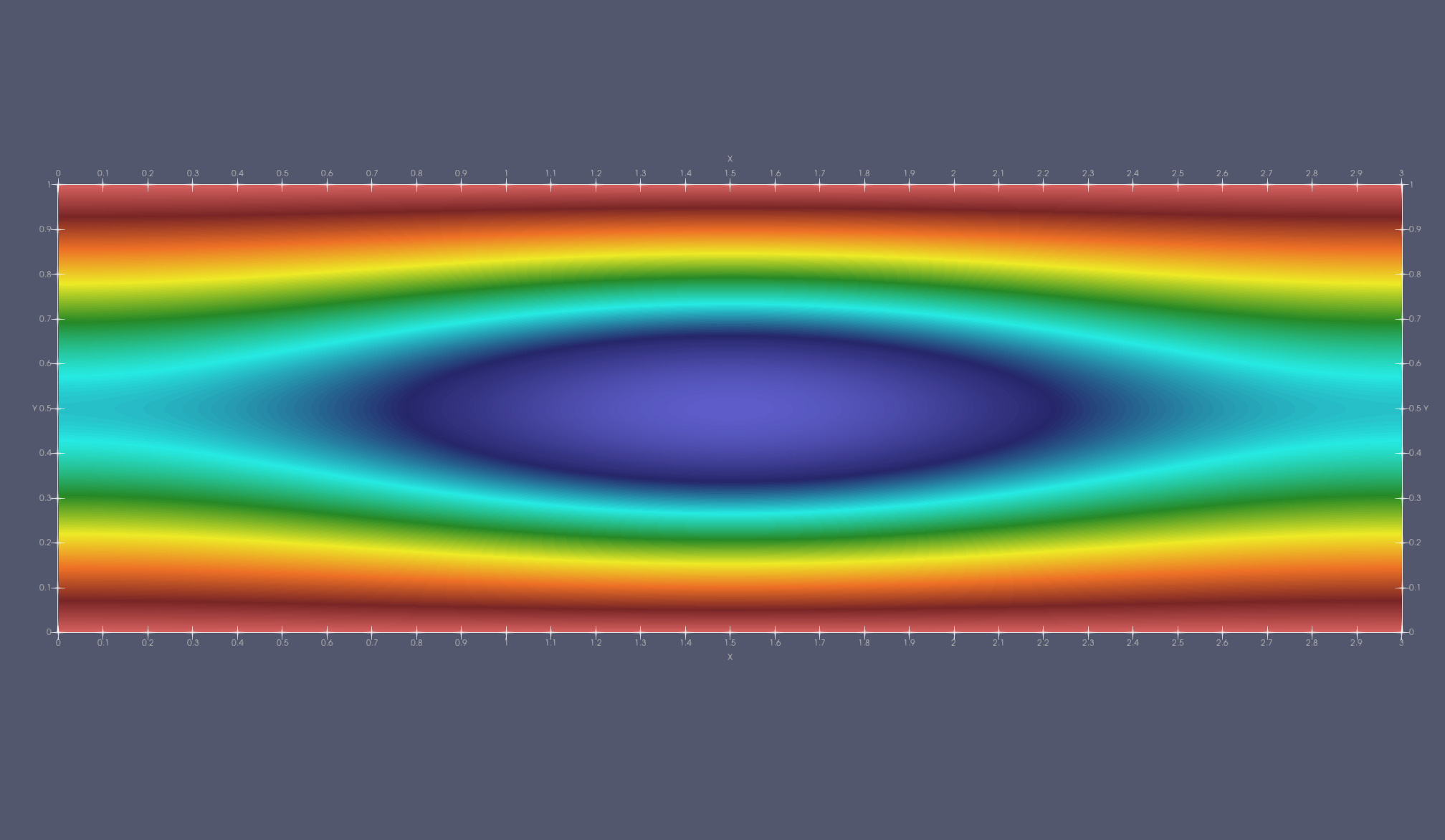}{\figWidth}};
  \draw(9.1,3.) node[anchor=south west,xshift=0pt,yshift=+0pt] {\trimfig{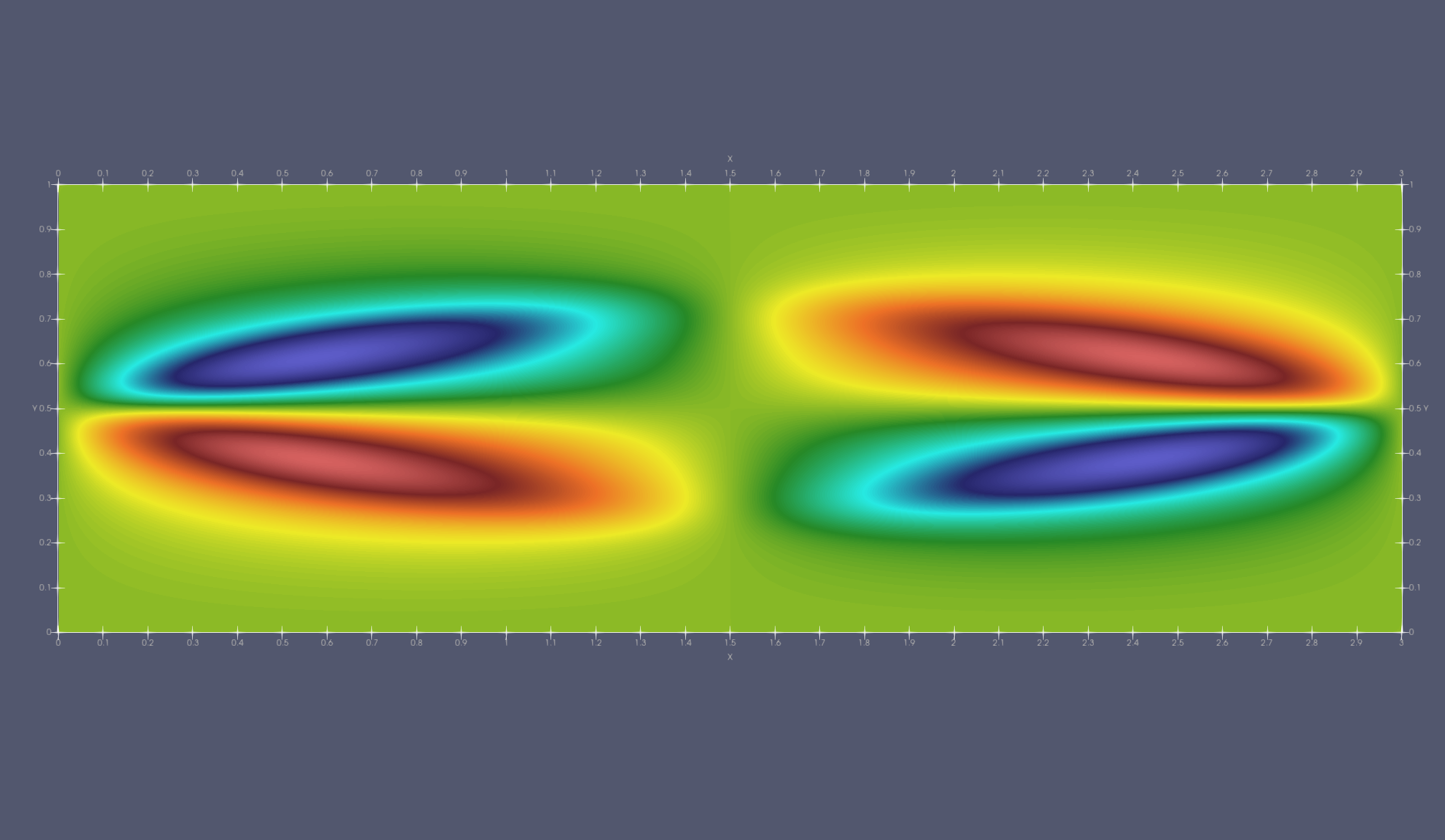}{\figWidth}};
    \draw(-.4,6.) node[draw,fill=white,anchor=west,xshift=4pt,yshift=0pt] {\scriptsize $\Psi$};
    \draw(9.2,6.) node[draw,fill=white,anchor=west,xshift=4pt,yshift=0pt] {\scriptsize $\Phi$};
  \draw(-.5,-.75) node[anchor=south west,xshift=0pt,yshift=+0pt] {\trimfig{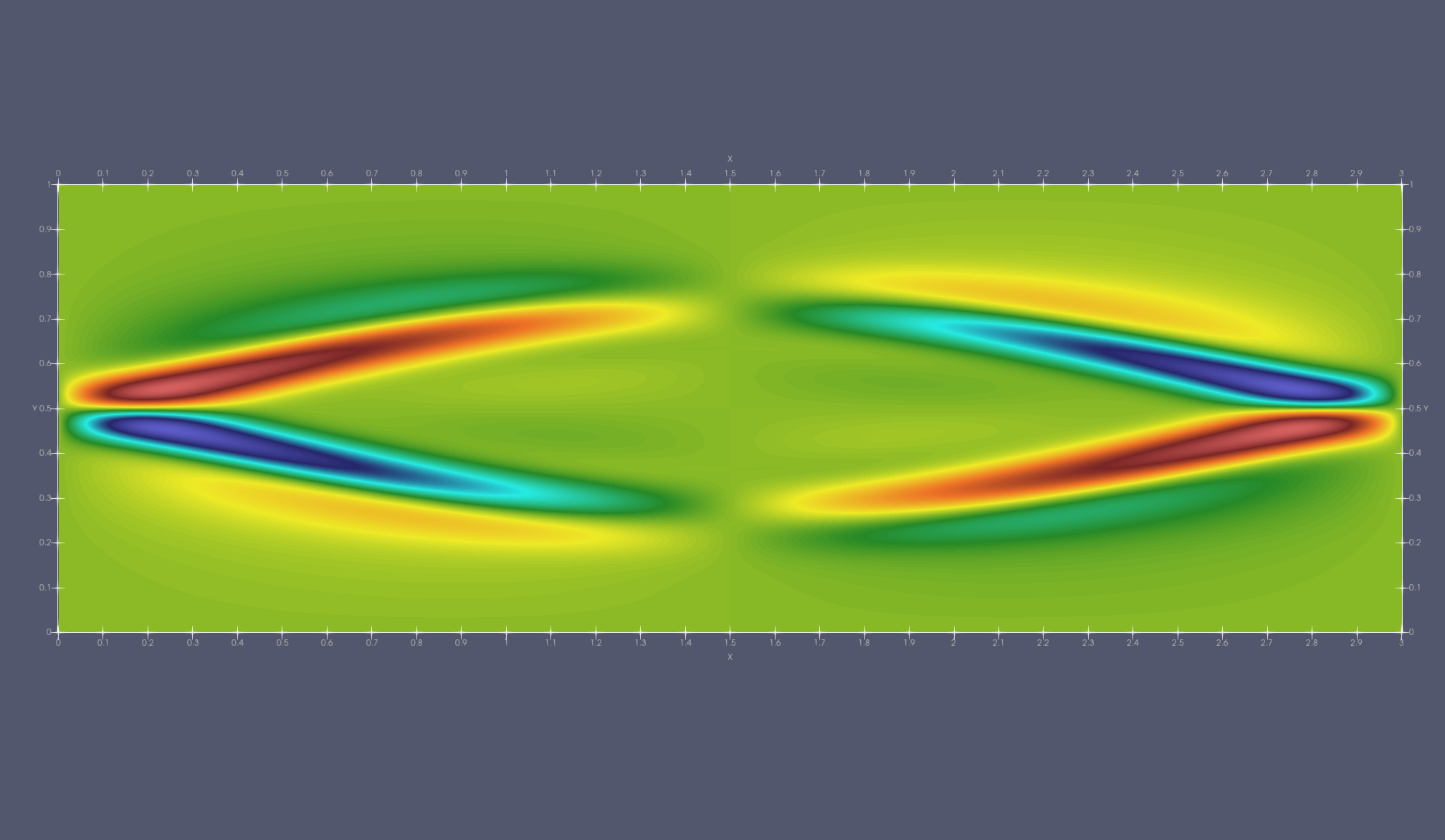}{\figWidth}};
  \draw(9.1,-.75) node[anchor=south west,xshift=0pt,yshift=+0pt] {\trimfig{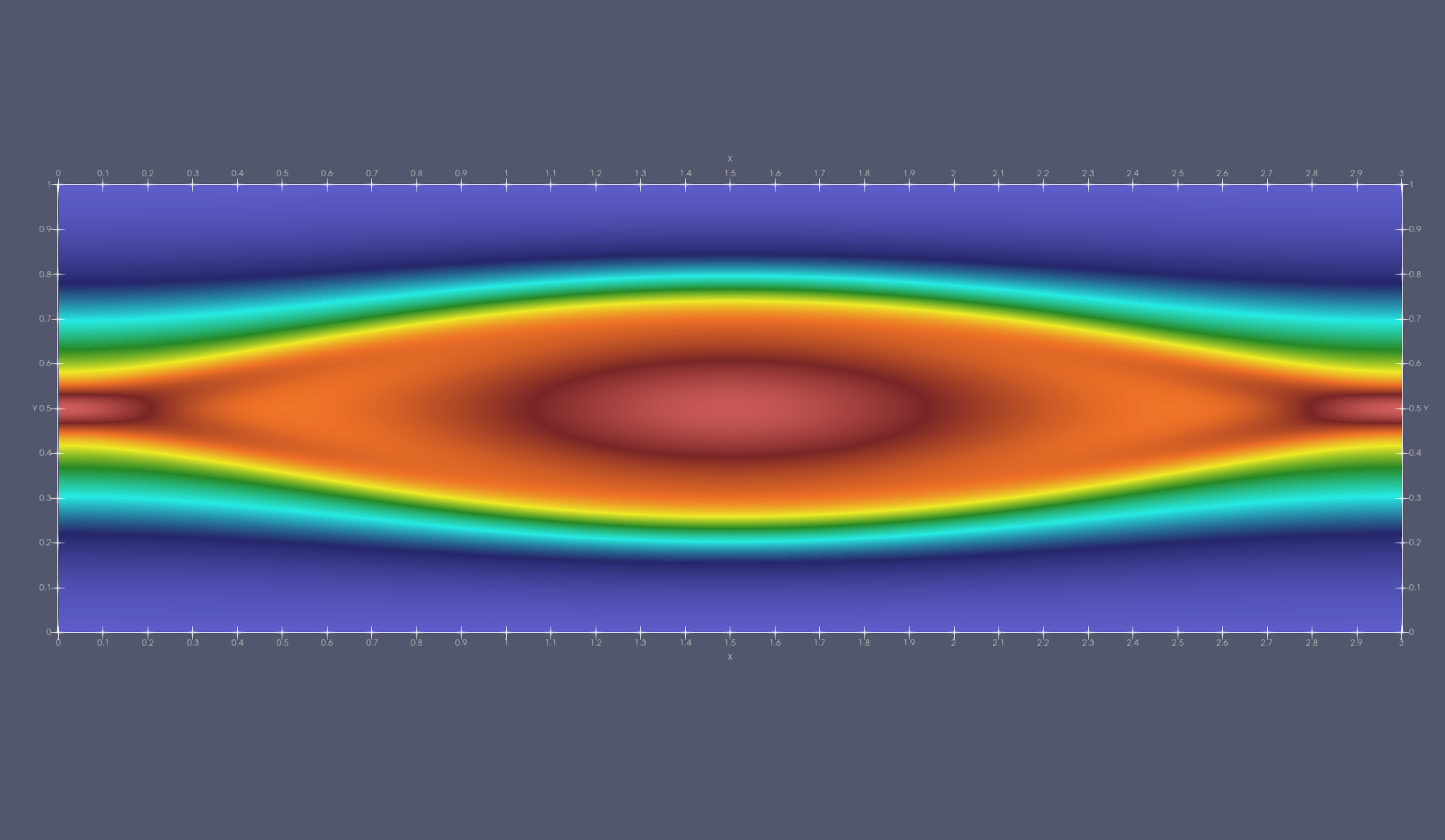}{\figWidth}};
    \draw(-.4,2.2) node[draw,fill=white,anchor=west,xshift=4pt,yshift=0pt] {\scriptsize $\omega$};
    \draw(9.2,2.2) node[draw,fill=white,anchor=west,xshift=4pt,yshift=0pt] {\scriptsize $J$};
%
\end{tikzpicture}
} 
\end{center}
  \caption{Tearing mode. The poloidal flux $\Psi$, streaming function $\Phi$, vorticity $\omega$ and parallel current $J$ at $t = 250$ are presented.
 Here $\nu=\eta=$ $10^{-3}$.
  The computational grid is $96\times96$ and the polynomial degree used is $p=2$. The time step is $\dt=5$ and DIRK2 is used as the time integrator.
  }
  \label{fig:tearModeO2}
\end{figure}
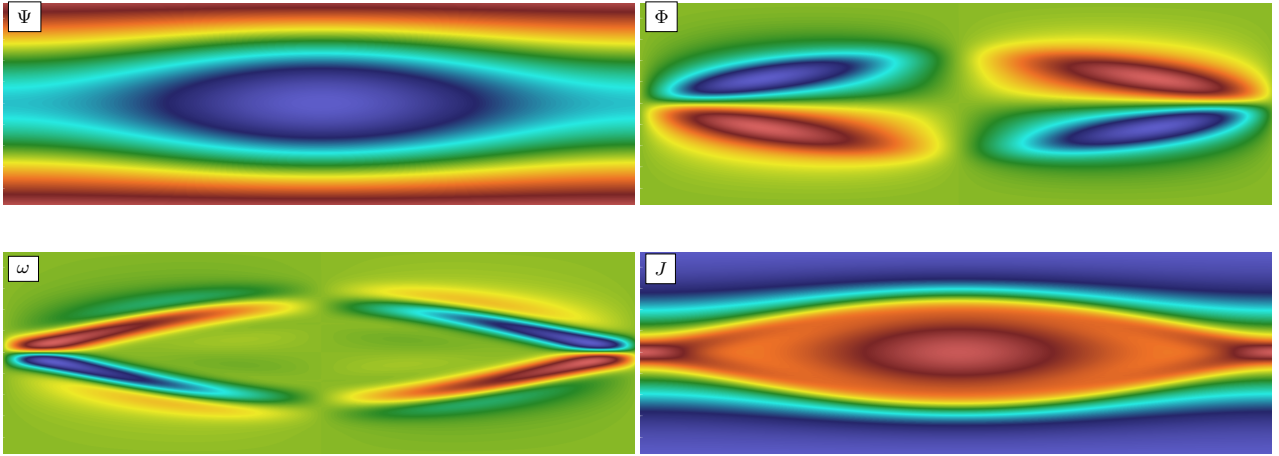
}

The problem is first solved on a uniform mesh to compare with the results given in~\cite{chacon2002}.
In Figure~\ref{fig:tearModeO2}, the solutions of the tearing mode at $t=250$ are presented.
In this test, a computational grid of $96\times96$ is used with a large time step $\dt=5$.
The test uses $p=2$ and the DIRK2 time integrator.
For such a resolution, the preconditioner performs very well with  large time steps
(the time step corresponds to roughly $160 \dt_{CFL}$ based on the definition given in~\cite{chacon2002}).
In particular, the averaged number of Newton iterations is 2 and the averaged total Krylov linear solver iterations are about 5.
Both the numerical solutions and performance of the solver are in excellent agreement with the results given in~\cite{chacon2002}.

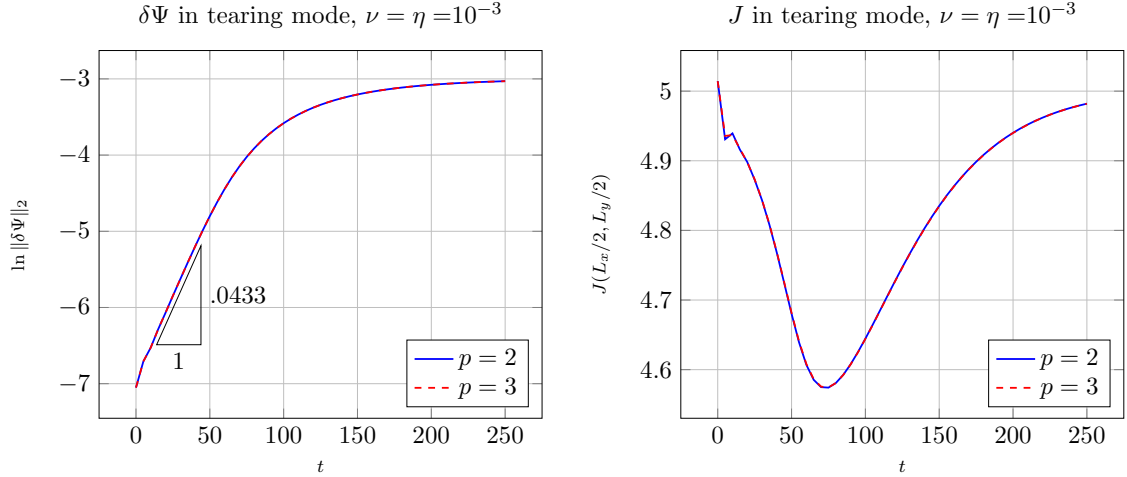
\begin{figure}[htb]
\begin{center}
\resizebox{13cm}{!}{
\begin{tikzpicture}
\useasboundingbox (0,-.3) rectangle (15,6.4);  
\begin{axis}[
title={$\delta\Psi$ in tearing mode, $\nu=\eta=$$10^{-3}$},
xlabel={$t$},
ylabel={$\ln \|\delta\Psi\|_2$},
label style={font=\scriptsize},
grid=both,
legend pos=south east
]
\logLogSlopeTriangle{0.23}{0.1}{0.2}{.0433}{black};
\addplot[color=blue,thick] table[x={t},y={dPsi}] {figures/tear/dPsi-o2.dat};
\addplot[color=red,thick, dashed] table[x={t},y={dPsi}] {figures/tear/dPsi-o3.dat};
\legend{$p=2$, $p=3$}
\end{axis}
  \begin{scope}[xshift=9cm]
\begin{axis}[
title={$J$ in tearing mode, $\nu=\eta=$$10^{-3}$},
xlabel={$t$},
ylabel={$J({L_x}/2, {L_y}/2)$},
label style={font=\scriptsize},
grid=both,
legend pos=south east
]
\addplot[color=blue,thick] table[x={t},y={J}] {figures/tear/J-o2.dat};
\addplot[color=red,thick, dashed] table[x={t},y={J}] {figures/tear/J-o3.dat};
\legend{$p=2$, $p=3$}
\end{axis}
 \end{scope}
%
\end{tikzpicture}
}
\end{center}
\caption{Time histories of the $L_2$-norm of the magnetic perturbation (global measure, left plot) and the current at the grid center point (local measure, right plot) for the tearing instability.
The corresponding growth rate of $\gamma=0.0433$ is also presented.
A computational grid of $96\times96$ is used with different polynomial degrees.
 The time step is $\dt=5$ with DIRK2 for $p=2$ and DIRK3 for $p=3$.
}
  \label{fig:tearModeCurve}
\end{figure}

As a part of our verification tests, we also perform the same scaling tests as presented in~\cite{chacon2002}. The first verification test is the time history of the magnetic perturbation
$\log \| \delta \Psi\|_2$ (global measure) and the current density at the midpoint of the computational domain (local measure).
We present the results of $p=2$ with the DIRK2 time integrator and $p=3$ with the DIRK3 time integrator in Figure~\ref{fig:tearModeCurve}.
We note that both curves match well with each other, as well as with the second-order large time-step solution given in~\cite{chacon2002}.
The kink in the current plot of Figure~\ref{fig:tearModeCurve} is due to the large time step taken ($\dt = 5$).
A simple linear regression is used to estimate the growth rate of the tearing mode as $0.0433$, which is also comparable with the reported growth rate in~\cite{chacon2002}.

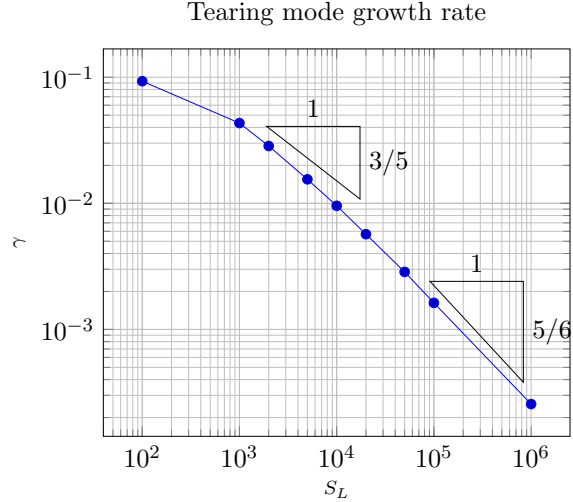
\begin{figure}[htb]
\begin{center}
\resizebox{6.cm}{!}{
\begin{tikzpicture}
\useasboundingbox (0,-.3) rectangle (6.5,6.4);  
\begin{loglogaxis}[
title={Tearing mode growth rate},
xlabel={$S_L$},
ylabel={$\gamma$},
label style={font=\scriptsize},
grid=both,
legend pos=south east
]
\logLogSlopeTriangleInv{0.55}{0.2}{0.8}{3/5}{black};
\logLogSlopeTriangleInv{0.9}{0.2}{0.4}{5/6}{black};
\addplot coordinates {
        (1e2,0.093)
    (1e3, 0.0433)
    (2e3, 0.0285)
    (5e3, 0.0155)
   (1e4,0.00956)
   (2e4, 0.005692)
   (5e4, 0.00286)
   (1e5, 0.001624)
   (1e6,0.000256)
};
\end{loglogaxis}
%
\end{tikzpicture}
}
\end{center}
\caption{Scaling of the tearing mode growth rate $\gamma$ verse the Lundquist number $S_L$ for a fixed Reynolds
number $Re$ = $10^3$. Two theoretical scalings ($S_L^{-3/5}$ and $S_L^{-5/6}$) are shown for comparison.
}
  \label{fig:tearModeGrowth}
\end{figure}

The growth rate is further studied by varying the Lundquist number. An identical scaling test is performed  in Fig.~3 of~\cite{chacon2002}.
The Reynolds number is fixed at $10^3$ while the Lundquist number varies between $10^2$ to $10^6$.
The scaling result is found to be comparable to Ref.~\cite{chacon2002}. The growth rate is close to $S_L^{-3/5}$ in the small Lundquist number cases and approaches to
 $S_L^{-5/6}$ when the Lundquist number increases.
 We find the SUPG stabilization on the $\Psi$ equation to be necessary when the Lundquist number is greater than or equal to
 $10^5$.
 Note the Lundquist number for this case is simply defined as $1/\eta$.

\begin{table}[hbt]\tableFont 
\begin{center}
\caption{Preconditioner performance study for the tearing mode. $Re=S_L= $ $10^4$.
DIRK2 is used that has two solves per time step. A fixed relative tolerance of $10^{-4}$ and a fixed relative solution tolerance of $10^{-6}$
are used as the Newton convergence criterion.
The steps are averaged over 10 time steps and $p=2$, $\dt=1$. Dofs stands for total degrees of freedoms in one scalar component.
 \label{tableTear}}
 \medskip
\begin{tabular}{|c|c|c|c|c|c|c|c|c|} \hline
  \multicolumn{8}{|c|}{Tearing mode} \\ \hline
Grid & dofs & $\dt/\dt_{CFL}$ & CPUs &  Newton/solve     &  Krylov/solve   &     Time(s)/$\dt$    & Time(s)/Krylov  \\[3pt] \hline
  $96	  \times32$ &12480 & 32 & 1 &  2.0 &  4.5     &  11.5   &    1.3    \\ \hline
  $192  \times64$& 49536 & 64 & 4  &    2.0 & 4.9  &  16.0 &    1.6 \\ \hline
  $384 \times128$ & 197376 & 128& 16  & 2.0 & 6.5 &  20.5 &  1.6 \\ \hline
  $768 \times256$ & 787968 & 256 & 64  & 2.0 &  8.5&  30.1  &  1.8 \\ \hline
  $1536 \times512$ &3.1M &  512 & 256 &  2.0	   &   15.5 &   38.7 &  1.2   \\ \hline
  $3072\times1024$ & 12.6M & 1024 & 1024  & 2.0  &   29.5 &   58.5  &   1.0 \\ \hline
  $6144\times2048$ & 50.3M & 2048 & 4096  & 3.0 &   35.5	 &     72.3  &  1.0 \\ \hline
\end{tabular}
\end{center}
\end{table}

The last important test on the uniform meshes is the parallel weak scaling test.
The Reynolds number and Lundquist number are taken as $10^4$ as representative, and the general trend of the scaling results is very similar for other choices of Reynolds/Lundquist numbers.
The weak scaling result is reported in Table~\ref{tableTear}.
Note the number of dofs on each processor is fixed to 12480.
We choose a uniform mesh such that the grid spacings along $x$- and $y$-directions are identical,
which we found gives a more consistent performance than the case of different grid spacings considered in~\cite{chacon2002}.
The reported dofs are the number of dofs of a scalar unknown.
The time step is fixed to $\dt = 1$ and thus the problem becomes more stiff as we refine the mesh.
The problem has been normalized such that the shear Alfven speed is 1.
As a result, the CFL number of the mesh $6144\times2048$ is about $2048$.
The number of Newton iterations, Krylov iterations and CPU time per time step are reported in Table~\ref{tableTear}.
The dominant computational cost of the current solver stems from inverting the preconditioner (representing 80-90\% of the total cost).
The computational cost is found to be   proportional   to  the number of Krylov iterations per solve.
The wall-clock time per Krylov solve (Time(s)/Krylov)   is found to stay more or less  constant,
which indicates a scalable parallel implementation (attributed to the domain decomposition of MFEM, algebraic multigrid of Hypre, etc.).
The more interesting and exciting result is the algorithmic scalability of the overall algorithm.
The number of Krylov iterations grows weakly with respect to the number of dofs.
Note that the Krylov iterations grow in the last few {\red resolutions}, largely due to  growth of the condition number of the system.
In fact, for the last resolution of $6144\times2048$, the solution update in the last few Newton iterations is close to ``noise''.
To avoid solving for noise, we add a fixed relative solution tolerance of $10^{-6}$, i.e., the iteration stops when the relative solution update is less than  $10^{-6}$.
This solution tolerance is only necessary in the finest mesh resolutions.
The overall performance of the current work is very comparable with~\cite{chacon2002}, in which the solver shows perfect scaling until the very last few refined meshes.
Note the convergence criteria used in the current work is tighter compared to other works such as~\cite{muralikrishnan2020multilevel},
where comparable finite-element schemes are used to solve extended MHD.
A looser tolerance can improve the scaling result significantly, particularly with fine meshes.
Overall,  the weak scaling performance is excellent.
The total dofs from a single processor to 4096 processors grow by a factor of 4096, while the
computational time only grows by a factor of 6.3.

We would like to comment that, despite the growth of iterations in the last few resolutions, this kind of uniform resolution is not necessary in practice.
AMR significantly reduces the total dofs and guarantees a smaller workload on the preconditioner.
This will be discussed in the next section.

\subsubsection{AMR results}

In the second part of this numerical example, we examine the performance of the fully implicit solver with dynamic AMR.
We will consider accuracy and performance with respect to the uniform mesh solver,
and we will compare our solver against existing ones for the same system~\cite{philip2008implicit, baty2019finmhd}.

We choose a simple tearing mode  test problem to perform all the AMR performance studies in this section, while a more interesting and challenging test will be considered in the next section.
Our setup follows the same test given in the finite-difference-based AMR solver of~\cite{philip2008implicit}.
The test case uses relatively large resistivity and viscosity  $\nu=\eta=$ $10^{-3}$.
As indicated by the uniform mesh results in Figure~\ref{fig:tearModeO2}, the solutions are rather smooth and the current sheet does not have very sharp structures.
Despite its simplicity, this is a very useful test to verify  accuracy improvements in the AMR solutions.
The error estimator is chosen as a combination of the error in $\Psi$ and $\omega$
  \begin{equation}
\label{eqn:errorTear}
e_K:= \alpha \, \eta_K(\Psi) +\eta_K(\omega),
\end{equation}
where $\alpha=0.1$. We choose the refinement ratio $\beta_r = 0.1$ and the coarsening ratio $\beta_c = 0.001$ as described in Section~\ref{sec:refine}.

Note that the indicator here uses the error estimator as the refinement criterion.
As a comparison, the AMR solver in~\cite{philip2008implicit} uses an {\it ad hoc} refinement/coarsening indicator,
while the work in~\cite{baty2019finmhd} chooses the Hessian matrix of certain solution as the indicator, which  according to Ref.~\cite{baty2019finmhd} is an indirect reflection of  some local error estimator.
In general, an indicator based on the error estimator is preferred since it needs less tuning compared to the approach in~\cite{philip2008implicit}
and it can be easily extensible to other test problems.
The AMR approach in the current work is non-conforming and it is octree-based using the quadrilateral elements as the building block.
The octree-based approach only changes the mesh very locally in the re-gridding, and the parallel communication and rebalancing is very minimal.
 As a result, the overhead of the AMR stage of the current solver needs less than 1\% of the total computational time.
This contrasts with the approach described in~\cite{philip2008implicit, baty2019finmhd}. The former uses a patch-based AMR package while the latter uses
an unstructured grid.
But, as we previously found in~\cite{peng2020}, due to the conforming constraint,
 the conforming unstructured adaptive grid is generally less effective compared to the octree-based adaptive grid.
The work of~\cite{baty2019finmhd} is based on FreeFem++~\cite{freefem}, and the adaptive mesh stage there regenerates the entire unstructured mesh every time and performs
a global projection, which can be a lot more expensive than our AMR algorithm.

Owing to the implicit scheme and physics-based preconditioning,
the time step can be kept  fixed throughout the entire simulation.
We use the polynomials of $p= 3$ in all the AMR tests and fix time step to 0.5.
Note in \cite{philip2008implicit} the time step is fixed to be grid-dependent at 140$\dt_{CFL}$, which is comparable to the time step used here.
On the other hand, \cite{baty2019finmhd} uses a much smaller time step since it does not focus on preconditioning  and   needs to respect the CFL constraint.
In the current test, the solver starts with a uniform mesh of $96 \times 32$ and uses three levels of refinement.
Every 10 time steps, the solver checks the error in the elements and mark those for the refinement and coarsening.
For better performance, the refinement and coarsening stages ensure the added and removed cells do not lead to a jump more than one level across neighboring cells.

{
\newcommand{\figWidth}{9.5cm}
\newcommand{\trimfig}[2]{\trimFig{#1}{#2}{.22}{.2}{.5}{.5}}
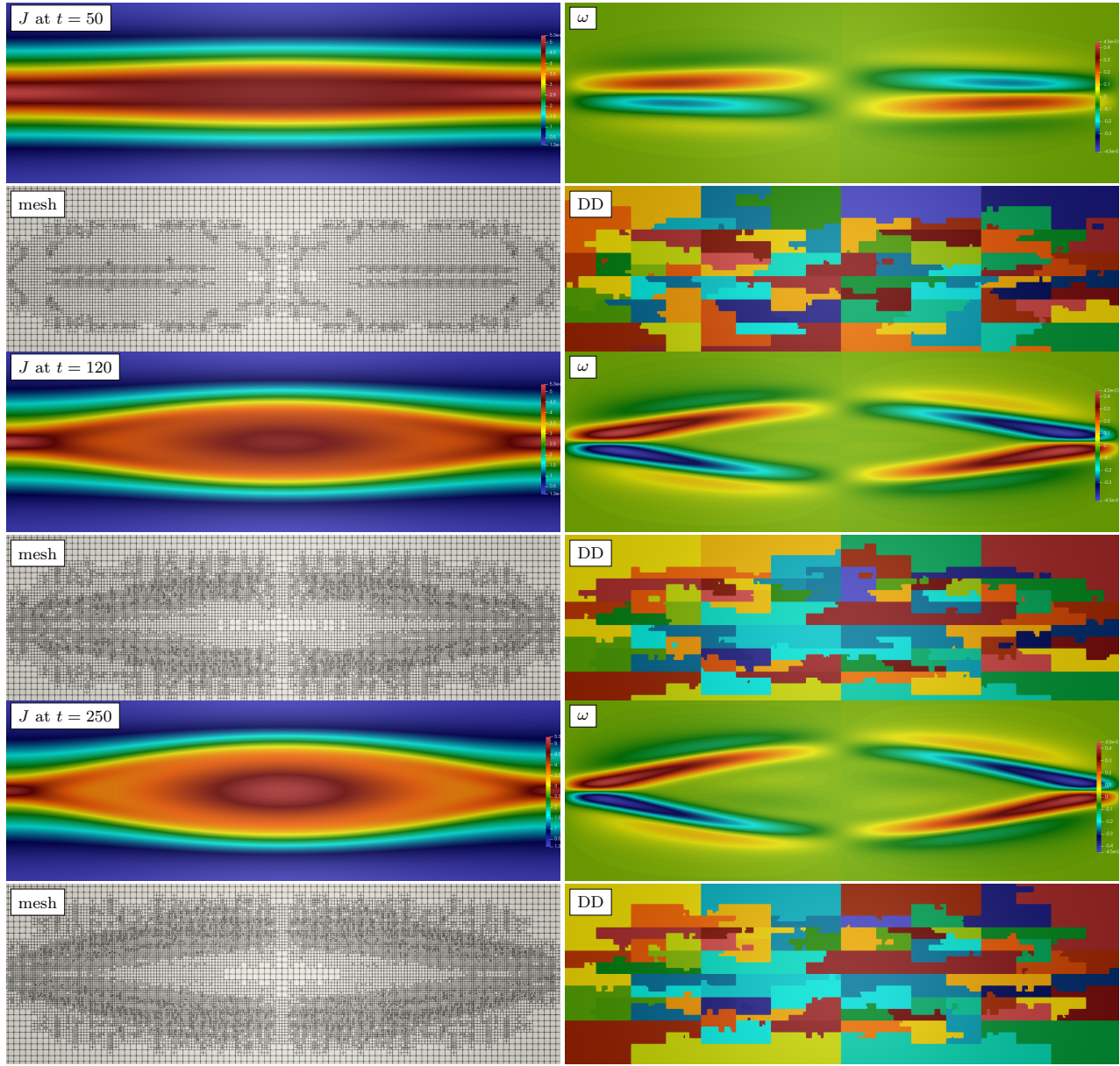
\begin{figure}[htb]
\begin{center}
\resizebox{16cm}{!}{
\begin{tikzpicture}[scale=1]
  \useasboundingbox (0.0,0) rectangle (18,18);  
    \begin{scope}[yshift=12.4cm]
  \draw(-.5,2.) node[anchor=south west,xshift=0pt,yshift=+0pt] {\trimfig{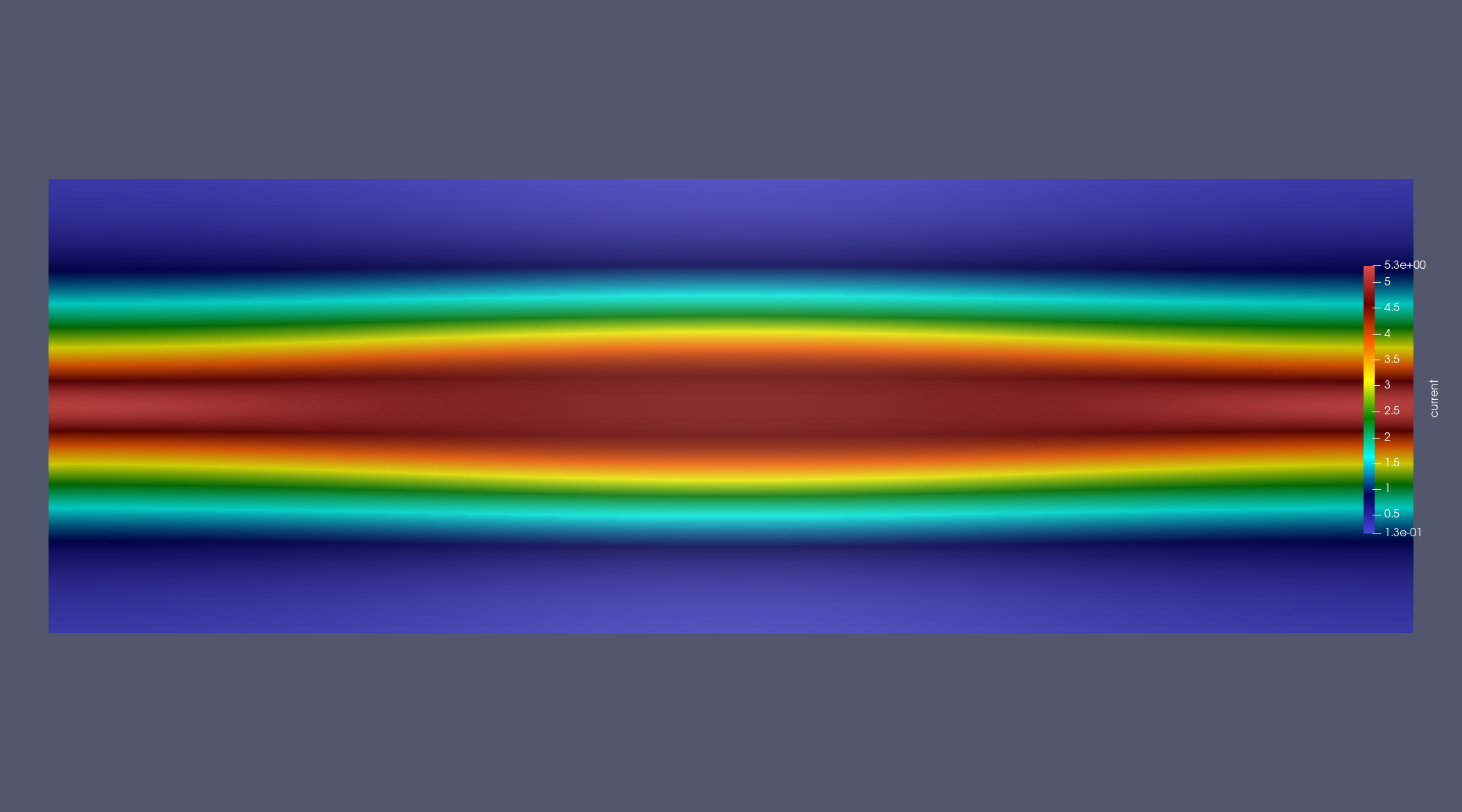}{\figWidth}};
  \draw(9.1,2.) node[anchor=south west,xshift=0pt,yshift=+0pt] {\trimfig{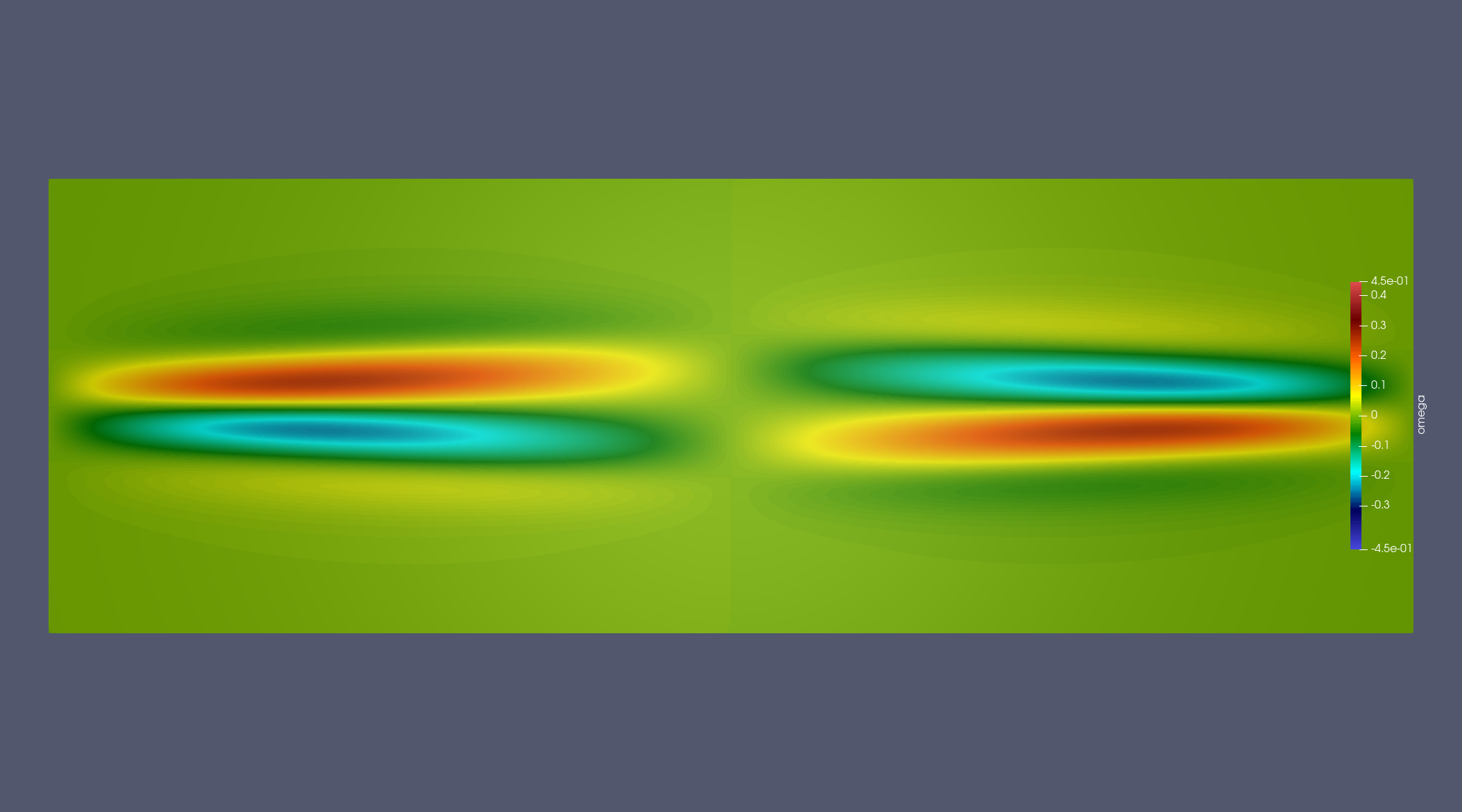}{\figWidth}};
  \draw(-.4,5.) node[draw,fill=white,anchor=west,xshift=4pt,yshift=0pt] {\scriptsize $J$ at $t=50$};
  \draw(9.2,5.) node[draw,fill=white,anchor=west,xshift=4pt,yshift=0pt] {\scriptsize $\omega$};
  \draw(-.5,-1.15) node[anchor=south west,xshift=0pt,yshift=+0pt] {\trimfig{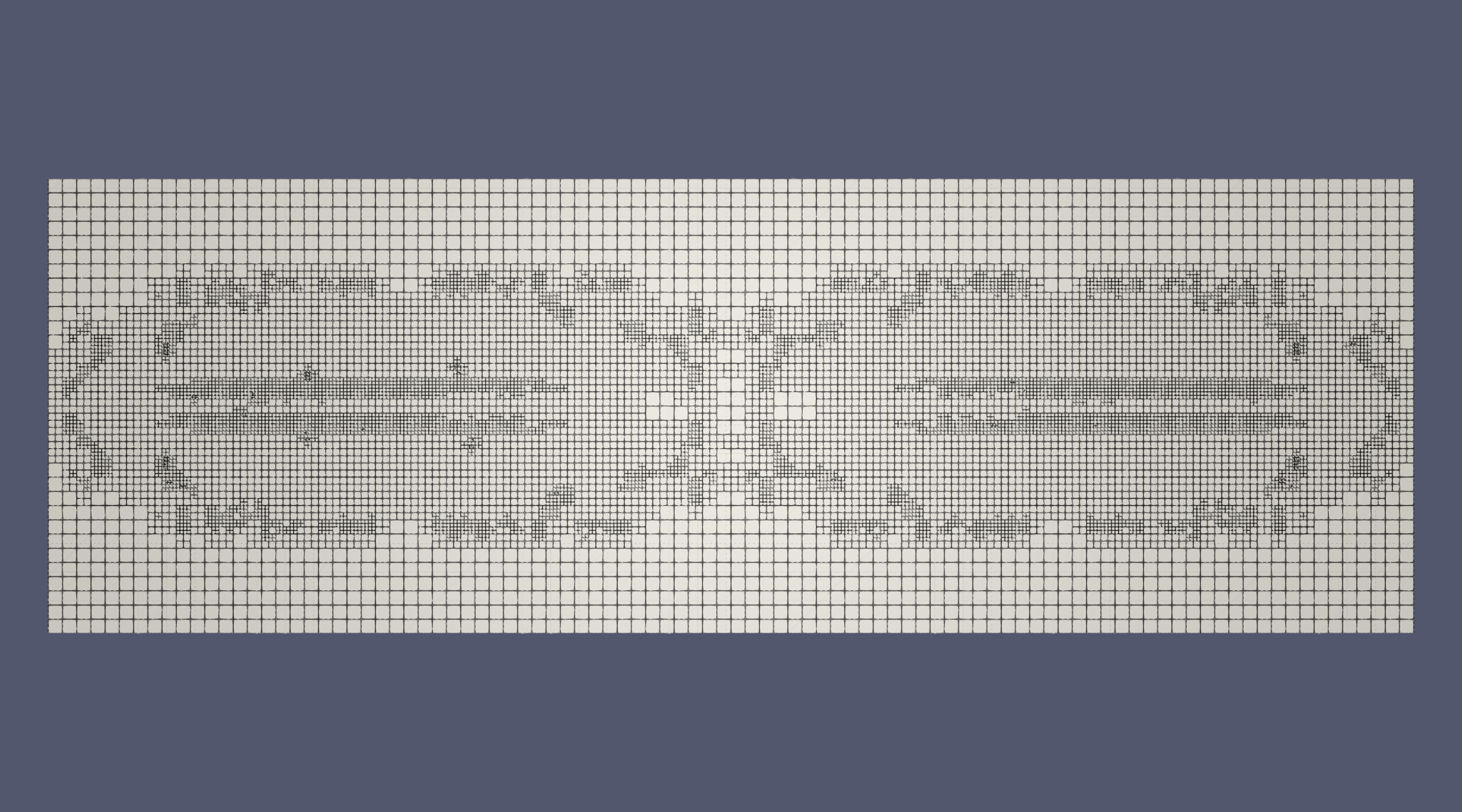}{\figWidth}};
  \draw(9.1,-1.15) node[anchor=south west,xshift=0pt,yshift=+0pt] {\trimfig{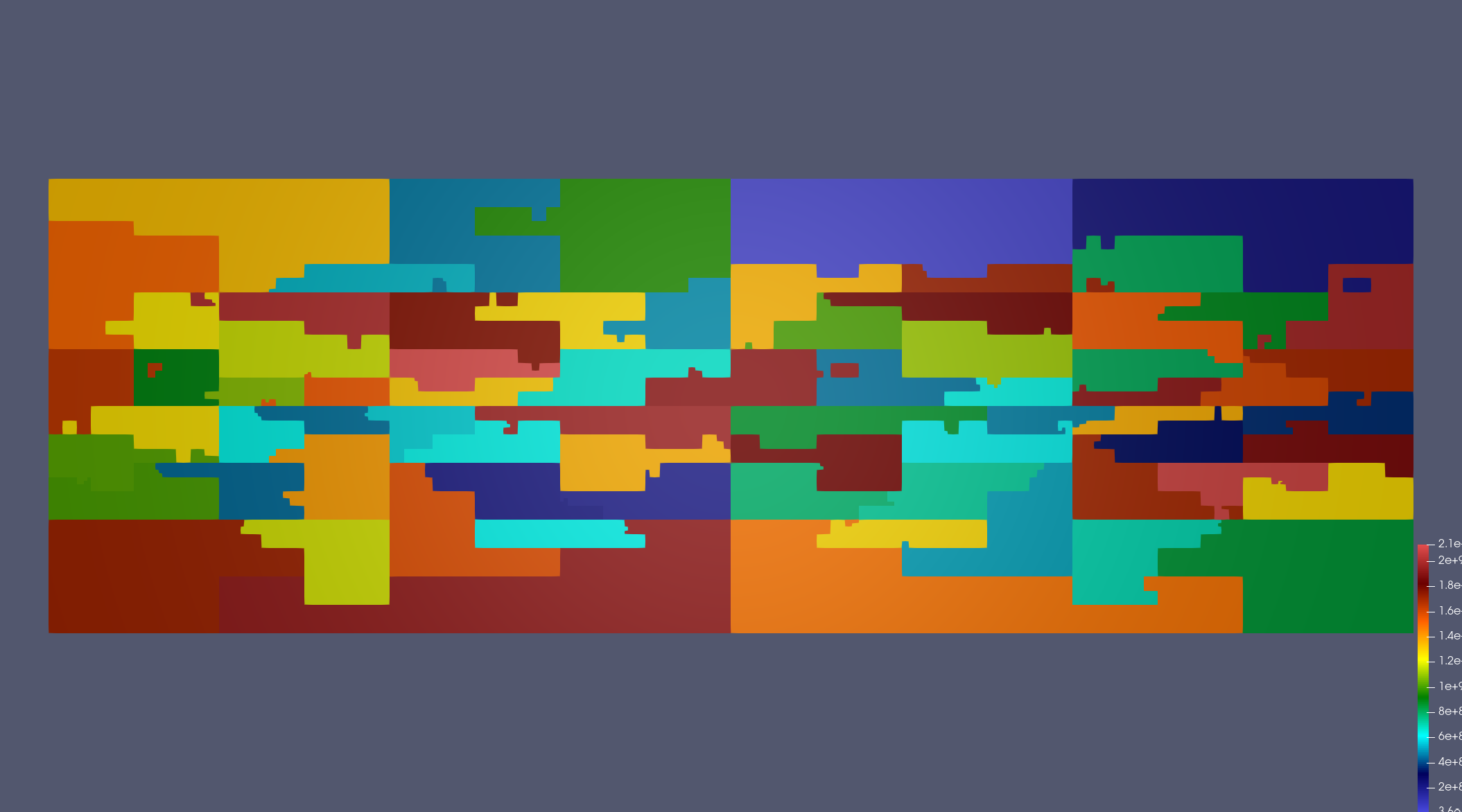}{\figWidth}};
    \draw(-.4,1.8) node[draw,fill=white,anchor=west,xshift=4pt,yshift=0pt] {\scriptsize mesh};
    \draw(9.2,1.8) node[draw,fill=white,anchor=west,xshift=4pt,yshift=0pt] {\scriptsize DD};
     \end{scope}
     \begin{scope}[yshift=6.cm]
       \draw(-.5,2.4) node[anchor=south west,xshift=0pt,yshift=+0pt] {\trimfig{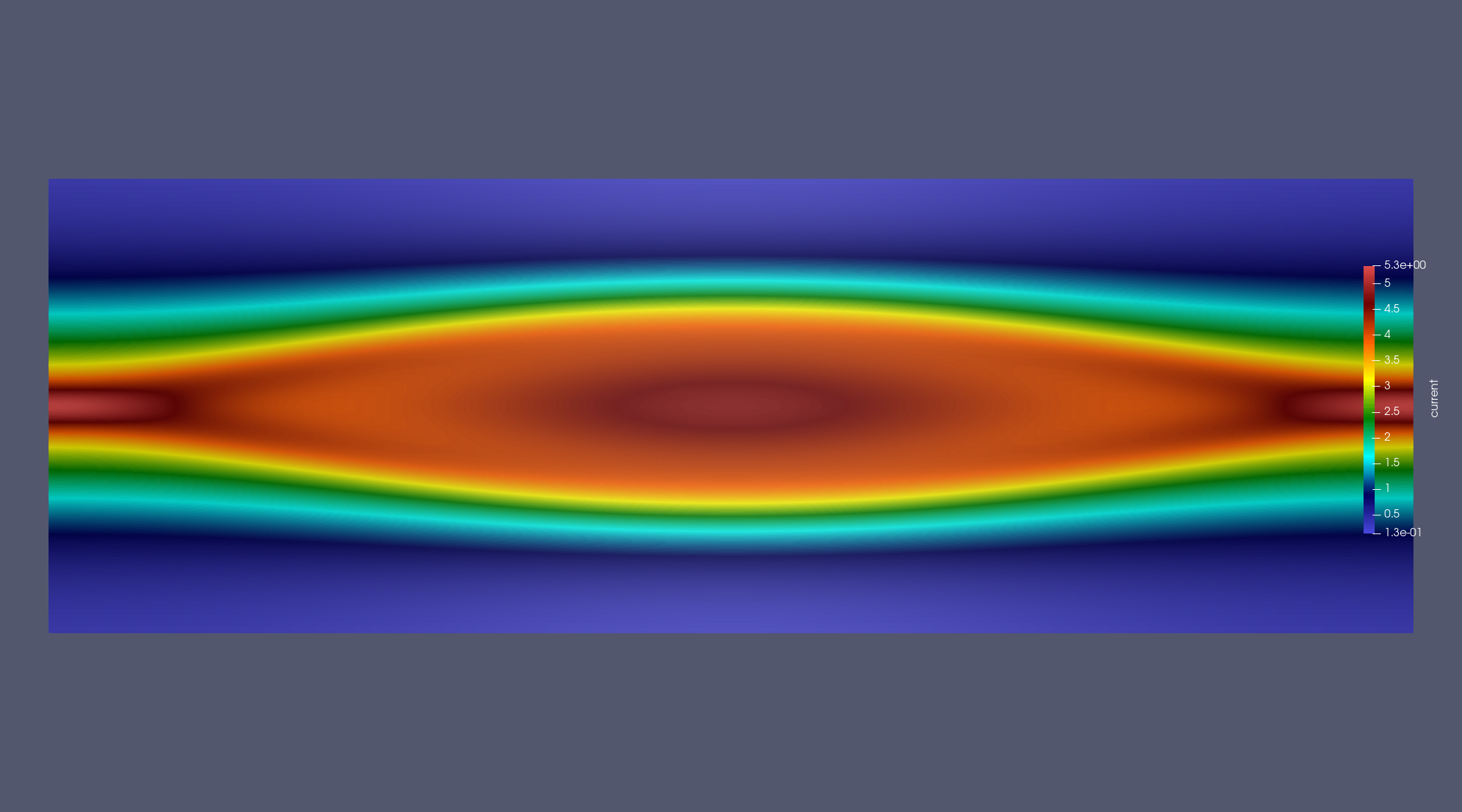}{\figWidth}};
  \draw(9.1,2.4) node[anchor=south west,xshift=0pt,yshift=+0pt] {\trimfig{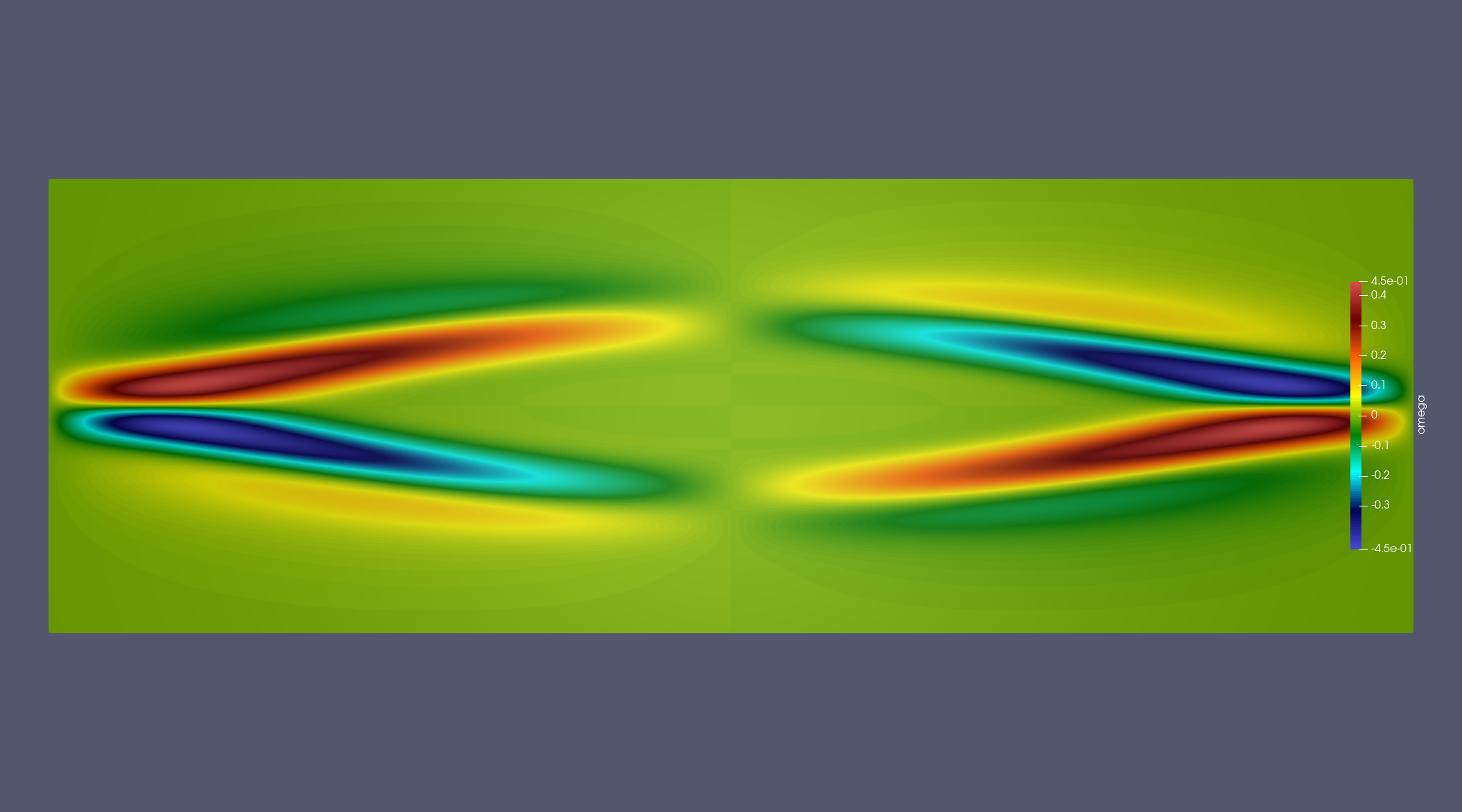}{\figWidth}};
  \draw(-.4,5.4) node[draw,fill=white,anchor=west,xshift=4pt,yshift=0pt] {\scriptsize $J$ at $t=120$};
  \draw(9.2,5.4) node[draw,fill=white,anchor=west,xshift=4pt,yshift=0pt] {\scriptsize $\omega$};
  \draw(-.5,-.75) node[anchor=south west,xshift=0pt,yshift=+0pt] {\trimfig{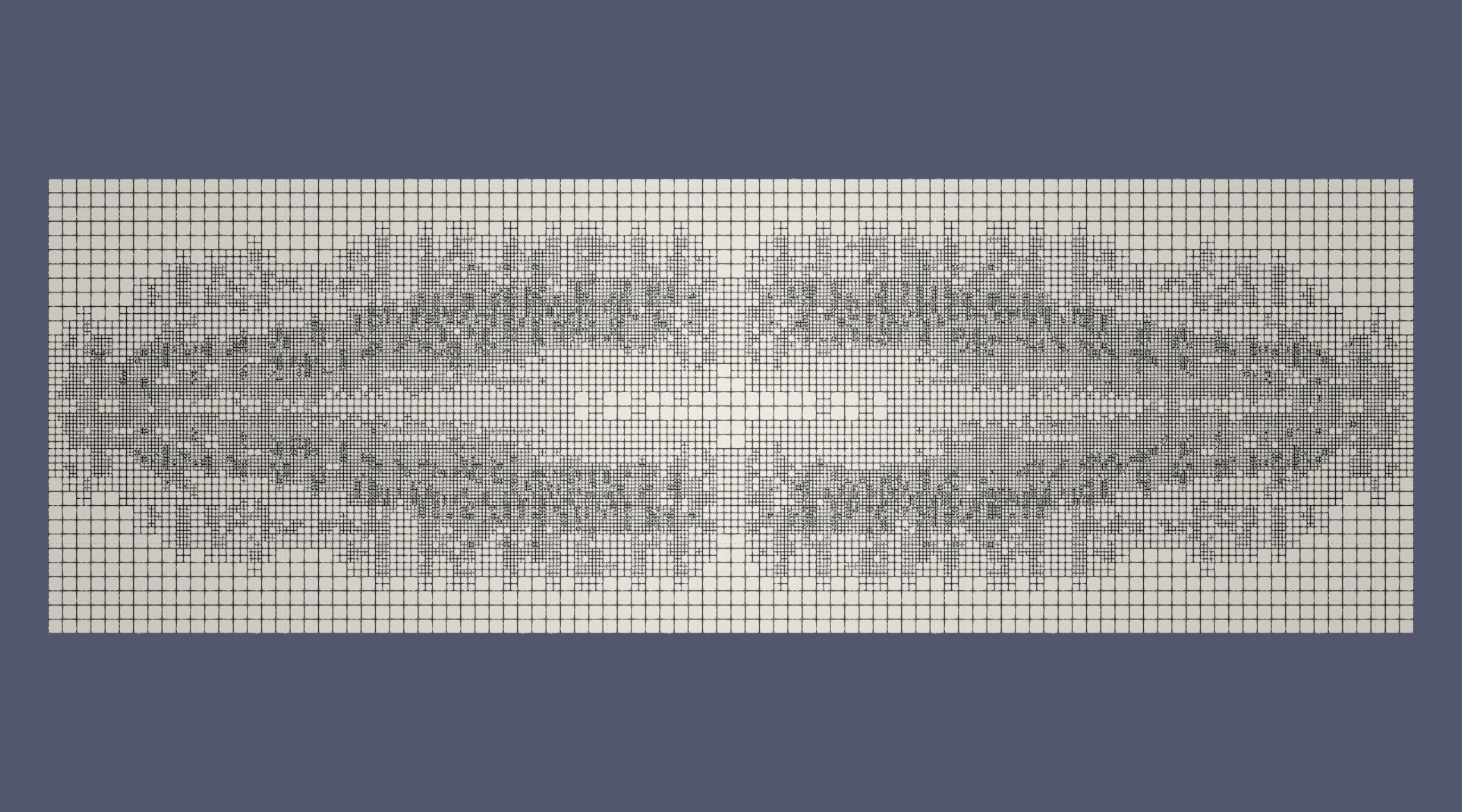}{\figWidth}};
  \draw(9.1,-.75) node[anchor=south west,xshift=0pt,yshift=+0pt] {\trimfig{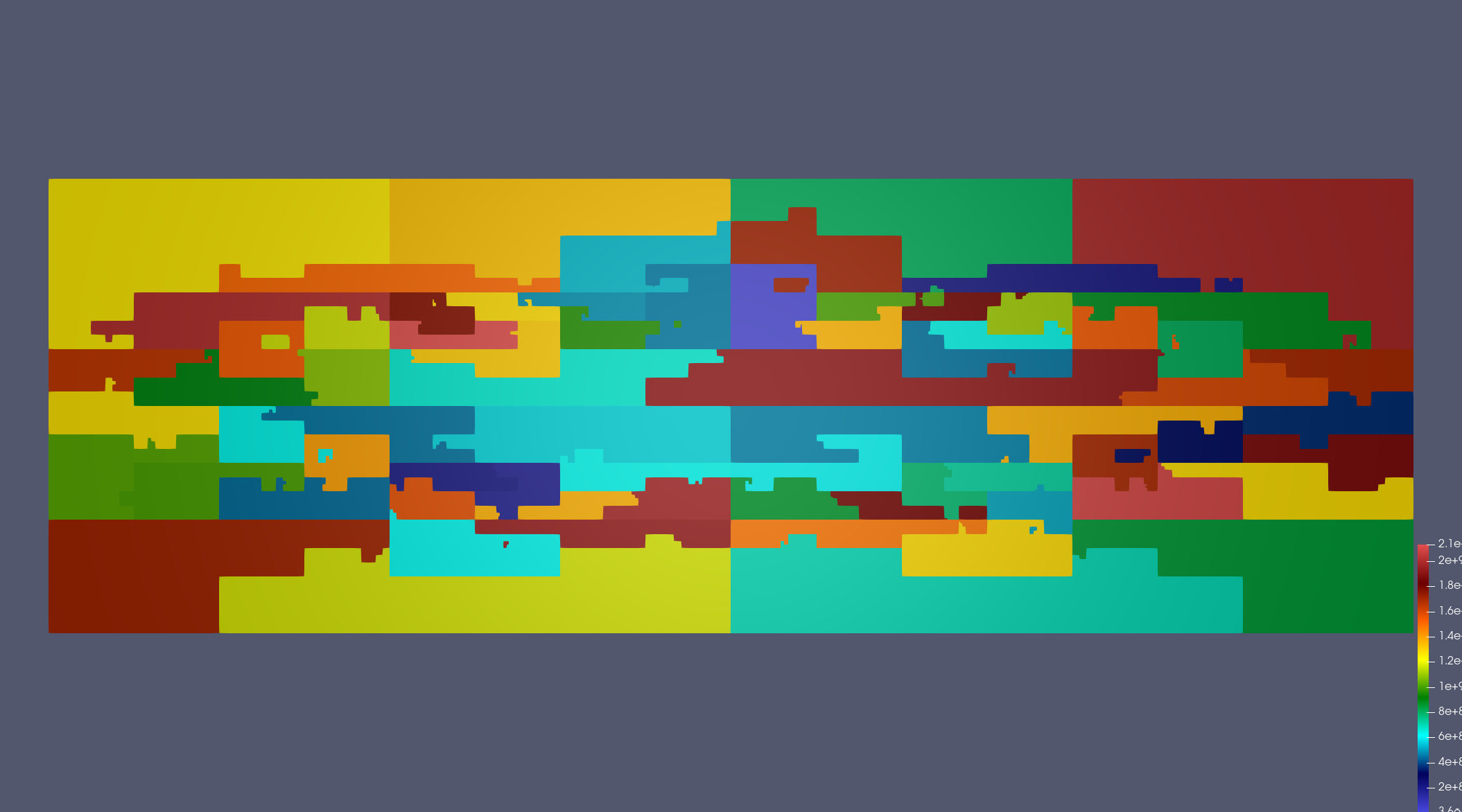}{\figWidth}};
    \draw(-.4,2.2) node[draw,fill=white,anchor=west,xshift=4pt,yshift=0pt] {\scriptsize mesh};
    \draw(9.2,2.2) node[draw,fill=white,anchor=west,xshift=4pt,yshift=0pt] {\scriptsize DD};
      \end{scope}
     \begin{scope}[yshift=0cm]
       \draw(-.5,2.4) node[anchor=south west,xshift=0pt,yshift=+0pt] {\trimfig{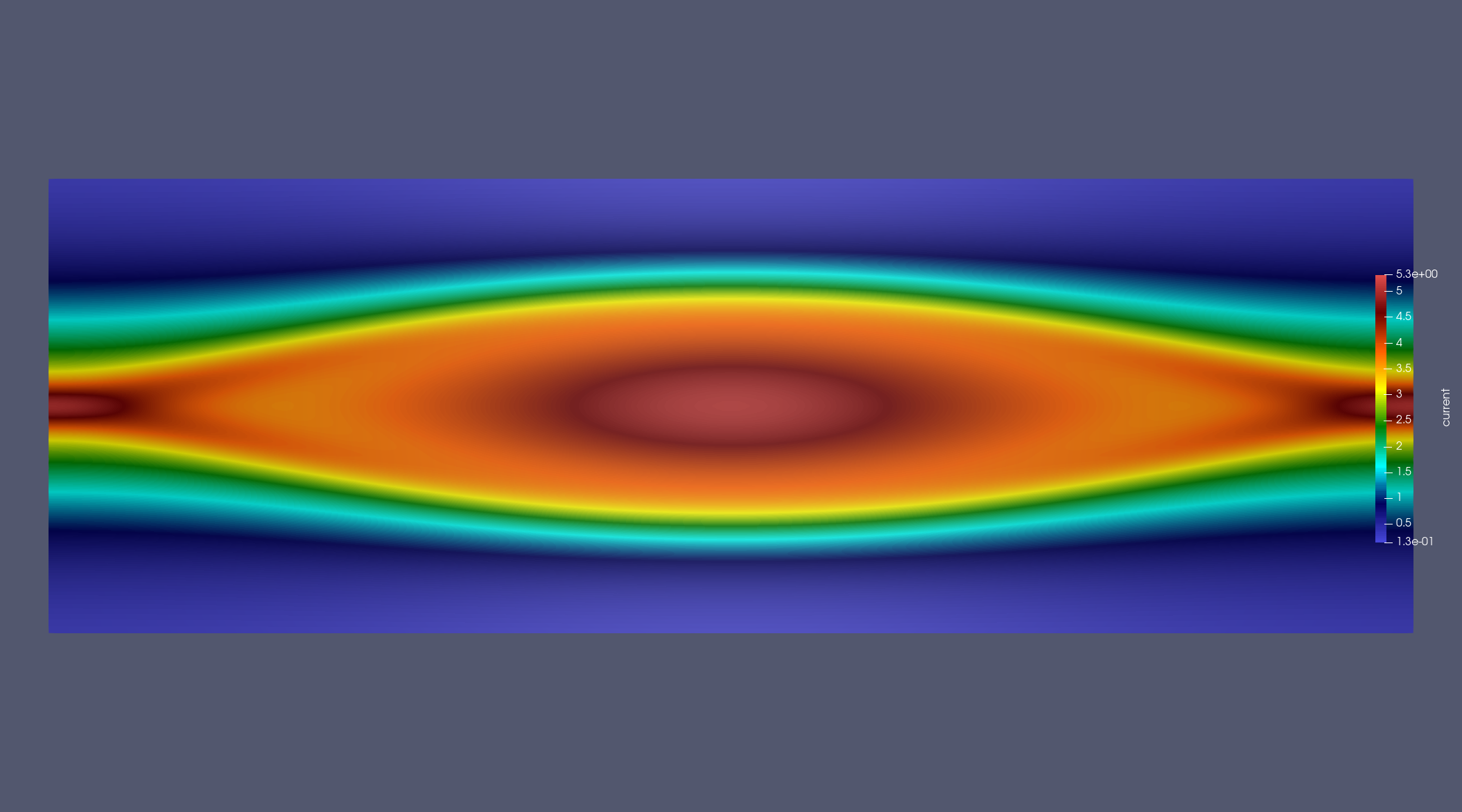}{\figWidth}};
  \draw(9.1,2.4) node[anchor=south west,xshift=0pt,yshift=+0pt] {\trimfig{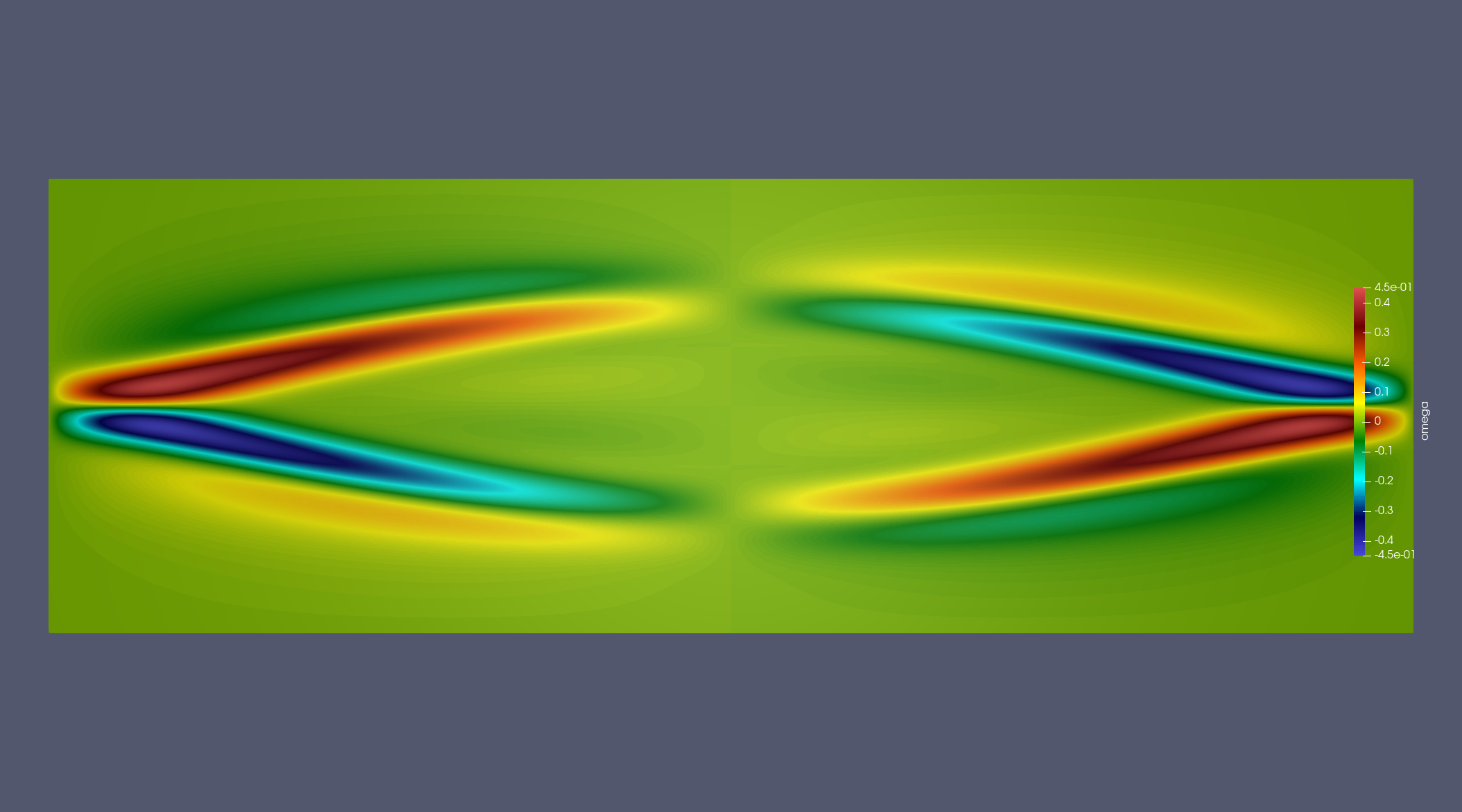}{\figWidth}};
  \draw(-.4,5.4) node[draw,fill=white,anchor=west,xshift=4pt,yshift=0pt] {\scriptsize $J$ at $t=250$};
  \draw(9.2,5.4) node[draw,fill=white,anchor=west,xshift=4pt,yshift=0pt] {\scriptsize $\omega$};
  \draw(-.5,-.75) node[anchor=south west,xshift=0pt,yshift=+0pt] {\trimfig{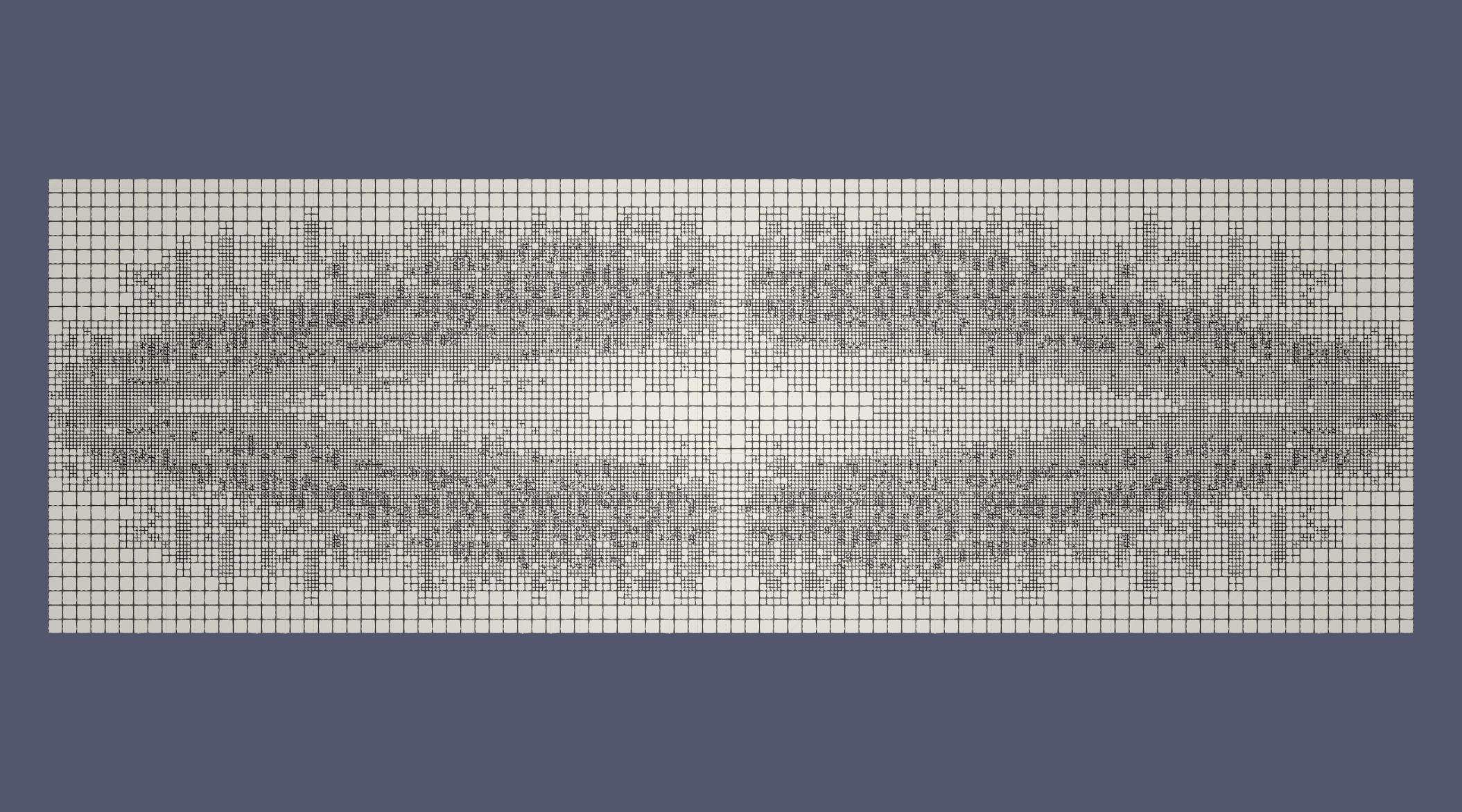}{\figWidth}};
  \draw(9.1,-.75) node[anchor=south west,xshift=0pt,yshift=+0pt] {\trimfig{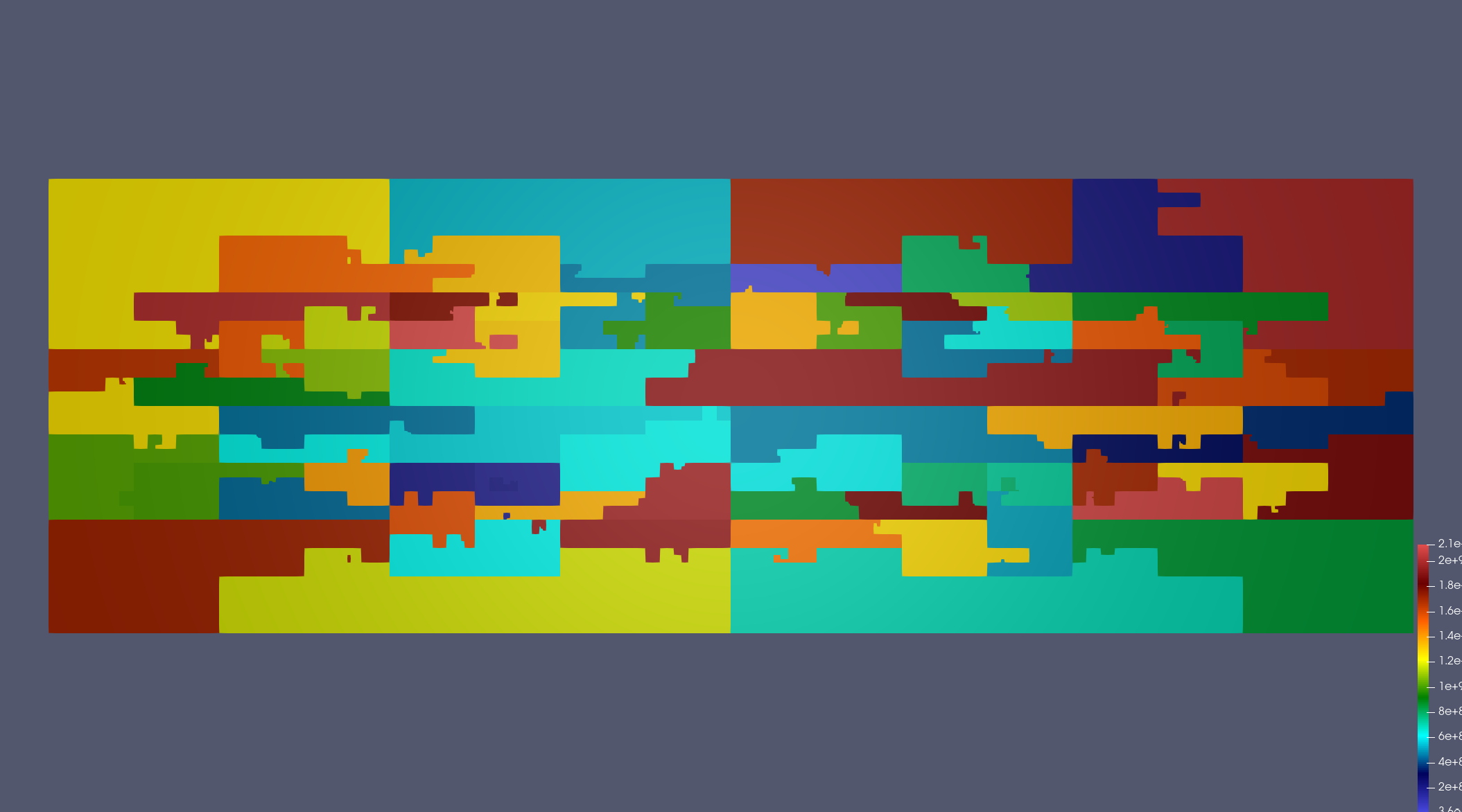}{\figWidth}};
    \draw(-.4,2.2) node[draw,fill=white,anchor=west,xshift=4pt,yshift=0pt] {\scriptsize mesh};
    \draw(9.2,2.2) node[draw,fill=white,anchor=west,xshift=4pt,yshift=0pt] {\scriptsize DD};
      \end{scope}
      %
\end{tikzpicture}
} 
\end{center}
  \caption{Tearing mode. Parallel current $J$, Vorticity $\omega$, adaptive meshes and corresponding domain decompositions at various times are presented.
 Here $\nu=\eta=$ $10^{-3}$.
  The base grid is $96\times32$ with three levels of refinements and the polynomial degree used is $p=3$. The time step is $\dt=0.5$ and DIRK2 is used as the time integrator. 72 CPUs are used in this test.
  }
  \label{fig:tearModeAMR}
\end{figure}
}

The solutions, adaptive meshes and domain partitioning are presented at several times in Figure~\ref{fig:tearModeAMR}.
Note the final result at $t=250$ matches well with the uniform solution result in Figure~\ref{fig:tearModeO2}.
The adaptive meshes are found to follow the solution structures very well.
For instance, at $t = 50$, the mesh shows two levels of refinement. The mesh at level 1 largely follows the initial perturbation $\delta \Psi$ and the mesh at level 2 has some local structures.
It is interesting to observe that the structures in $\omega$ are very well captured in the level 2 mesh at $t = 50$.
This indicates the proposed error estimator is effective in capturing the features of the solutions.
As time evolves, the adaptive mesh changes slowly, and, eventually, all three levels of refinement are fully utilized at $t = 120$ and $t = 250$.
The mesh still follows the structures in $\Psi$ and $\omega$ well. Even with continuous re-gridding and load balancing,
the overall meshes still keep a symmetric structure and are largely comparable with the adaptive mesh in~\cite{philip2008implicit}.
72 CPUs are used to perform the AMR test. The domain partitioning is presented along with the solution to show the evolution of partitioning.
To better visualize the partitioning, a random number is assigned to each partition for easily distinguishing different sub-domains. It is observed that the partitioning forms connected sub-domains.

Next, we examine the algorithmic  performance of the AMR solver.
The averaged number of dofs for a scalar unknown in the AMR simulation is about 213329,
while the total number of dofs for a scalar unknown on the uniform mesh of the same finer resolution needs 1771776.
It follows that the AMR solver only needs 12\% of the total number of dofs on the uniform fine mesh.
This is comparable with the performance of the same AMR test in~\cite{philip2008implicit}, in which 14\% of the total number of dofs on the fine mesh is reported.
Note since a high-order finite-element method is used, the dofs in our implementation is more than 10 times more than the dofs in~\cite{philip2008implicit} for
a similar mesh resolution. Thus, the computational time is considerably longer than the time reported in~\cite{philip2008implicit}.
The uniform mesh of $768 \times 256$ needs  24 Krylov iterations on average per solve, and takes 106250s on 72 CPUs.
As a comparison, the AMR solver needs 29 Krylov iterations   and  12451s on 72 CPUs, or 11.7\% of the computational time of the uniform grid solver.

{
\newcommand{\figWidth}{9.5cm}
\newcommand{\trimfig}[2]{\trimFig{#1}{#2}{.22}{.2}{.5}{.5}}
\begin{figure}[htb]
\begin{center}
\resizebox{16cm}{!}{
\begin{tikzpicture}[scale=1]
  \useasboundingbox (0.0,0) rectangle (18,12);  
     \begin{scope}[yshift=0.cm]
       \draw(-.5,2.4) node[anchor=south west,xshift=0pt,yshift=+0pt] {\trimfig{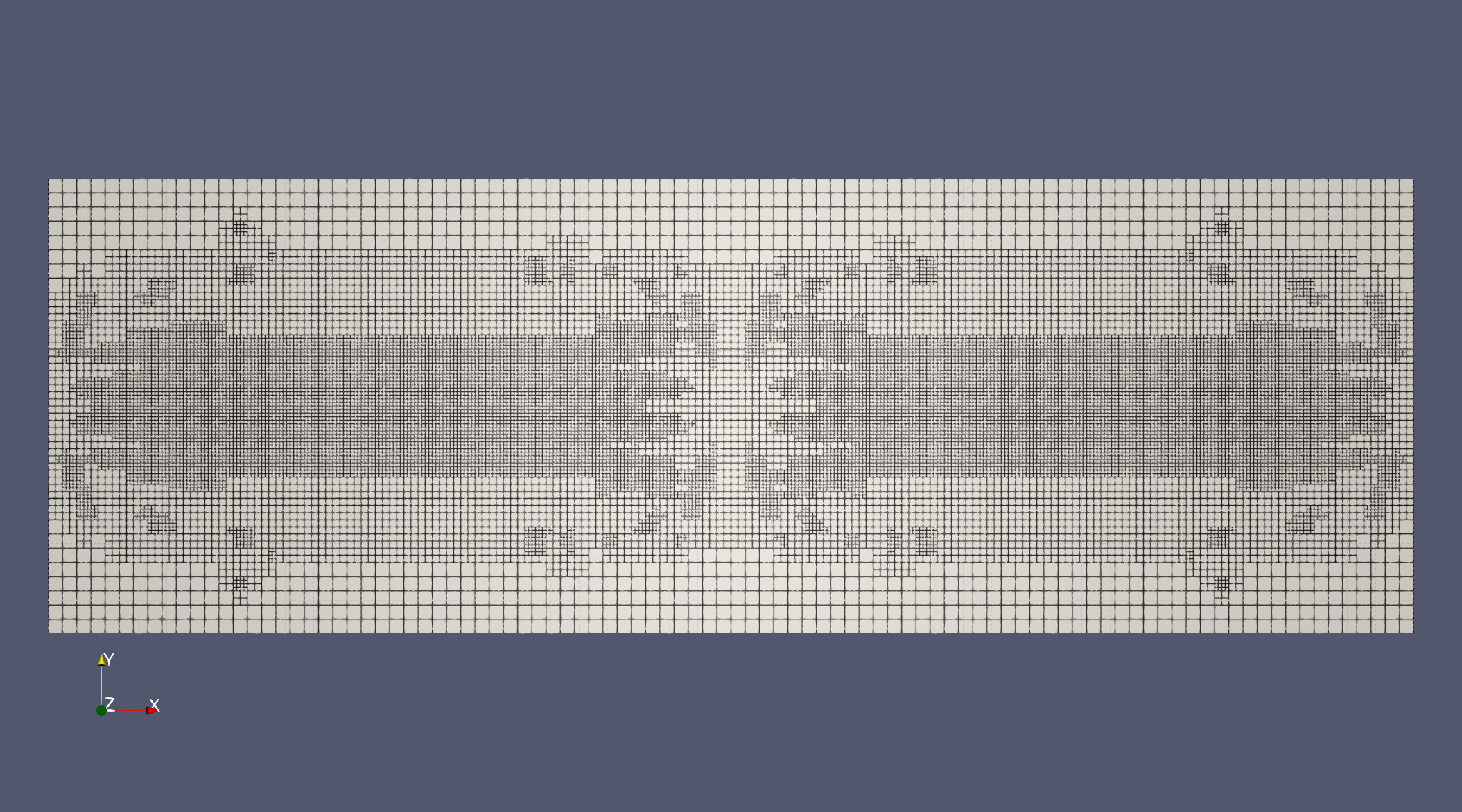}{\figWidth}};
  \draw(9.1,2.4) node[anchor=south west,xshift=0pt,yshift=+0pt] {\trimfig{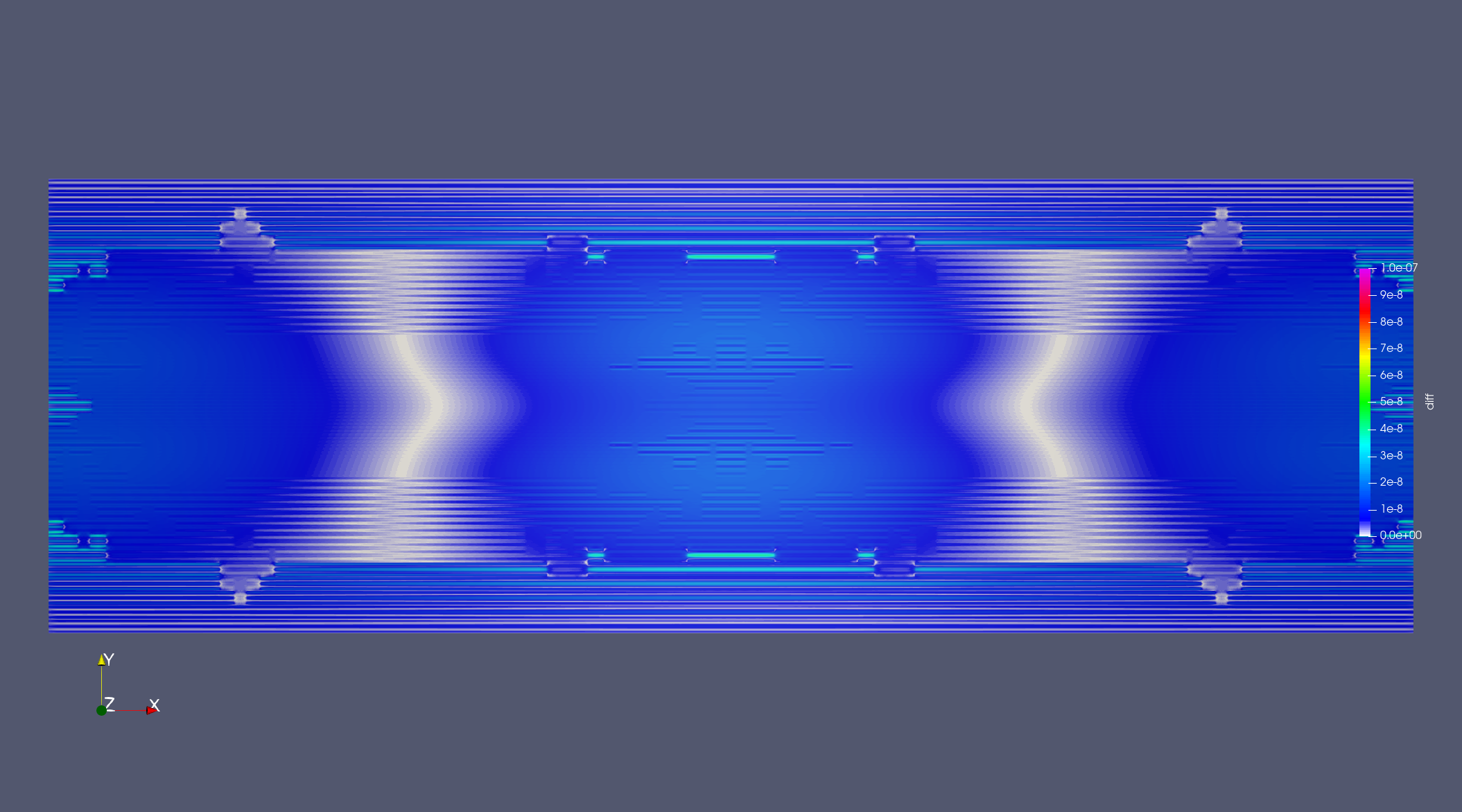}{\figWidth}};
  \draw(-.4,5.4) node[draw,fill=white,anchor=west,xshift=4pt,yshift=0pt] {\scriptsize 2-level mesh};
  \draw(9.2,5.4) node[draw,fill=white,anchor=west,xshift=4pt,yshift=0pt] {\scriptsize Error};
  \draw(-.5,-.75) node[anchor=south west,xshift=0pt,yshift=+0pt] {\trimfig{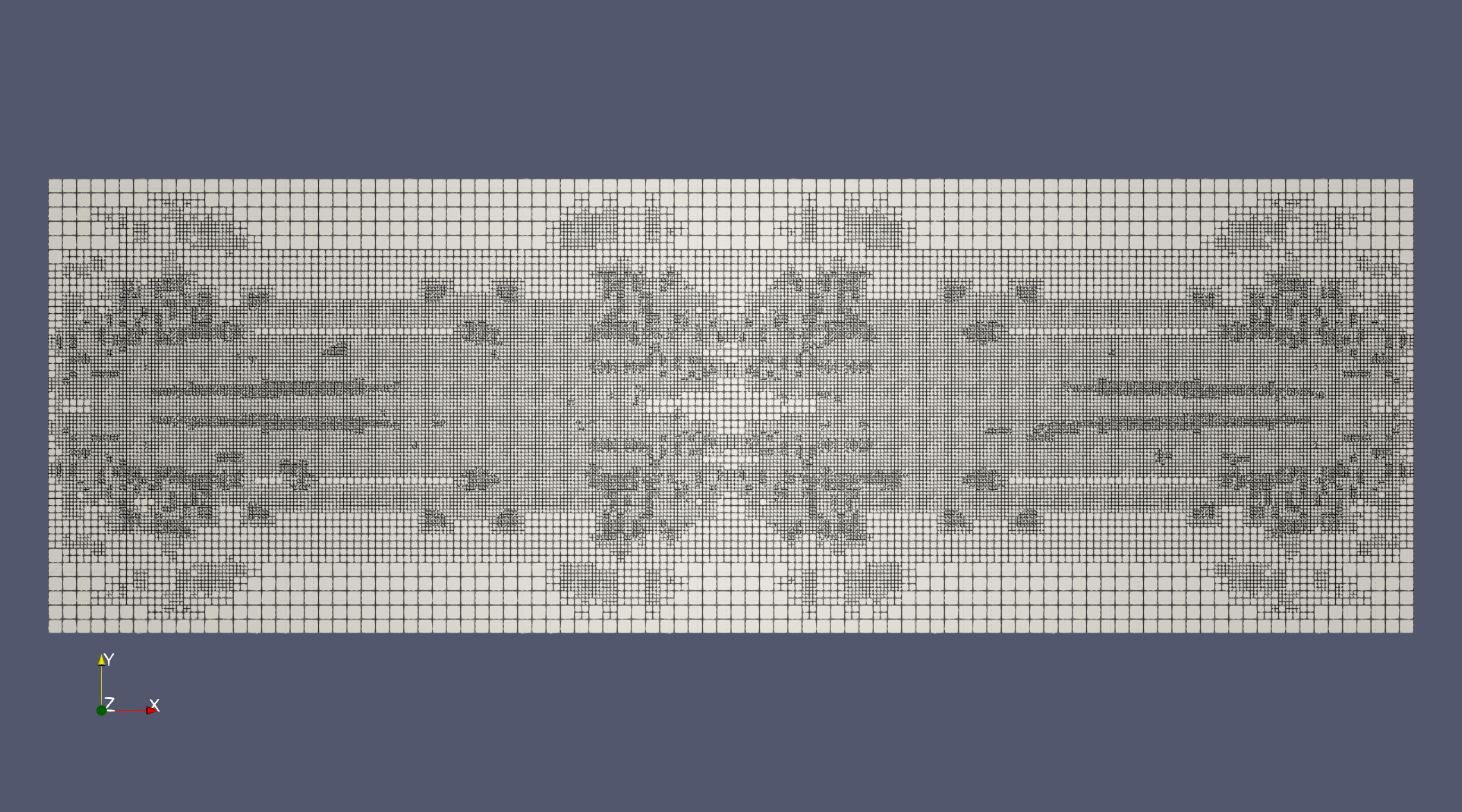}{\figWidth}};
  \draw(9.1,-.75) node[anchor=south west,xshift=0pt,yshift=+0pt] {\trimfig{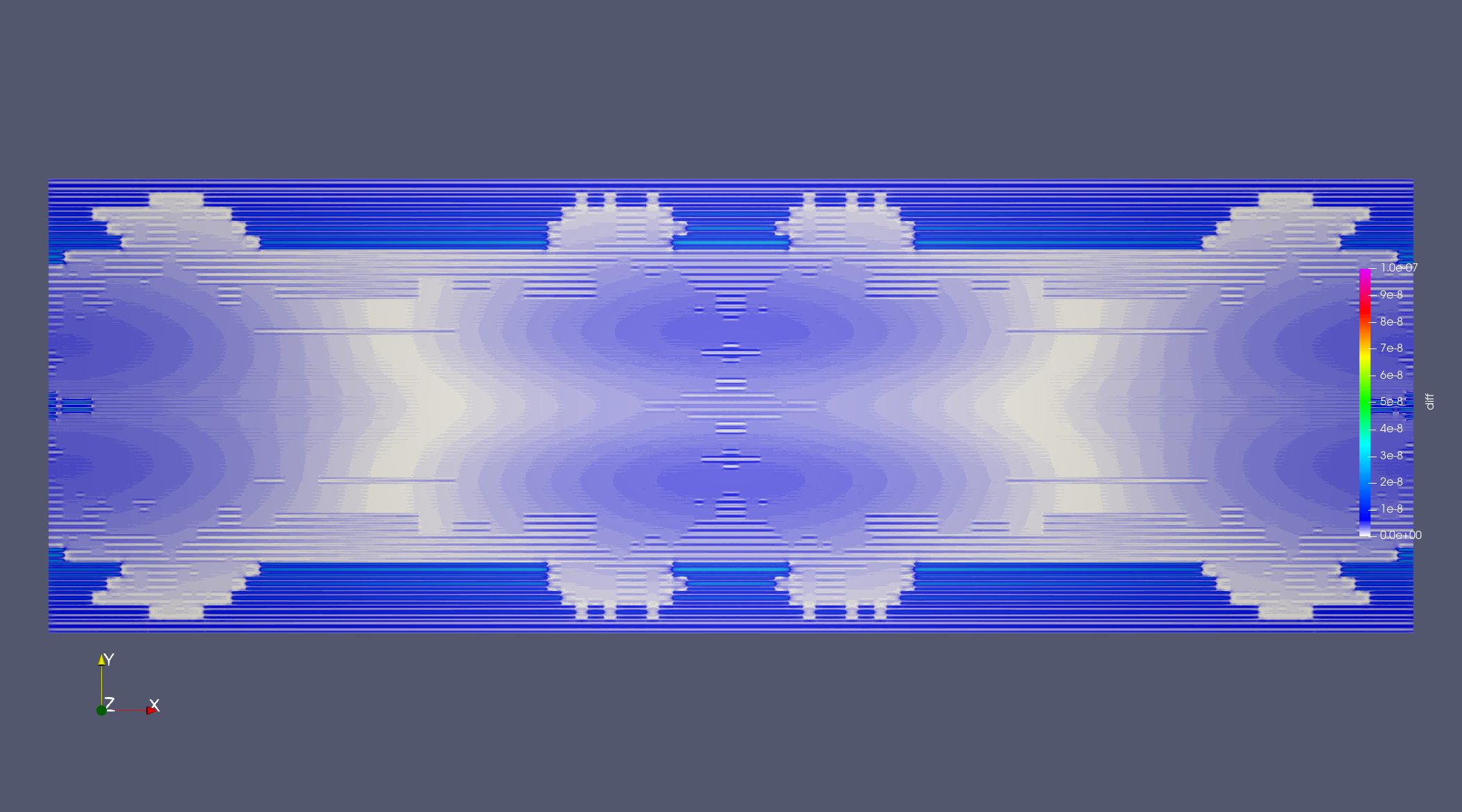}{\figWidth}};
    \draw(-.4,2.2) node[draw,fill=white,anchor=west,xshift=4pt,yshift=0pt] {\scriptsize 3-level mesh};
    \draw(9.2,2.2) node[draw,fill=white,anchor=west,xshift=4pt,yshift=0pt] {\scriptsize Error};
      \end{scope}
     \begin{scope}[yshift=6.4cm]
       \draw(-.5,2.4) node[anchor=south west,xshift=0pt,yshift=+0pt] {\trimfig{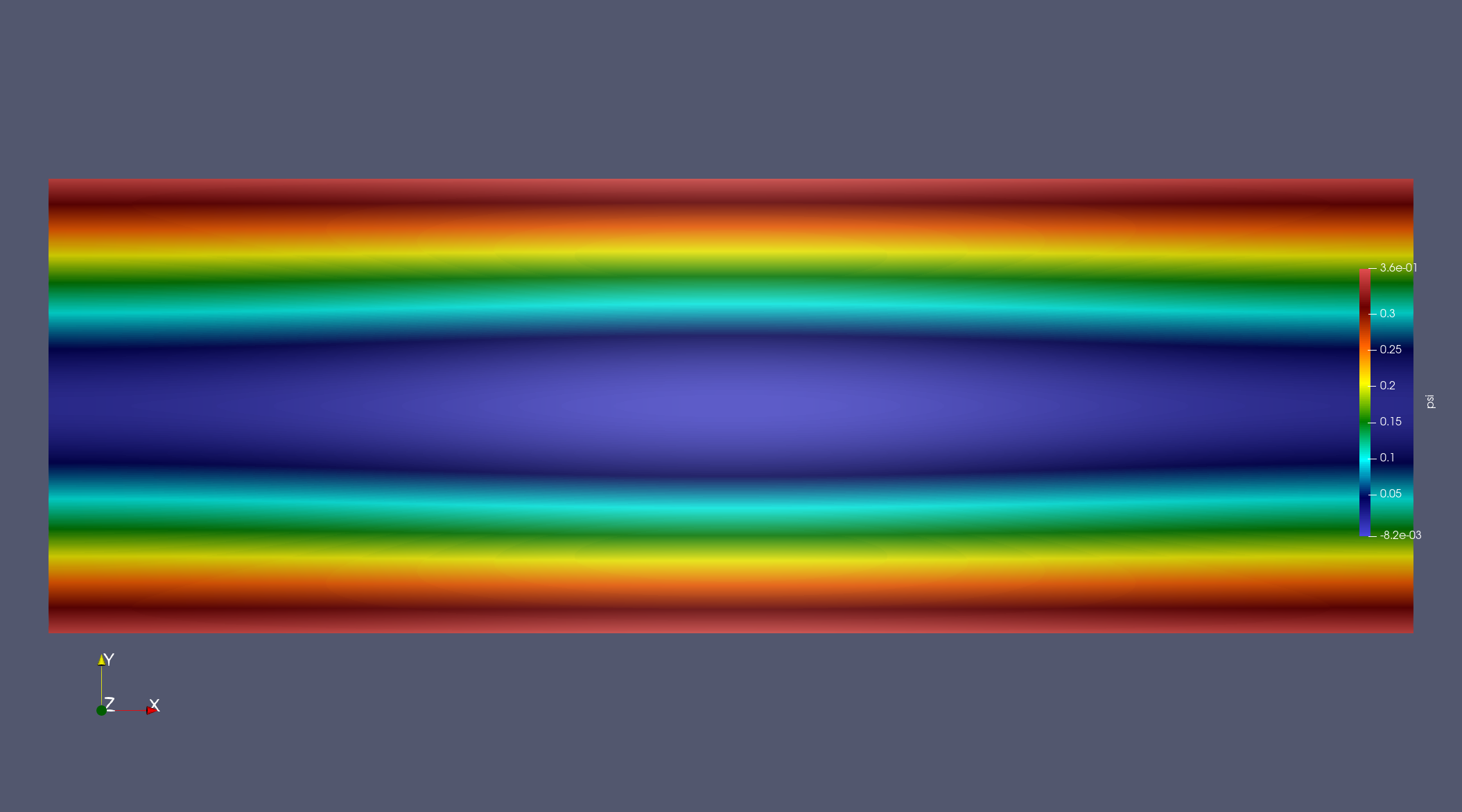}{\figWidth}};
  \draw(9.1,2.4) node[anchor=south west,xshift=0pt,yshift=+0pt] {\trimfig{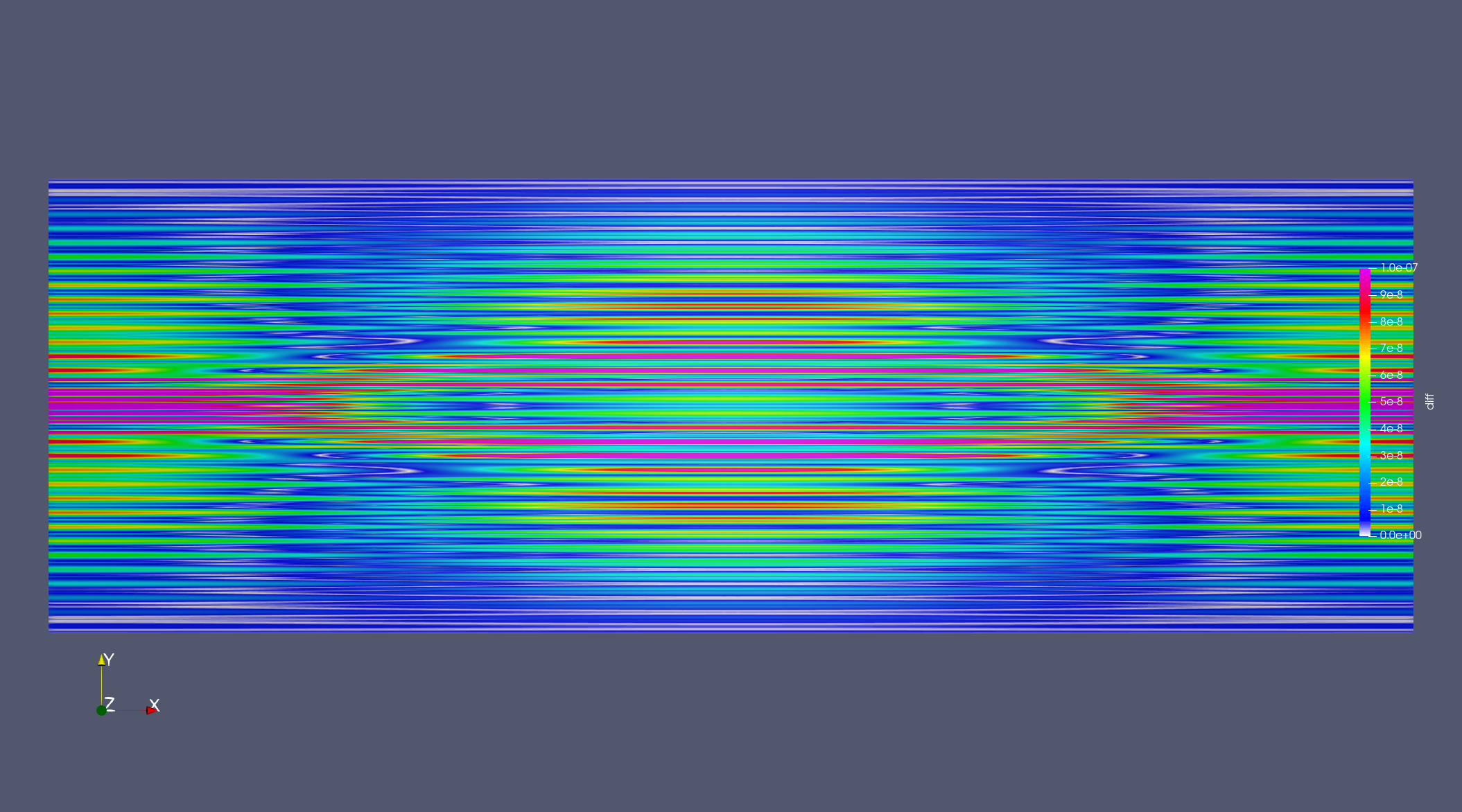}{\figWidth}};
  \draw(-.4,5.4) node[draw,fill=white,anchor=west,xshift=4pt,yshift=0pt] {\scriptsize $\Psi$};
  \draw(9.2,5.4) node[draw,fill=white,anchor=west,xshift=4pt,yshift=0pt] {\scriptsize Error on uniform mesh};
  \draw(-.5,-.75) node[anchor=south west,xshift=0pt,yshift=+0pt] {\trimfig{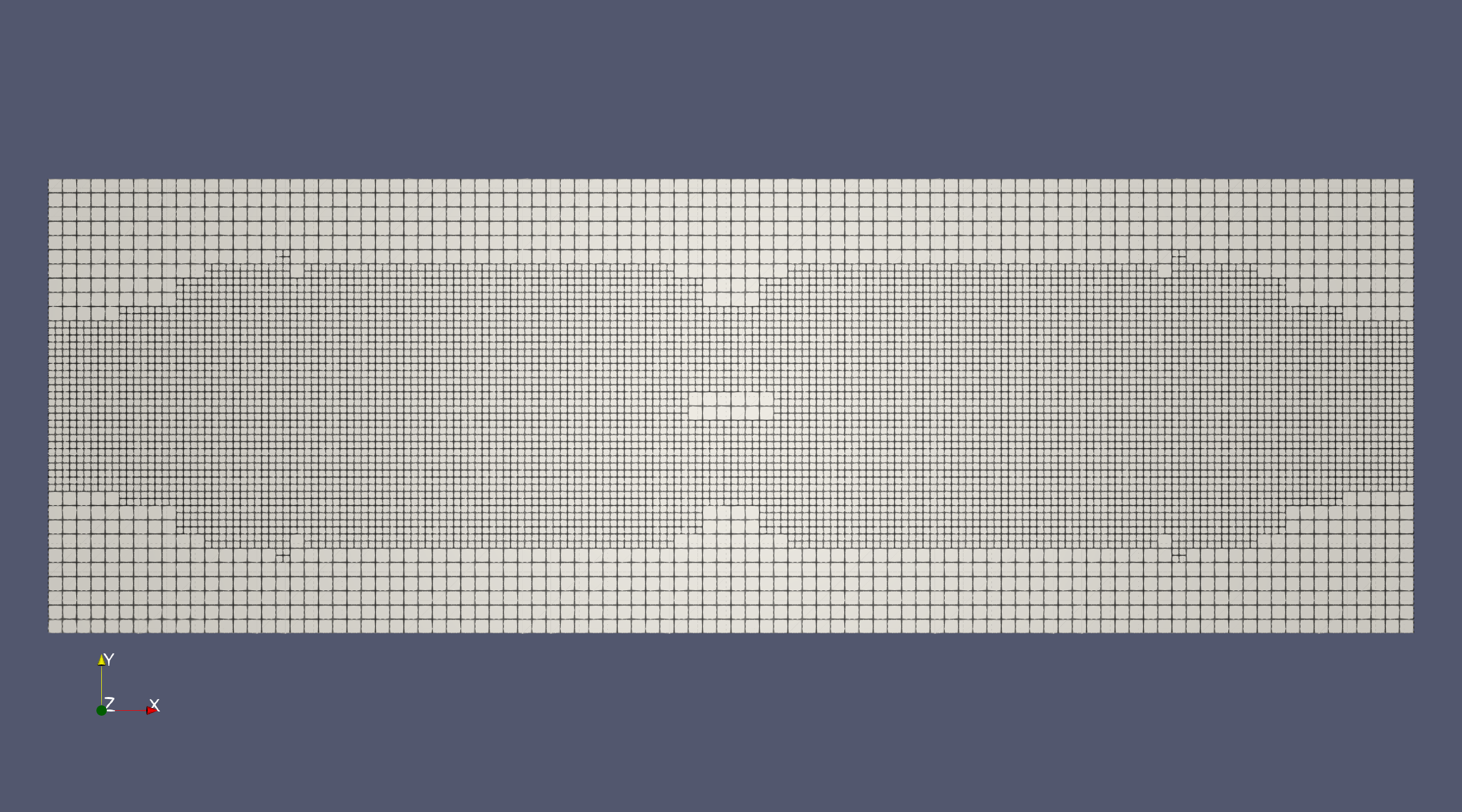}{\figWidth}};
  \draw(9.1,-.75) node[anchor=south west,xshift=0pt,yshift=+0pt] {\trimfig{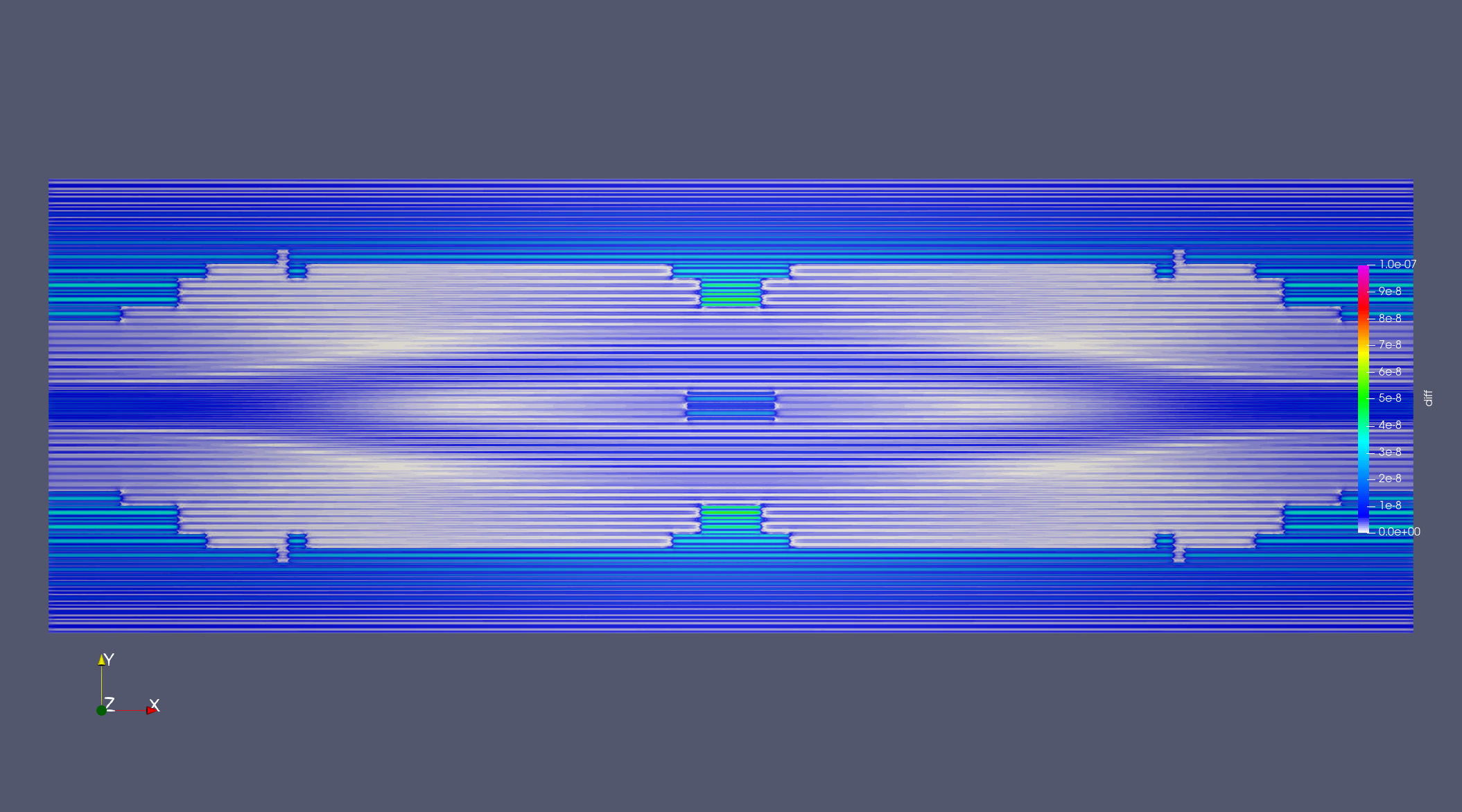}{\figWidth}};
    \draw(-.4,2.2) node[draw,fill=white,anchor=west,xshift=4pt,yshift=0pt] {\scriptsize 1-level mesh};
    \draw(9.2,2.2) node[draw,fill=white,anchor=west,xshift=4pt,yshift=0pt] {\scriptsize Error};
      \end{scope}
      %
\end{tikzpicture}
} 
\end{center}
  \caption{Tearing mode. Errors on different meshes are presented for comparison. Top row: the solution $\Psi$ at $t = 50$ and the point value errors on the mesh of $96\times32$.
  Second row: the mesh using one  refinement level and the error.
  Third row: the mesh using two  refinement levels and the error.
   Bottom row: the mesh using three  refinement levels and the error.
   The errors are computed as point value difference between the solution and highly resolved solution on a uniform grid of   $1536 \times512$.
  To visualize the errors, they are all evaluated at the nodes of the grid  $1536 \times512$.
  For easy comparison, all the error plots share the same colormap in the range of $[0, 10^{-7}]$.
    }
  \label{fig:tearAMRerror}
\end{figure}
}

The final examination focuses on the potential accuracy improvement of AMR.
While it is challenging to compare two finite-element solutions on two different meshes,
MFEM provides limited support for such a task.
We rely on the {\tt gslib} tool on MFEM, which provides a high-order interpolation on any given point in the domain.
We use this tool to pick 40000 points in the entire domain to compare the two interpolated solutions on the coarse mesh and the finer mesh
and collect  its averaged difference and maximal difference.
The reference solution is computed on  a $1536 \times512$ grid and we choose $\Psi$ to compute the errors.
We focus on a given time ($t = 50$) as the demonstration and compare four different meshes (a very coarse uniform mesh of $96\times32$,
and adaptive meshes with a base mesh of $96\times32$ with one, two and three levels of refinement, respectively).
The solutions and their differences are presented in Figure~\ref{fig:tearAMRerror}, and the corresponding difference values are reported in Table~\ref{tableTearAMR}.
To easily visualize the differences in Figure~\ref{fig:tearAMRerror}, they are only evaluated at the node point of the  $1536 \times512$ grid during  visualization.
Nevertheless, this provides us a good indication of where  error concentrates in the solutions.

\begin{table}[hbt]\tableFont 
\begin{center}
\caption{Maximum and average differences on 40000 points between solutions on coarse meshes and the reference solutions.
 \label{tableTearAMR}}
 \medskip
\begin{tabular}{|c|c|c|c|} \hline
  \multicolumn{4}{|c|}{Tearing mode AMR test at $t=50$} \\ \hline
Grid & Average dofs &   Maximum error &  Average error         \\[3pt] \hline
  $96\times32$ &27936 & \num{2.69}{-7} & \num{3.09}{-8}      \\ \hline
  1-level mesh & 63247 & \num{5.35}{-8} & \num{8.13}{-9}     \\ \hline
   2-level mesh & 157432 & \num{3.92}{-8} &   \num{9.38}{-9}      \\ \hline
  3-level mesh& 225487 &  \num{2.64}{-8}  & \num{3.36}{-9}  \\ \hline
\end{tabular}
\end{center}
\end{table}

We first note the uniform mesh has a very large error in the center region, which is largely proportional to the solution $\Psi$.
The adaptive mesh with one-level refinement improves the error significantly by effectively locating the region of larger error and reduce the error in the center region.
The adaptive mesh with two-level refinement has a mixed result, in which some part of the error increases and some part of the error decreases compared to the one-level mesh.
It is however still noticeably better than the error on the uniform mesh. Such a mixed result is likely the outcome of frequent refinement and coarsening steps.
The adaptive mesh with three-level refinement is found to be better than all the other meshes.
It is also observed that the errors closely follow the pattern of the refined regions in all the meshes,
which indicates the effectiveness of the AMR solver.
For a quantitative  comparison of  performance, we present the average and maximum errors in Table~\ref{tableTear}.
First note the error in the current work is much smaller than the errors reported in~\cite{philip2008implicit} (Fig.10 in~\cite{philip2008implicit} indicates a relative error of $10^{-3}$ to  $10^{-4}$
compared to the relative error in the level of  $10^{-8}$ to   $10^{-9}$ in the current work of the same test). 
We note that the maximum and averaged errors among all the solutions decay as the number of  refinement levels increases.
It is also important to stress that the proposed error estimator~\eqref{eqn:errorTear} is used in all the tests considered in this study.
Although it is not an optimal error estimator, it provides a satisfactory result based on the above accuracy study.
(The optimal error estimator here refers to the error estimator using a true solution or a reference solution.)
We conclude that the AMR solver provides a more accurate solution with increasing levels of refinement.

\subsection{Island coalescence}
\label{sec:ic}
The final example we consider is the island coalescence, originally considered in studies such as~\cite{finn1977coalescence, biskamp1980coalescence}.
The problem shows true time and space multiscale behaviors when the Lundquist number increases including plasmoid development,
 and therefore serves as a great test to challenge our solver.

The coalescence instability, which drives this example, is an ideal MHD instability,  driven by the attractive force between parallel currents.
For low and moderately high Lundquist number, the simulation can be divided into two phases~\cite{biskamp1980coalescence}. In the first phase,
the current-density ropes are freely accelerated toward each other and a current sheet starts to form, while in the second phase the current sheet at the reconnection layer stays somewhat stationary and
gets slowly damped until the two islands fully merge. 
For high-Lundquist numbers, the current sheet tears after the aspect ratio of the current sheet becomes sufficient large, and results in  plasmoids forming along the sheet~\cite{biskamp1996magnetic}.
The plasmoids can further excite more plasmoids and eventually lead to very complicated turbulence-like local structures.

The physics properties described above indicate the numerical solutions for the test can be very challenging.
The first challenge is due to the stiffness of the Alfv\'en wave. The earlier stage of the dynamics is very slow, and one would like to step over the CFL constraint from the Alfv\'en speed using implicit time stepping.
The second challenge is the evolving small spatial scales due to the formulation of a localized current sheet. In particular, we found that the current sheet has to be well-resolved for two reasons:
(i) an under-resolved current sheet may lead to spurious oscillations in the solution, resulting in the failure of the solver;
(ii) although one may reduce the time step so that the implicit solver may still converge, we found those oscillations may lead to  unphysical solutions.
For instance, a large ``plasmoid'' may form at the reconnection site for small Lundquist numbers.
We observe one giant plasmoid forms for Lundquist number of $5\times10^4$ when it is under-resolved, and similar unphysical solutions were observed previously in~\cite{adler2013island, toth2008hall}.
Note the current sheet thickness scales as $\sqrt{\eta}$ (according to the Sweet-Parker model), becoming thinner as the Lundquist number increases.
The third challenge is due to  large flow speed along the current sheet at the peak reconnection time. As a result, it needs  adequate stabilization.
Last, the violent plasmoid instability poses a significant challenge to the solver. The spatial structure becomes more complicated than a simple current sheet,
which requires the dynamic AMR to  capture fully those small structures.
The dynamics also happens much faster than the coalescence instability, requiring the time step to be reduced.
All those challenges will be carefully studied and addressed  here.

 Many of the previous studies applied various numerical schemes with some preconditioning techniques to this problem; see~\cite{knoll2006coalescence, philip2008implicit, shadid2010towards, lin2010parallel, shadid2010current, adler2011efficiency,  cyr2013new, adler2013island, muralikrishnan2020multilevel} for instance.
 Most of those studies however focus on the moderately high Lundquist number cases (the maximal Lundquist number considered is typically under $10^{5}$).
 Among them,~\cite{adler2011efficiency, adler2013island} also develop an implicit AMR FEM solver, but
 owing to low Lundquist numbers in their tests the plasmoid instability is not observed.
 One of the highest Lundquist number considered in previous studies is~\cite{shadid2010current} in which $S_L$ up to $10^{9}$ is studied for this example using a resistive MHD model, while~\cite{muralikrishnan2020multilevel}
 considers a simulation of $S_L=$ $10^{7}$ for the same example.
 The approach proposed in~\cite{shadid2010current, muralikrishnan2020multilevel} does not use AMR to capture plasmoids.
Instead, \cite{shadid2010current} relied on a locally packed structured grid and~\cite{muralikrishnan2020multilevel} used  a very high-order HDG scheme with stabilization to resolve plasmoids
 and currents sheets. Those techniques are quite different from the approach proposed here.

The same problem setup as in~\cite{knoll2006coalescence, philip2008implicit} is adopted. The initial condition is a Fadeev magnetic equilibrium~\cite{fadeev1965self} given by
\begin{align*}
    \Psi_0 = -\lambda \ln \left[\cosh (\frac y  \lambda) + \epsilon \cos (\frac x \lambda) \right], \qquad \omega_0=0, \qquad \Phi_0= 0,
\end{align*}
balanced with a source term of $E_0 = \eta \grad^2 \Psi_0$.
The perturbation exciting the instability is
\[
   \delta \Psi = 10^{-3} \cos(\pi y/2) \cos (\pi x),
\]
The computational domain is $[-1, 1]\times[-1, 1]$ and the boundary conditions are the PEC wall (Dirichlet) along the $y$ boundaries and periodic in $x$.
Other parameters are
    $\epsilon=0.2$, and   $\nu=\eta=10^{-4}$,
unless otherwise noted.
The definitions of Reynolds and Lundquist numbers also follow~\cite{knoll2006coalescence, philip2008implicit, shadid2010towards}, i.e., $Re =1/\nu$ and $S_L=1/\eta$.
The reference length scale and Alfv\'en speed are both 1. Thus, the time scale for the Alfv\'en wave is $\tau_A = 1$.
Note unlike the computations in~\cite{knoll2006coalescence, shadid2010towards, shadid2010current}, our computational domain is the entire domain and we do not assume symmetry in either direction.
This is important to capture the full solutions structures when plasmoids start to appear   and break the symmetry.
The readers are referred to~\cite{knoll2006coalescence} for more discussions on physics.

The example is again split into two parts in the following discussion. The first part focuses on the results on static meshes,
while, in the second part, the solver is truly pushed to the limit with dynamic adaptivity, studying the cases of high Lundquist numbers and particular the one with violent plasmoid instabilities.

\subsubsection{Static-mesh results}
\label{sec:ic-steady}
{
\newcommand{\figWidth}{9.2cm}
\newcommand{\trimfig}[2]{\trimFig{#1}{#2}{.1}{.1}{.45}{.45}}
\newcommand{\trimfigb}[2]{\trimFig{#1}{#2}{.3}{.3}{0}{.0}}
\begin{figure}[htb]
\begin{center}
\resizebox{16cm}{!}{
\begin{tikzpicture}[scale=1]
  \useasboundingbox (0.0,0) rectangle (18,6.8);  
  \draw(-.5,3.) node[anchor=south west,xshift=0pt,yshift=+0pt] {\trimfig{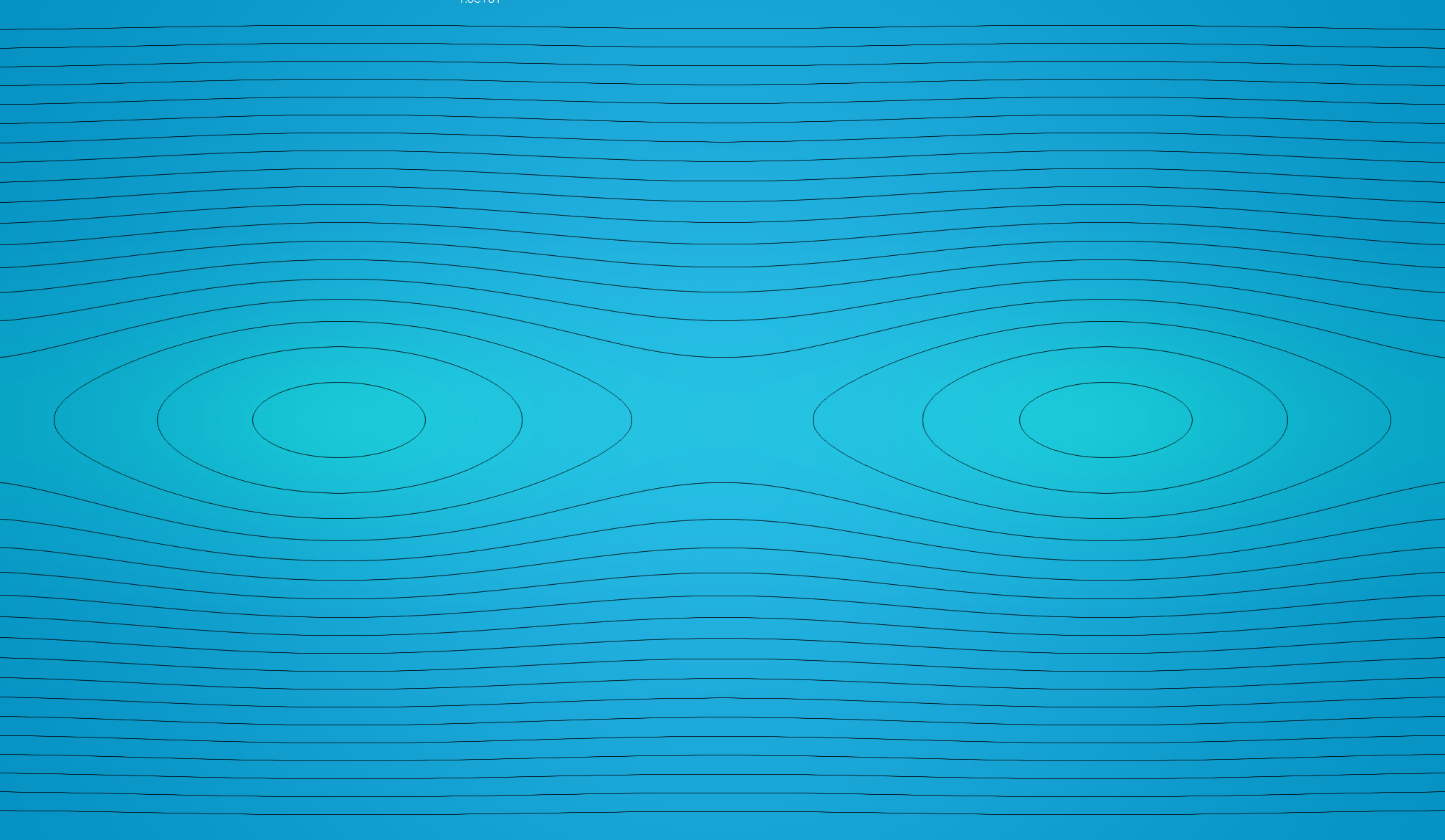}{\figWidth}};
  \draw(9.1,3.) node[anchor=south west,xshift=0pt,yshift=+0pt] {\trimfig{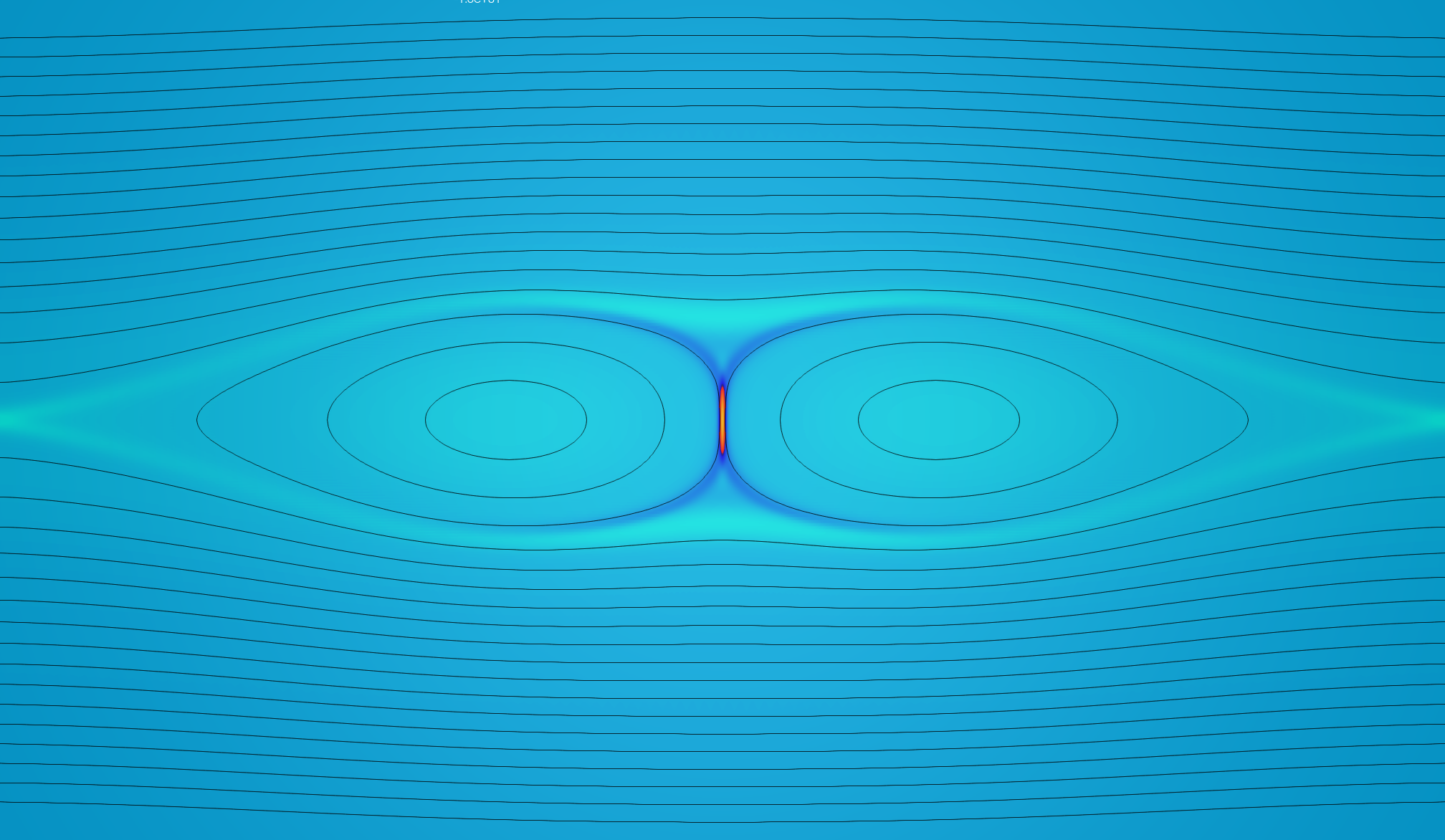}{\figWidth}};
    \draw(-.4,6.3) node[draw,fill=white,anchor=west,xshift=4pt,yshift=0pt] {\scriptsize $t=0$};
    \draw(9.2,6.3) node[draw,fill=white,anchor=west,xshift=4pt,yshift=0pt] {\scriptsize $t=6$};
  \draw(-.5,-.75) node[anchor=south west,xshift=0pt,yshift=+0pt] {\trimfig{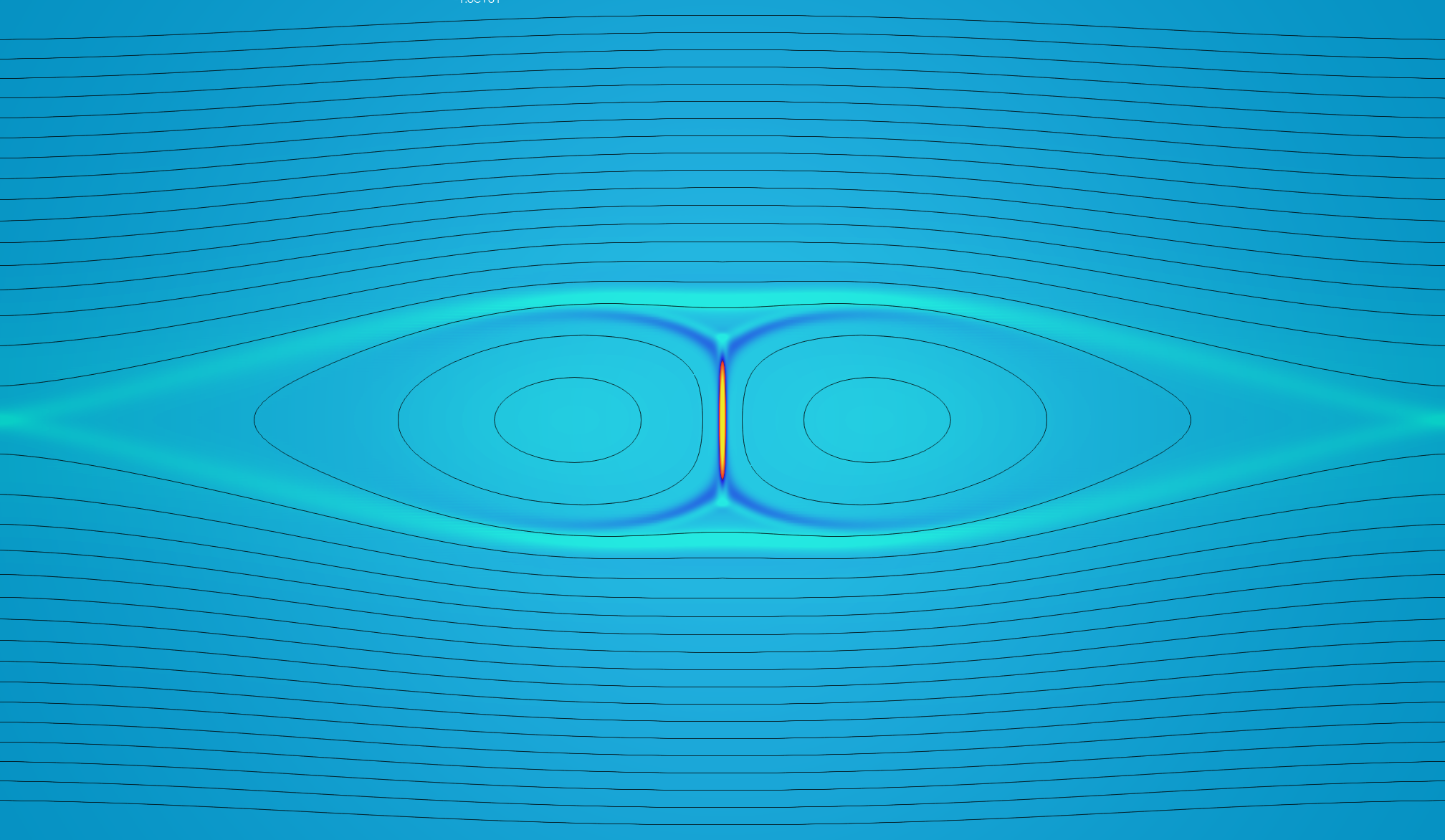}{\figWidth}};
  \draw(9.1,-.75) node[anchor=south west,xshift=0pt,yshift=+0pt] {\trimfig{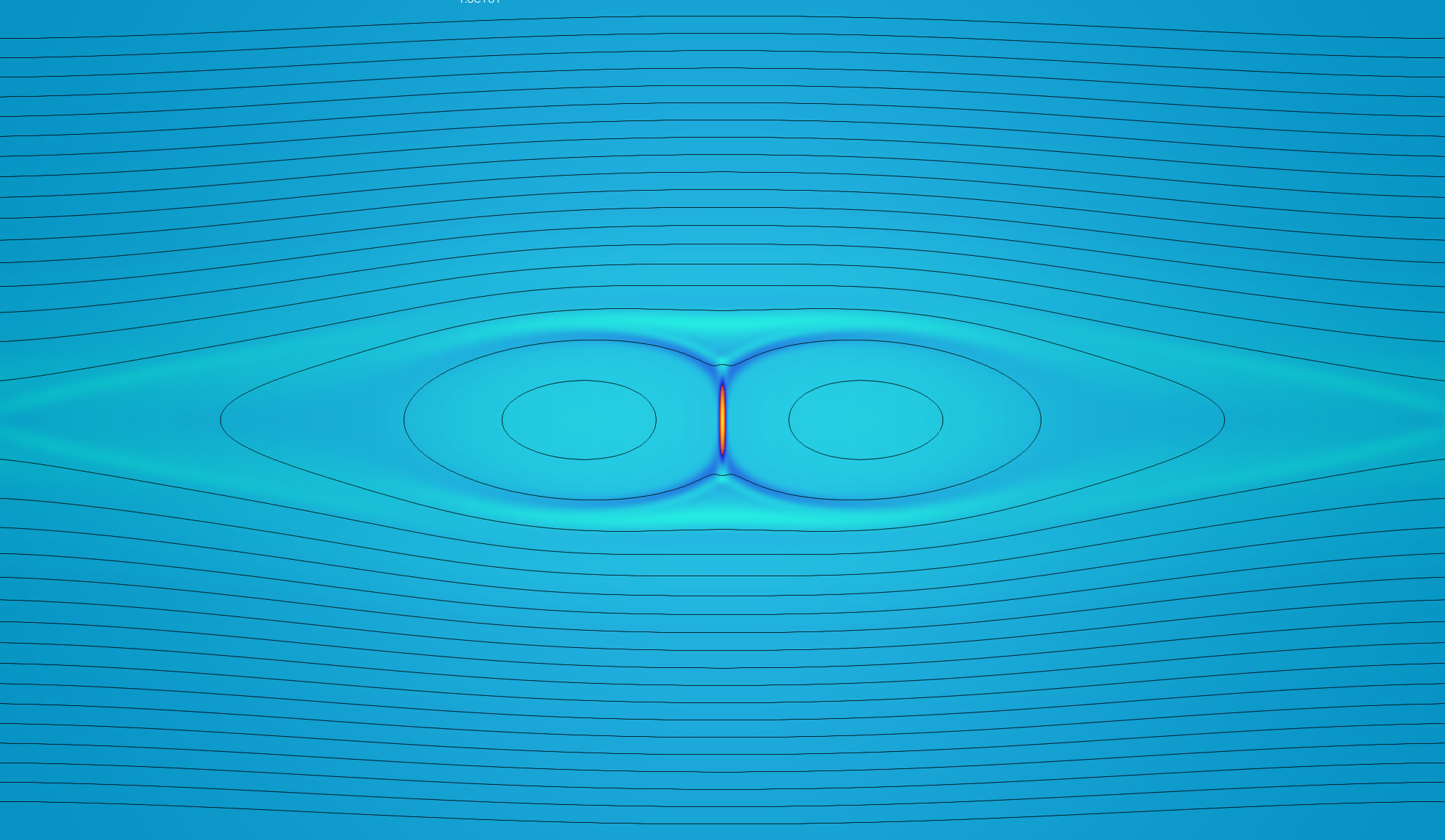}{\figWidth}};
    \draw(-.4,2.5) node[draw,fill=white,anchor=west,xshift=4pt,yshift=0pt] {\scriptsize $t=7$};
    \draw(9.2,2.5) node[draw,fill=white,anchor=west,xshift=4pt,yshift=0pt] {\scriptsize $t=8$};
%
\end{tikzpicture}
} 
\end{center}
  \caption{Island coalescence. Solutions of $\nu=\eta=$ $10^{-4}$ at different times are presented.
  The parallel current $J$ is presented as pseudocolor  and the magnetic flux $\Psi$ is presented as black contours.
The colormap and contour values are fixed over time.
  The presented computation uses a uniform grid of $512\times512$ with $p=2$ and DIRK2 with a fixed time step of $\dt=0.1$.
  }
  \label{fig:icOverT}
\end{figure}
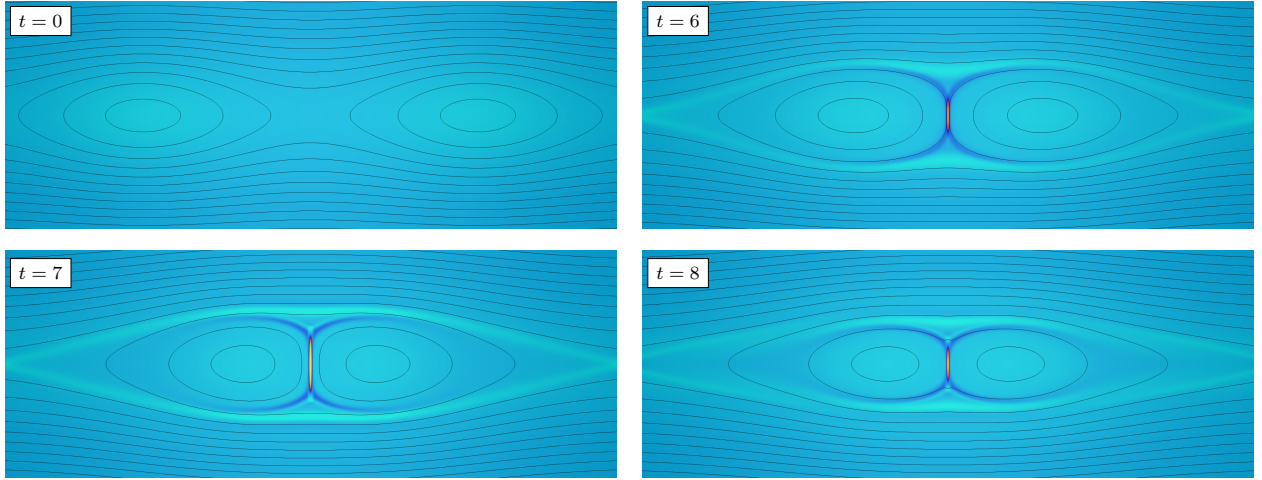
}

The problem is first solved on a uniform mesh of  $512\times512$ with the results presented in Figure~\ref{fig:icOverT}.
The test uses $\nu=\eta=$ $10^{-4}$,  $p=2$ and the DIRK2 time integrator with a large time step $\dt=0.1$.
Again, the preconditioner performs very well with a large time step.
In particular, the average Newton iteration is 2 and the average total Krylov linear solver is about 11.
The numerical solutions  are very comparable with the results given in Ref.~\cite{shadid2010towards} but with better performance.
Starting from time of $t = 5$, a current sheet starts to form in the center and the number of iterations grows from 5 iterations to 10 to 20 iterations.

{
\newcommand{\figWidth}{9.2cm}
\newcommand{\trimfig}[2]{\trimFig{#1}{#2}{.1}{.1}{.45}{.45}}
\begin{figure}[htb]
\begin{center}
\resizebox{16cm}{!}{
\begin{tikzpicture}[scale=1]
  \useasboundingbox (0.0,0) rectangle (18,3.);  
  \draw(-.5,-.75) node[anchor=south west,xshift=0pt,yshift=+0pt] {\trimfig{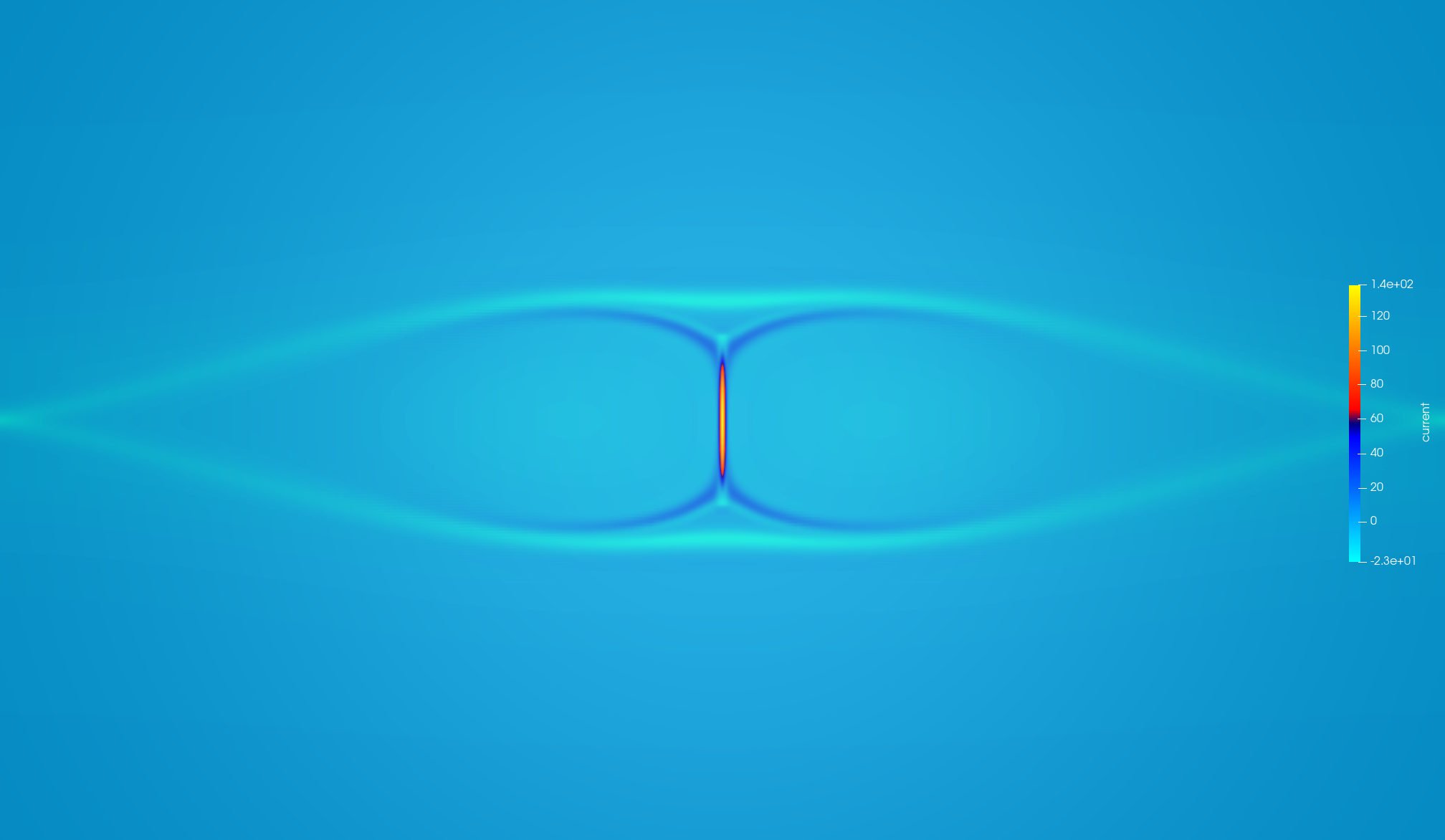}{\figWidth}};
  \draw(9.1,-.75) node[anchor=south west,xshift=0pt,yshift=+0pt] {\trimfig{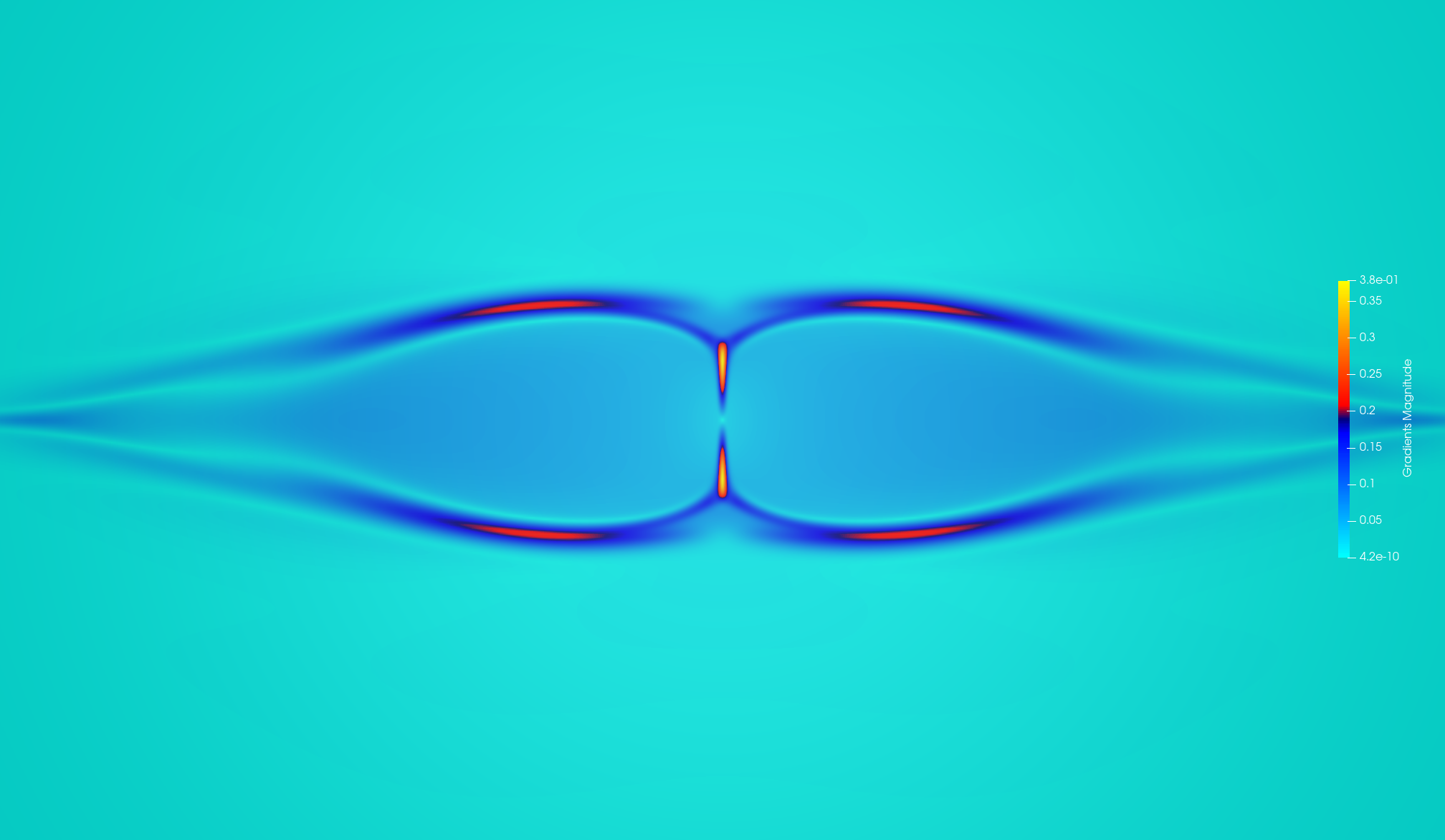}{\figWidth}};
    \draw(-.4,2.5) node[draw,fill=white,anchor=west,xshift=4pt,yshift=0pt] {\scriptsize $J$};
    \draw(9.2,2.5) node[draw,fill=white,anchor=west,xshift=4pt,yshift=0pt] {\scriptsize $\|\vv\|$};
%
\end{tikzpicture}
} 
\end{center}
  \caption{Island coalescence.
The current and flow speed of $\nu=\eta=$ $10^{-4}$ at first peak time $t=6.9$ are presented.
The peak current at the center is about 140 and the highest flow speed is about 0.38 (with normalized Alfv\'en speed of 1).
  The presented computation uses a uniform grid of $512\times512$ with $p=2$ and DIRK2 with a fixed time step of $\dt=0.1$.
  }
  \label{fig:icPeak}
\end{figure}
}

The solution at peak reconnection   is presented in Figure~\ref{fig:icPeak}.
With this Lundquist number, the  current sheet structure follows the Sweet-Parker model.
The current sheet is found to be well resolved and the largest current density in the current sheet is about 140.
The largest flow speed in the outgoing jet regions of the  reconnection diffusion region is about 0.38 of the Alfv\'en speed.
Note that, for this test, since the Lundquist and Reynolds numbers are relatively small, the solvers  work well without the stabilization term  and produce good results.

{
\newcommand{\figWidth}{10cm}
\newcommand{\trimfig}[2]{\trimFig{#1}{#2}{.0}{.0}{.0}{.0}}
\begin{figure}[htb]
\begin{center}
\resizebox{16cm}{!}{
\begin{tikzpicture}[scale=1]
  \useasboundingbox (0.0,0) rectangle (18,4.7);  
  \draw(3.9,-1) node[anchor=south west,xshift=0pt,yshift=+0pt] {\trimfig{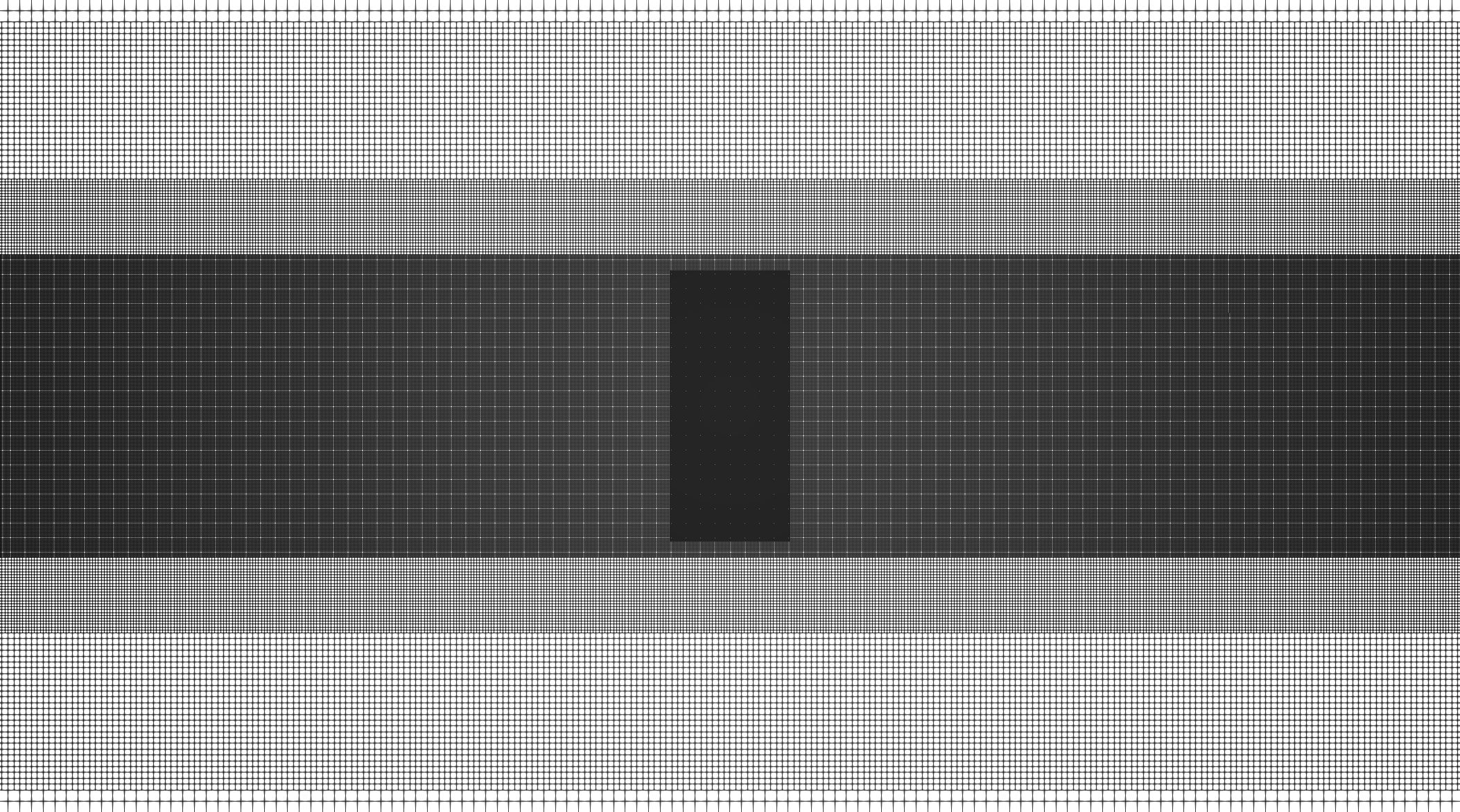}{\figWidth}};
%
\end{tikzpicture}
} 
\end{center}
  \caption{Locally packed mesh for the island coalescence scaling tests. The grid is visualized between $y \in [-0.5, 0.5]$. The base uniform grid is refined by four levels gradually from the external region to the center. Here the finest grid corresponds to a uniform grid of $2048\times2048$.
  }
  \label{fig:icFixedMesh}
\end{figure}
}

We further test the solver by varying the Reynolds and Lundquist numbers following~\cite{knoll2006coalescence, shadid2010towards} and study the reconnection rate at the reconnection X-point,
 i.e., $\xv = [0, 0]$. We simulate the same test in~\cite{knoll2006coalescence, shadid2010towards} by varying the Lundquist number from $10^2$ to $2\times10^5$.
In this test, the Reynolds number is always chosen to be identical to the Lundquist number.
We use a locally packed grid for all the simulations given in Figure \ref{fig:icFixedMesh}.
The mesh has a uniform background mesh of $512\times512$ and is gradually refined towards the center region.
The final level is packed around the current sheet and has an effective grid spacing identical to the $2048\times 2048$ mesh.
This is a meshing technique commonly employed for this test when dynamic AMR is not utilized; see~\cite{shadid2010towards} for instance.
Such a mesh indeed saves computational cost considerably. For the third-order polynomials used, the packed mesh has 2.8M dofs in a scalar variable
while the uniform mesh of $2048\times 2048$ has 38M dofs.
Thus, the packed mesh only needs 7.4\% of the comparable uniform mesh.

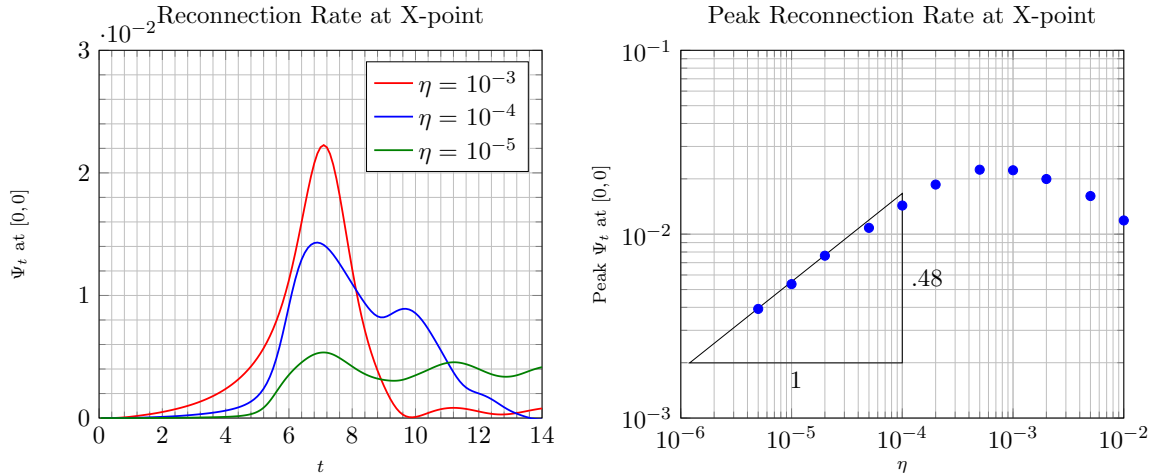
\begin{figure}[htb]
\begin{center}
\resizebox{13cm}{!}{
\begin{tikzpicture}
\useasboundingbox (0,-.3) rectangle (15,6.4);  
\begin{axis}[
title={Reconnection Rate at X-point},
xlabel={$t$},
ylabel={$\Psi_t$ at $[0, 0]$},
label style={font=\scriptsize},
grid=both,
grid style={draw=gray!50},
minor tick num=4,
legend pos=north east,
xmin=0,   xmax=14,
ymin=0,   ymax=0.03,
]
\addplot[color=red,thick] table[x={t},y={dPsidt}] {figures/icUniform/1e-3dpsidt.dat};
\addplot[color=blue,thick] table[x={t},y={dPsidt}] {figures/icUniform/1e-4dpsidt.dat};
\addplot[color=Green,thick] table[x={t},y={dPsidt}] {figures/icUniform/1e-5dpsidt.dat}; 
\legend{$\eta=$ $10^{-3}$, $\eta=$ $10^{-4}$, $\eta=$ $10^{-5}$}
\end{axis}
  \begin{scope}[xshift=9cm]
\begin{loglogaxis}[
title={Peak Reconnection Rate at X-point},
xlabel={$\eta$},
ylabel={Peak $\Psi_t$ at $[0, 0]$},
label style={font=\scriptsize},
grid=both,
grid style={draw=gray!50},
legend pos=south east,
 xmin=1e-6, xmax=1e-2,
 ymin=1e-3, ymax=1e-1,
]
\addplot [color=blue, mark=*,only marks] coordinates {
    (1e-2, 0.0118771)
    (5e-3, 0.0161228)
    (2e-3, 0.0199592)
    (1e-3, 0.0222687)
   (5e-4,0.0224286)
   (2e-4, 0.0186182)
   (1e-4, 0.0143167)
   (5e-5, 0.0108235) 
   (2e-5,0.00763906)
   (1e-5,0.00534996) 
   (5e-6,0.00392301)
};
\logLogSlopeTriangle{0.5}{0.48}{0.15}{.48}{black};
\end{loglogaxis}
 \end{scope}
%
\end{tikzpicture}
}
\end{center}
\caption{Left plot: time histories of the reconnection rate, defined as $\partial \Psi/\partial t$,  at the X-point for different resistivity. Right plot: scaling result between the peak reconnection rate at X-point  and resistivity.
A locally packed grid described in Figure~\ref{fig:icFixedMesh} is used in all the tests. The scaling results match well with~\cite{knoll2006coalescence, shadid2010towards}.
The linear regression of the smallest three resistivity data points indicate the scaling of reconnection rate is $\Psi_t\sim \eta^{0.48}$.
}
  \label{fig:icScaling}
\end{figure}

The reconnection rate results are presented in Figure~\ref{fig:icScaling}.
In all these tests, a smaller time step of 0.025 is used. When a large time step of $0.1$ is used, we found the reconnection rate may not be accurate for the large Lundquist number runs,
which has been also described in~\cite{adler2013island}.
Our results match very well with~\cite{knoll2006coalescence, shadid2010towards}.
The shape of the reconnection rate in the left plot of Figure~\ref{fig:icScaling} matches well with the result presented in~\cite{shadid2010towards}.
The runs with $\eta=$ $10^{-5}$ are found to have a slightly smaller peak reconnection rate than those in Fig.~17 of~\cite{shadid2010towards}.
The peak reconnection rate also matches well with~\cite{knoll2006coalescence, shadid2010towards}.
The linear regression of the smallest three resistivity data points in the scaling plot indicates the scaling of reconnection rate satisfies $\Psi_t\sim \eta^{0.48}$,
which is identical to the scaling result found in~\cite{shadid2010towards} and comparable with the Sweet-Parker slow reconnection rate of $\Psi_t\sim \sqrt{\eta}$.
We found that the stabilization terms are necessary for all the simulations with $\eta\le5\times 10^{-5}$. 
The above scaling test validates our solver and also indicates that SUPG is not affecting the quality of the physics results.


\begin{table}[hbt]\tableFont 
\begin{center}
\caption{Preconditioner performance study with different  grid resolutions for the island coalescence. $Re=S_L= 10^4$.
DIRK2 is used with two solves per time step. A fixed relative tolerance of $10^{-4}$ and a fixed relative solution tolerance of $10^{-6}$
are used as the Newton convergence criterion.
The steps are averaged over 10 time steps and $p=2$, $\dt=0.1$. Dofs stands for the total degrees of freedoms for one scalar component.
 \label{tableIC}}
 \medskip
\begin{tabular}{|c|c|c|c|c|c|c|c|c|} \hline
  \multicolumn{8}{|c|}{Island  Coalescence} \\ \hline
Grid & dofs & $\dt/\dt_{CFL}$ & CPUs &  Newton/solve     &  Krylov/solve   &     Time/$\dt$   &     Time/Krylov      \\[3pt] \hline
  $64\times64$ &16512 & 3.2& 1 &  2.0 &  3.9 &            8.9   & 1.1      \\ \hline
  $128\times128$& 65792 & 6.4 & 4  &    2.0 & 3.9 &     16.5  & 2.1   \\ \hline
   $256\times256$ & 262656 & 12.8 & 16  & 2.0 & 3.9 &  19.3  & 2.4    \\ \hline
  $512\times512$ & 1M & 25.6& 64  & 2.2 &  4.0 &  23.3    & 2.9 \\ \hline
 $1024\times1024$ &4.2M &  51.2 & 256 &  2.1	 &   5.3 &   21.3  & 2.0    \\ \hline
   $2048\times2048$ & 16.8M & 102.4 & 1024  & 3.0 &	11.2 &   41.1   & 1.8   \\ \hline
    $4096\times4096$ & 67.1M & 204.8 & 4096  & 3.1 & 19.5	 &     64.5  & 1.6   \\ \hline
\end{tabular}
\end{center}
\end{table}

The last uniform-mesh test  is the weak parallel scaling test.
The Reynolds number and Lundquist number are chosen as $10^4$.
The weak scaling result is reported in Table~\ref{tableIC}.
The result is very comparable to the tearing-mode test.
The weak scaling is almost optimal until we reach the finest resolutions.
We also found the performance to be quite comparable with the finite difference version~\cite{chacon2002}.
(Note the original work~\cite{chacon2002} does not perform a weak scaling test with as many as CPUs as the current work.
But a similar weak scaling test has been performed using the code developed in~\cite{chacon2002}   during the development of the current work.)
Overall, the performance of the solver is satisfactory.
The total dofs from a single processor to 4096 processors grows by a factor of 4096 while the
computational time only grows by a factor of 7.2.
We also would like to highlight the current solver is  better than the weak scaling result reported in Table 7 of~\cite{shadid2010towards}.
The setup between two weak scaling tests are almost identical but the number of Krylov iterations grows much faster for the 
purely algebraic-based multilevel smoother in~\cite{shadid2010towards}.
 
\subsubsection{AMR results}
\label{sec:ic-amr}
We consider next numerical results using the implicit dynamic AMR solver.
The simulation considered here is a first step to demonstrate viability of dynamically resolving current sheets and plasmoid structures in  fusion devices
using AMR.
Note some plasmoid related simulations for practical devices can be found in~\cite{ebrahimi2016large, ali2019effects}.

In the following AMR tests, the error estimator~\eqref{eqn:error} is used with $\alpha = 0.1$.
The error estimator is chosen since the solutions such as Figure~\ref{fig:icOverT} indicate structures are mostly observed in the current and vorticity.
We choose the refinement ratio $\beta_r = 0.1$ and the coarsening ratio $\beta_c = 0.001$ as described in Section~\ref{sec:refine}.
The refinement and coarsening steps are called every 10 time steps but are switched to every 4 time steps later at   $t = 5.5$ when the simulation becomes more dynamic.
This is particularly necessary  when plasmoids form and collide into each other.
The base computational grid is $128\times128$, and all the simulation starts with the same base mesh but with different refinement levels.
Our coarsening algorithm is implemented such that it will not coarsen the element beyond the base mesh.
We found that coarsening beyond the base level may introduce a lot of oscillations during the projection.
All the AMR tests use a polynomial order of three.
The number of processors used are between 180 to 3600 processors for all the runs presented.

{
\newcommand{\figWidth}{9.2cm}
\newcommand{\trimfig}[2]{\trimFig{#1}{#2}{.0}{.0}{.3}{.3}}
\begin{figure}[htb]
\begin{center}
\resizebox{16cm}{!}{
\begin{tikzpicture}[scale=1]
  \useasboundingbox (0.0,0) rectangle (18,13.3);  
    \begin{scope}[yshift=7cm]
  \draw(-.5,2.7) node[anchor=south west,xshift=0pt,yshift=+0pt] {\trimfig{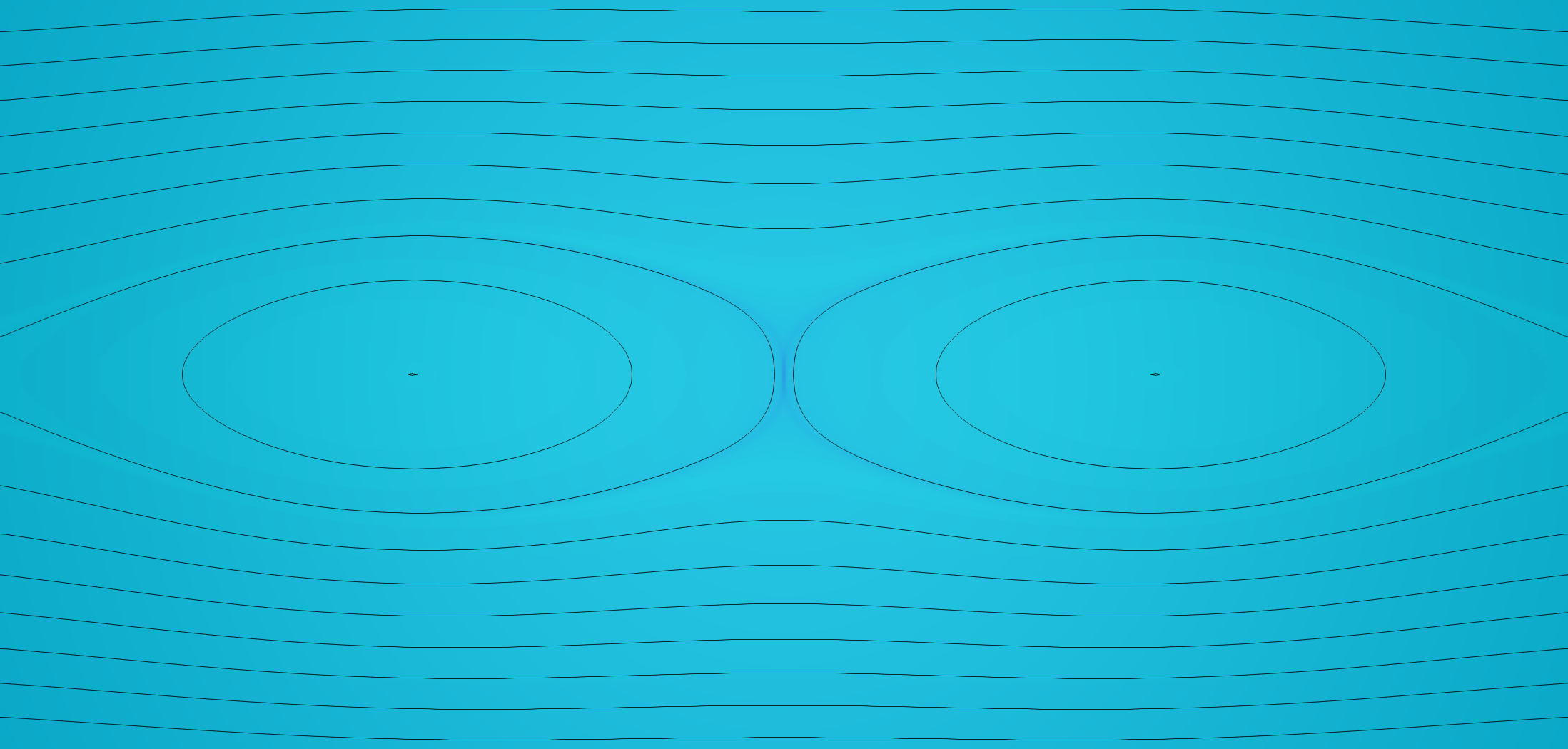}{\figWidth}};
  \draw(9.1,2.7) node[anchor=south west,xshift=0pt,yshift=+0pt] {\trimfig{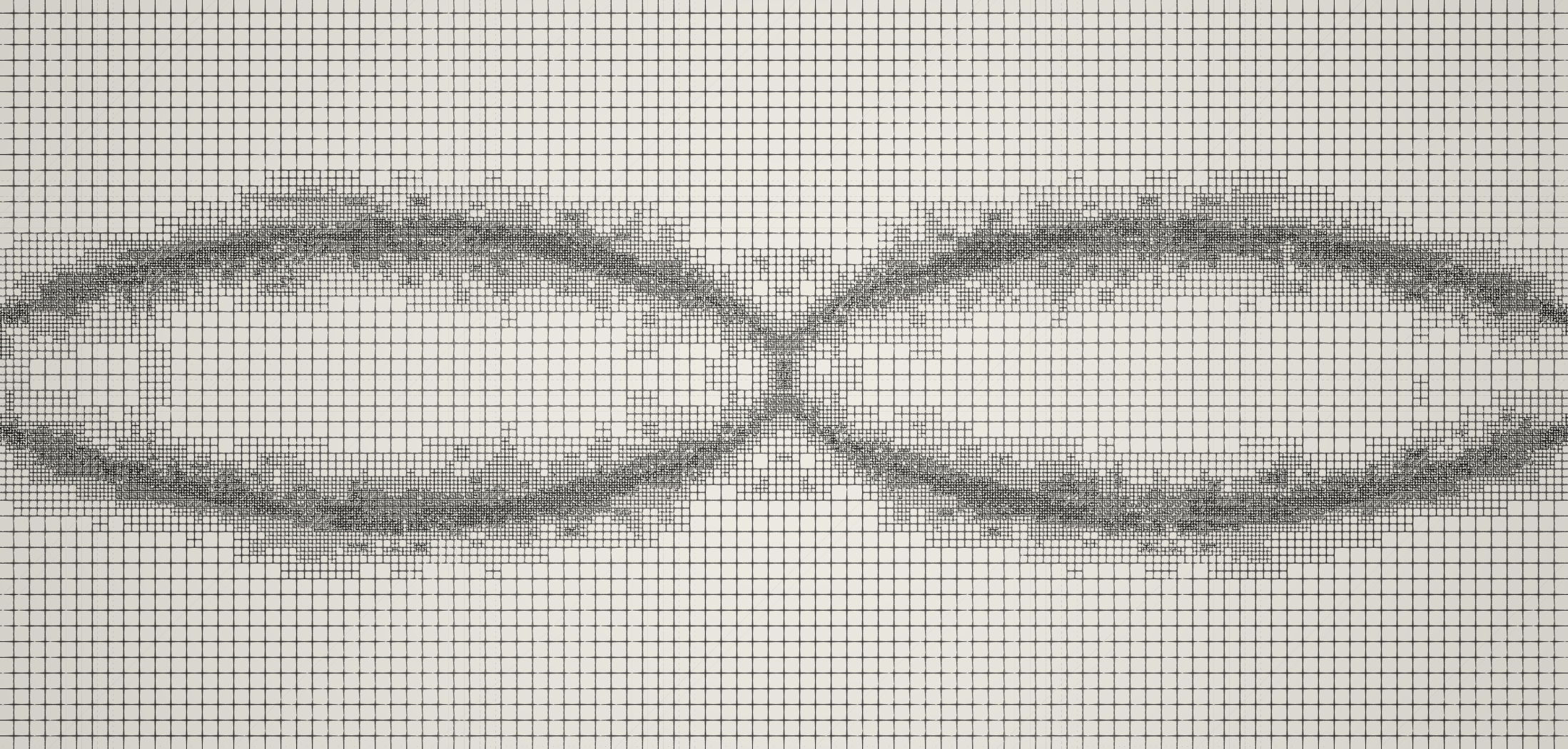}{\figWidth}};
    \draw(-.4,6) node[draw,fill=white,anchor=west,xshift=4pt,yshift=0pt] {\scriptsize $t=5$};
    \draw(9.2,6) node[draw,fill=white,anchor=west,xshift=4pt,yshift=0pt] {\scriptsize mesh};
  \draw(-.5,-.8) node[anchor=south west,xshift=0pt,yshift=+0pt] {\trimfig{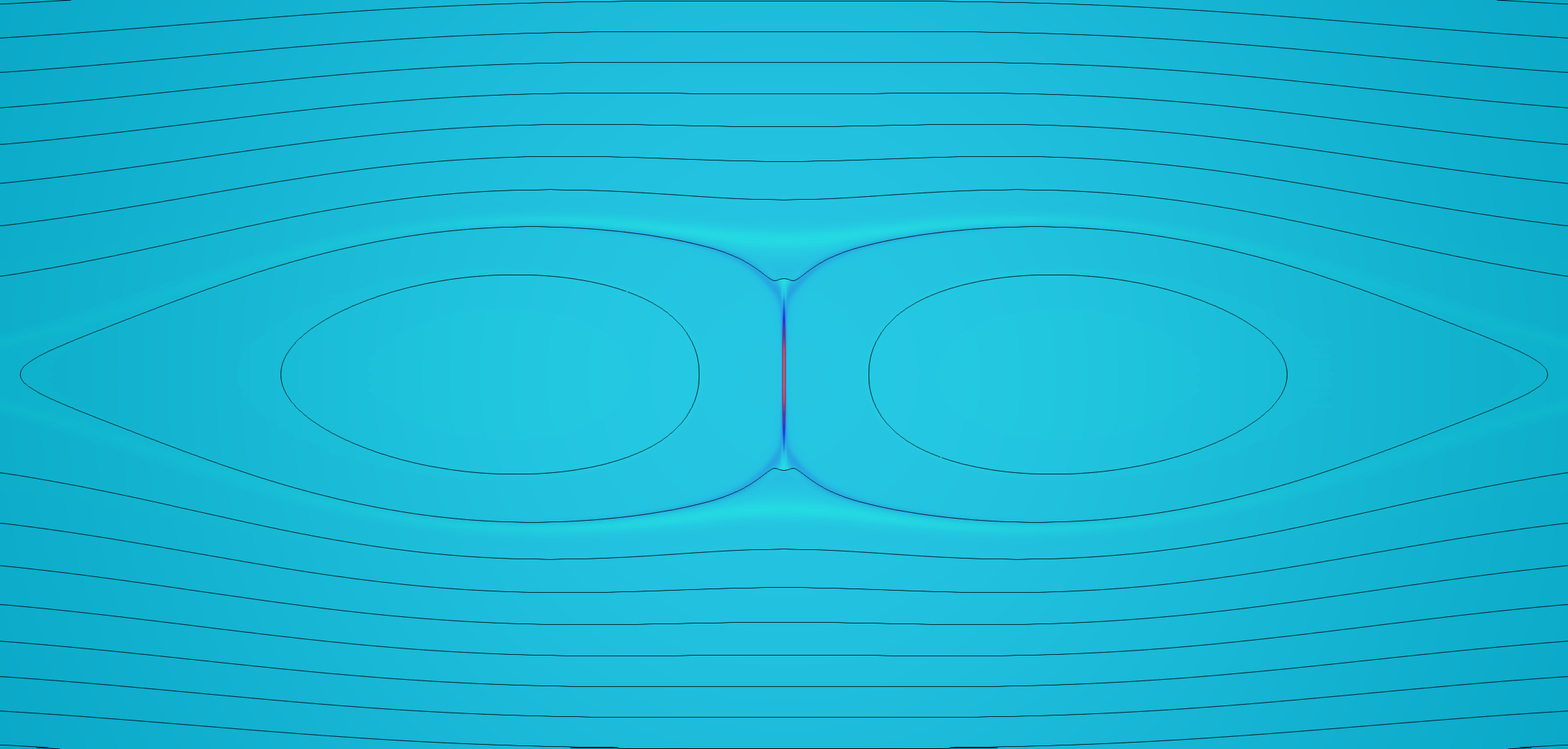}{\figWidth}};
  \draw(9.1,-.8) node[anchor=south west,xshift=0pt,yshift=+0pt] {\trimfig{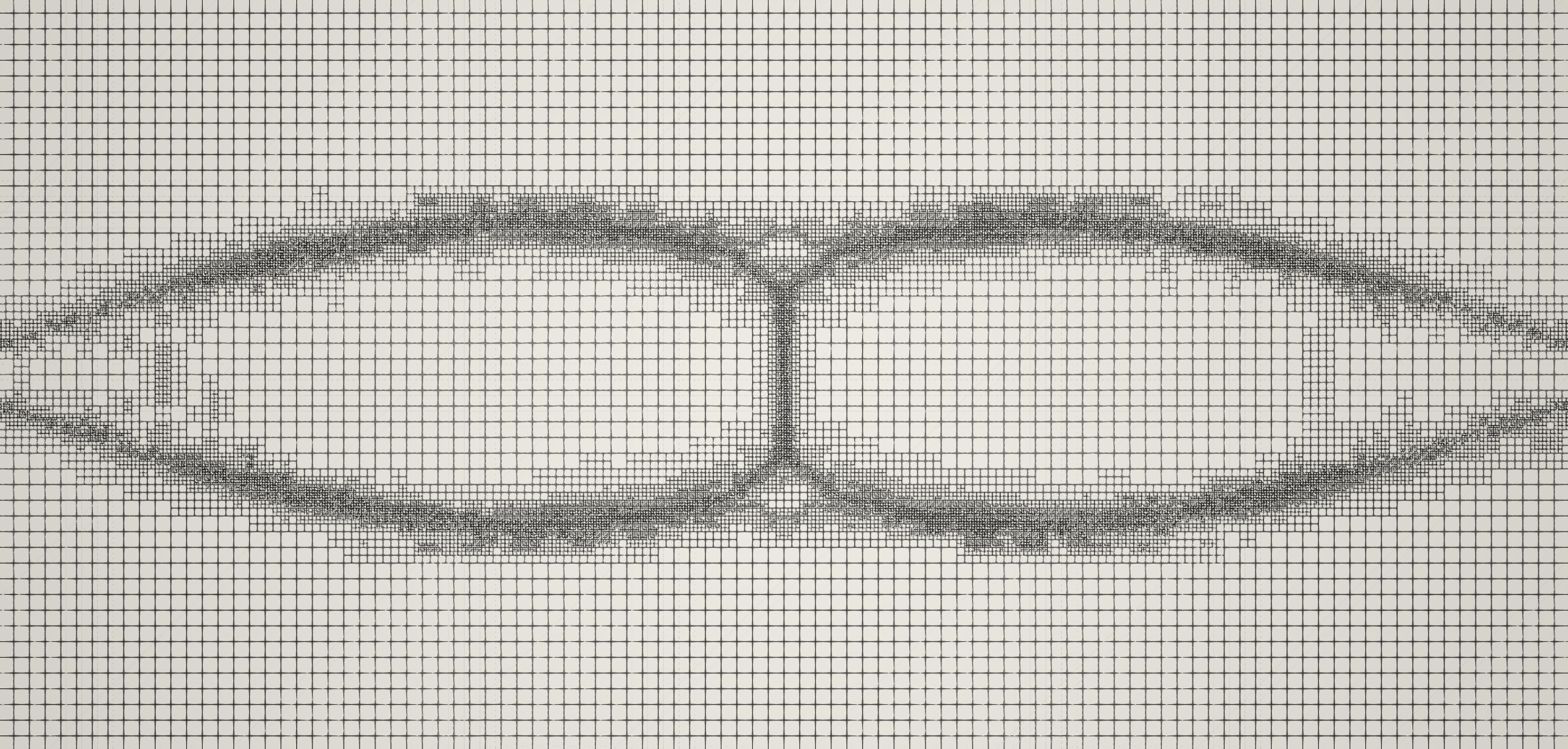}{\figWidth}};
    \draw(-.4,2.5) node[draw,fill=white,anchor=west,xshift=4pt,yshift=0pt] {\scriptsize $t=6$};
    \draw(9.2,2.5) node[draw,fill=white,anchor=west,xshift=4pt,yshift=0pt] {\scriptsize mesh};
        \end{scope}
      \draw(-.5,2.7) node[anchor=south west,xshift=0pt,yshift=+0pt] {\trimfig{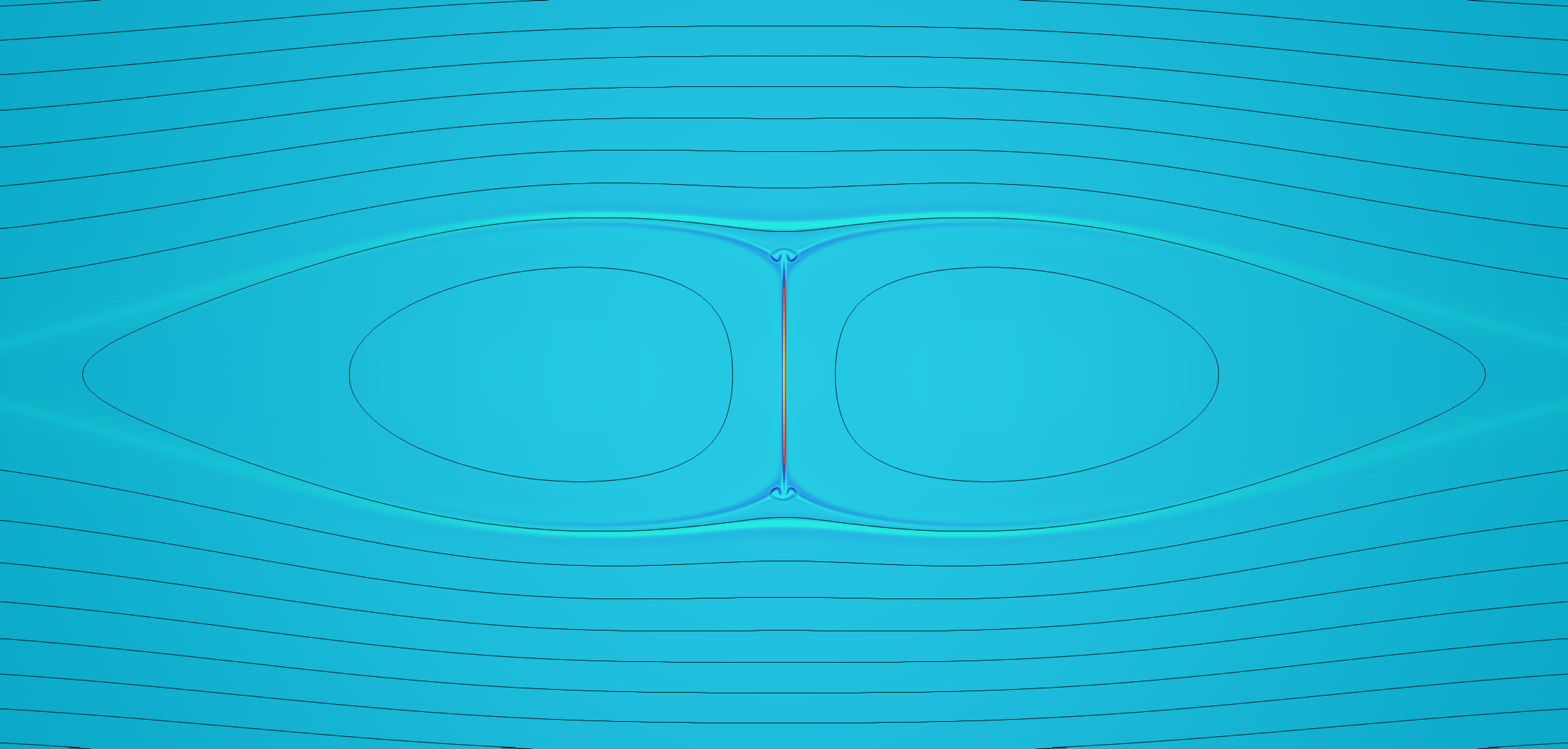}{\figWidth}};
  \draw(9.1,2.7) node[anchor=south west,xshift=0pt,yshift=+0pt] {\trimfig{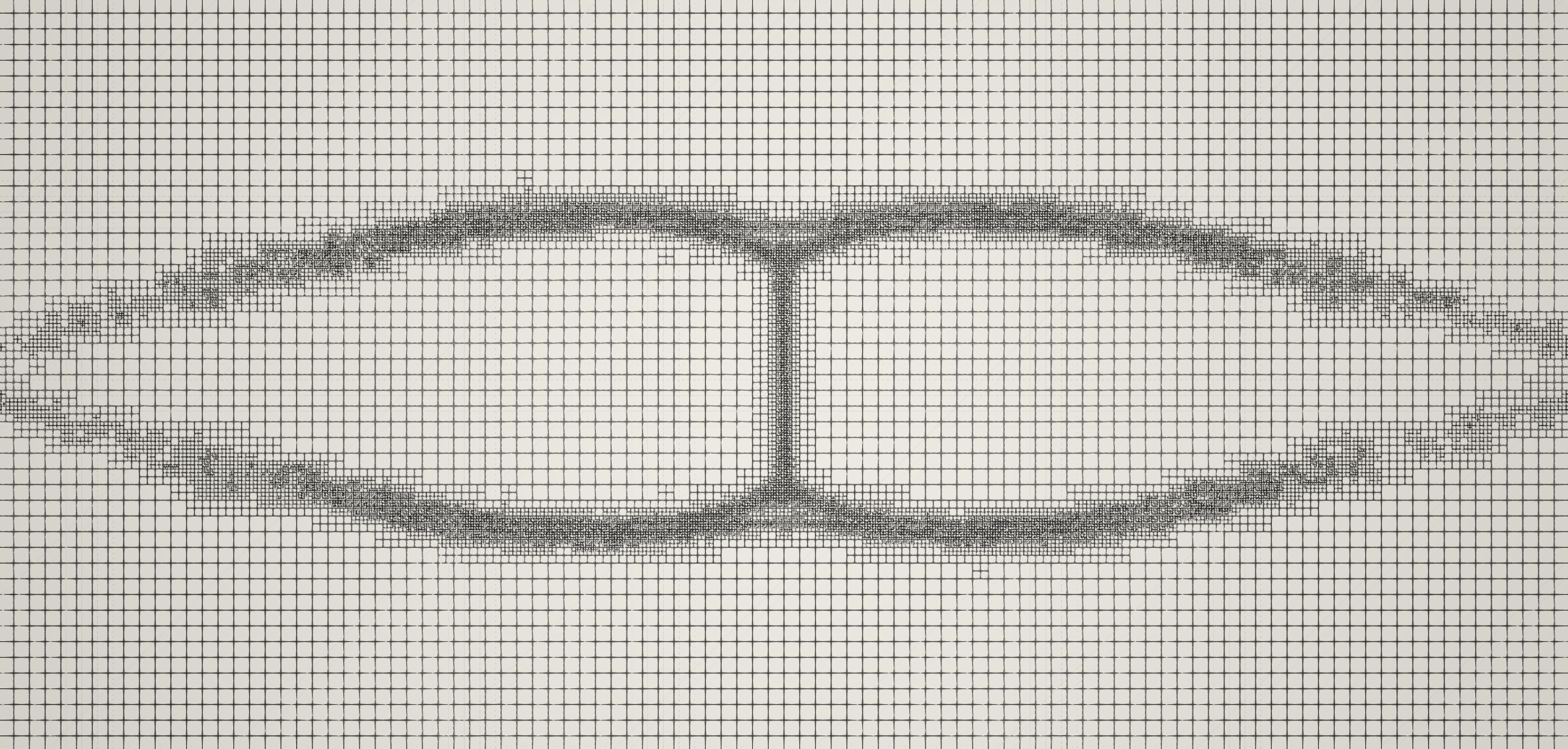}{\figWidth}};
    \draw(-.4,6) node[draw,fill=white,anchor=west,xshift=4pt,yshift=0pt] {\scriptsize $t=7$};
    \draw(9.2,6) node[draw,fill=white,anchor=west,xshift=4pt,yshift=0pt] {\scriptsize mesh};
  \draw(-.5,-.8) node[anchor=south west,xshift=0pt,yshift=+0pt] {\trimfig{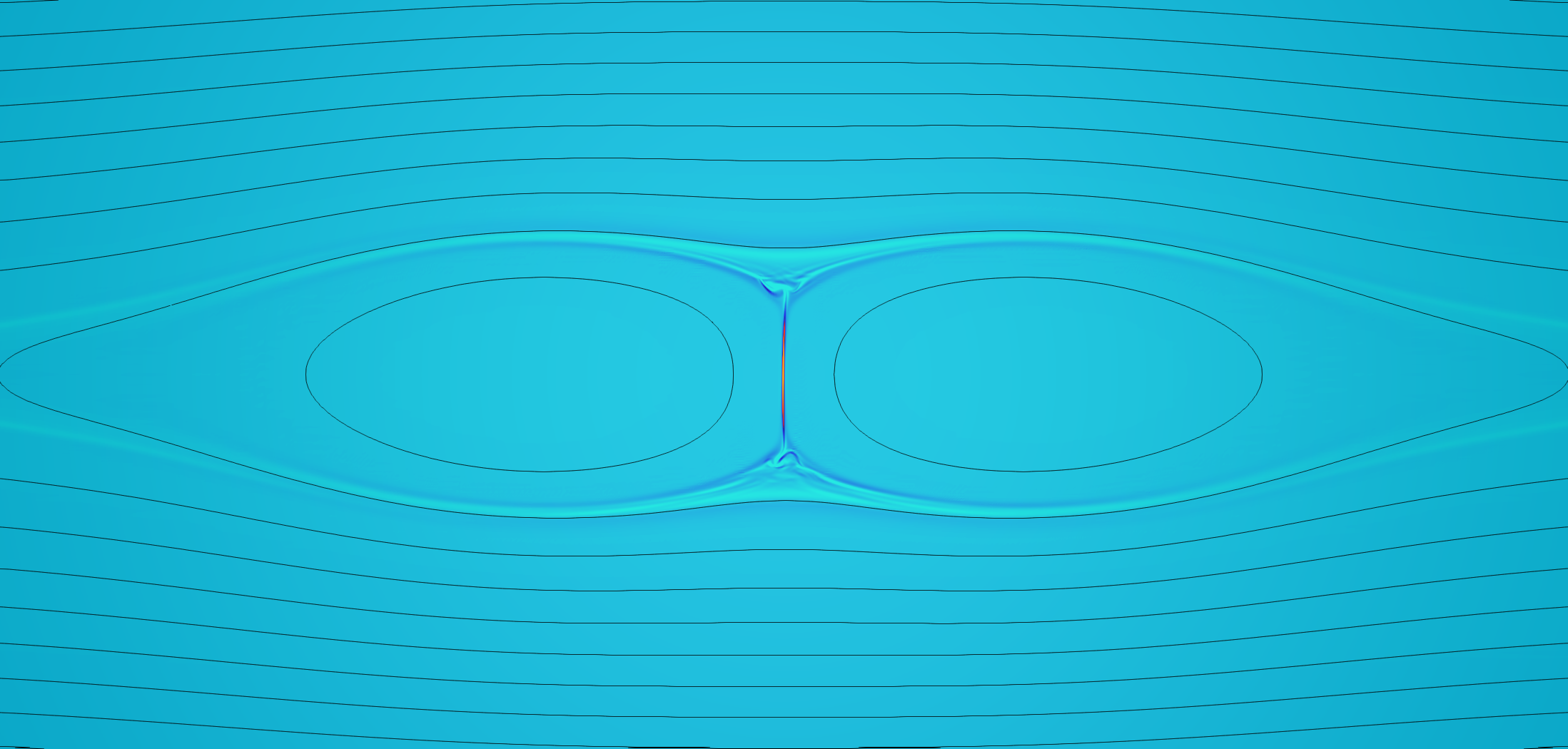}{\figWidth}};
  \draw(9.1,-.8) node[anchor=south west,xshift=0pt,yshift=+0pt] {\trimfig{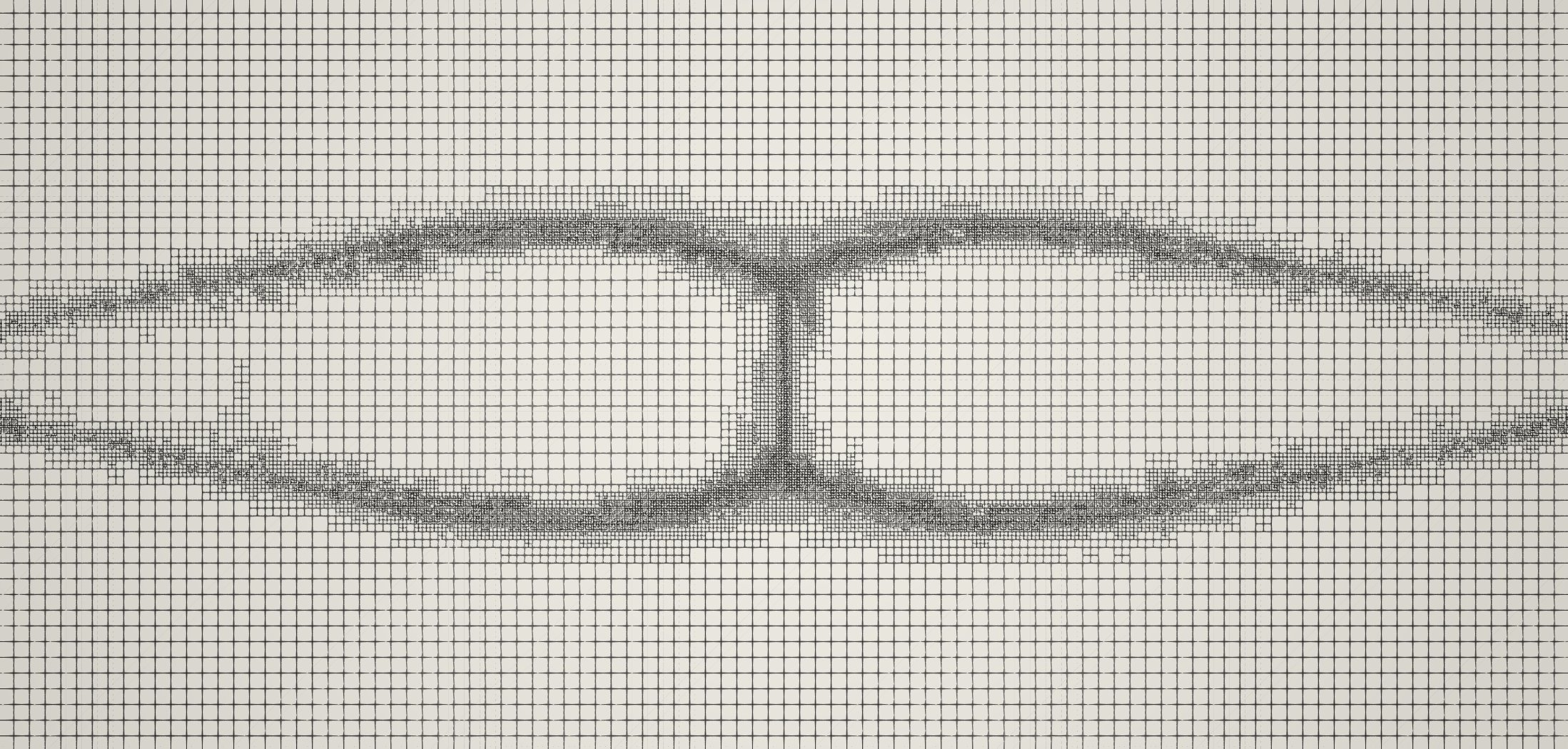}{\figWidth}};
    \draw(-.4,2.5) node[draw,fill=white,anchor=west,xshift=4pt,yshift=0pt] {\scriptsize $t=8$};
    \draw(9.2,2.5) node[draw,fill=white,anchor=west,xshift=4pt,yshift=0pt] {\scriptsize mesh};
%
\end{tikzpicture}
} 
\end{center}
  \caption{Island coalescence  of $\nu=\eta=$ $10^{-5}$. Solutions at different times and the corresponding adaptive meshes are  presented.
  The parallel current $J$ is presented as pseudocolor plots and the magnetic flux $\Psi$ is presented as black contours.
The colormap and contour values are fixed over time.
   Note the adaptive meshes capture the interesting structure in the current very well.
  }
  \label{fig:icAMROverT}
\end{figure}
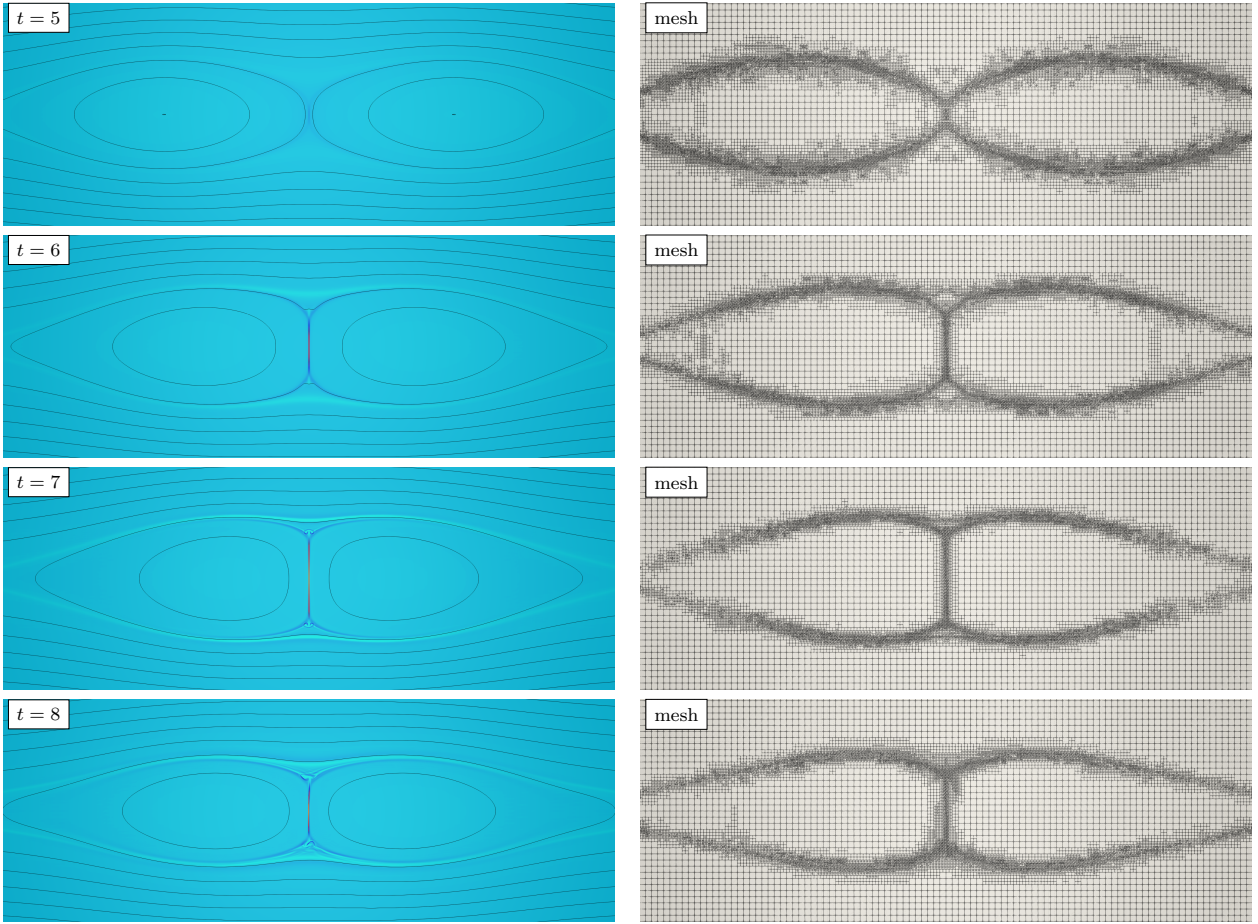
}

We first consider the case of $\nu=\eta=$ $10^{-5}$.
The solutions and the corresponding adaptive meshes at different times are presented in  Figure~\ref{fig:icAMROverT}.
The time step is fixed to $\dt=0.05$ throughout the simulation.
The number of refinement levels is four, and thus the adaptive grid is effectively comparable to the grid  in Figure~\ref{fig:icFixedMesh}.
Throughout the entire simulation,  the adaptive meshes capture the interesting structure in the current very well
and the meshes are very localized, with the base mesh covering the majority of the computational domain.
Therefore, the adaptive mesh saves a lot of computational cost even compared to an optimized locally packed grid like the one in Figure~\ref{fig:icFixedMesh}.
In fact, the average number of dofs in the entire run is 472961, which is 17\% of the locally packed grid  in Figure~\ref{fig:icFixedMesh} and 1.2\% of the effective uniform mesh.
The adaptive meshes are largely symmetric even at a later time.
At $t=8$, the solution is found to be slightly asymmetric, which is also observed in the fixed-mesh solution.
It is also interesting to observe the AMR algorithm starts to capture this asymmetry in the mesh at $t=8$.
The test indicates the AMR algorithm performs as expected, and the techniques such as refinement, coarsening and the error estimator are very effective
and suitable for  practical simulations like this.


{
\newcommand{\figWidth}{9.2cm}
\newcommand{\trimfig}[2]{\trimFig{#1}{#2}{.0}{.0}{.4}{.4}}
\newcommand{\trimfigz}[2]{\trimwb{#1}{#2}{.16}{.2}{0}{.05}}
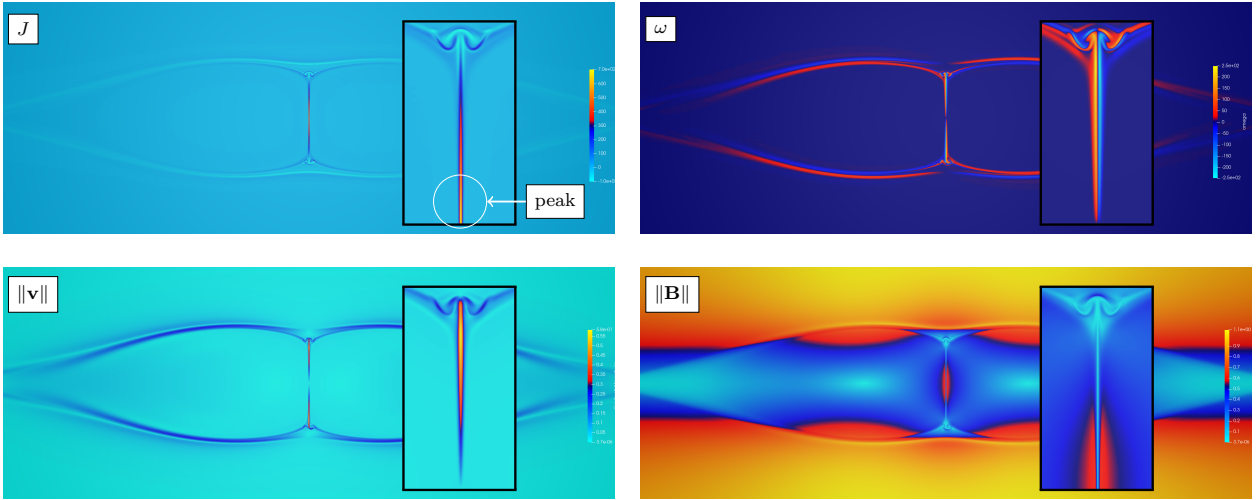
\begin{figure}[htb]
\begin{center}
\resizebox{16cm}{!}{
\begin{tikzpicture}[scale=1]
  \useasboundingbox (0.0,0) rectangle (18,7.);  
   \begin{scope}[yshift=4cm]
  \draw(-.5,-.75) node[anchor=south west,xshift=0pt,yshift=+0pt] {\trimfig{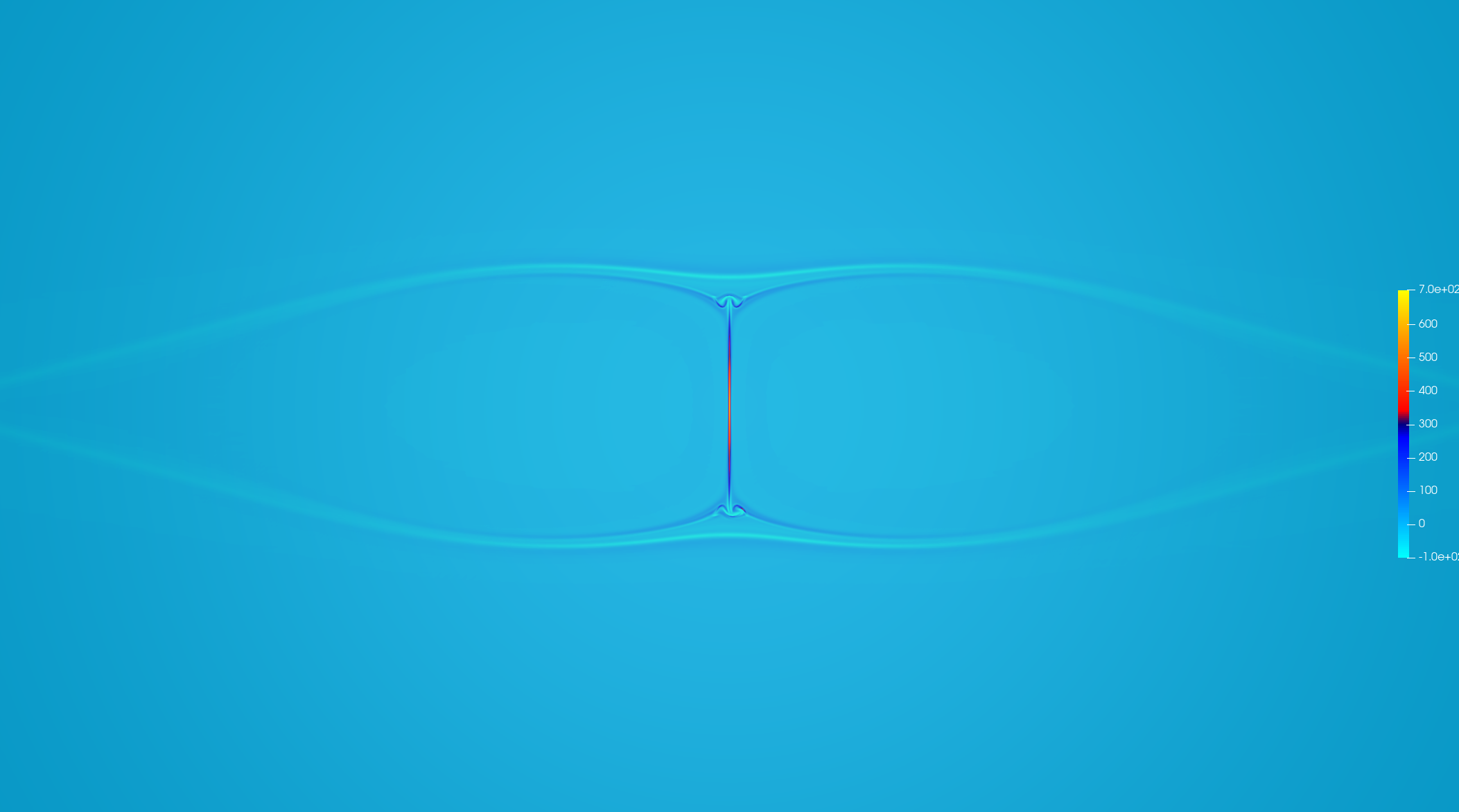}{\figWidth}};
  \draw(9.1,-.75) node[anchor=south west,xshift=0pt,yshift=+0pt] {\trimfig{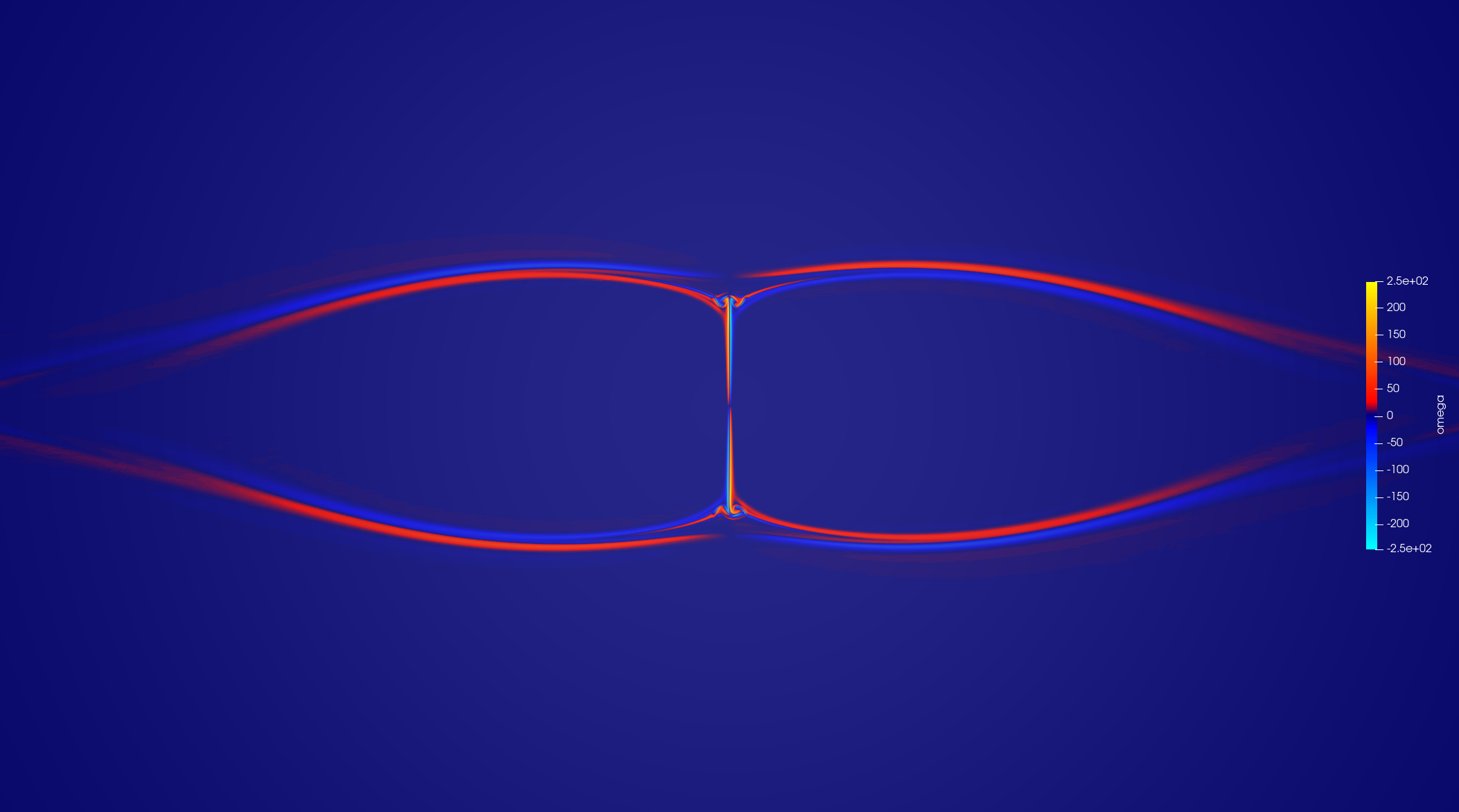}{\figWidth}};
  \draw(5.3,-.6) node[anchor=south west,xshift=-4pt,yshift=+0pt] {\trimfigz{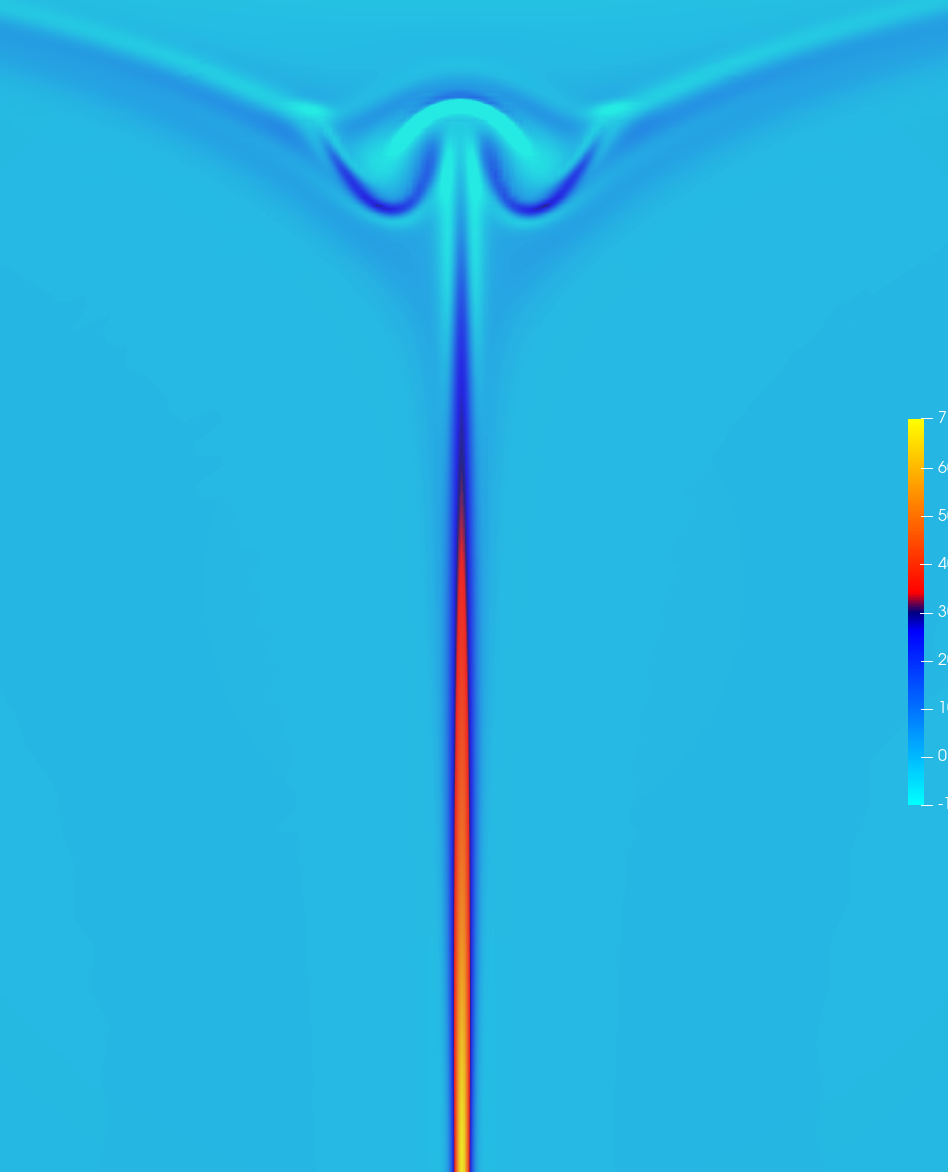}{1.65cm}};
    \draw(14.9,-.6) node[anchor=south west,xshift=-4pt,yshift=+0pt] {\trimfigz{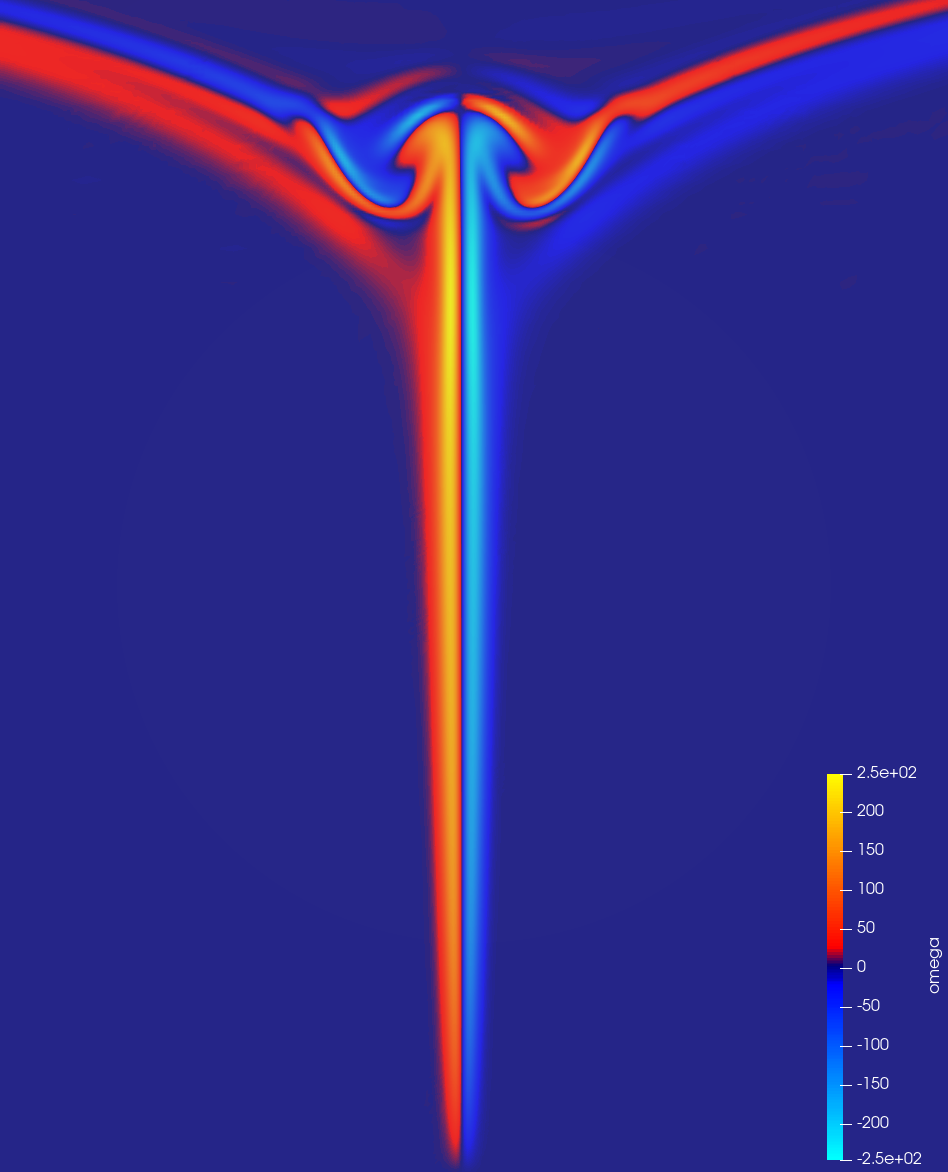}{1.65cm}};
    \draw(-.4,2.5) node[draw,fill=white,anchor=west,xshift=4pt,yshift=0pt] {\scriptsize $J$};
    \draw(9.2,2.5) node[draw,fill=white,anchor=west,xshift=4pt,yshift=0pt] {\scriptsize $\omega$};
\node [circle, minimum size=.8cm, draw, color=white] (b1) at (6.55,-.1) {};
\node[draw, fill=white, anchor=west,xshift=4pt,yshift=0pt] (b2) at (7.4,-.1) {\scriptsize peak};
\draw [<-, thick, white] (b1) -- (b2);
   \end{scope}

  \draw(-.5,-.75) node[anchor=south west,xshift=0pt,yshift=+0pt] {\trimfig{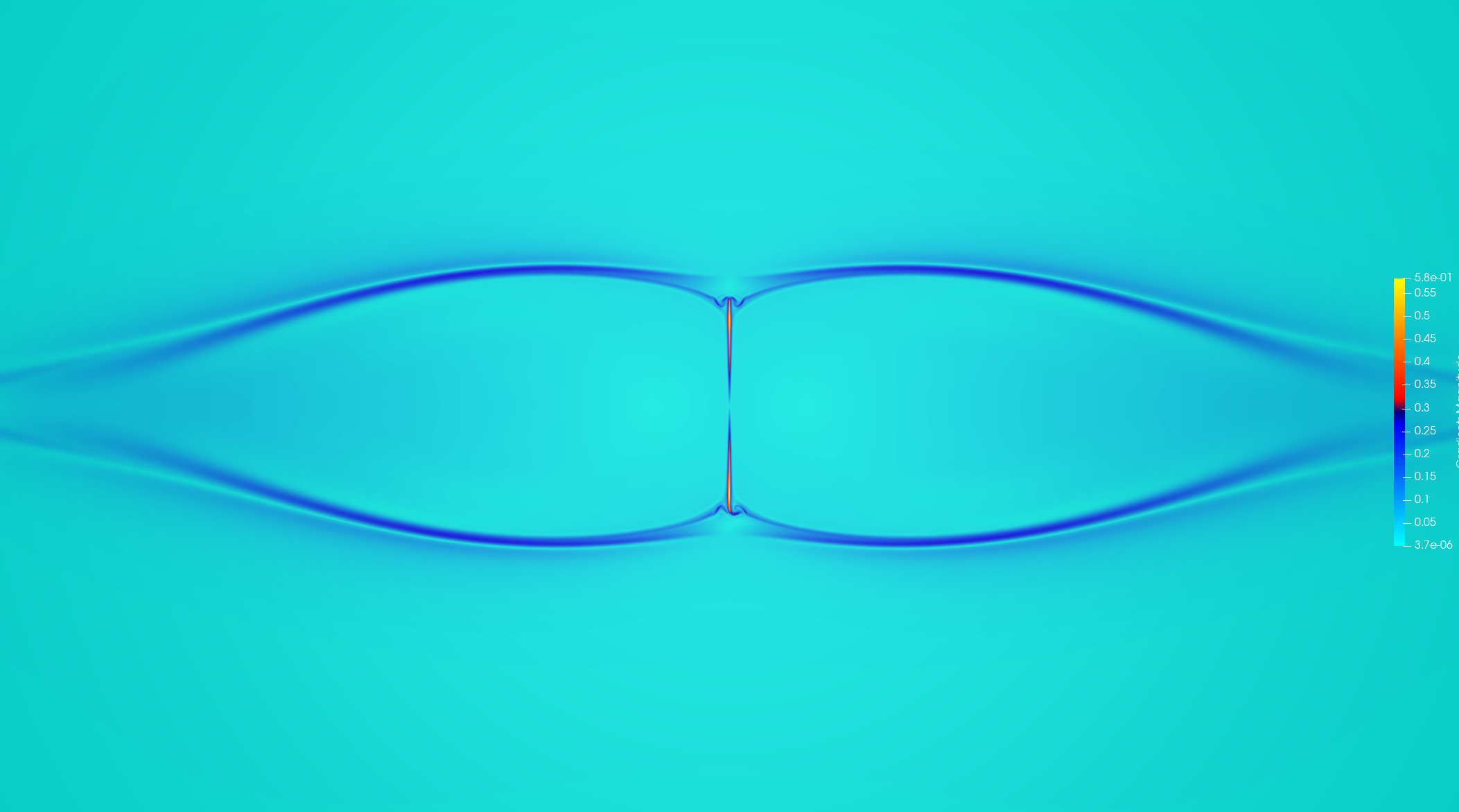}{\figWidth}};
  \draw(9.1,-.75) node[anchor=south west,xshift=0pt,yshift=+0pt] {\trimfig{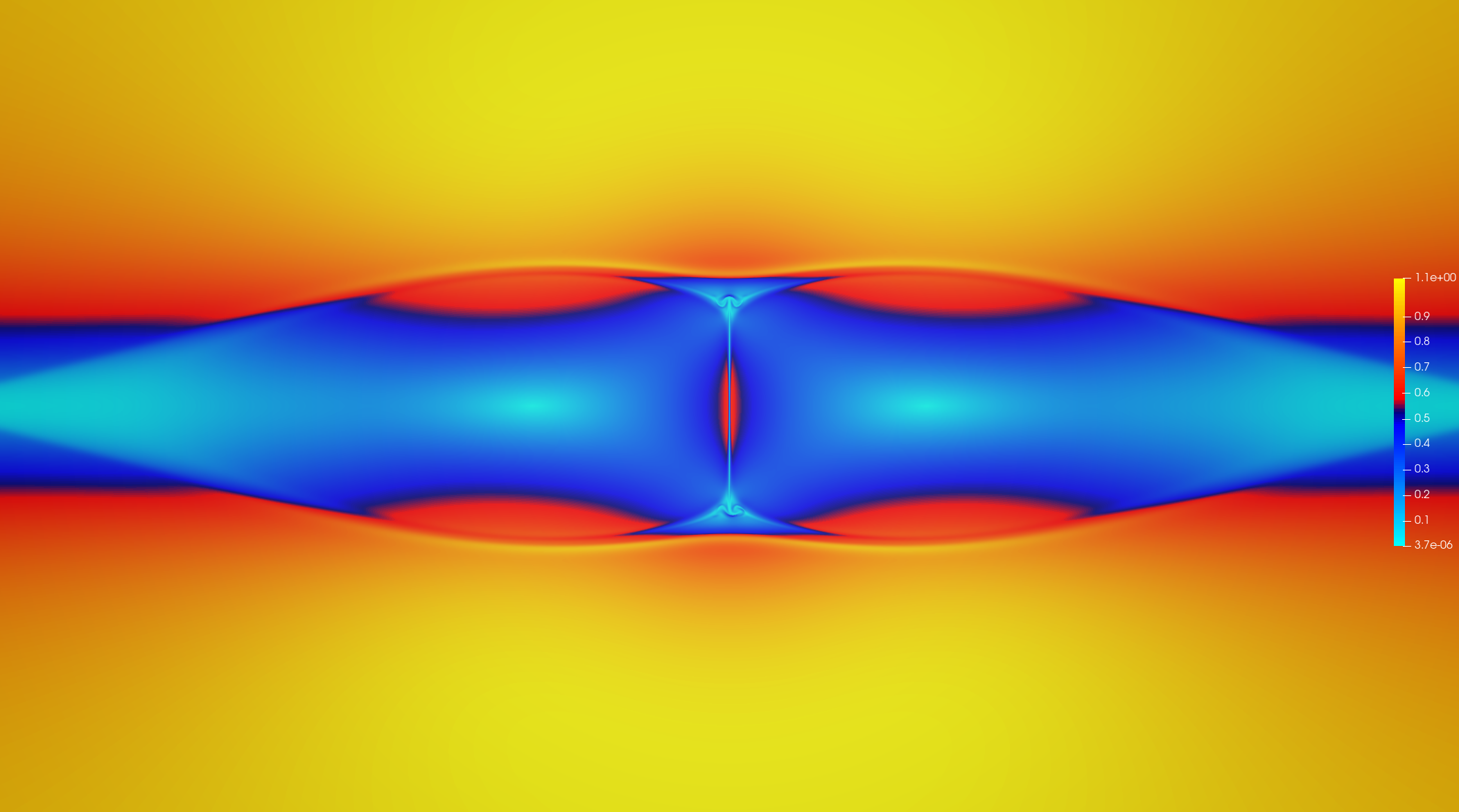}{\figWidth}};
   \draw(5.3,-.6) node[anchor=south west,xshift=-4pt,yshift=+0pt] {\trimfigz{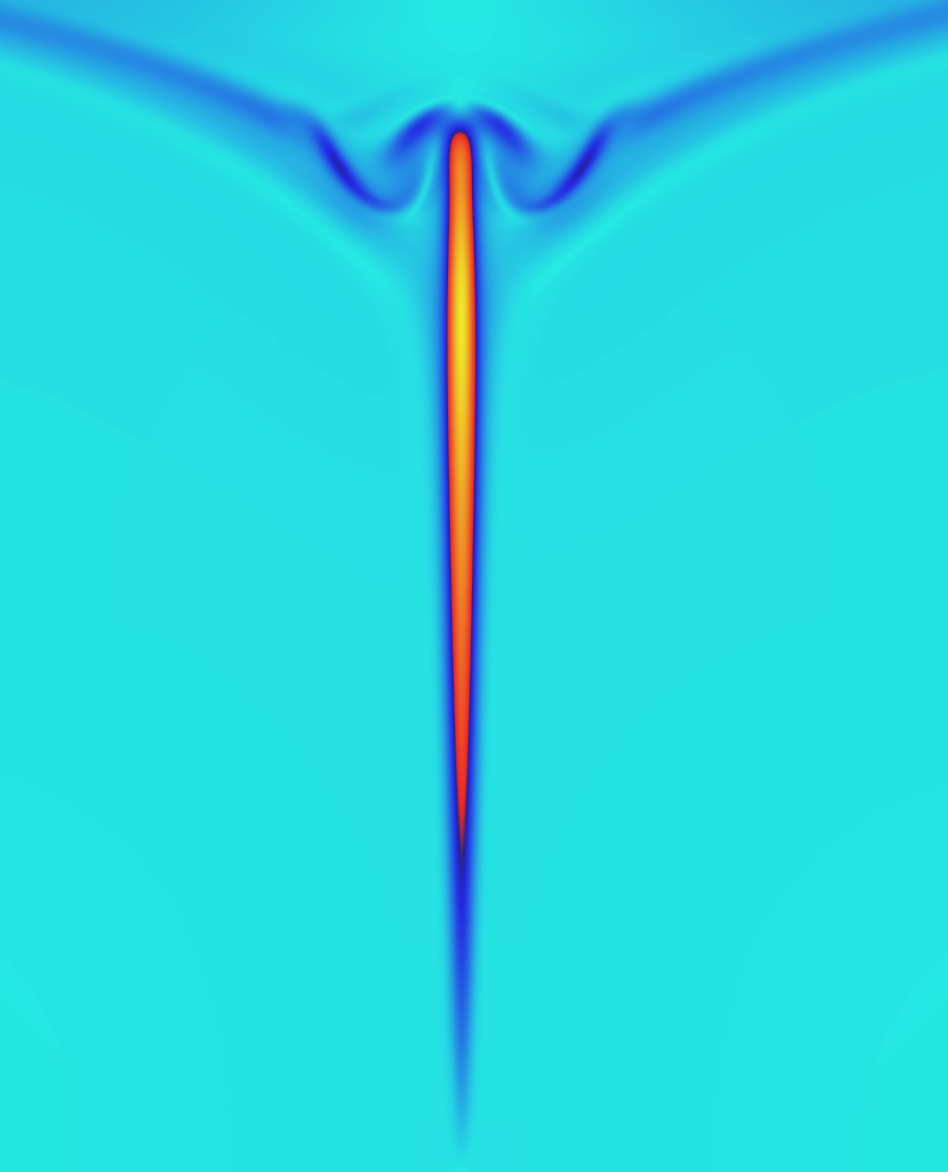}{1.65cm}};
   \draw(14.9,-.6) node[anchor=south west,xshift=-4pt,yshift=+0pt] {\trimfigz{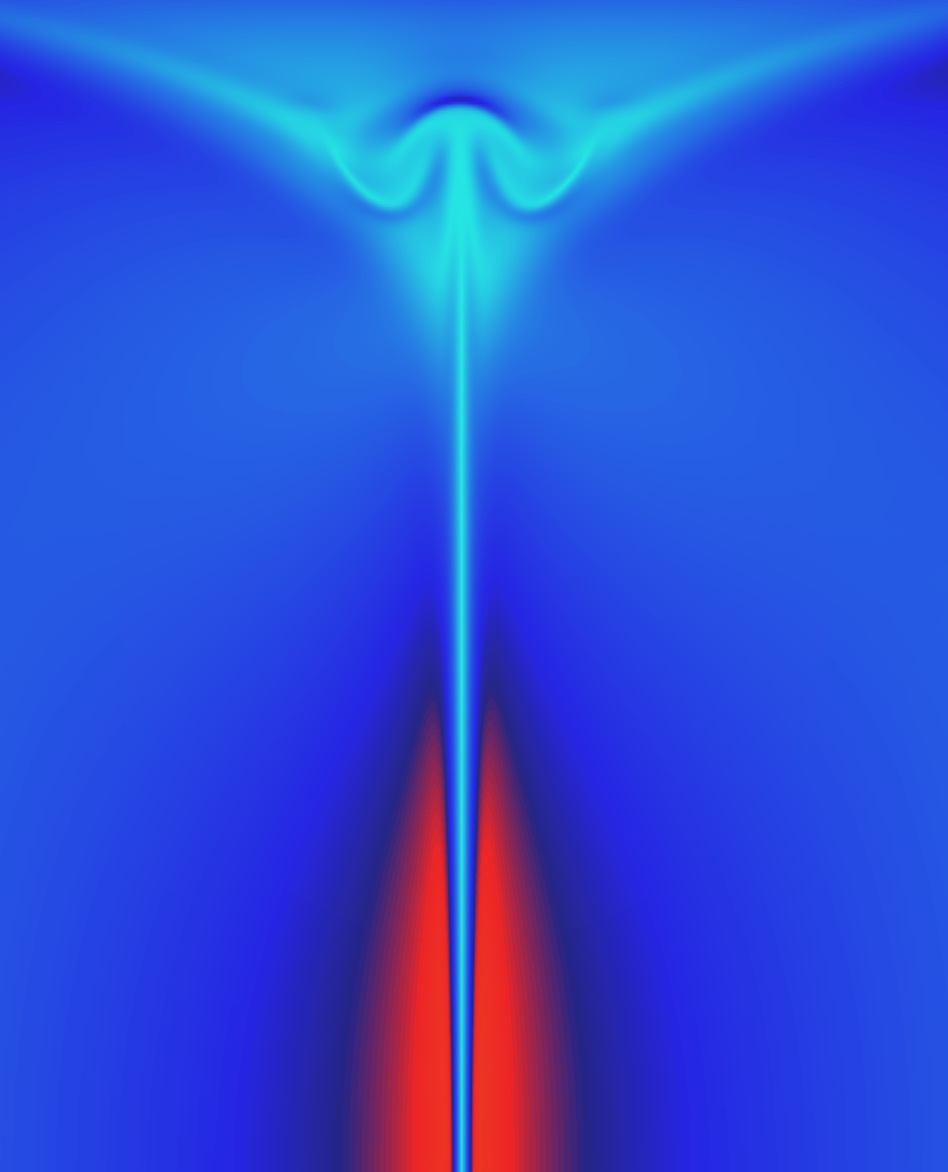}{1.65cm}};
    \draw(-.4,2.5) node[draw,fill=white,anchor=west,xshift=4pt,yshift=0pt] {\scriptsize $\|\vv\|$};
    \draw(9.2,2.5) node[draw,fill=white,anchor=west,xshift=4pt,yshift=0pt] {\scriptsize $\|\Bv\|$};
%
\end{tikzpicture}
} 
\end{center}
  \caption{Island coalescence  of $\nu=\eta=$ $10^{-5}$.
The current, vorticity, flow speed and magnetic field (with their zoomed-in views around the center sheet)  are presented at the peak time of $t=7.25$.
The zoomed-in views are focused on the top half part of the current sheet region.
The peak current at the center is about 700 and the highest flow speed is about 0.58 (with a normalized Alfv\'en speed of 1).
  }
  \label{fig:ic1e-5full}
\end{figure}
}

The  solution of the above run at the first reconnection peak time is presented in Figure~\ref{fig:ic1e-5full}.
With a Lundquist number of $10^{5}$, the structure still follows the current sheet in the Sweet-Parker model.
The current sheet is found to be well resolved and the largest current density in the current sheet is about 700.
The largest speed in the top and bottom jet regions of the current sheet is about 0.58.
The aspect ratio of this peak current sheet is about 70. Plasmoids are not observed in the entire simulation since the aspect ratio is not large enough.
(Plasmoids will appear for aspect ratios  greater than 100~\cite{biskamp1996magnetic}).

{Figure~\ref{fig:ic1e-5full} also illustrates the importance of including the stabilization terms in the physics-based preconditioners for these strongly hyperbolic operators.}
The stabilization terms are needed for two reasons. The obvious one is inversion of the advection-diffusion operator
$\Ac_{\nu}+\tau_\nu T$ in the preconditioner.
Without the SUPG term, the scheme becomes unstable and the inversion of the operator $\Ac_{\nu}$ through AMG will fail.
The second subtle reason is  the inversion of the Schur complement $    \Sc = \Ac_{\eta} + \tau_\eta S  -  N_{\Bv_0}
    \Dc_{\nu}^{-1}  N_{\Bv_0}$.
We note that around the jet region the magnetic-field strength is very small.
 Therefore, the Schur complement  is dominated by the advection-diffusion operator $\Ac_{\eta}$, which
is dominated by the strong advective term.

{
\newcommand{\figWidth}{9.2cm}
\newcommand{\figWidthc}{6cm}
\newcommand{\trimfig}[2]{\trimFig{#1}{#2}{.3}{.3}{.52}{.52}}
\newcommand{\trimfigb}[2]{\trimFig{#1}{#2}{.3}{.3}{.0}{.0}}
\newcommand{\trimfigd}[2]{\trimFig{#1}{#2}{.4}{.2}{.0}{.0}}
\newcommand{\trimfigc}[2]{\trimFig{#1}{#2}{1}{1}{.0}{.0}}
\begin{figure}[htb]
\begin{center}
\resizebox{16cm}{!}{
\begin{tikzpicture}[scale=1]
  \useasboundingbox (0.0,0) rectangle (18,8.5);  
     \begin{scope}[yshift=3.3cm]
  \draw(-.5,2.) node[anchor=south west,xshift=0pt,yshift=+0pt] {\trimfig{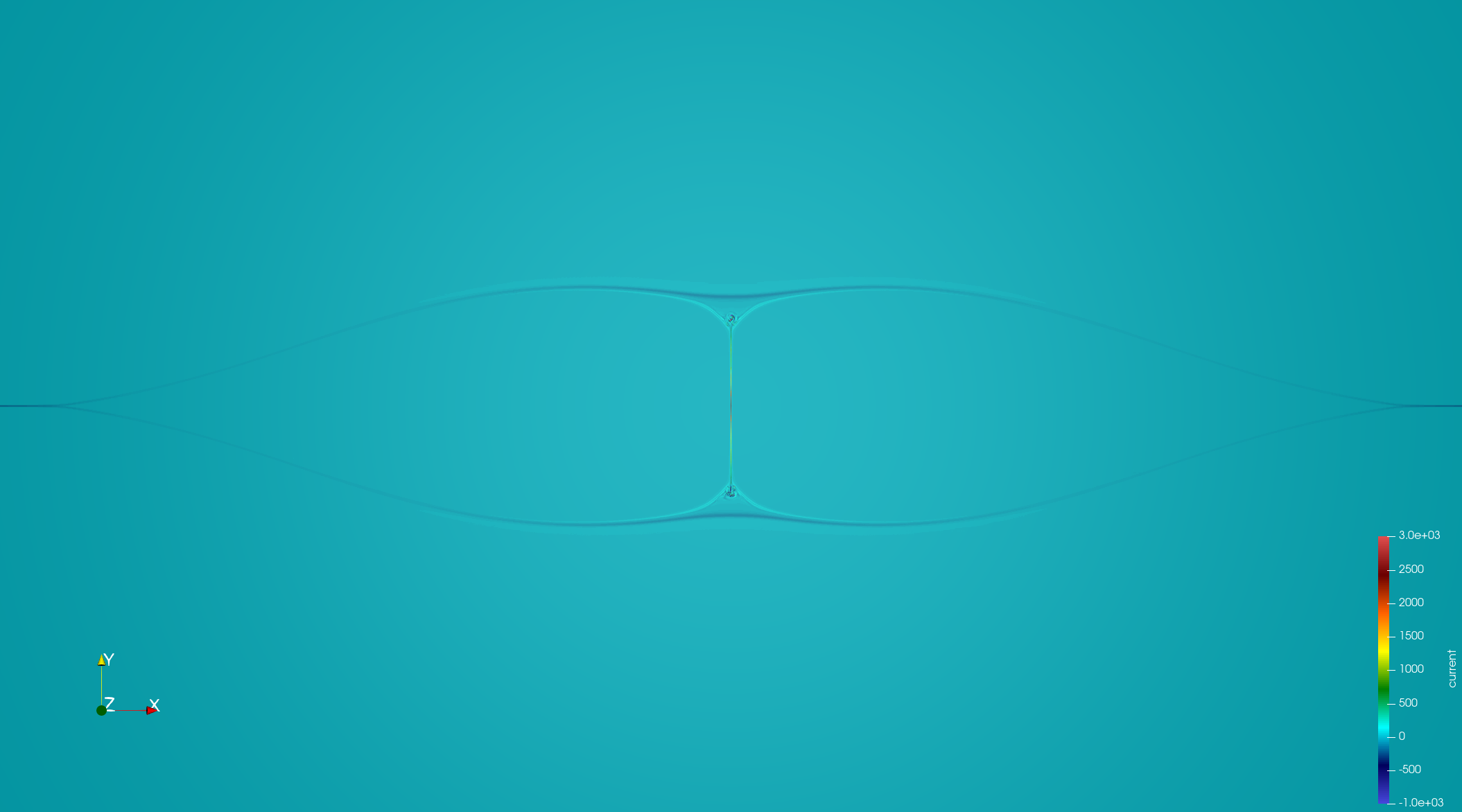}{\figWidth}};
  \draw(8.9,2.) node[anchor=south west,xshift=0pt,yshift=+0pt] {\trimfig{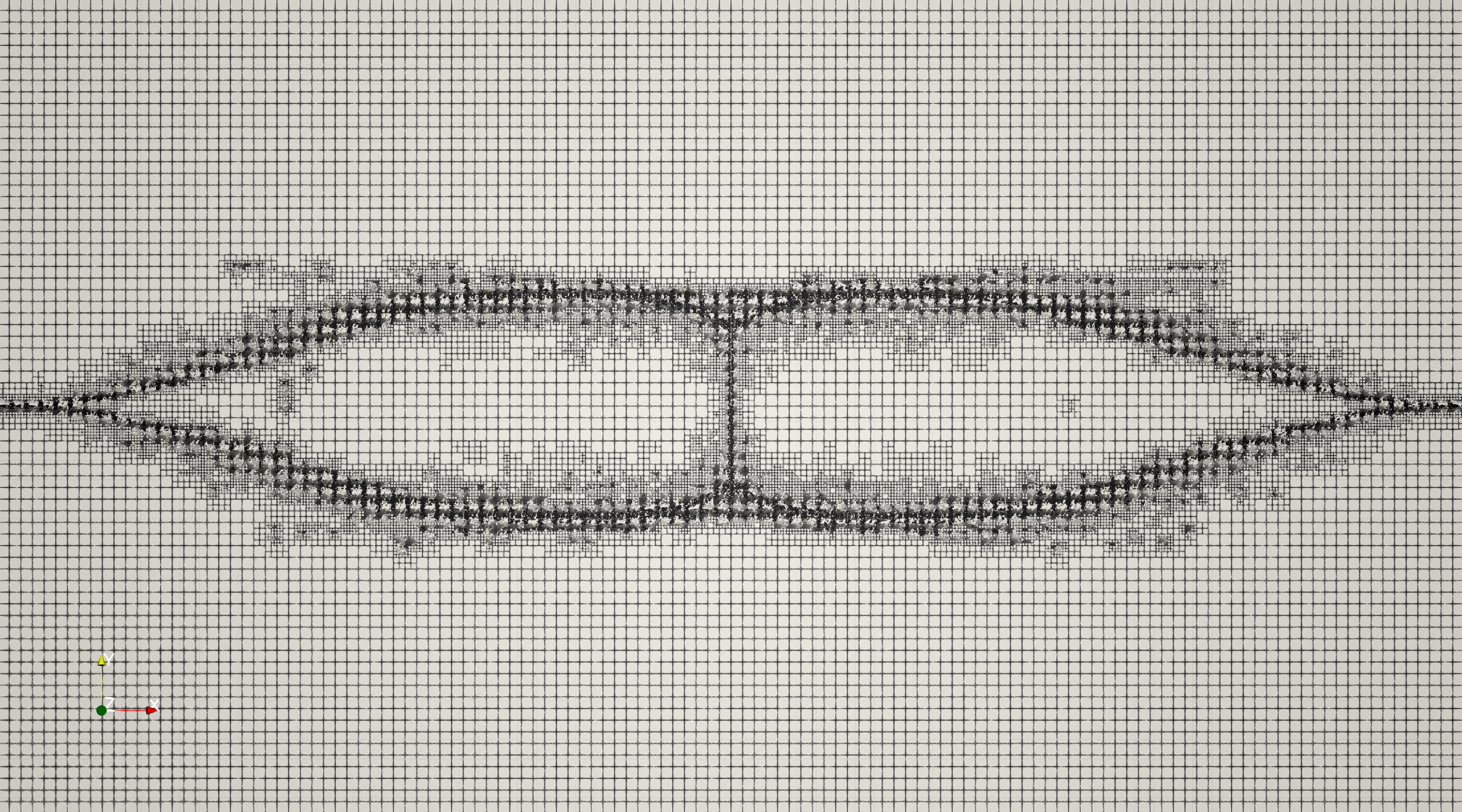}{\figWidth}};
  \draw(-.4,5.) node[draw,fill=white,anchor=west,xshift=4pt,yshift=0pt] {\scriptsize $t=6.5$};
  \draw(9,5.) node[draw,fill=white,anchor=west,xshift=4pt,yshift=0pt] {\scriptsize mesh};
  \draw(-.5,-4.) node[anchor=south west,xshift=0pt,yshift=+0pt] {\trimfigb{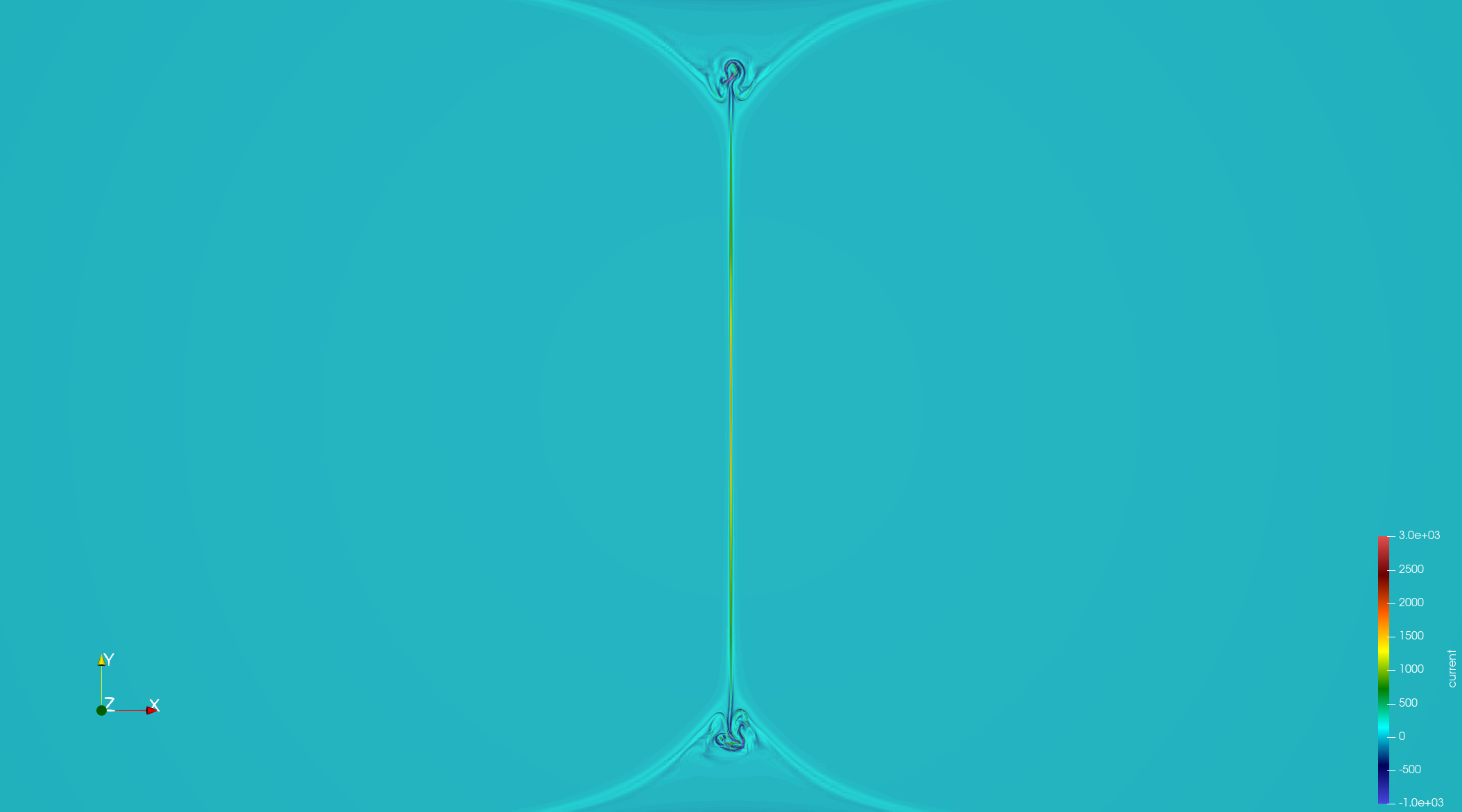}{\figWidth}};
  \draw(8.9,-4.) node[anchor=south west,xshift=0pt,yshift=+0pt] {\trimfigd{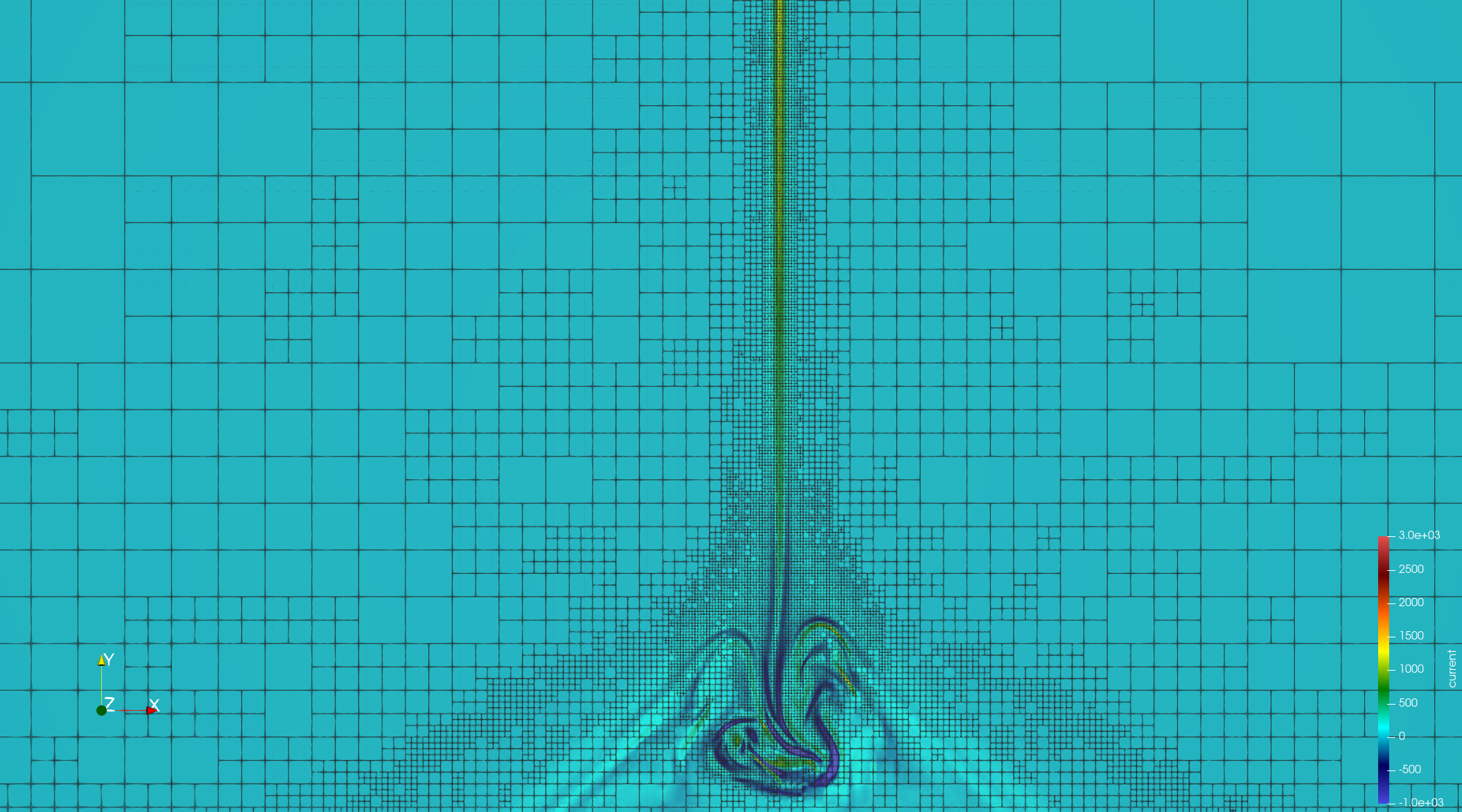}{\figWidth}};
    \draw(-.4,1.7) node[draw,fill=white,anchor=west,xshift=4pt,yshift=0pt] {\scriptsize zoomed-in 1};
    \draw(9.,1.7) node[draw,fill=white,anchor=west,xshift=4pt,yshift=0pt] {\scriptsize zoomed-in 2};
     \node [draw, minimum height=1cm,minimum width=1.5cm, draw, color=white, dashed] (b1) at (4.27,-3) {};
\node[draw, fill=white, anchor=west,xshift=4pt,yshift=0pt] (b2) at (5.4,-3) {\tiny zoomed-in 2};
\draw [<-, thick, white] (b1) -- (b2);
     \node [draw, minimum height=1.6cm,minimum width=2.6cm, draw, color=white,dashed] (b1) at (4.27,3.7) {};
\node[draw, fill=white, anchor=west,xshift=4pt,yshift=0pt] (b2) at (6,2.5) {\tiny zoomed-in 1};
\draw [<-, thick, white] (b1) -- (b2);
      \end{scope}
      %
\end{tikzpicture}
} 
\end{center}
  \caption{Island coalescence of $\nu = \eta= $ $10^{-6}$. Top row: current at $t = 6.5$ and the corresponding adaptive mesh. Here 6 levels of refinement are used and $p=3$.
  The mesh captures the interesting structure in the current very well.
  Second row: the zoomed-in views around the current sheet region. The left plot focuses on the center region while the right plot zooms in further to visualize the bottom
  jet and current sheet. Note the jet and the current sheet are very well captured by AMR.
   }
  \label{fig:ic1e-6}
\end{figure}
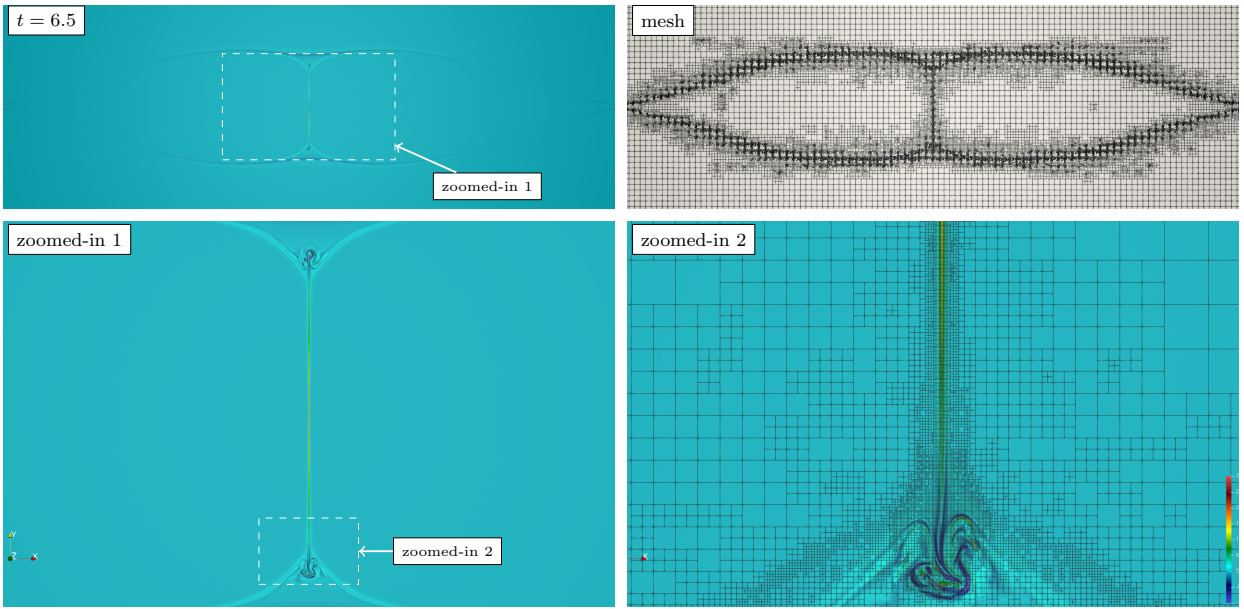
}

Next we consider the case of $\nu = \eta= $ $10^{-6}$.
In this test, we use 6  levels of refinement on top of the $128\times128$ base mesh, $p=3$ and an initial time step of $0.05$.
Unlike the previous case,  plasmoids do develop  in the simulation.
To accommodate the fast dynamics of plasmoids, we use an approach similar to~\cite{shadid2010current,muralikrishnan2020multilevel} that  reduce time step dynamically when the local dynamics become too fast.
Specifically, if the maximum number of Newton iterations is reached, we reduce the time step by a factor of two. 
After the time step is reduced, the simulation runs with the new time step. If the solver successfully runs with the same time step for 10 steps, we start to slowly increase the time step by 10\%.
The time step will continuously increase every 10 time steps until the solver fails again or reaches the original target maximum time step.
This approach is found to be sufficient to run through the entire simulation.
The smallest time step  for this run is found to be 0.003125.
The computational cost is  significantly reduced compared to the uniform mesh case.
The average number of dofs in this run is 1.30M, while a uniform mesh of the same resolution needs  604M dofs.
As a result, the implicit adaptive solver  needs only 0.21\% of the total dofs compared to a comparable uniform mesh.

The solutions and mesh with their zoomed-in views at $t = 6.5$ are depicted in Figure~\ref{fig:ic1e-6}.
The zoomed-in views show that the current sheet and jet regions are very well resolved, indicating the error estimator is very effective.
The time slice of $t = 6.5$ is right before the first plasmoid appears at $t = 6.525$.
At $t = 6.5$, the aspect ratio is estimated as 200 and the appearance of the plasmoid is consistent with the estimate given in~\cite{biskamp1996magnetic}.
The highest current density in the center peak region is about 2400.
In the next time step, the first plasmoid appears right above the central peak region.

{
\newcommand{\figWidth}{9.2cm}
\newcommand{\figWidthc}{6cm}
\newcommand{\trimfig}[2]{\trimFig{#1}{#2}{.3}{.3}{.52}{.52}}
\newcommand{\trimfigb}[2]{\trimFig{#1}{#2}{.3}{.3}{.0}{.0}}
\newcommand{\trimfigd}[2]{\trimFig{#1}{#2}{.35}{.25}{.0}{.0}}
\newcommand{\trimfigc}[2]{\trimFig{#1}{#2}{1}{1}{.0}{.0}}
\begin{figure}[htb]
\begin{center}
\resizebox{16cm}{!}{
\begin{tikzpicture}[scale=1]
  \useasboundingbox (0.0,0) rectangle (18,12.4);  
  \draw(-.5,-1) node[anchor=south west,xshift=0pt,yshift=+0pt] {\trimfigc{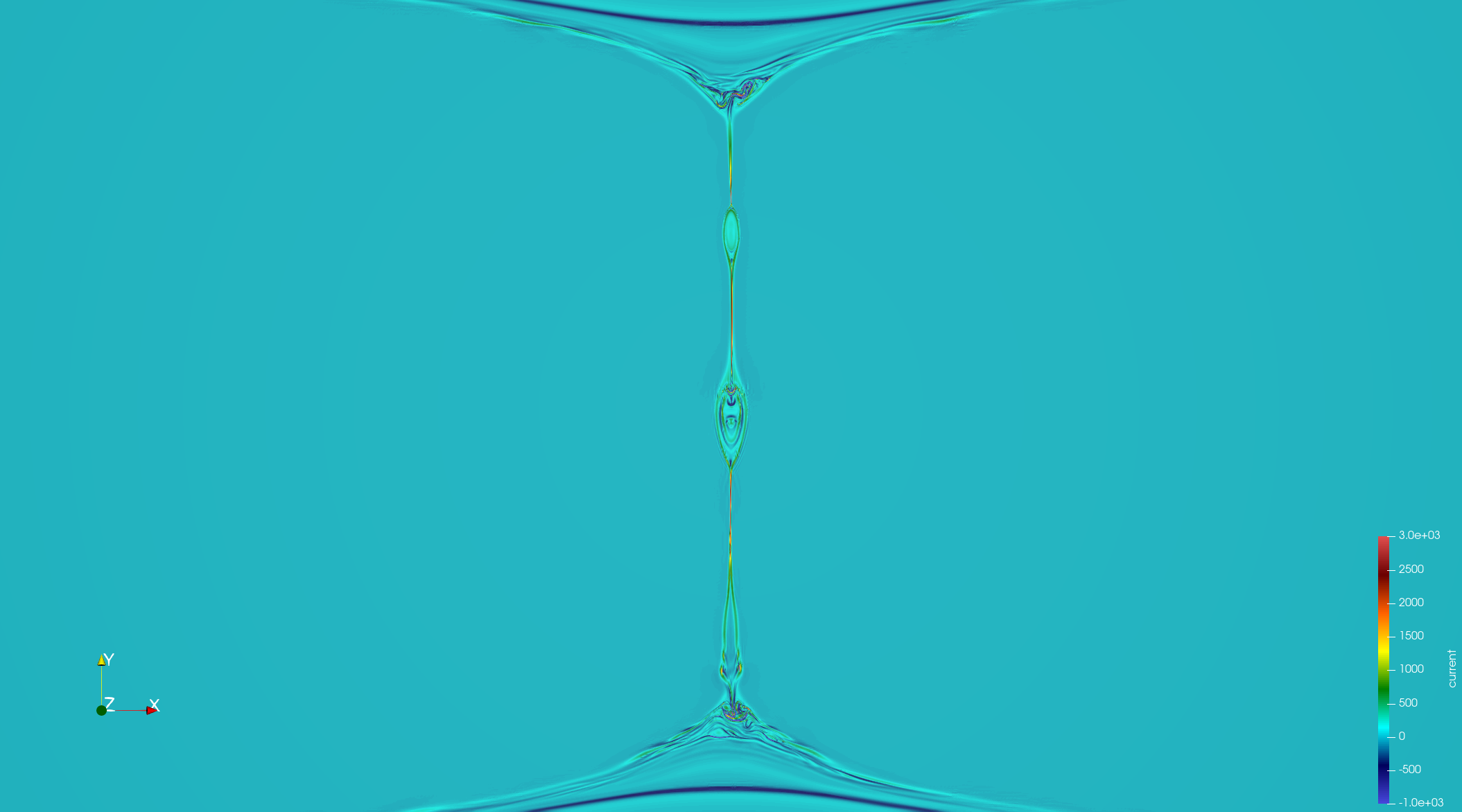}{\figWidthc}};
  \draw(5.8,-1) node[anchor=south west,xshift=0pt,yshift=+0pt] {\trimfigc{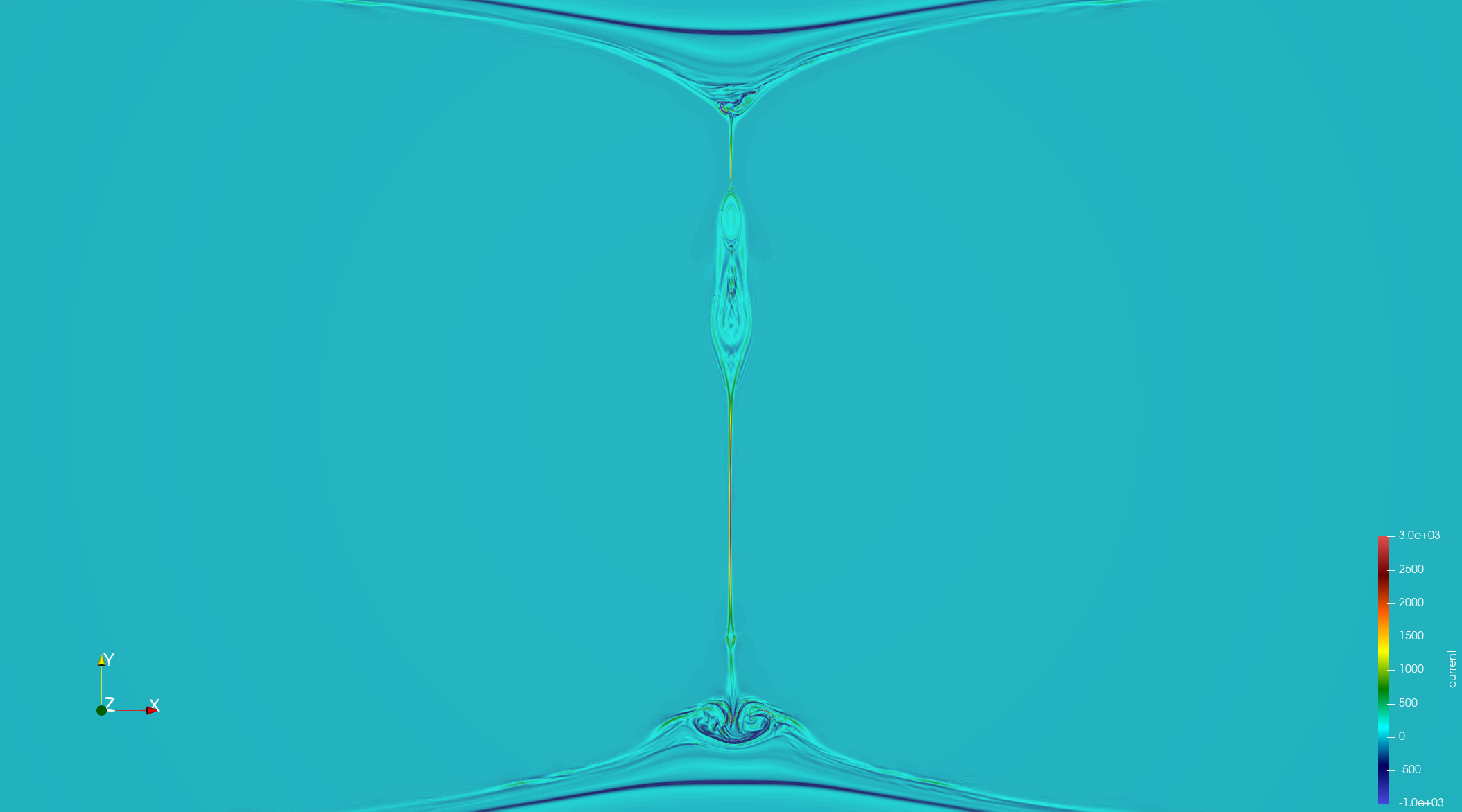}{\figWidthc}};
    \draw(12.1,-1) node[anchor=south west,xshift=0pt,yshift=+0pt] {\trimfigc{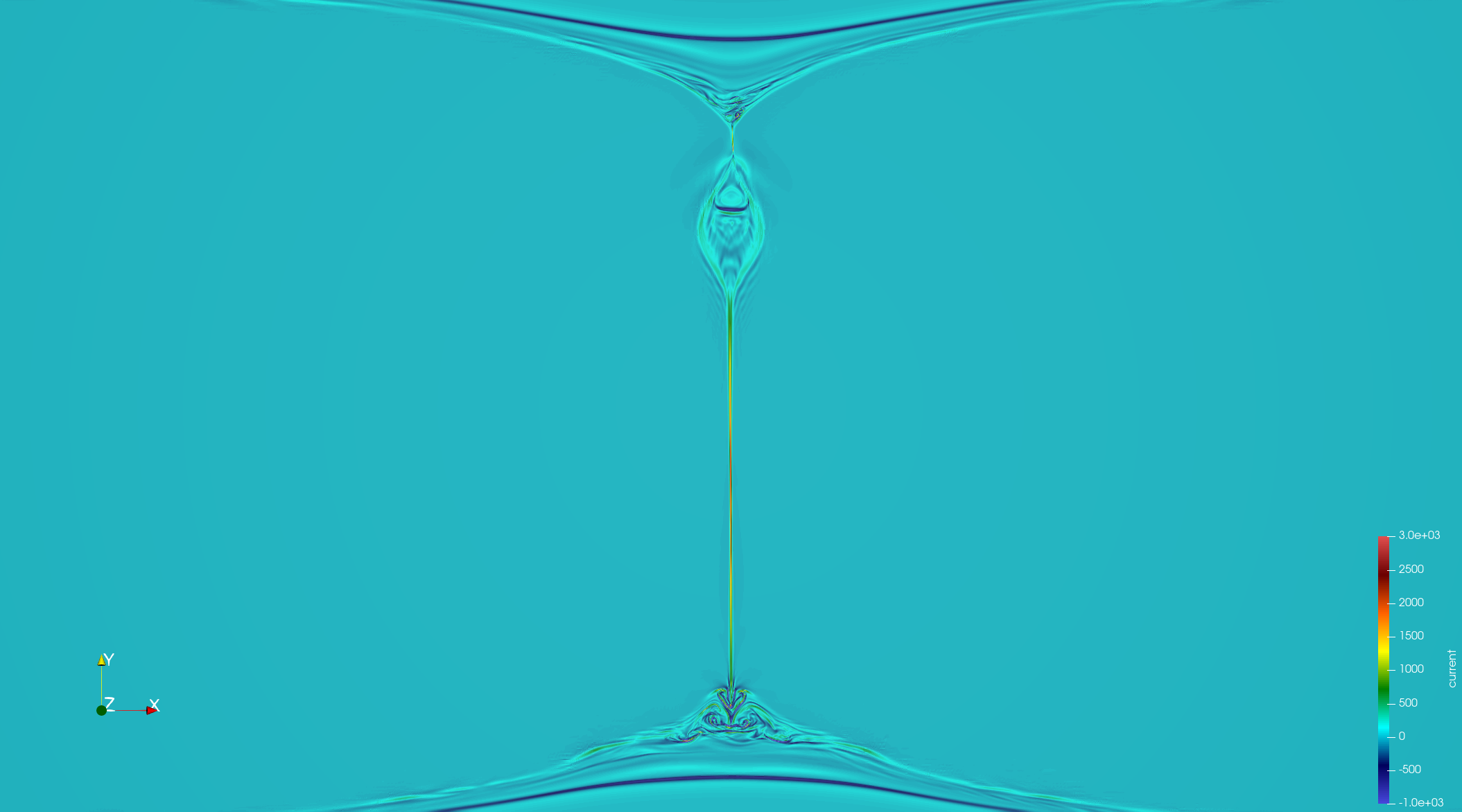}{\figWidthc}};
    \draw(-.4,5.4) node[draw,fill=white,anchor=west,xshift=4pt,yshift=0pt] {\scriptsize $t=7.28$};
    \draw(5.9,5.4) node[draw,fill=white,anchor=west,xshift=4pt,yshift=0pt] {\scriptsize $t=7.36$};
    \draw(12.2,5.4) node[draw,fill=white,anchor=west,xshift=4pt,yshift=0pt] {\scriptsize $t=7.43$};
    \node [circle, minimum size=1.3cm, draw, color=white] (b1) at (15.25,3.8) {};
\node[draw, fill=white, anchor=west,xshift=4pt,yshift=0pt] (b2) at (15.5,2.5) {\tiny Merged plasmoid};
\draw [<-, thick, white] (b1) -- (b2);
    \begin{scope}[yshift=6.7cm]
  \draw(-.5,-1) node[anchor=south west,xshift=0pt,yshift=+0pt] {\trimfigc{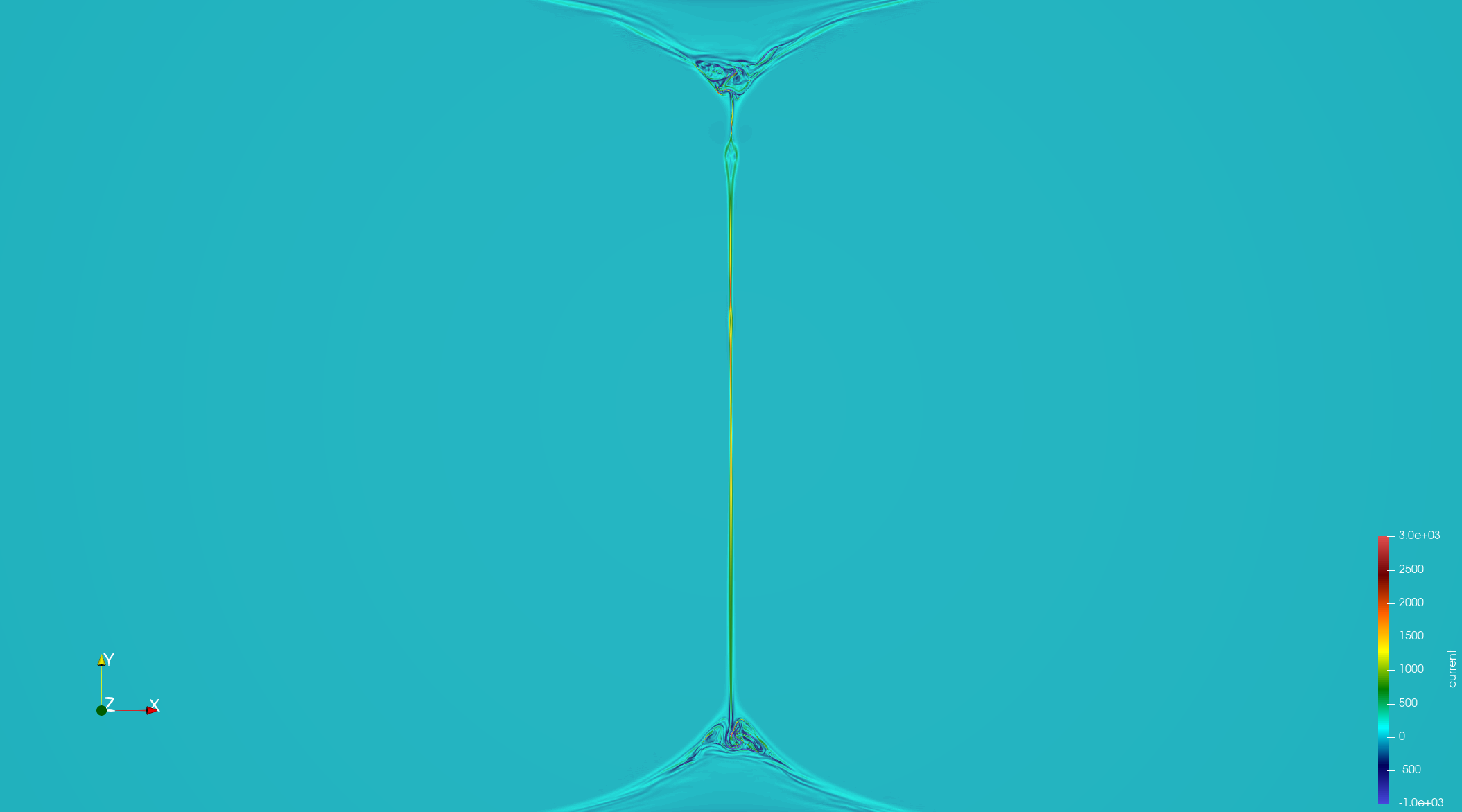}{\figWidthc}};
  \draw(5.8,-1) node[anchor=south west,xshift=0pt,yshift=+0pt] {\trimfigc{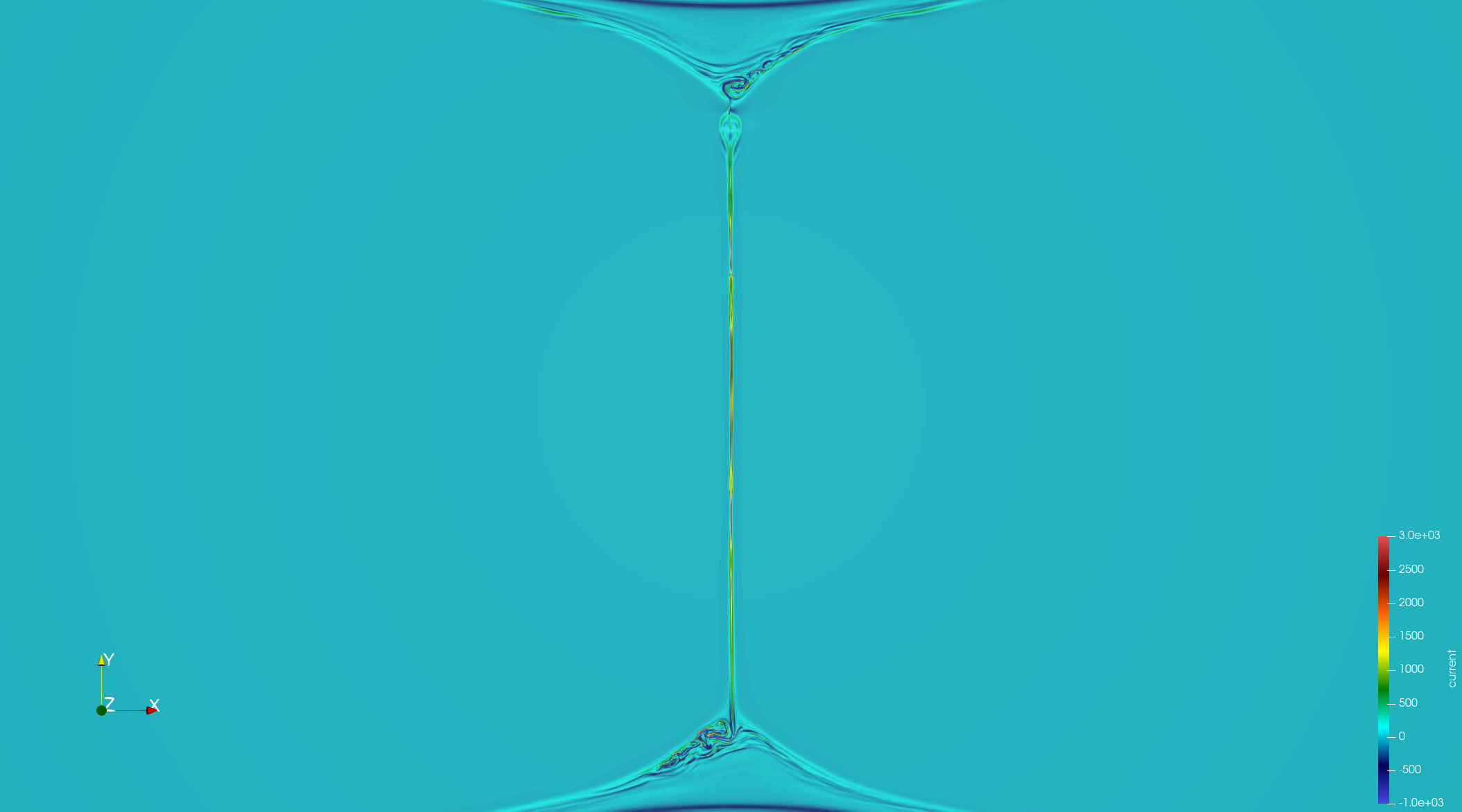}{\figWidthc}};
    \draw(12.1,-1) node[anchor=south west,xshift=0pt,yshift=+0pt] {\trimfigc{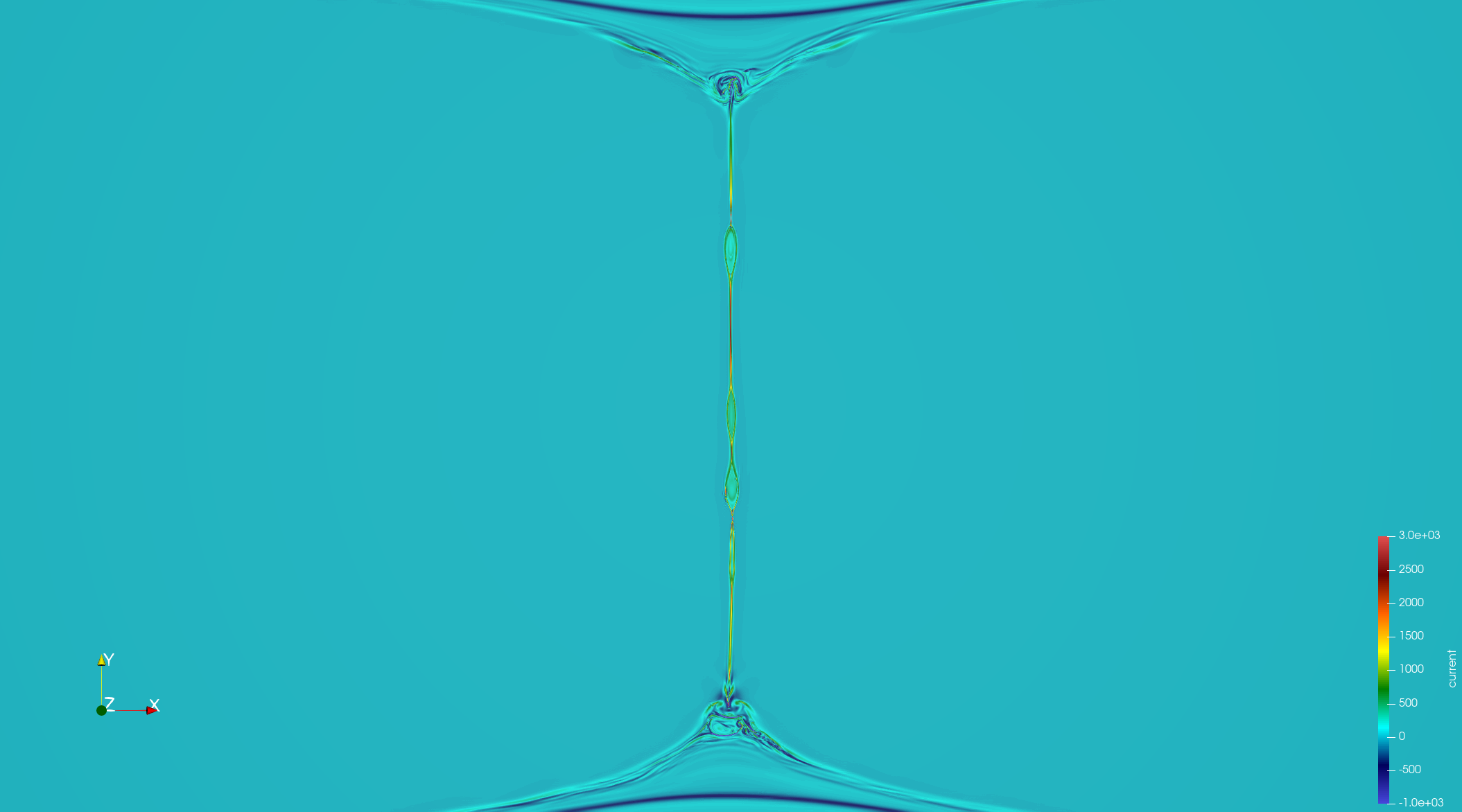}{\figWidthc}};
    \draw(-.4,5.4) node[draw,fill=white,anchor=west,xshift=4pt,yshift=0pt] {\scriptsize $t=6.86$};
    \draw(5.9,5.4) node[draw,fill=white,anchor=west,xshift=4pt,yshift=0pt] {\scriptsize $t=7.08$};
    \draw(12.2,5.4) node[draw,fill=white,anchor=west,xshift=4pt,yshift=0pt] {\scriptsize $t=7.21$};
         \node [circle, minimum size=.5cm, draw, color=white] (b1) at (2.66,4.3) {};
         \node [circle, minimum size=.5cm, draw, color=white] (b3) at (2.66,3.05) {};
\node[draw, fill=white, anchor=west,xshift=4pt,yshift=0pt] (b2) at (3.4,3.67) {\tiny Plasmoids};
\draw [<-, thick, white] (b1) -- (b2);
\draw [<-, thick, white] (b3) -- (b2);
      \end{scope}
      %
\end{tikzpicture}
} 
\end{center}
  \caption{Island coalescence of $\nu = \eta= $ $10^{-6}$.  The current sheet and plasmoid evolution around the center region after the sheet becomes unstable.
  A giant plasmoid forms in the very end after several plasmoids merge together.
   }
  \label{fig:ic1e-6time}
\end{figure}
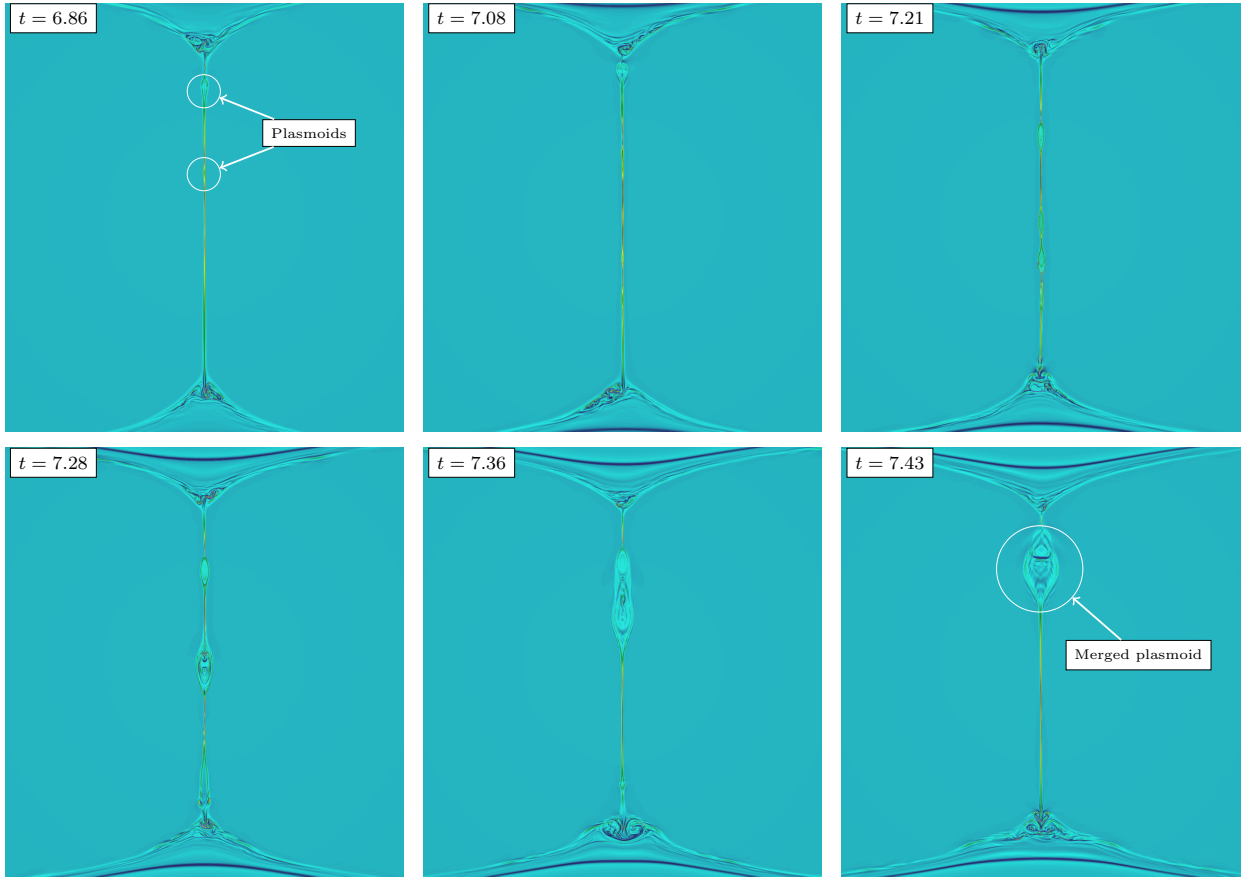
}

The time evolution of the current sheet is presented in Figure~\ref{fig:ic1e-6time}.
We select a few time slides to demonstrate the dynamics of the plasmoids.   Between $t = 6.5$ and $t = 7.0$,
the plasmoid dynamics are rather simple. There are a few isolated plasmoids (typically two to three) appearing along the current sheet.
The plasmoids then propagate towards the top or bottom jet regions.
When the plasmoids hit the jet regions, the plasmoids collapse and separate into two small bumps to propagate away along the current layers to the left and right, respectively.
At this stage, the movement of the plasmoids is rather straightforward, and they are driven by the flow velocity,
i.e., the plasmoids appearing on the top part of the current sheet move upward and those appearing on the bottom move downward.
At the later time, starting around $t =7.08$,  there are multiple plasmoids appearing.
For instance, at $t = 7.21$, a few plasmoids  with different sizes appear at the same time.
It is observed that the plasmoids merge into a larger plasmoid, which starts to move upward at $t = 7.28$.
Eventually, all the plasmoids merge together and form a giant plasmoid in the center at $t = 7.43$.
At this time, the current sheet has passed the peak time and the plasmoid dynamics becomes less violent until the second peak starts to form.
One important takeaway here is the simulation in the high Lundquist number breaks symmetry in both $x$ and $y$ directions.
Therefore, this test suggests it is necessary to perform the simulation in the entire domain instead of assuming symmetry as in previous studies~\cite{knoll2006coalescence, shadid2010towards}.

{
\newcommand{\figWidth}{4.6cm}
\newcommand{\figWidthc}{6cm}
\newcommand{\trimfig}[2]{\trimFig{#1}{#2}{.2}{.2}{.1}{.1}}
\newcommand{\trimfigb}[2]{\trimFig{#1}{#2}{.9}{.9}{.0}{.0}}
\newcommand{\trimfigc}[2]{\trimFig{#1}{#2}{1}{1}{.0}{.0}}
\begin{figure}[htb]
\begin{center}
\resizebox{16cm}{!}{
\begin{tikzpicture}[scale=1]
  \useasboundingbox (0.0,0) rectangle (18,5.7);  
\begin{scope}[yshift=0cm]
  \draw(-.5,-1) node[anchor=south west,xshift=0pt,yshift=+0pt] {\trimfigc{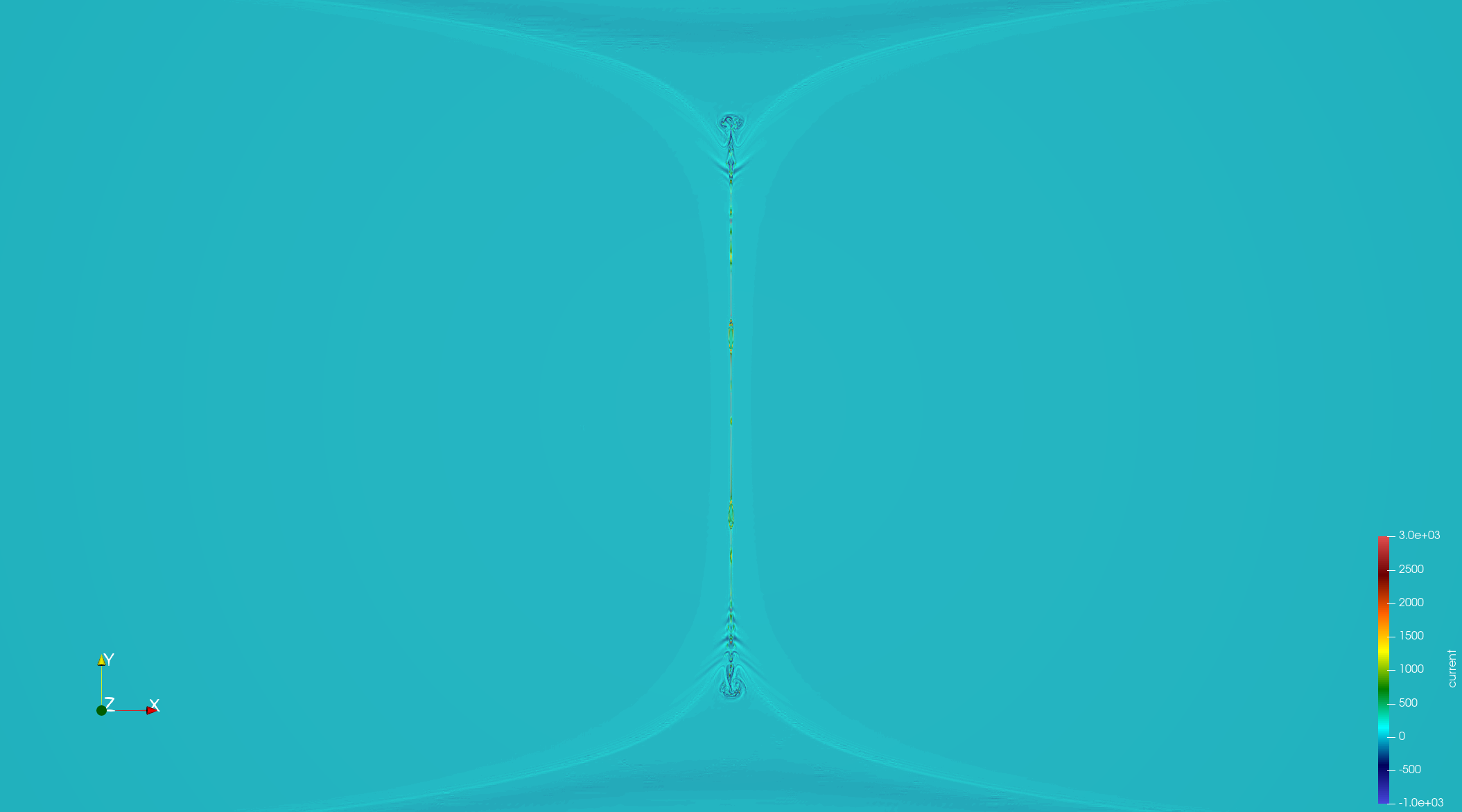}{\figWidthc}};
  \draw(5.8,-1) node[anchor=south west,xshift=0pt,yshift=+0pt] {\trimfigc{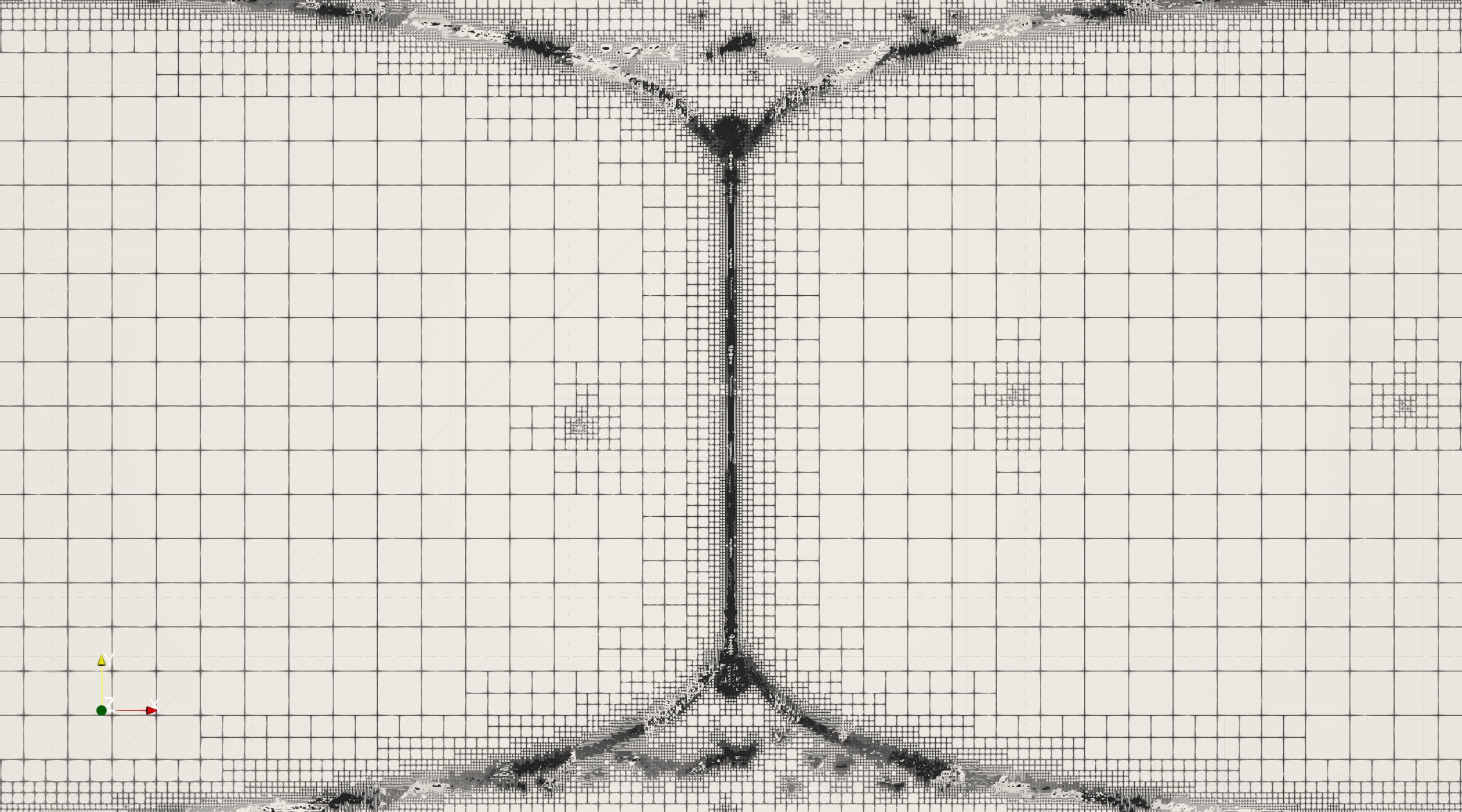}{\figWidthc}};
    \draw(12.1,-.5) node[anchor=south west,xshift=0pt,yshift=+0pt] {\trimfigb{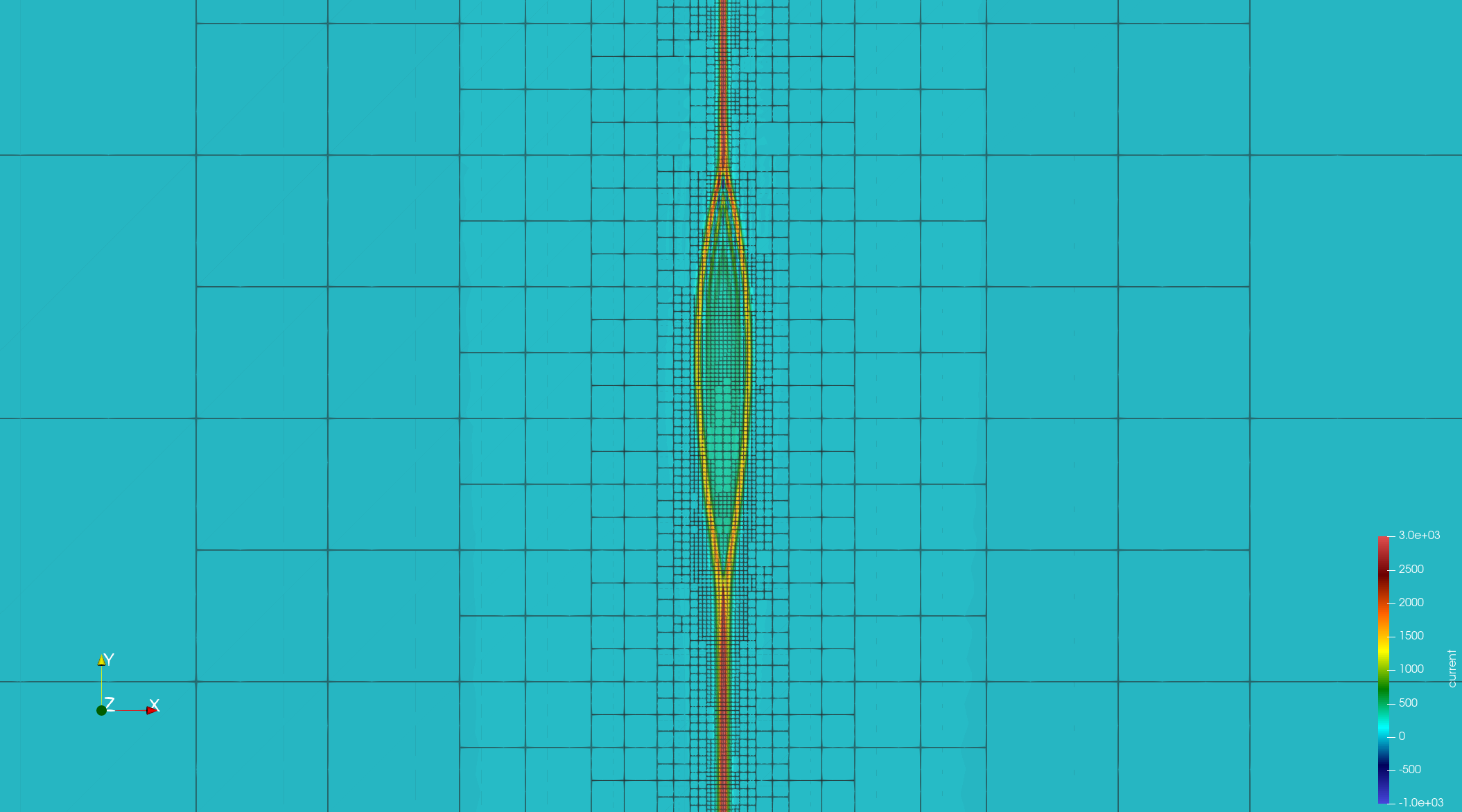}{\figWidthc}};
    \draw(-.4,5.4) node[draw,fill=white,anchor=west,xshift=4pt,yshift=0pt] {\scriptsize $t=6$};
    \draw(5.9,5.4) node[draw,fill=white,anchor=west,xshift=4pt,yshift=0pt] {\scriptsize mesh};
    \draw(12.2,5.) node[draw,fill=white,anchor=west,xshift=4pt,yshift=0pt] {\scriptsize zoomed-in};
 \node [draw, minimum size=.5cm, draw, color=white] (b1) at (2.67,2.9) {};
\node[draw, fill=white, anchor=west,xshift=4pt,yshift=0pt] (b2) at (3.25,2.9) {\tiny zoomed-in};
\draw [<-, thick, white] (b1) -- (b2);
\end{scope}
\end{tikzpicture}
} 
\end{center}
  \caption{Island coalescence for $\nu = \eta = 10^{-7}$. 7 levels of refinement are used with $p=3$. Left: the current sheet at the center.
  Middle: the corresponding adaptive mesh. Right: the zoomed-in view around the central plasmoid.
  }
  \label{fig:ic1e-7}
\end{figure}
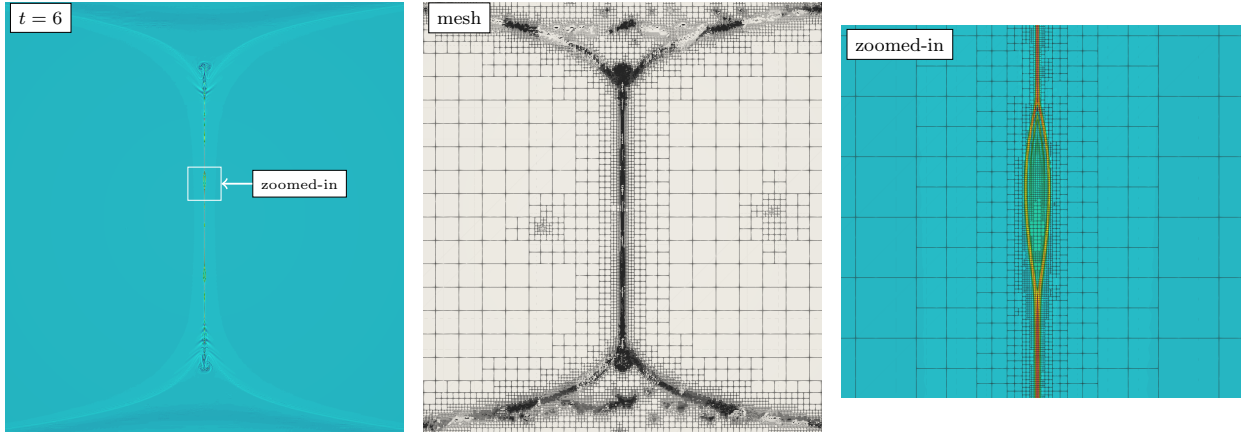
}

In the final test, we consider the most challenging case of $\nu = \eta= $ $10^{-7}$.
In this test, we use 7  levels of refinement.
The current sheet is found to be much thinner and the plasmoid dynamics are much more violent.
As a result, the smallest time step is found to be 0.00091 for this run.
The computational cost is again  significantly reduced compared to the uniform mesh.
The average number of dofs in this run is 1.77M, while a uniform mesh   needs  2416M dofs.
As a result, the implicit adaptive solver only needs 0.07\% of the total dofs compared to a comparable uniform mesh.
The current density and mesh at $t = 6$ is presented at Figure~\ref{fig:ic1e-7}. From the zoomed-in view, we found the plasmoid in the center is very well captured by AMR.
There are 8 plasmoids already along this current sheet and the maximum current density values is 7086, which is found  between two isolated plasmoids.
The plasmoids first appear around $t = 5.75$ with  current sheet aspect ratio of  250 and the maximum current density in the center around 3000.

{
\newcommand{\figWidth}{4.6cm}
\newcommand{\figWidthc}{6cm}
\newcommand{\trimfig}[2]{\trimFig{#1}{#2}{.2}{.2}{.1}{.1}}
\newcommand{\trimfigb}[2]{\trimFig{#1}{#2}{.9}{.9}{.0}{.0}}
\newcommand{\trimfigc}[2]{\trimFig{#1}{#2}{1}{1}{.0}{.0}}
\begin{figure}[htb]
\begin{center}
\resizebox{16cm}{!}{
\begin{tikzpicture}[scale=1]
  \useasboundingbox (0.0,0) rectangle (18,17.6);  
  \draw(-.5,-1) node[anchor=south west,xshift=0pt,yshift=+0pt] {\trimfig{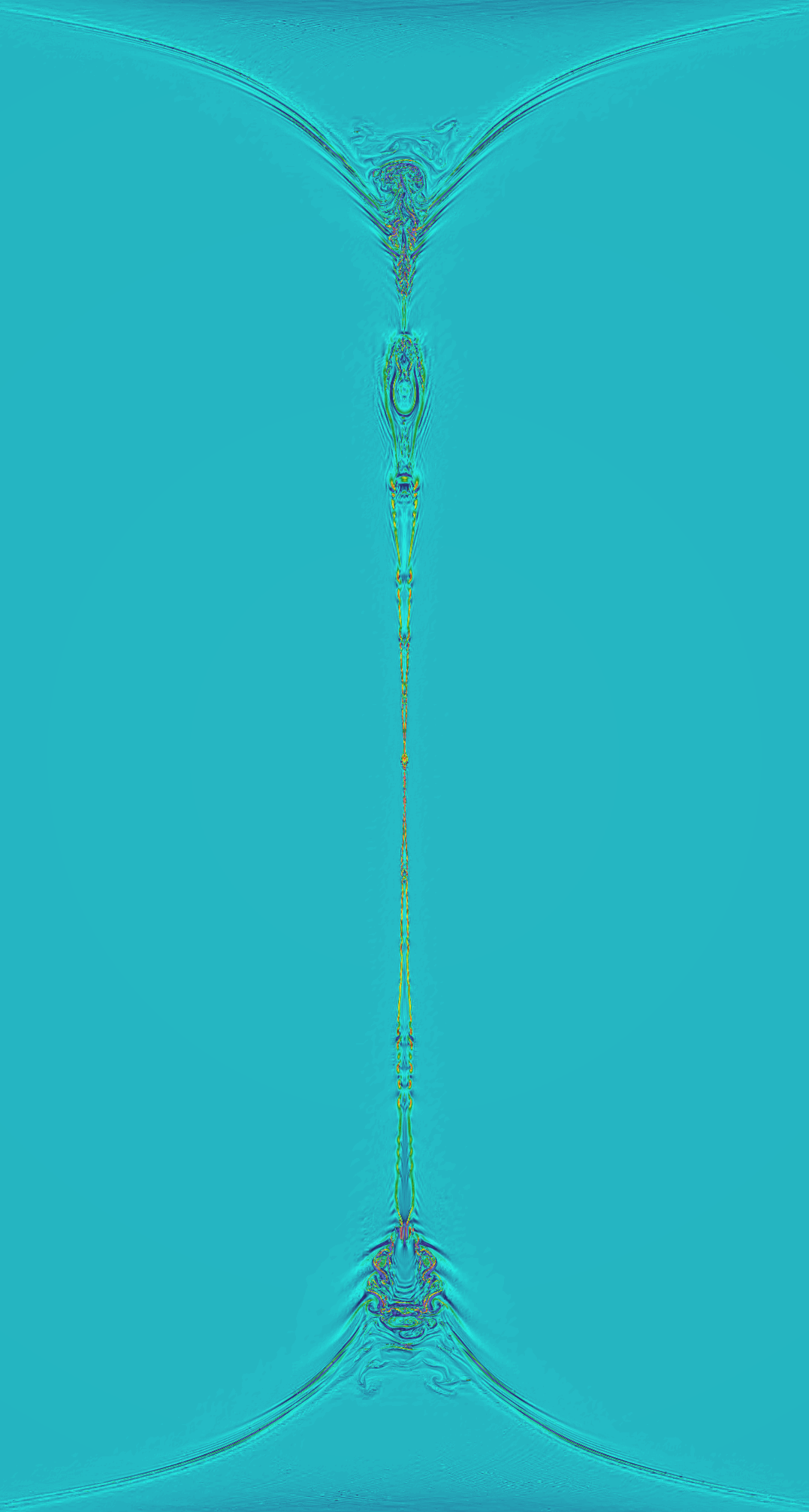}{\figWidth}};
  \draw(4.2,-1) node[anchor=south west,xshift=0pt,yshift=+0pt] {\trimfig{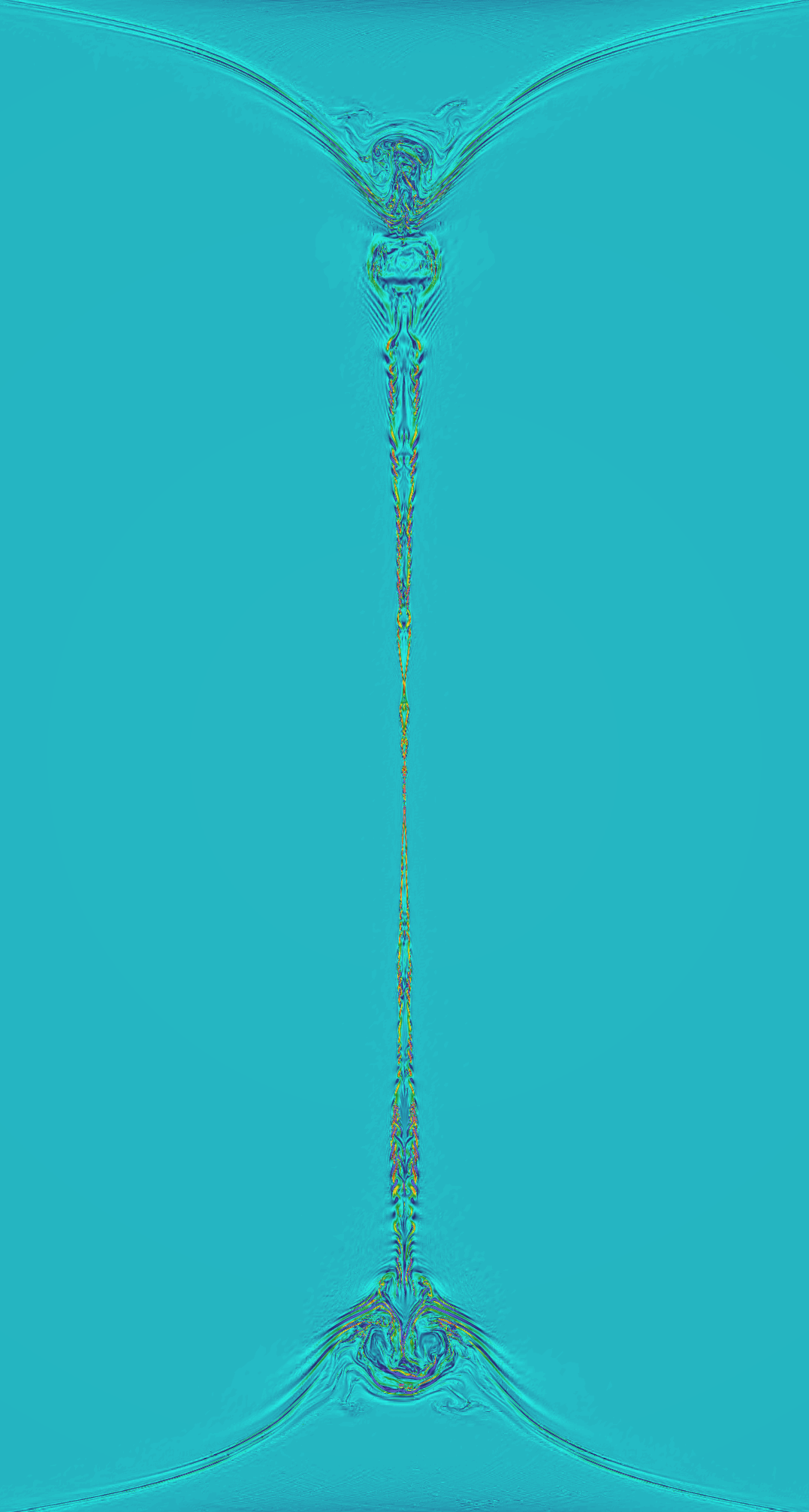}{\figWidth}};
  \draw(8.9,-1) node[anchor=south west,xshift=0pt,yshift=+0pt] {\trimfig{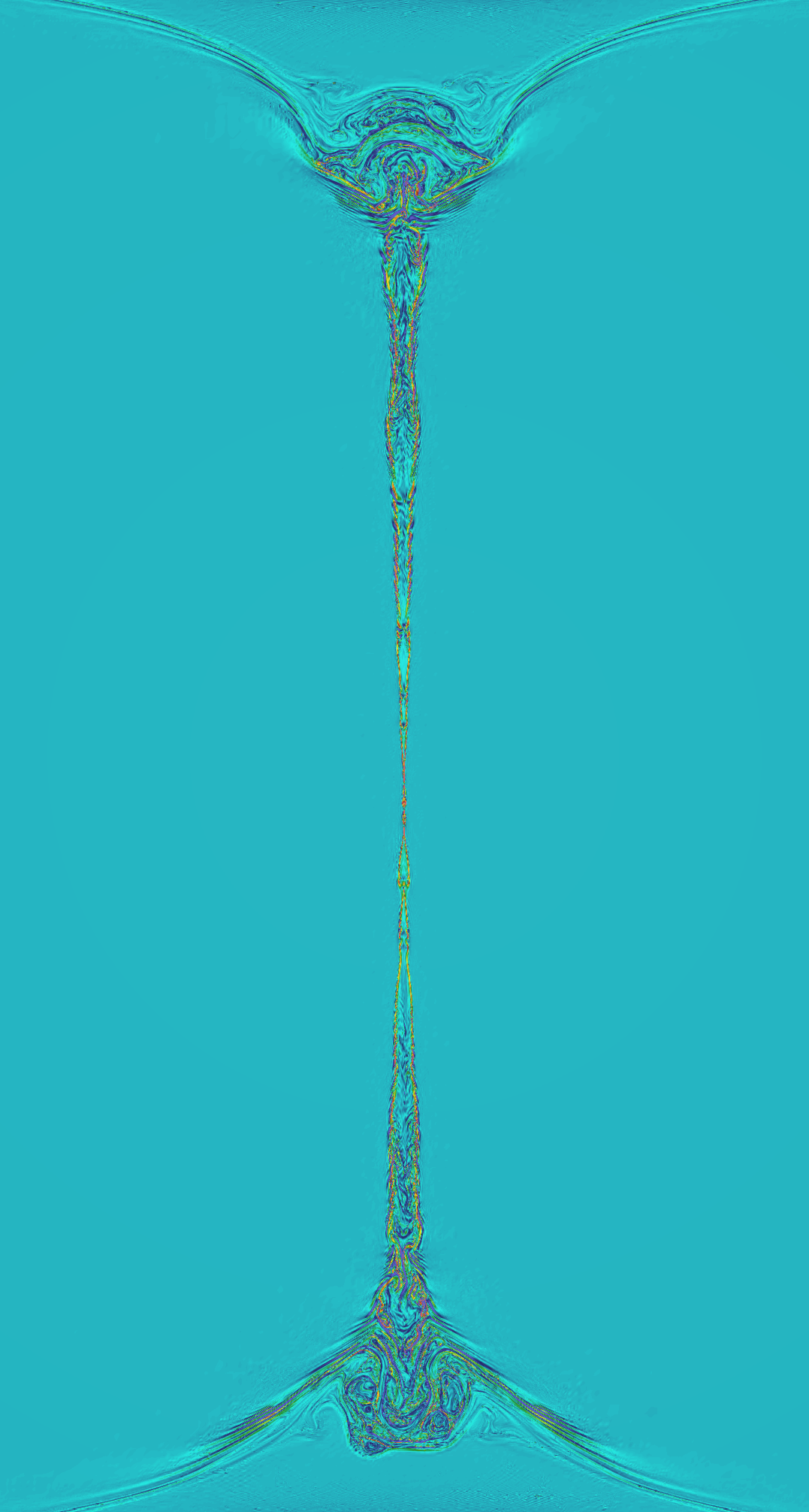}{\figWidth}};
  \draw(13.6,-1) node[anchor=south west,xshift=0pt,yshift=+0pt] {\trimfig{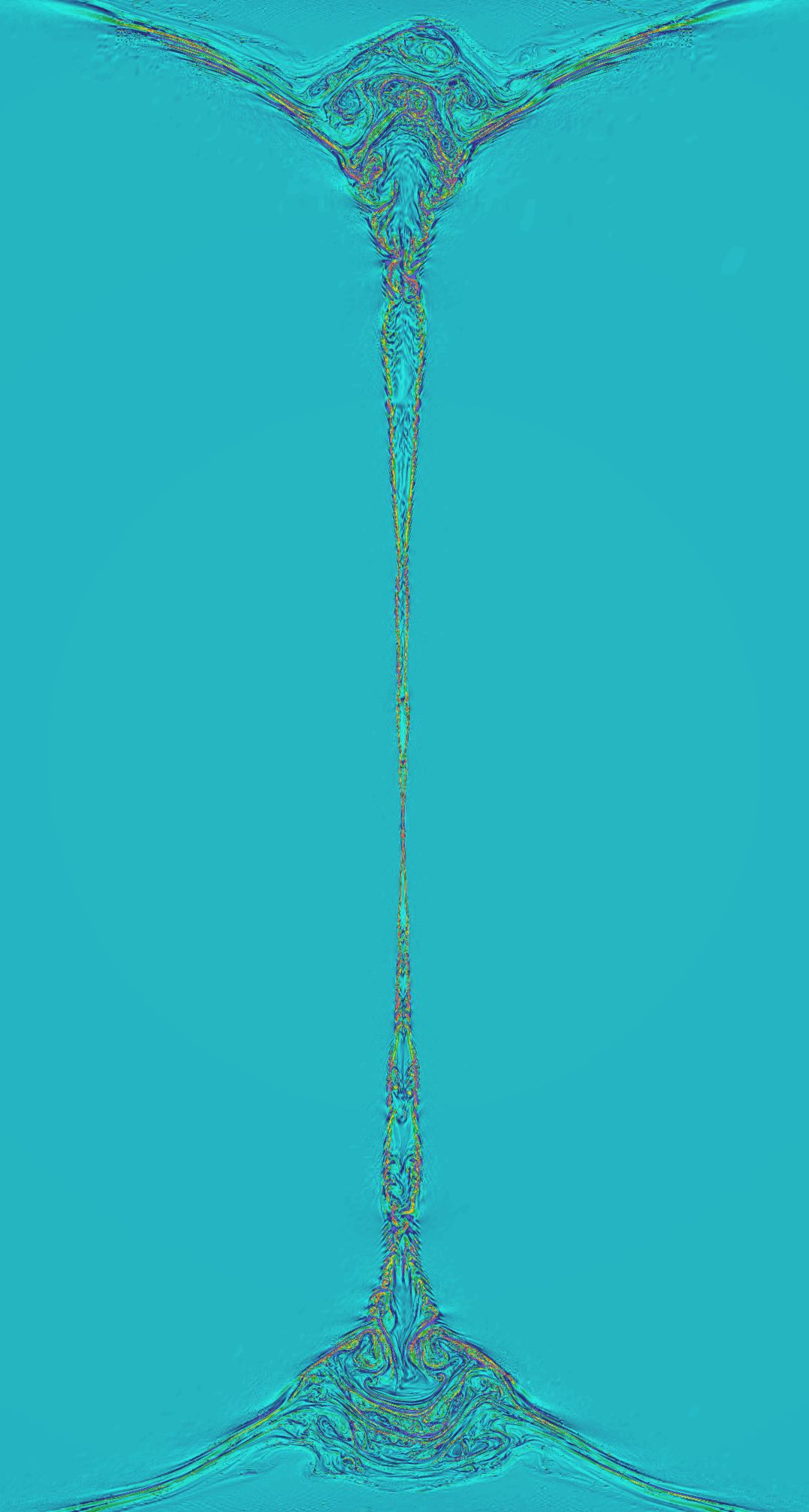}{\figWidth}};
    \draw(-.4,8) node[draw,fill=white,anchor=west,xshift=4pt,yshift=0pt] {\scriptsize $t=6.31$};
    \draw(4.3,8) node[draw,fill=white,anchor=west,xshift=4pt,yshift=0pt] {\scriptsize $t=6.36$};
    \draw(9,8) node[draw,fill=white,anchor=west,xshift=4pt,yshift=0pt] {\scriptsize $t=6.44$};
     \draw(13.7,8) node[draw,fill=white,anchor=west,xshift=4pt,yshift=0pt] {\scriptsize $t=6.56$};
    \begin{scope}[yshift=9.3cm]
  \draw(-.5,-1) node[anchor=south west,xshift=0pt,yshift=+0pt] {\trimfig{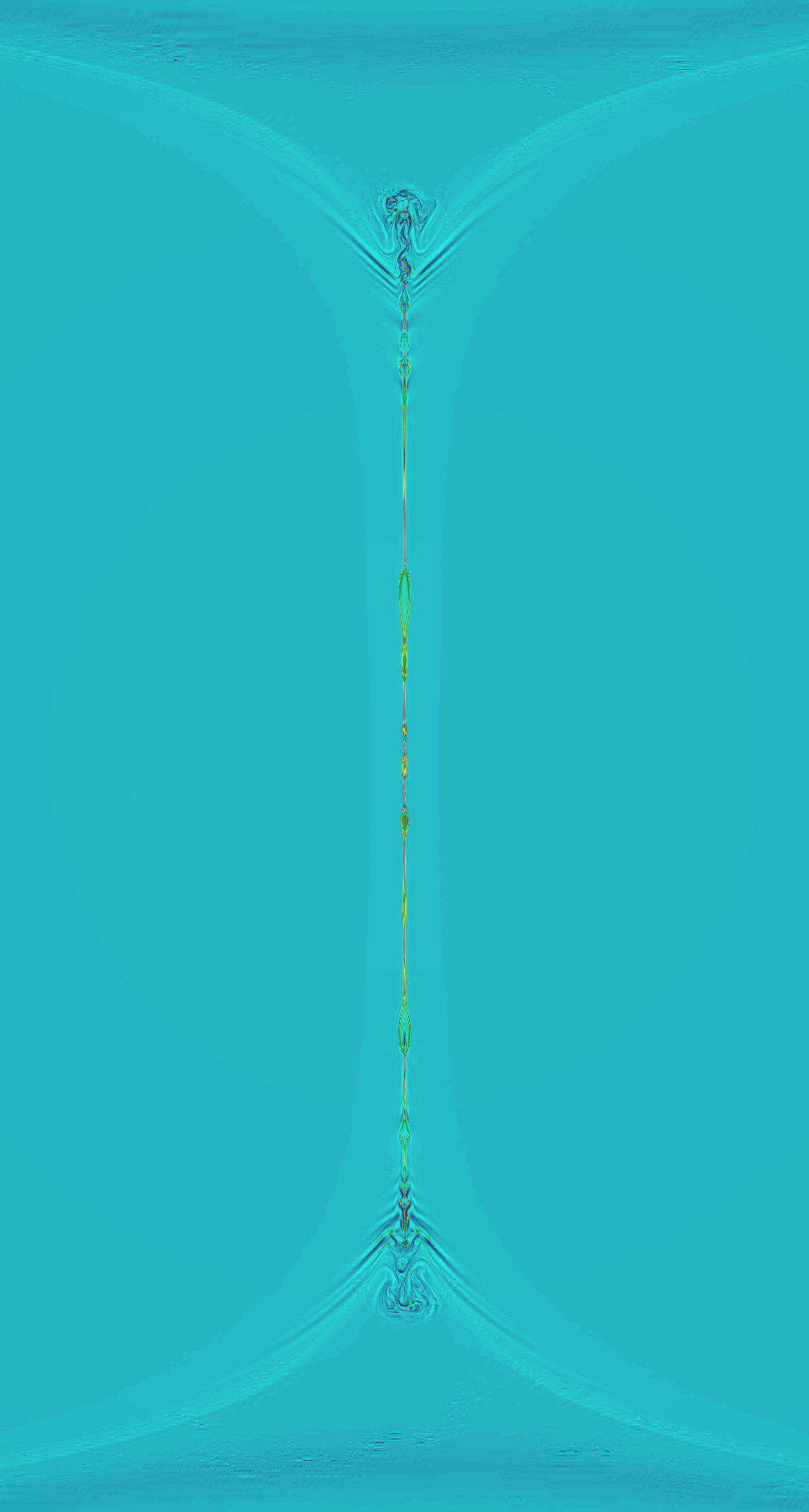}{\figWidth}};
  \draw(4.2,-1) node[anchor=south west,xshift=0pt,yshift=+0pt] {\trimfig{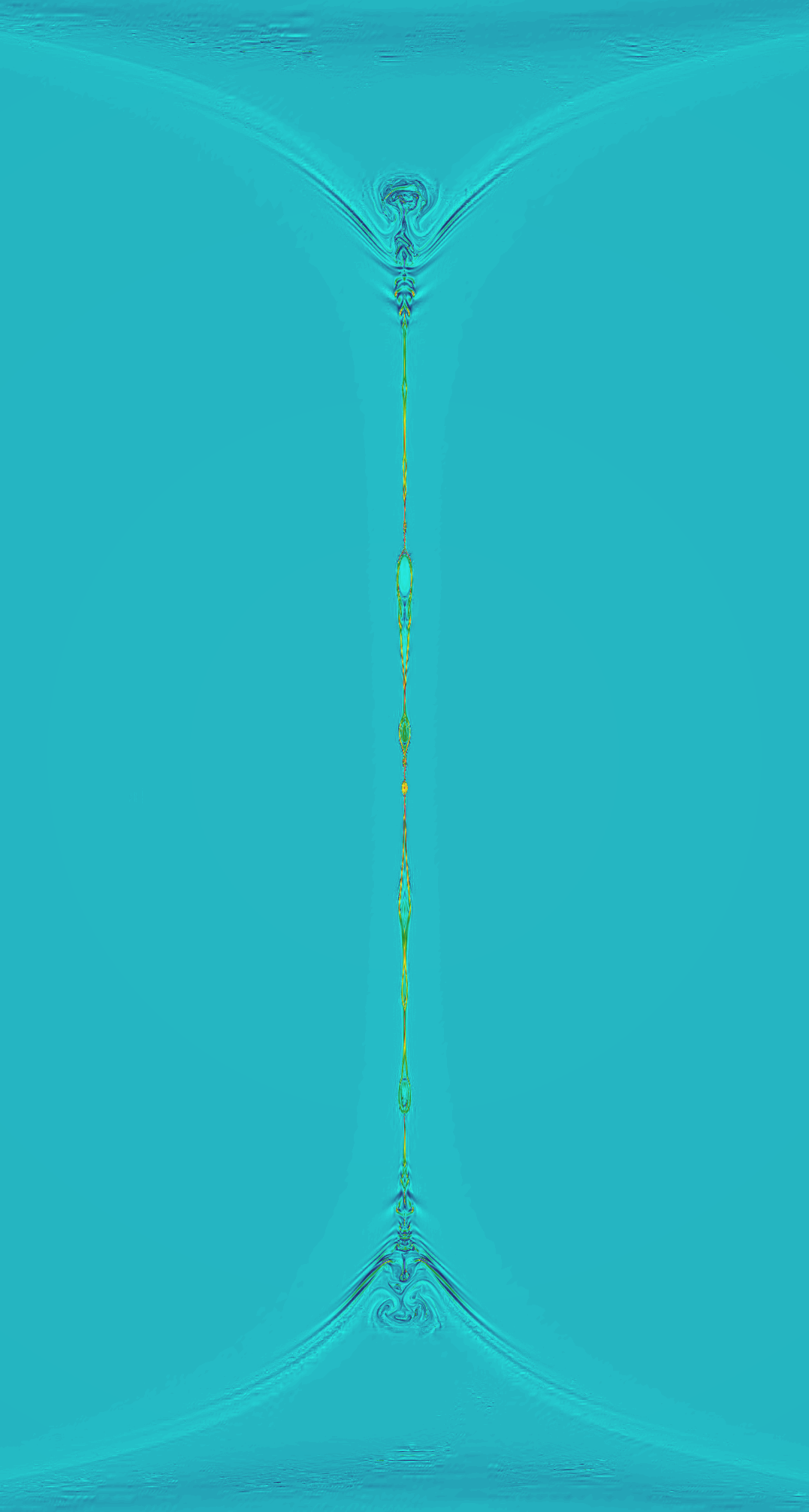}{\figWidth}};
  \draw(8.9,-1) node[anchor=south west,xshift=0pt,yshift=+0pt] {\trimfig{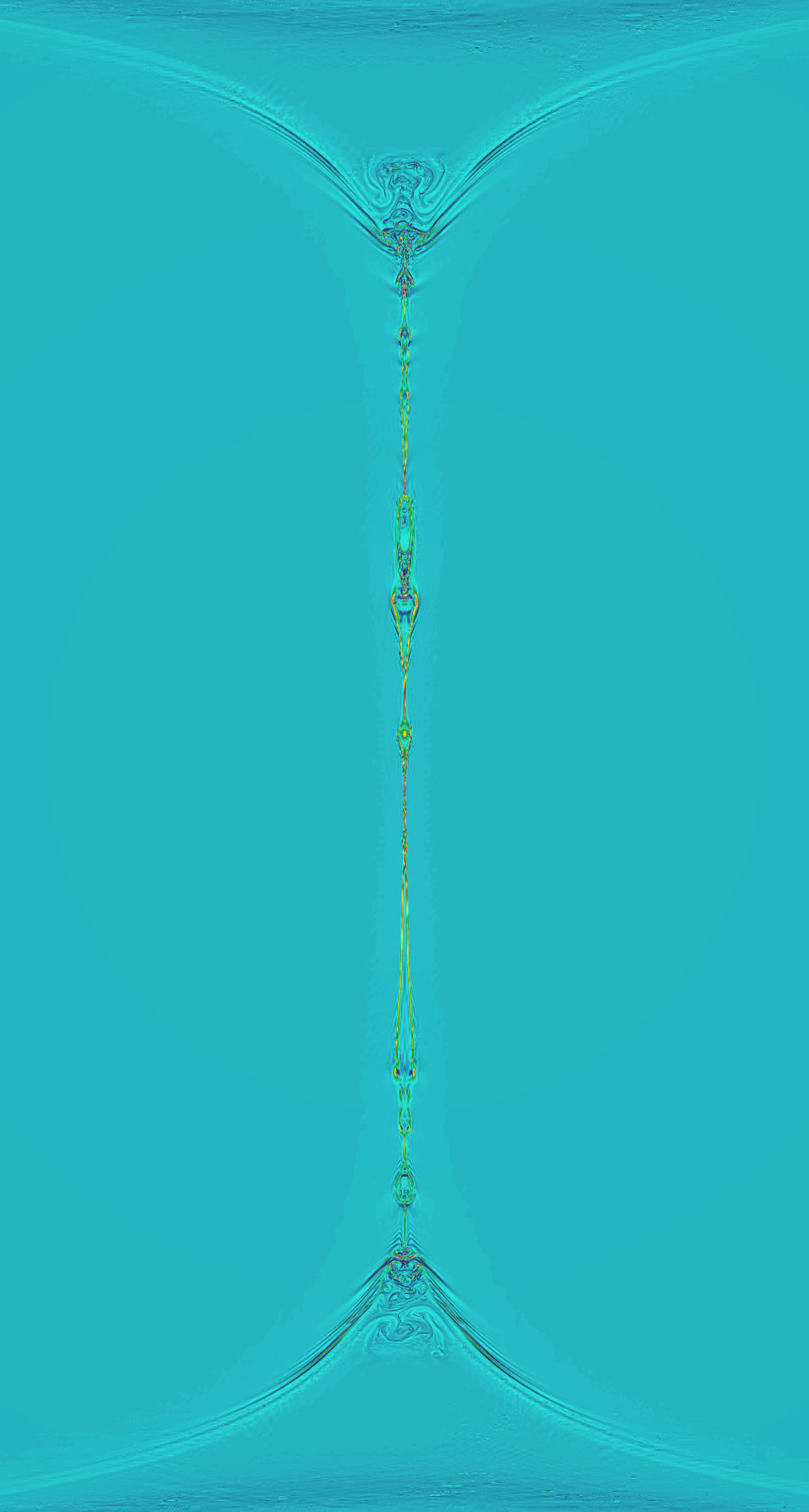}{\figWidth}};
  \draw(13.6,-1) node[anchor=south west,xshift=0pt,yshift=+0pt] {\trimfig{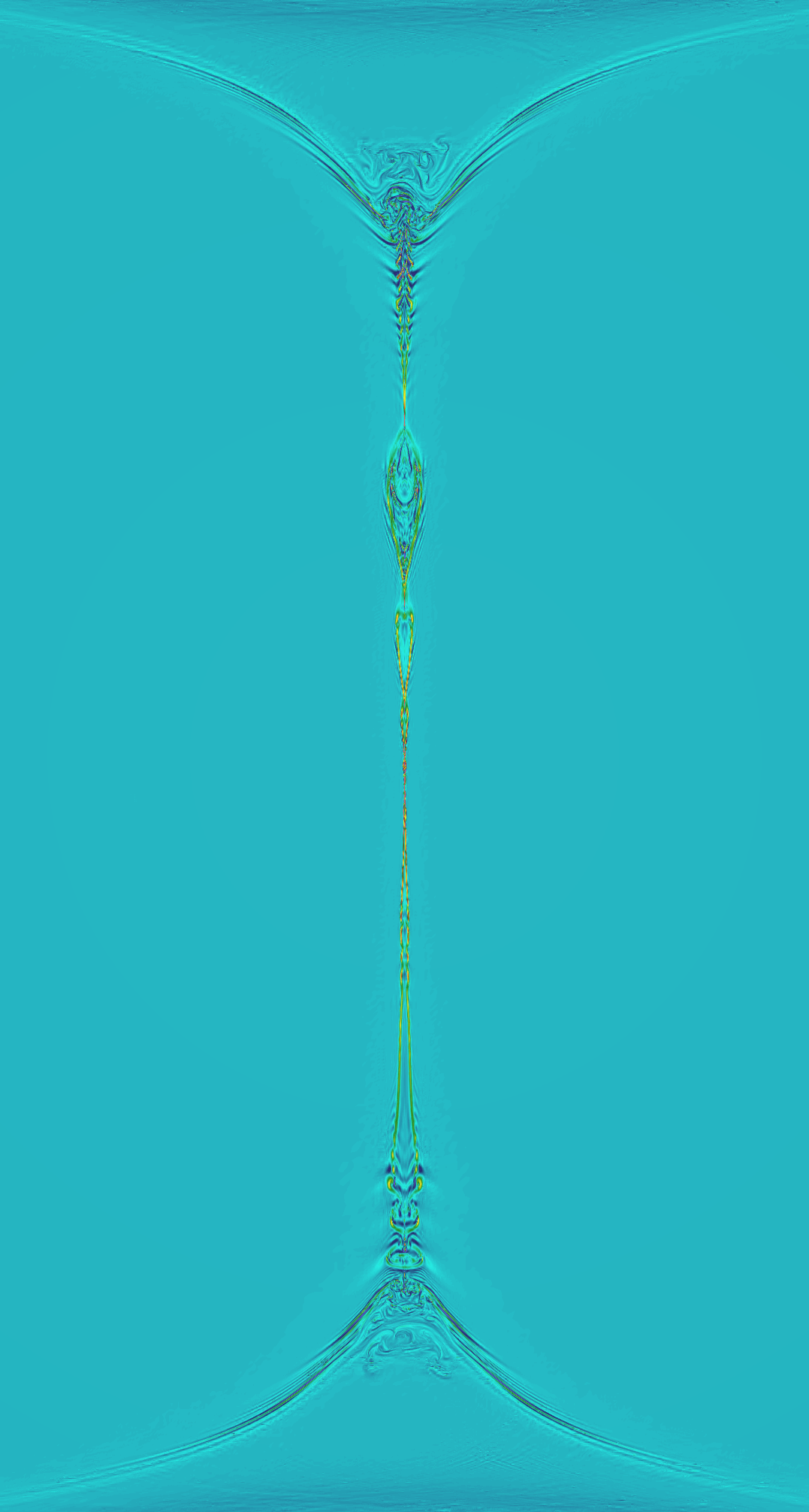}{\figWidth}};
    \draw(-.4,8) node[draw,fill=white,anchor=west,xshift=4pt,yshift=0pt] {\scriptsize $t=6.08$};
    \draw(4.3,8) node[draw,fill=white,anchor=west,xshift=4pt,yshift=0pt] {\scriptsize $t=6.13$};
    \draw(9,8) node[draw,fill=white,anchor=west,xshift=4pt,yshift=0pt] {\scriptsize $t=6.21$};
     \draw(13.7,8) node[draw,fill=white,anchor=west,xshift=4pt,yshift=0pt] {\scriptsize $t=6.26$};%
      \end{scope}
\end{tikzpicture}
} 
\end{center}
  \caption{Island coalescence for $\nu = \eta = 10^{-7}$. Evolution around the current sheet regions. The dynamics become very fast and the current sheet breaks into many smaller structures.
  }
  \label{fig:ic1e-7time}
\end{figure}
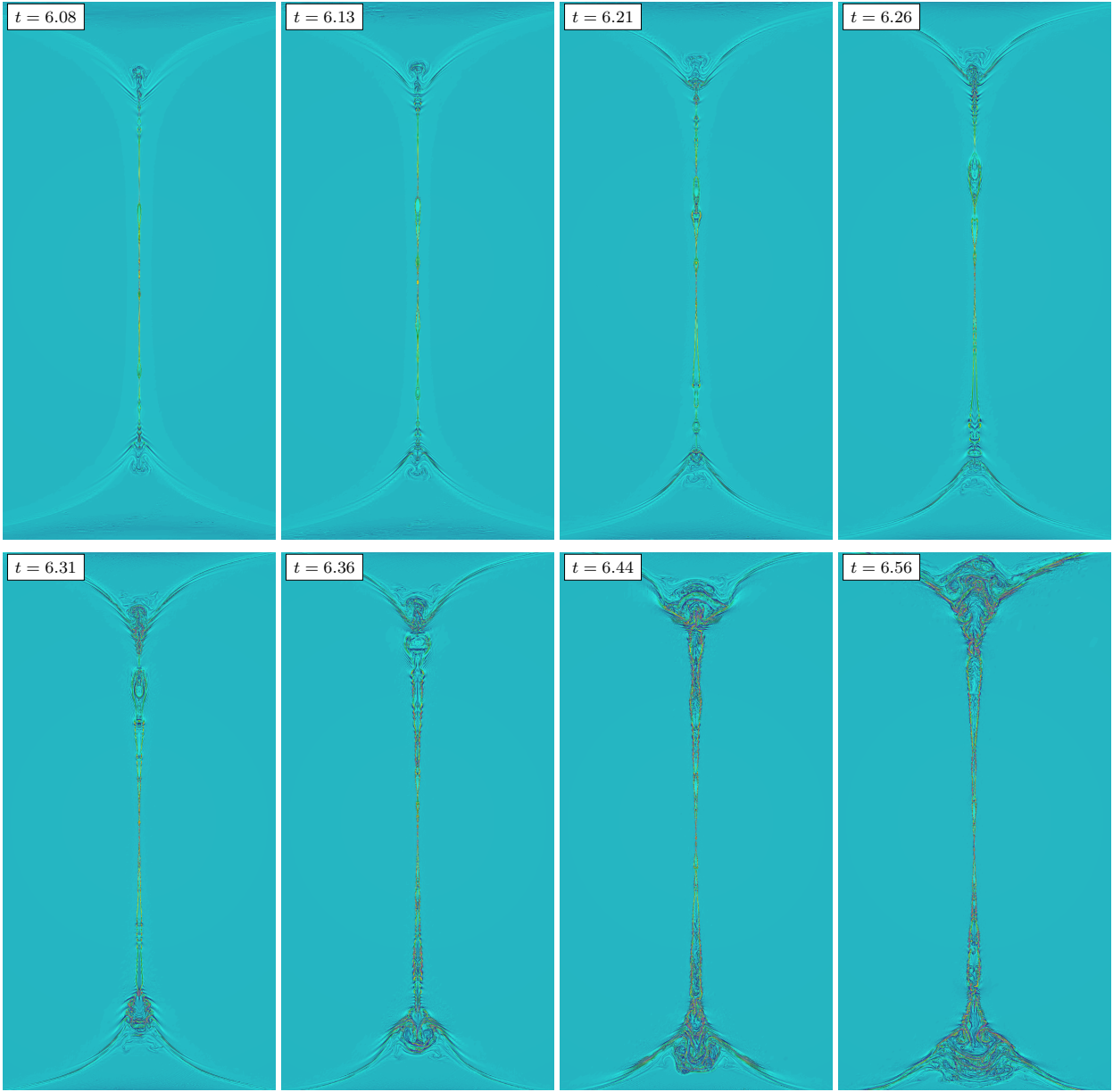
}

The time evolution of the current sheet is presented in Figure~\ref{fig:ic1e-7time}.
We select a few time slices to demonstrate the dynamics of the plasmoids.
At an earlier stage, the plasmoid dynamics are similar to the dynamics of $\nu = \eta = 10^{-6}$, i.e.,
there are many plasmoids of different sizes forming along the current sheet and they move and merge together, roughly between $t = 5.75$ and 6.26.
However, this dynamics changes at a later time.
Starting at $t = 6.31$, the current sheet continuously generates small plasmoids along both directions.
When those plasmoids move up or down, they deform into an asymmetric drop-like shape, with the front speed faster than the tail speed.
As a result, the tail is caught up by  plasmoids generated at a later time,  leading to their merging.
We find  that  this process occurs continuously, and as a result a stable X-shape structure forms in the center of the current sheet,
which is very similar to the structure in the Petschek reconnection model (as marked in Figure~\ref{fig:ic1e-7time}). Note such a structure is also observed at the plot of $t = 7.28$ in Figure~\ref{fig:ic1e-6time} (along the bottom current sheet)
for the case of $\nu = \eta = 10^{-6}$.
Similar results were previously observed in~\cite{mei2012numerical, baty2020petschek} for  moderately high Lundquist number simulations. 
Discussions on forming of such a structure can be found in~\cite{mei2012numerical}.
We defer a deeper investigation into high Lundquist number cases to a future physics study.

\begin{table}[hbt]\tableFont 
\begin{center}
\caption{Comparison between average dofs in AMR meshes and dofs in the corresponding uniform meshes of comparable fine grid resolutions
} \label{tableICAMR}
 \medskip
\begin{tabular}{|c|c|c|c|} \hline
  \multicolumn{4}{|c|}{Average scalar dofs in island coalescence tests} \\ \hline
$S_L$ &  AMR &    uniform meshes  &  AMR/uniform meshes       \\[3pt] \hline
$10^5$ & 0.47M & 38M & 1.2\%      \\ \hline
 $10^6$ & 1.30M & 604M & 0.21\%    \\ \hline
 $10^{7}$ & 1.77M & 2416M &  0.07\%      \\ \hline
\end{tabular}
\end{center}
\end{table}

Based on the above tests, we conclude the proposed implicit AMR solver is suitable for simulations of ultra high Lundquist numbers.
It is found that the implicit AMR solver is robust and efficient for different Lundquist number regimes, ranging from no plasmoids and simple dynamics to many colliding plasmoids
and very complicated dynamics.
The highest Lundquist number tested with the proposed solver is $10^{8}$, and in a future physics study we plan to  explore this limit further.
The savings in  dofs have been summarized in Table~\ref{tableICAMR},
demonstrating the ability of the approach to evolve previously inaccessible dissipation regimes.

\section{Conclusions} \label{sec:conclusions}
We have developed a fully implicit, stabilized finite-element formulation for 2D incompressible reduced resistive MHD,  incorporating
important techniques of physics-based preconditioning and adaptive mesh refinement.
An SUPG-based stabilization approach   is added into the continuous finite-element formulation to handle  high Lundquist numbers.
The proposed physics-based preconditioner handles the stiff hyperbolic Alfv\'en wave, the diffusive terms and the stabilization terms in the formulation.
An octree-based parallel AMR algorithm is employed, for which we propose an error estimator and a proper refining/coarsening strategy.
The algorithm is implemented in the scalable finite-element framework, MFEM, with which we achieve good performance up to thousands of cores.

The performance of the proposed algorithm is carefully studied with several numerical examples.
The major points of interest include the accuracy of  the discretization,  the efficiency and validation of the associated solver, and the performance improvement of the AMR algorithm (both in accuracy and computational cost).
Our study confirms the expected high-order convergence rates in both space and time.
Weak parallel scaling results up to 4096 CPUs for moderately high Lundquist-number simulations  demonstrate excellent algorithmic  and parallel scalability.
Results using the implicit AMR solver for low Lundquist-number (tearing mode) to moderately-high and high Lundquist-number  problems (island coalescence)
demonstrate the AMR algorithm improves the accuracy of numerical solutions and reduces computational cost significantly over the implicit solver on a uniform mesh.
In particular, the AMR algorithm is capable of capturing the vast scale separations and dramatic dynamics in the solution when the thin current sheet breaks into small plasmoids.

The proposed solver provides a unique and promising tool to study the physics  of high  Lundquist numbers, such as  magnetic reconnection  when plasmoid instabilities become violent.
Those results will be investigated  in a future physics study.
Another important future direction is to adapt the solver for tokamak simulations of practical interest.

\section*{Acknowledgement}
We would like to thank the MFEM, PETSc and deal.II teams for a number of helpful discussions.
{\red We also thank Hubert Baty to point out a missing term in the $J$-$\omega$ formulation in Section 2.}
We also extend gratitude to Barry Smith for the helpful discussion on JFNK and Sriramkrishnan Muralikrishnan for the helpful discussion on island coalescence simulations.
This research used resources provided by the Los Alamos National Laboratory Institutional Computing Program, which is supported by the U.S. Department of Energy National Nuclear Security Administration under Contract No.~89233218CNA000001.
In particular, most of the scaling tests are performed on the LANL Grizzly supercomputer.
This research also used resources of the National Energy Research Scientific Computing Center (NERSC), a U.S. Department of Energy Office of Science User Facility located at Lawrence Berkeley National Laboratory, operated under Contract No. DE-AC02-05CH11231 using NERSC awards FES-ERCAP0016552 and FES-ERCAP0016553, in some long runs of the final example.

\bibliographystyle{elsart-num}
\bibliography{journal-ISI,ref}

\begin{thebibliography}{10}
\expandafter\ifx\csname url\endcsname\relax
  \def\url#1{\texttt{#1}}\fi
\expandafter\ifx\csname urlprefix\endcsname\relax\def\urlprefix{URL }\fi

\bibitem{biskamp1997nonlinear}
D.~Biskamp, Nonlinear magnetohydrodynamics, no.~1, Cambridge University Press,
  1997.

\bibitem{jardin2012review}
S.~C. Jardin, Review of implicit methods for the magnetohydrodynamic
  description of magnetically confined plasmas, Journal of Computational
  Physics 231~(3) (2012) 822--838.

\bibitem{chacon2002}
L.~Chac{\'o}n, D.~A. Knoll, J.~Finn, An implicit, nonlinear reduced resistive
  mhd solver, J. Comput. Phys. 178~(1) (2002) 15--36.

\bibitem{shadid2010towards}
J.~N. Shadid, R.~P. Pawlowski, J.~W. Banks, L.~Chac{\'o}n, P.~T. Lin, R.~S.
  Tuminaro, Towards a scalable fully-implicit fully-coupled resistive mhd
  formulation with stabilized fe methods, Journal of Computational Physics
  229~(20) (2010) 7649--7671.

\bibitem{chacon20032d}
L.~Chac{\'o}n, D.~A. Knoll, A 2d high-$\beta$ hall mhd implicit nonlinear
  solver, Journal of Computational Physics 188~(2) (2003) 573--592.

\bibitem{chacon2004non}
L.~Chac{\'o}n, A non-staggered, conservative, $\nabla\cdot \bv= 0$,
  finite-volume scheme for 3d implicit extended magnetohydrodynamics in
  curvilinear geometries, Computer Physics Communications 163~(3) (2004)
  143--171.

\bibitem{chacon2016scalable}
L.~Chac{\'o}n, A.~Stanier, A scalable, fully implicit algorithm for the reduced
  two-field low-$\beta$ extended mhd model, Journal of Computational Physics
  326 (2016) 763--772.

\bibitem{shadid2016scalable}
J.~N. Shadid, R.~P. Pawlowski, E.~C. Cyr, R.~S. Tuminaro, L.~Chac{\'o}n, P.~D.
  Weber, Scalable implicit incompressible resistive mhd with stabilized fe and
  fully-coupled newton--krylov-amg, Computer Methods in Applied Mechanics and
  Engineering 304 (2016) 1--25.

\bibitem{cyr2013new}
E.~C. Cyr, J.~N. Shadid, R.~S. Tuminaro, R.~P. Pawlowski, L.~Chac{\'o}n, A new
  approximate block factorization preconditioner for two-dimensional
  incompressible (reduced) resistive mhd, SIAM Journal on Scientific Computing
  35~(3) (2013) B701--B730.

\bibitem{phillips2014block}
E.~G. Phillips, H.~C. Elman, E.~C. Cyr, J.~N. Shadid, R.~P. Pawlowski, A block
  preconditioner for an exact penalty formulation for stationary mhd, SIAM
  Journal on Scientific Computing 36~(6) (2014) B930--B951.

\bibitem{phillips2016block}
E.~G. Phillips, J.~N. Shadid, E.~C. Cyr, H.~C. Elman, R.~P. Pawlowski, Block
  preconditioners for stable mixed nodal and edge finite element
  representations of incompressible resistive mhd, SIAM Journal on Scientific
  Computing 38~(6) (2016) B1009--B1031.

\bibitem{lee2019analysis}
J.~J. Lee, S.~J. Shannon, T.~Bui-Thanh, J.~N. Shadid, Analysis of an {HDG}
  method for linearized incompressible resistive mhd equations, SIAM Journal on
  Numerical Analysis 57~(4) (2019) 1697--1722.

\bibitem{muralikrishnan2020multilevel}
S.~Muralikrishnan, S.~Shannon, T.~Bui-Thanh, J.~N. Shadid, A multilevel block
  preconditioner for the hdg trace system applied to incompressible resistive
  mhd, arXiv preprint arXiv:2012.07648.

\bibitem{hu2017stable}
K.~Hu, Y.~Ma, J.~Xu, Stable finite element methods preserving $\nabla\cdot \bv=
  0$ exactly for mhd models, Numerische Mathematik 135~(2) (2017) 371--396.

\bibitem{ma2016robust}
Y.~Ma, K.~Hu, X.~Hu, J.~Xu, Robust preconditioners for incompressible mhd
  models, Journal of Computational Physics 316 (2016) 721--746.

\bibitem{li2021constrained}
L.~Li, D.~Zhang, W.~Zheng, A constrained transport divergence-free finite
  element method for incompressible mhd equations, Journal of Computational
  Physics 428 (2021) 109980.

\bibitem{christlieb2014finite}
A.~J. Christlieb, J.~A. Rossmanith, Q.~Tang, Finite difference weighted
  essentially non-oscillatory schemes with constrained transport for ideal
  magnetohydrodynamics, Journal of Computational Physics 268 (2014) 302--325.

\bibitem{christlieb2016high}
A.~J. Christlieb, X.~Feng, D.~C. Seal, Q.~Tang, A high-order
  positivity-preserving single-stage single-step method for the ideal
  magnetohydrodynamic equations, Journal of Computational Physics 316 (2016)
  218--242.

\bibitem{donea2003finite}
J.~Donea, A.~Huerta, Finite element methods for flow problems, John Wiley \&
  Sons, 2003.

\bibitem{biskamp1996magnetic}
D.~Biskamp, Magnetic reconnection in plasmas, Astrophysics and Space Science
  242~(1) (1996) 165--207.

\bibitem{bhattacharjee2009fast}
A.~Bhattacharjee, Y.-M. Huang, H.~Yang, B.~Rogers, Fast reconnection in
  high-lundquist-number plasmas due to the plasmoid instability, Physics of
  Plasmas 16~(11) (2009) 112102.

\bibitem{huang2010scaling}
Y.-M. Huang, A.~Bhattacharjee, Scaling laws of resistive magnetohydrodynamic
  reconnection in the high-lundquist-number, plasmoid-unstable regime, Physics
  of Plasmas 17~(6) (2010) 062104.

\bibitem{lynch2016reconnection}
B.~Lynch, J.~Edmondson, M.~Kazachenko, S.~Guidoni, Reconnection properties of
  large-scale current sheets during coronal mass ejection eruptions, The
  Astrophysical Journal 826~(1) (2016) 43.

\bibitem{baty2020petschek}
H.~Baty, Petschek-type reconnection in the high-lundquist-number regime during
  nonlinear evolution on the tilt instability, arXiv preprint arXiv:2005.04221.

\bibitem{philip2008implicit}
B.~Philip, L.~Chac{\'o}n, M.~Pernice, Implicit adaptive mesh refinement for 2d
  reduced resistive magnetohydrodynamics, Journal of Computational Physics
  227~(20) (2008) 8855--8874.

\bibitem{adler2011efficiency}
J.~Adler, T.~A. Manteuffel, S.~F. McCormick, J.~Nolting, J.~W. Ruge, L.~Tang,
  Efficiency based adaptive local refinement for first-order system
  least-squares formulations, SIAM Journal on Scientific Computing 33~(1)
  (2011) 1--24.

\bibitem{baty2019finmhd}
H.~Baty, Finmhd: An adaptive finite-element code for magnetic reconnection and
  formation of plasmoid chains in magnetohydrodynamics, The Astrophysical
  Journal Supplement Series 243~(2) (2019) 23.

\bibitem{strauss1976nonlinear}
H.~R. Strauss, Nonlinear, three-dimensional magnetohydrodynamics of noncircular
  tokamaks, The Physics of Fluids 19~(1) (1976) 134--140.

\bibitem{strauss1998adaptive}
H.~Strauss, D.~Longcope, An adaptive finite element method for
  magnetohydrodynamics, Journal of Computational Physics 147~(2) (1998)
  318--336.

\bibitem{lankalapalli2007adaptive}
S.~Lankalapalli, J.~E. Flaherty, M.~S. Shephard, H.~Strauss, An adaptive finite
  element method for magnetohydrodynamics, Journal of Computational Physics
  225~(1) (2007) 363--381.

\bibitem{gunzburger2012finite}
M.~D. Gunzburger, Finite element methods for viscous incompressible flows: a
  guide to theory, practice, and algorithms, Elsevier, 2012.

\bibitem{cyr2012stabilization}
E.~C. Cyr, J.~N. Shadid, R.~S. Tuminaro, Stabilization and scalable block
  preconditioning for the navier--stokes equations, Journal of Computational
  Physics 231~(2) (2012) 345--363.

\bibitem{owren1995alternative}
B.~Owren, H.~Simonsen, Alternative integration methods for problems in
  structural dynamics, Computer Methods in Applied Mechanics and Engineering
  122~(1-2) (1995) 1--10.

\bibitem{manteuffel2018nonsymmetric}
T.~A. Manteuffel, J.~Ruge, B.~S. Southworth, Nonsymmetric algebraic multigrid
  based on local approximate ideal restriction (lair), SIAM Journal on
  Scientific Computing 40~(6) (2018) A4105--A4130.

\bibitem{dembo1982inexact}
R.~S. Dembo, S.~C. Eisenstat, T.~Steihaug, Inexact newton methods, SIAM Journal
  on Numerical analysis 19~(2) (1982) 400--408.

\bibitem{knoll2004jacobian}
D.~A. Knoll, D.~E. Keyes, Jacobian-free newton--krylov methods: a survey of
  approaches and applications, Journal of Computational Physics 193~(2) (2004)
  357--397.

\bibitem{ainsworth2011posteriori}
M.~Ainsworth, J.~T. Oden, A posteriori error estimation in finite element
  analysis, Vol.~37, John Wiley \& Sons, 2011.

\bibitem{arndt2020deal}
D.~Arndt, W.~Bangerth, B.~Blais, T.~C. Clevenger, M.~Fehling, A.~V. Grayver,
  T.~Heister, L.~Heltai, M.~Kronbichler, M.~Maier, et~al., The deal. ii
  library, version 9.2, Journal of Numerical Mathematics 1~(ahead-of-print).

\bibitem{kelly1983posteriori}
D.~Kelly, J.~De~SR~Gago, O.~Zienkiewicz, I.~Babuska, A posteriori error
  analysis and adaptive processes in the finite element method: Part i-error
  analysis, International journal for numerical methods in engineering 19~(11)
  (1983) 1593--1619.

\bibitem{zienkiewicz1992superconvergent1}
O.~C. Zienkiewicz, J.~Z. Zhu, The superconvergent patch recovery and a
  posteriori error estimates. part 1: The recovery technique, International
  Journal for Numerical Methods in Engineering 33~(7) (1992) 1331--1364.

\bibitem{zienkiewicz1992superconvergent2}
O.~C. Zienkiewicz, J.~Z. Zhu, The superconvergent patch recovery and a
  posteriori error estimates. part 2: Error estimates and adaptivity,
  International Journal for Numerical Methods in Engineering 33~(7) (1992)
  1365--1382.

\bibitem{peng2020}
Z.~Peng, Q.~Tang, X.-Z. Tang, An adaptive discontinuous petrov--galerkin method
  for the grad--shafranov equation, SIAM Journal on Scientific Computing 42~(5)
  (2020) B1227--B1249.

\bibitem{bonilla2020monotonicity}
J.~Bonilla, S.~Badia, Monotonicity-preserving finite element schemes with
  adaptive mesh refinement for hyperbolic problems, Journal of Computational
  Physics (2020) 109522.

\bibitem{anderson2020mfem}
R.~Anderson, J.~Andrej, A.~Barker, J.~Bramwell, J.-S. Camier, J.~Cerveny,
  V.~Dobrev, Y.~Dudouit, A.~Fisher, T.~Kolev, et~al., {MFEM}: a modular finite
  element methods library, Computers \& Mathematics with Applications 81 (2021)
  42--74.

\bibitem{balay2019petsc}
S.~Balay, S.~Abhyankar, M.~Adams, J.~Brown, P.~Brune, K.~Buschelman, L.~Dalcin,
  A.~Dener, V.~Eijkhout, W.~Gropp, et~al., {PETSc users manual}.

\bibitem{falgout2002hypre}
R.~D. Falgout, U.~M. Yang, {hypre: A library of high performance
  preconditioners}, in: International Conference on Computational Science,
  Springer, 2002, pp. 632--641.

\bibitem{cerveny2019nonconforming}
J.~Cerveny, V.~Dobrev, T.~Kolev, Nonconforming mesh refinement for high-order
  finite elements, SIAM Journal on Scientific Computing 41~(4) (2019)
  C367--C392.

\bibitem{moreau2013}
R.~J. Moreau, Magnetohydrodynamics, Vol.~3, Springer Science \& Business Media,
  2013.

\bibitem{ben1999conservative}
N.~Ben~Salah, A.~Soulaimani, W.~G. Habashi, M.~Fortin, A conservative
  stabilized finite element method for the magneto-hydrodynamic equations,
  International Journal for Numerical Methods in Fluids 29~(5) (1999) 535--554.

\bibitem{glasser2004sel}
A.~H. Glasser, X.~Tang, The sel macroscopic modeling code, Computer physics
  communications 164~(1-3) (2004) 237--243.

\bibitem{lin2020krylov}
P.~T. Lin, J.~N. Shadid, P.~H. Tsuji, Krylov smoothing for fully-coupled amg
  preconditioners for vms resistive mhd, in: Numerical Methods for Flows,
  Springer, 2020, pp. 277--286.

\bibitem{freefem}
F.~Hecht, New development in freefem++, J. Numer. Math. 20~(3-4) (2012)
  251--265.

\bibitem{finn1977coalescence}
J.~M. Finn, P.~Kaw, Coalescence instability of magnetic islands, The Physics of
  Fluids 20~(1) (1977) 72--78.

\bibitem{biskamp1980coalescence}
D.~Biskamp, H.~Welter, Coalescence of magnetic islands, Physical Review Letters
  44~(16) (1980) 1069.

\bibitem{adler2013island}
J.~H. Adler, M.~Brezina, T.~A. Manteuffel, S.~F. McCormick, J.~W. Ruge,
  L.~Tang, Island coalescence using parallel first-order system least squares
  on incompressible resistive magnetohydrodynamics, SIAM Journal on Scientific
  Computing 35~(5) (2013) S171--S191.

\bibitem{toth2008hall}
G.~T{\'o}th, Y.~Ma, T.~I. Gombosi, Hall magnetohydrodynamics on block-adaptive
  grids, Journal of Computational Physics 227~(14) (2008) 6967--6984.

\bibitem{knoll2006coalescence}
D.~Knoll, L.~Chac{\'o}n, Coalescence of magnetic islands, sloshing, and the
  pressure problem, Physics of Plasmas 13~(3) (2006) 032307.

\bibitem{lin2010parallel}
P.~T. Lin, J.~N. Shadid, R.~S. Tuminaro, M.~Sala, G.~L. Hennigan, R.~P.
  Pawlowski, A parallel fully coupled algebraic multilevel preconditioner
  applied to multiphysics pde applications: Drift-diffusion,
  flow/transport/reaction, resistive mhd, International Journal for Numerical
  Methods in Fluids 64~(10-12) (2010) 1148--1179.

\bibitem{shadid2010current}
J.~Shadid, R.~Pawlowski, L.~Chac{\'o}n, D.~Knoll, Current sheet break-up via
  fast plasmoid formation in the island coalescence problem the ultra-high
  lundquist number regime (s$\sim10^9$), in: APS Division of Plasma Physics
  Meeting Abstracts, Vol.~52, 2010, pp. CP9--152.

\bibitem{fadeev1965self}
V.~Fadeev, I.~Kvabtskhava, N.~Komarov, Self-focusing of local plasma currents,
  Nuclear fusion 5~(3) (1965) 202.

\bibitem{ebrahimi2016large}
F.~Ebrahimi, R.~Raman, Large-volume flux closure during plasmoid-mediated
  reconnection in coaxial helicity injection, Nuclear Fusion 56~(4) (2016)
  044002.

\bibitem{ali2019effects}
A.~Ali, P.~Zhu, Effects of plasmoid formation on sawtooth process in a tokamak,
  Physics of Plasmas 26~(5) (2019) 052518.

\bibitem{mei2012numerical}
Z.~Mei, C.~Shen, N.~Wu, J.~Lin, N.~Murphy, I.~Roussev, Numerical experiments on
  magnetic reconnection in solar flare and coronal mass ejection current
  sheets, Monthly Notices of the Royal Astronomical Society 425~(4) (2012)
  2824--2839.

\end{thebibliography}

\end{document}